\documentclass[lettersize,journal]{IEEEtran}
\usepackage{amsmath,amssymb}
\usepackage{textcomp}
\usepackage{stfloats}
\usepackage{url}
\usepackage{verbatim}
\usepackage{graphicx}
\usepackage{gensymb}

\usepackage{microtype}                 
\usepackage{times}                     
\usepackage{cite}                      
\usepackage{tabu}                      
\usepackage{booktabs}                  
\usepackage{mathtools}
\usepackage{subfigure}
\usepackage{multirow}
\usepackage{caption}
\usepackage{colortbl}
\usepackage{arydshln,xcolor,array}
\usepackage{comment}
\usepackage{enumerate}
\usepackage{float}
\usepackage[export]{adjustbox}
\usepackage[linewidth=1pt]{mdframed}

\usepackage[ruled,linesnumbered]{algorithm2e}
\usepackage{tabularx}
\usepackage{wrapfig}
\usepackage{breakurl}

\usepackage[spaces,hyphens]{xurl} 
\usepackage[colorlinks,allcolors=blue]{hyperref}

\def\BibTeX{{\rm B\kern-.05em{\sc i\kern-.025em b}\kern-.08em
    T\kern-.1667em\lower.7ex\hbox{E}\kern-.125emX}}
\usepackage{balance}

\begin{document}

\title{LFS-Aware Surface Reconstruction from Unoriented 3D Point Clouds}

\author{Rao Fu, Kai Hormann and Pierre Alliez
\thanks{Rao Fu is with INRIA, 06902 Sohia Antipolis, France, and also with Geometry Factory, 06560 Antibes, France (e-mail: rao.fu@inria.com).}
\thanks{Kai Hormann is with Universit\`a della Svizzera italiana (USI), 6900 Lugano, Switzerland (e-mail: kai.hormann@usi.ch).}
\thanks{Pierre Alliez is with INRIA, 06902 Sohia Antipolis, France (e-mail: pierre.alliez@inria.fr).}

\thanks{The source code will be publicly available.  Code: \url{https://github.com/bizerfr/LFS-Aware-Reconstruction}}

}

\markboth{SUBMITTED TO xxxxxxxx,~Vol.~xx, No.~x, March~2024}%
{How to Use the IEEEtran \LaTeX \ Templates}

\maketitle

\begin{abstract}
We present a novel approach for generating isotropic surface triangle meshes directly from unoriented 3D point clouds, with the mesh density adapting to the estimated local feature size (LFS). Popular reconstruction pipelines first reconstruct a dense mesh from the input point cloud and then apply remeshing to obtain an isotropic mesh. The sequential pipeline makes it hard to find a lower-density mesh while preserving more details. Instead, our approach reconstructs both an implicit function and an LFS-aware mesh sizing function directly from the input point cloud, which is then used to produce the final LFS-aware mesh without remeshing. We combine local curvature radius and shape diameter to estimate the LFS directly from the input point clouds. Additionally, we propose a new mesh solver to solve an implicit function whose zero level set delineates the surface without requiring normal orientation. The added value of our approach is generating isotropic meshes directly from 3D point clouds with an LFS-aware density, thus achieving a trade-off between geometric detail and mesh complexity. Our experiments also demonstrate the robustness of our method to noise, outliers, and missing data and can preserve sharp features for CAD point clouds. 
\end{abstract}

\begin{IEEEkeywords}
\textbf{I.3.5} [Computing Methodologies]: Computer Graphics -- Computational Geometry and Object Modeling;
\end{IEEEkeywords}

\section{Introduction}
\IEEEPARstart{T}{he} increasing popularity of 3D scanners has facilitated the capture of large-scale 3D point cloud data that are required to generate accurate 3D models. This has fueled the interest in surface reconstruction that refers to recovering a continuous surface (here a triangle surface mesh) from a 3D point cloud. 
Surface mesh reconstruction is relevant for numerous applications in various fields, such as computer animation \cite{agudo2018shape}, computer-aided design \cite{lv2021voxel}, and compression \cite{ahn2012efficient}. Arbitrarily reconstructing a surface mesh is often insufficient, as additional properties are often sought. In this work, we focus on generating a mesh with well-shaped (isotropic) elements and variable mesh sizing in accordance with the local feature size (LFS). Such meshes exhibit a satisfactory balance between the quality of the elements,  level of complexity, and geometric fidelity. They are relevant for downstream applications, such as simulation and visualization.

Surface reconstruction has been explored through various methods, including explicit interpolation with Delaunay triangulations or Voronoi diagrams, as well as implicit functions coupled with iso-surfacing techniques. However, when the goal is to obtain isotropic meshes with adaptive sizing that ensures both isotropic triangle elements and adjustments to surface features like curvature or local feature size, the conventional pipeline involves first reconstructing a dense mesh from point clouds and subsequently applying remeshing to increase the mesh quality. This cascaded approach separates reconstruction and remeshing, making it hard to produce such a mesh that preserves more details while using fewer triangles. Furthermore, obtaining these adaptive isotropic meshes is complex, requiring a delicate balance between mesh complexity and reconstruction accuracy.

\begin{figure*}[htb!]
  \begin{center}
\begin{minipage}[t]{0.185\linewidth}
\includegraphics[width=\linewidth]{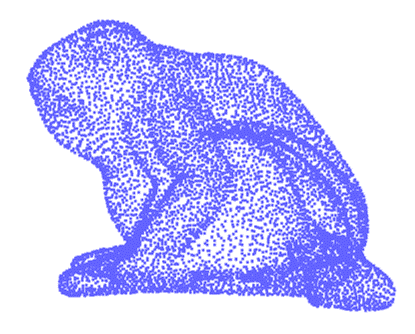}
\captionof{subfigure}{Input point cloud}
\end{minipage}
\begin{minipage}[t]{0.185\linewidth}
\includegraphics[width=\linewidth]{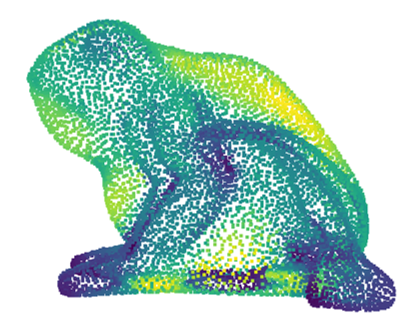}
\captionof{subfigure}{Estimated LFS}
\end{minipage}
\begin{minipage}[t]{0.185\linewidth}
\includegraphics[width=\linewidth]{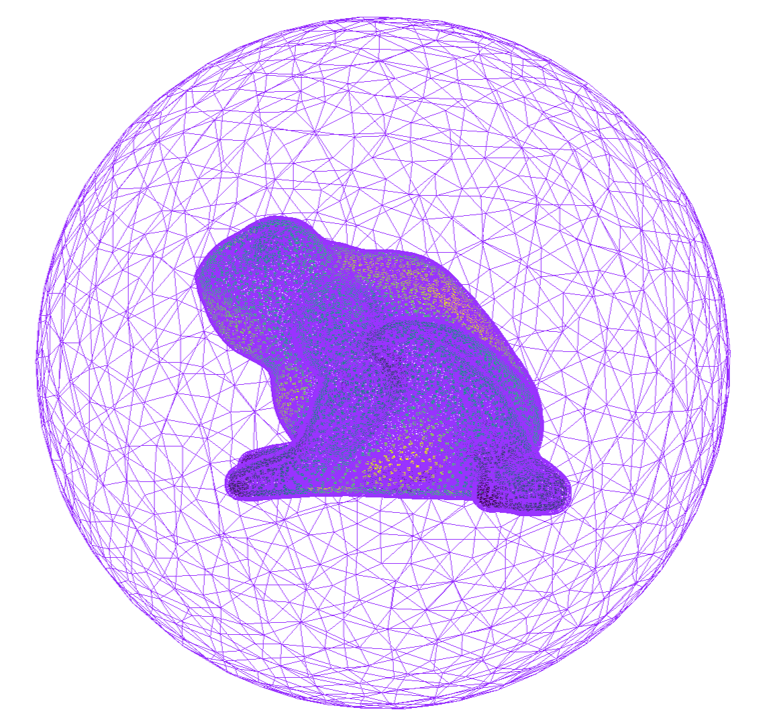}
\captionof{subfigure}{Multi-domain}
\end{minipage}
\begin{minipage}[t]{0.178\linewidth}
\includegraphics[width=\linewidth]{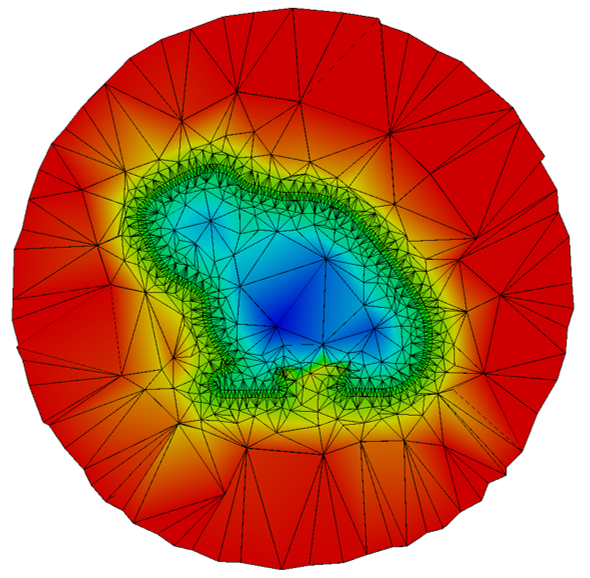}
\captionof{subfigure}{Implicit function}
\end{minipage}
\begin{minipage}[t]{0.185\linewidth}
\includegraphics[width=\linewidth]{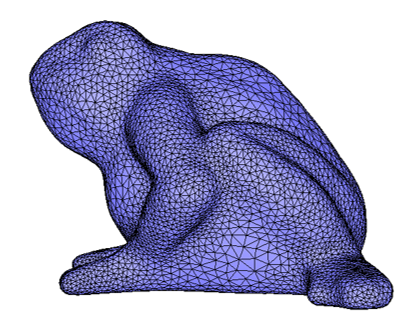}
\captionof{subfigure}{LFS-aware mesh}
\end{minipage}
\captionof{figure}{A step-by-step reconstruction example. 
(a) Our algorithm takes the unoriented point sets as input. (b) First, we estimate the local feature size (LFS) directly on the inputs. (c) Second, we construct a reach-aware multi-domain. We show the boundary of the multi-domain: an envelope domain embedded in a sphere domain. (d) Third, we use a new mesh solver to solve an implicit function defined on the multi-domain whose zero level set is the target surface. We present a clip view of the implicit function. (e) We extract the LFS-aware mesh from the solved implicit function.
}
\label{fig:teaser}
\end{center}
\end{figure*}

We tackle instead reconstruction and remeshing altogether by reconstructing an adaptive isotropic mesh directly from an unoriented 3D point cloud. Our method is motivated by the desire to apply LFS-aware meshing via Delaunay refinement \cite{jamin2015cgalmesh} on an implicit function whose zero level set delineates the reconstructed surface. LFS captures important local topological information: curvature, thickness, and separation.  We guide the size of triangle elements using the LFS to preserve fine details---obtaining lower reconstruction errors than remeshing as a postprocessing step. Fig.~\ref{fig:teaser} presents a step-by-step reconstruction example of our algorithm.

We summarize our contributions as follows:
\begin{enumerate}
\item We introduce a new method to estimate the local feature size from a 3D point cloud, which avoids constructing the medial axis via the Voronoi diagram; the latter approach being sensitive to noise and sampling conditions. Instead, our approach estimates LFS via jet fitting and analyzing a Lipschitz distance function along random rays. Such an LFS estimation approach is reliable and robust to low noise and non-uniform sampling. 
\item We introduce a novel approach for solving a signed implicit function in three main steps. First, we generate and discretize in tetrahedra a multi-domain composed of a thin envelope around the input points and the complement in a loose bounding sphere of the input. The envelope is derived from an unsigned implicit function with a level set of the estimated reach that casts off outliers. Second, we solve a signed implicit function from the signing guess of edges in a least-squares sense, which is capable of being resilient to outliers and filling large holes. Finally, we construct a robust implicit signed function from the solved implicit signed function that offers robustness to noise.  
\item We conduct extensive experiments on synthetic and real-world 3D point clouds, including the AIM@Shape dataset \cite{falcidieno2004aim} and the ABC dataset \cite{Koch_2019_CVPR}. Our method generates output meshes that are valid (i.e., intersection-free), with adaptive sizing and smaller reconstruction errors than previous approaches. 
\end{enumerate}

\section{Related Work}

The proposed approach involves LFS estimation, surface reconstruction, and iso-surfacing and remeshing. We now briefly review each component.

\subsection{LFS Estimation} 
The local feature size (LFS) on a 3D shape is the distance from a query point to its closest point on the shape's medial axis. The reach of a shape refers to the minimum of the LFS \cite{aamari2019estimating}. There are two main approaches to estimating the LFS: indirect and direct. Indirect methods commonly involve computing the medial axis \cite{tagliasacchi20163d}, defined as the locus of centers of spheres that touch the shape's boundary at two or more unique points. Amenta and Bern \cite{amenta1998surface} proved that the Voronoi poles provide a good estimation of the medial axis if the point set used to generate the corresponding Voronoi diagram is noise-free and satisfies the $\epsilon$-sampling conditions. Several approaches have been proposed based on this idea \cite{amenta1998surface, amenta2001power, boissonnat2001natural, dey2006normal}, which estimate the medial axis and then estimate the LFS by finding the nearest valid Voronoi pole. The limitations of these methods are sensitivity to noise and non-uniform sampling, which significantly affect the accuracy of the estimated medial axis. Other approaches \cite{Ma20113DMA, rebain2019lsmat,lee2021imat} attempt to gain robustness by directly computing a medial axis in an optimization scheme, free of the construction of the Voronoi diagram. However, some of them require oriented normals as a prior.
Direct methods avoid estimating the medial axis explicitly \cite{shapira2008consistent, rolland2013robust}. Instead, they estimate the shape diameter function from the input shape. While direct methods are less sensitive to noise, estimating the shape diameter function on a 3D point cloud is challenging, because finding the antipodal point given a ray cast from an input point is ill-posed. Furthermore, the shape diameter function only measures the thickness of the shape, while LFS also captures separation and curvature. Departing from previous approaches, our LFS estimation approach relies on both the shape diameter function and curvature estimation via jet fitting \cite{cazals2005estimating}. The shape diameter function is estimated by analyzing the unsigned distance function computed from the input point cloud. 

\subsection{Surface Reconstruction}  
Many surface reconstruction approaches have been proposed over the years, and we review the learning-based and non-learning-based separately.

\subsubsection{Learning-based}  Learning-based methods have drawn lots of attention in recent years. The occupancy network and its convolutional variant \cite{mescheder2019occupancy, peng2020convolutional} predict an occupancy probability for the grid and then extract the mesh based on the prediction. Point2Surf \cite{erler2020points2surf} learns an implicit signed distance function from a local patch of the point cloud. The Shape-As-Points method \cite{peng2021shape} proposes a differentiable formulation of Poisson surface reconstruction. However, these methods are constrained by the dataset and lack scalability. To solve the scalability issue, the POCO method \cite{boulch2022poco} uses convolutions and computes latent vectors at each point to deduce an occupancy for large-scale point clouds. To solve the generality issue, the neural kernel reconstruction methods \cite{huang2022neural,huang2023neural} solve for an implicit function using optimization with the data-driven kernel function and thus gain generalizability and scalability. There are also other approaches that use neural networks to learn implicit representations \cite{wang2023alto, ben2022digs, wang2023neural, liu2021deep, ye2022gifs}.

\begin{figure*}[htb!]
  \centering
  \includegraphics[width=\linewidth,trim={10 10 10 10},clip]{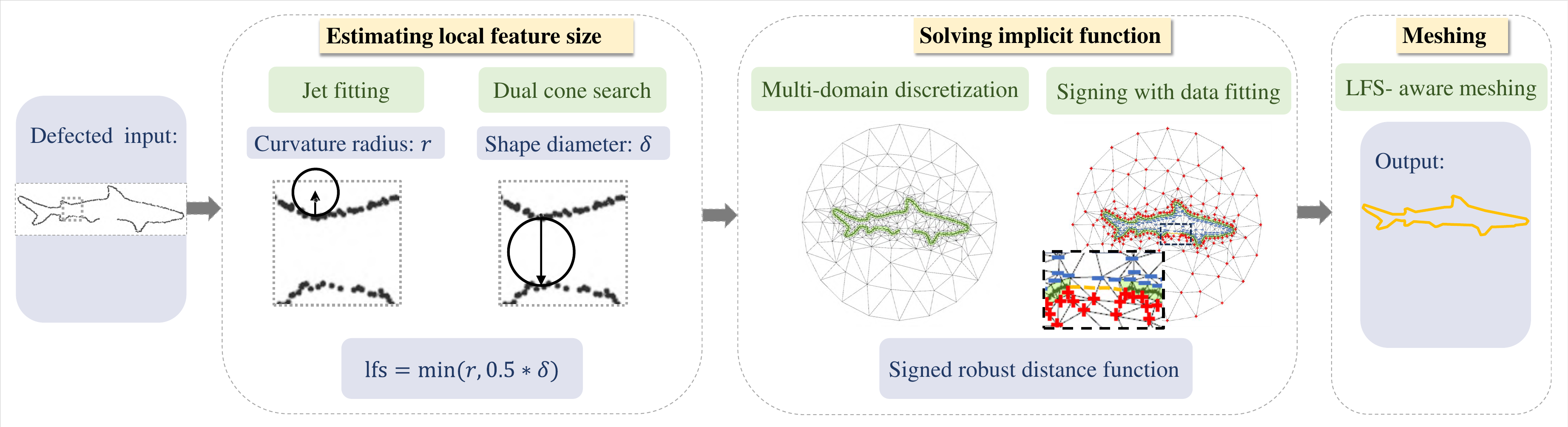}
  \caption{Overview. The input is a defective point cloud, with or without normals. The algorithm first estimates LFS as the minimum of the local curvature radius and half of the shape diameter. An implicit function is solved on a multi-domain discretization obtained by Delaunay refinement, so as to fill large holes. Yellow dashed lines delineate the filled holes, ``+''  denotes the positive vertices, and ``-'' denotes the negative vertices. Finally, the output LFS-aware mesh is generated by Delaunay refinement.}
  \label{fig-overview}
\end{figure*}

\subsubsection{Non-learning-based}  Non-learning-based methods can be categorized into Delaunay-based, implicit-based, primitive-based, and hybrid methods. Delaunay-based methods utilize the Delaunay triangulation or Voronoi diagram to generate a mesh that interpolates the input points. The popular crust and Power Crust approaches use the Voronoi diagram to compute the medial axis of a shape and then construct a mesh from the Delaunay triangulation of the crust or power diagram \cite{amenta1998surface, amenta2001power}. The Tight Cocone algorithm \cite{dey2003tight} extends the Cocone algorithm \cite{amenta2000simple} by introducing a tightness criterion that ensures that the mesh is close to the input point cloud. The advancing front algorithm \cite{cohen2004greedy} constructs a mesh by greedily advancing a front of Delaunay triangles. The restricted Delaunay method \cite{wang2022restricted} interpolates input points by film-sticking and sculpting.
While Delaunay-based methods provide provable guarantees under certain conditions, they are sensitive to noise and outliers. 
Implicit-based methods are devised to represent the reconstructed surface as an isolevel of an implicit function.  The function is often represented by a discretized 3D domain such as an octree, or defined using smoothness priors like radial basis functions (RBF), moving least squares (MLS) or kernel regression. 
Some implicit-based approaches require oriented normals as input, while others do not. More specifically, MLS-based methods \cite{hoppe1992surface, kolluri2008provably} rely on oriented normals, Poisson surface reconstruction approaches \cite{kazhdan2006poisson, kazhdan2013screened} diffuse the input oriented normals on an octree before solving for a Poisson equation. If normals are unoriented, normal-orienting methods \cite{hoppe1992surface, metzer2021orienting, xu2023globally} shall be applied.
Some approaches  deduce an implicit signed function from the unoriented point set \cite{mullen2010signing,giraudot2013noise}, and a spectral approach solves a generalized eigenvalue problem to obtain a signed function from unoriented normals and covariance matrices \cite{alliez2007voronoi}. VIPSS \cite{huang2019variational} solves for a signed function by Duchon's energy. PGR \cite{lin2022surface} re-formulates the Poisson reconstruction by incorporating the Gauss formula, thus leading to a dense system that lacks scalability, while iPSR \cite{hou2022iterative} runs Poisson reconstruction in an iterative manner and a recent work \cite{xiao2023point} orients point cloud normals by incorporating isovalue constraints to Poisson reconstruction. 
Some hybrid methods combine Delaunay triangulations and implicit functions. The AMLS approach \cite{dey2005adaptive} projects input points into an LFS-aware implicit surface and then uses the tight cone to interpolate the input points. A recent progressive approach interleaves refinement of a triangulated 3D domain with implicit mesh-based solvers \cite{zhao2021progressive}. RFEPS \cite{xu2022rfeps} preserves sharp features by using the restricted power diagram and a Poisson-reconstructed mesh.
Implicit-based methods are often more robust to noise than Delaunay-based methods but may require a dense mesh. They often require a fine discretization of the 3D domain to capture the surface details, resulting in a complex uniform output mesh, even in flat areas.
We refer to the surveys \cite{berger2017survey, huang2022surface, sulzer:hal-03968453} for a detailed overview. 

\subsection{Iso-surfacing and Remeshing} 
Iso-surfacing refers to extracting a sublevel from a given scalar field. Marching cubes \cite{lorensen1987marching} can extract a triangular mesh on a regular voxel grid or an octree. The Marching tetrahedra method \cite{doi1991efficient} simplifies the marching cubes algorithm by replacing voxels with tetrahedral cells. In order to preserve the sharp features, dual contouring \cite{ju2002dual} with its variants \cite{nielson2004dual, schaefer2004dual, shen2023flexible} optimizes vertex locations within the confines of individual cells using the quadratic error function. Besides, neural versions of marching cubes and dual contouring \cite{chen2021neural, chen2022neural} are also proposed to improve the performance. In order to obtain isotropic triangular meshes, GradNorm and its variant \cite{hass2020approximating, zong2023region} first tile space with fixed or adaptive tetrahedra and then extract a triangle mesh. 
Remeshing is used to improve mesh quality in terms of vertex sampling, regularity, and the shape of its elements \cite{alliez2008recent}. The remeshing process involves computing a new mesh approximating the original mesh while meeting specific quality requirements. It proceeds either in some parameter space or directly on the input mesh. Only a few remeshing approaches\cite{surazhsky2003explicit, su2019curvature, nivoliers2015anisotropic} generate output meshes that are adapted to the local curvature or estimated LFS with an estimation process applied to the mesh. In contrast to these methods, our method estimates LFS directly on the input 3D point cloud. LFS is then used for isocontouring the implicit function via Delaunay refinement, resulting in a lower mesh complexity for a given level of detail.

\section{Method}

The input of our algorithm is a 3D point cloud $\mathbf{P}=\{\mathbf{p}_i\in\mathbb{R}^3\}_{i=1}^N$ with or without normals $\mathbf{N}=\{\mathbf{n}_i\in\mathbb{R}^3\}_{i=1}^N$, where $N$ is the number of points. The output is an isotropic triangle surface mesh where the size of the triangle elements is controlled by the estimated LFS. Fig.~\ref{fig-overview} depicts an overview of our algorithm.

\subsection{LFS Estimation}

\begin{figure}[t!]
  \centering
  \includegraphics[width=\columnwidth,trim={20 30 20 30},clip]{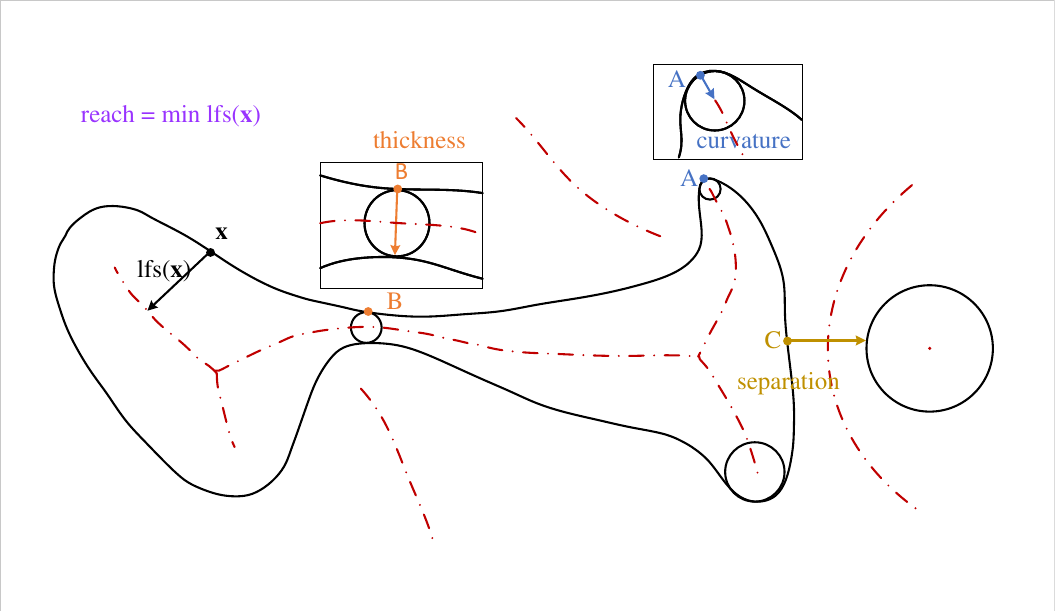}
  \caption{LFS captures the local curvature, thickness, and separation. The reach is the minimum of the LFS for the whole shape. The red dashed line depicts the medial axis.}
  \label{fig-lfs}
\end{figure}

Mathematically, LFS is defined as the distance from a query point on the surface to its closest point on the medial axis. Estimating the medial axis, e.g., from the Voronoi diagram \cite{dey2006normal} or finding the maximal empty ball \cite{Ma20113DMA}, is difficult as it is sensitive to noise and sampling density.
Instead, we estimate LFS directly from the input point set and bypass the construction of the medial axis. LFS captures a shape's local curvature, thickness, and separation (see Fig.~\ref{fig-lfs}). Given a query point $\mathbf{x}$ on a shape embedded in 3D space, we define LFS as
\begin{equation*}
  \textrm{lfs}(\mathbf{x}) = \min ( r(\mathbf{x}), 0.5 * \delta(\mathbf{x}) ),
\end{equation*}
where $r(\mathbf{x})$ denotes the estimated local curvature radius and $\delta(\mathbf{x})$ denotes the estimated shape diameter function \cite{shapira2008consistent, rolland2013robust}. The curvature radius captures the local curvature, while the shape diameter function captures the local thickness and separation of a shape. In addition, our surface reconstruction approach relies on the reach (i.e., the minimum LFS) to construct a reach-aware multi-domain discretization.

\subsubsection{Local curvature radius}
The local curvature radius is estimated by first fitting a differential jet locally\cite{cazals2005estimating}. Jet fitting constructs a local Monge coordinate system, whose basis ($\mathbf{d}_1$, $\mathbf{d}_2$, $\mathbf{n}$) is defined by three orthogonal directions: maximum principal curvature, minimum principal curvature, and normal. The jet fitting approach fits a Monge form as
\begin{equation*}
  z = \frac{1}{2}(k_1 x^2 + k_2 y^2) + \mathit{hot},
\end{equation*}
where $k_1$ and $k_2$ denote the local principal curvatures, $x$, $y$ and $z$ denote the coordinates, and $\mathit{hot}$ denotes a higher-order polynomial term. The curvature radius for a query point $\mathbf{x}$ is then defined as
\begin{equation*}
  r(\mathbf{x}) = 1 / |h(\mathbf{x})|,
\end{equation*}
where $|h(\mathbf{x})|$ is the curvature estimated via jet fitting with the maximum absolute value.

When the input 3D point cloud has no normal attributes, we directly derive the normals from the analytical jet surface. Note that the normals estimated via jet fitting are globally unoriented. We refer to \cite{cazals2005estimating} for more details.

\subsubsection{Shape Diameter} 
\label{shape_diameter}
The shape diameter function maps a query point $\mathbf{x}$ on the surface of a 3D solid shape to a scalar defined as the distance from $\mathbf{x}$ to its antipodal surface point with respect to the local normal direction. Previous approaches \cite{shapira2008consistent, rolland2013robust} only consider the local thickness of the shape as they only search along the inward normal direction. Thus, the local separation is ignored.  Furthermore, the original shape diameter function can only be estimated on a surface mesh where the antipodal points are well-defined given rays cast from a query point. Finding the antipodal points for an input 3D point cloud is hampered by the fact that the intersection point of a ray with a point cloud is ill-defined.

We address the two aforementioned issues by using a dual cone search algorithm that estimates an extended version of the shape diameter function $\delta(\mathbf{x})$, defined as
\begin{equation*}
  \delta(\mathbf{x}) = \min(\tau(\mathbf{x}), \sigma(\mathbf{x})),
\end{equation*}
where $\tau(\mathbf{x})$ denotes the thickness defined as the distance from $\mathbf{x}$ to its antipodal point along the inward normal direction, and $\sigma(\mathbf{x})$ denotes the separation defined as the distance from $\mathbf{x}$ to its antipode along the outward normal direction. It is worth noting that in our implementation, we conduct the search along the normal direction and its inverse direction, subsequently selecting the minimum of these two values. The final shape diameter value does not need any normal orientation.

\begin{figure}[t!]
  \centering
  \includegraphics[width=0.7\columnwidth,trim={20 20 20 20},clip]{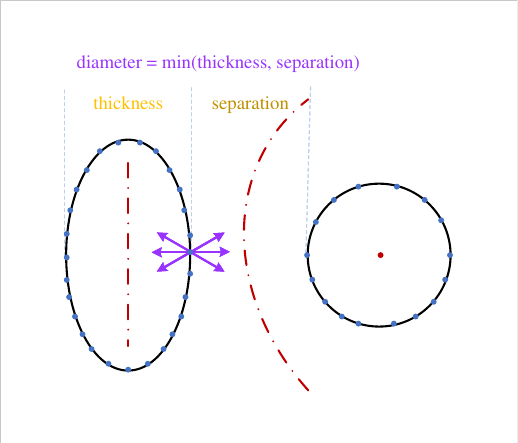}
  \caption{Dual cone search. 
           The red dashed line depicts the medial axis.
           The shape diameter is the minimum of the thickness and separation.}
  \label{fig-dual-cone}
\end{figure}

To find the antipodal points, we analyze the unsigned distance function to the input point cloud $\mathbf{P}$ along a ray $\mathbf{r}$.  The unsigned distance function is defined as 
\begin{equation*}
  d_u(\mathbf{\mathbf{x}})
  =\min_{\mathbf{p} \in \mathbf{P}}\{|\mathbf{x}-\mathbf{p}|\},
\end{equation*}
where $\mathbf{x} \in \mathbb{R}^3$ denotes a query point. If $d_u(\mathbf{x})$ is smaller than a small distance value referred to as $\epsilon$, we treat $\mathbf{x}$ as the antipodal point. We estimate $\epsilon$ from the unsigned robust distance function 
\begin{equation*}
\hat{d}_{u}(\mathbf{x}) 
= \sqrt{\frac{1}{k}\sum_{p \in N_k(\mathbf{x})} \|\mathbf{x} - \mathbf{p}\|^2},
\end{equation*}
where $N_k(\mathbf{x})$ denotes the $k$ nearest neighbors to a query point $\mathbf{x} \in \mathbb{R}^3$. The parameter $k$ is used to trade accuracy for robustness to noise and outliers \cite{mullen2010signing}. Then $\epsilon$ is estimated as the smallest $\hat{d}_{u}(\mathbf{x})$ computed at all input points, i.e., $\epsilon = \mathrm{min} \{ \hat{d}_{u}(\mathbf{p}) | \mathbf{p} \in \mathbf{P} \}$. Equipped with this, the brute-force approach is to sample points densely on $\mathbf{r}$ to find the antipodal points whose function value $d_u(\mathbf{x})$ is below $\epsilon$. However, as $d_u(\mathbf{x})$ is $1$-Lipschitz continuous, we utilize a Lipschitz-guided recursive dichotomic search to avoid dense sampling via using the Lipschitz continuity as a prior (see Appendix). In addition, to capture both thickness and separation, we cast random searching rays inside two opposite cones with $\mathbf{x}$ as the apex  with $\mathbf{n}$ and $-\mathbf{n}$ as their axes, because it is challenging for a single ray to find an antipodal point directly. More specifically, we cast $N_c$ random rays within each cone for a query point $\mathbf{x}$ to collect $k_c$ antipodal points and calculate the robust distance function from $\mathbf{x}$ to the collected $k_c$ antipodal points. Such a dual cone search may not find any antipodal points when the local point sampling is too sparse. In this case, we set the shape diameter as the diameter of the loose bounding sphere of the input point cloud. Fig.~\ref{fig-dual-cone} illustrates the dual cone search for estimating the shape diameter. 

Finally, the LFS value is taken as the minimum of the local curvature radius and half of the shape diameter. Then, we apply smoothing on the raw estimated LFS values.

\subsection{Implicit Function}
\label{implicit_function}

Our objective is to solve for an implicit signed function, that is piecewise-linear and defined on a 3D triangulation, whose zero level set corresponds to the reconstructed surface. Instead of solving for the implicit function using a single global solve step, our rationale is to compute its components sequentially. First, we construct a 3D multi-domain that discretizes an unsigned distance function to the input points (multi-domain). Next, we solve a least squares problem to obtain a signed implicit function from sign guesses estimated at the edges of the discretization (signing with data fitting). Finally, we utilize the signs of the solved implicit function to sign the unsigned robust distance function (signed robust distance function). We detail each step below.

\subsubsection{Multi-domain}

We construct a multi-domain discretization to represent the piecewise-linear implicit function. 
The multi-domain is the union of two 3D sub-domains: a thin envelope enclosing the input $\mathbf{P}$ and a loose bounding sphere containing $\mathbf{P}$ minus the above envelope. The thin envelope is made reach-aware to separate thin structures and preserve fine details, while the loose bounding sphere is added and discretized to help fill large holes with a smoothly graded 3D triangulation. 

The thin envelope domain is derived from an unsigned implicit function $I_u(\mathbf{x})$ and made read-aware by constraining $I_u(\mathbf{x})$ to be smaller than the estimated reach $I_R$, i.e., the minimum of LFS. The definition is
\begin{equation*}
I_{E}(\mathbf{x})=\{\mathbf{x} : I_{u}(\mathbf{x}) \leq I_R \},
\end{equation*}
and $I_{u}(\mathbf{x})$ is defined as
\begin{equation*}
  I_u(\mathbf{x}) = \frac{\sum_{\mathbf{p} \in N_{k}(\mathbf{x})} |(\mathbf{x} - \mathbf{p})^{T} \cdot \mathbf{n}| W(\mathbf{x})}{\sum_{\mathbf{p} \in N_{k}(\mathbf{x})} W(\mathbf{x})},
\end{equation*}
where $W(\mathbf{x}) = e^{-||\mathbf{x}-\mathbf{p}||^2/h^2}$ denotes a Gaussian weighted function. Note that $I_u(\mathbf{x})$ uses the absolute value of $(\mathbf{x} - \mathbf{p})^{T} \cdot \mathbf{n}$. Thus, normal orientation is not required here. 

The other domain is the complement between the envelope and a loose bounding sphere. The loose bounding sphere is $I_{S}(\mathbf{x})=\{\mathbf{x} : ||\mathbf{x} - \mathbf{c}|| \leq r\}$. $\mathbf{c}$ is the centroid of $\mathbf{P}$, and $r=\max_{\mathbf{p},\mathbf{q}}|\mathbf{p} - \mathbf{q}|$, where $\mathbf{p} \in \mathbf{P}$, $\mathbf{q} \in \mathbf{P}$ and $\mathbf{p} \neq \mathbf{q}$.

We utilize the Delaunay refinement paradigm \cite{jamin2015cgalmesh} to generate the multi-domain discretization. In order to generate well-shaped cells, we specify a constant cell radius-edge ratio in all sub-domains. We also specify a constant facet shape criterion for the boundary of the loose bounding sphere. We refer to \cite{jamin2015cgalmesh} for more details.

\subsubsection{Signing with Data Fitting}
\label{signing_with_data_fitting}

Our multi-domain is a reach-aware envelope $I_E(\mathbf{x})$ embedded in a 3D Delaunay triangulation $\mathbf{Tr}$ bounded by a sphere $I_{S}(\mathbf{x})$. The sublevel of the unsigned function inside $I_E(\mathbf{x})$ contains most of the inferred surface, but it remains to compute a signed function in order to fill holes and delineate the inferred surface as its zero level set. We proceed by solving for a signed function defined at the vertices of the multi-domain. 
We first estimate the sign guesses at the edges of the multi-domain. We then incorporate a data-fitting term to solve an implicit signed function value for each vertex in the least squares sense. 

The sign guesses are estimated as follows. Let $\text{sign}(\mathbf{e}_{mn})$ denote the binary sign guess for an edge $\mathbf{e}_{mn} \in \mathbf{Tr}$. If $\mathbf{e}_{mn}$ crosses the inferred surface, then $\text{sign}(\mathbf{e}_{mn})$ is set to $-1$ to reflect that it connects two vertices with opposite signs and set to $+1$ otherwise. An edge $\mathbf{e}_{mn}$ has three possible locations: outside, inside, and on the boundary of the envelope. If $\mathbf{e}_{mn}$ is outside or on the boundary of the envelope, $\text{sign}(\mathbf{e}_{mn})$ is directly set to $+1$ as $\mathbf{e}_{mn}$ does not cross the inferred surface. If $\mathbf{e}_{mn}$ is inside the envelope, we check whether $\mathbf{e}_{mn}$ connects two vertices with opposite signs. Note that the edges of slivers (flat cells) inside the envelope may not cross the inferred surface. We still perform the Lipschitz-guided recursive dichotomic search (see Appendix) on $\mathbf{e}_{mn}$ inside the envelope to check if the unsigned distance function $I_{u}(\mathbf{\mathbf{x}})$ is below a value $\epsilon$, which is set as half of the surface reach $I_R$. If $I_{u}(\mathbf{x})$ is below $\epsilon$, we set $\mathbf{e}_{mn}$ to $-1$, and $+1$ otherwise. Equipped with sign guesses at edges, we then solve for a signed implicit function $d_s(\cdot)$ at all vertices $\mathbf{V}$ of $\mathbf{Tr}$ by minimizing the objective function
\begin{equation}
\begin{aligned}
  E = \sum_{\mathbf{e}_{mn}}\left[d_s(\mathbf{v}_m) - \mathrm{sign}(\mathbf{e}_{mn})\cdot d_s(\mathbf{v}_n)\right]^2 
  \\ +\lambda \sum_{t}\sum_{\mathbf{p}_t}\Biggl[\sum_{i=0}^{3}\alpha_{i}(\mathbf{p}_t) \cdot d_s(\mathbf{v}_{t_i})\Biggr]^2, 
\end{aligned}
  \label{signing}
\end{equation}
where $d_s(\cdot)$ denotes the signed implicit function solved for at the vertices $\mathbf{V}$, $\alpha_i(\mathbf{p}_t)$ is the $i$-th barycentric coordinate for the input point $\mathbf{p}_t$ located in the tetrahedron $t$ with vertices $\mathbf{v}_{t_0},\dots,\mathbf{v}_{t_3}$.

The first term of the above objective function optimizes the sign consistency on the multi-domain. The second data fitting term favors the fact that the input points lie near the zero level set of the interpolated signed implicit function. The parameter $\lambda$ provides a means to balance between the two terms. For all shown experiments, $\lambda$ is set to $1.0$. 

\begin{figure}[t!]
\centering
\subfigure[]{
    \includegraphics[width=0.45\columnwidth,trim={20 20 20 20},clip]{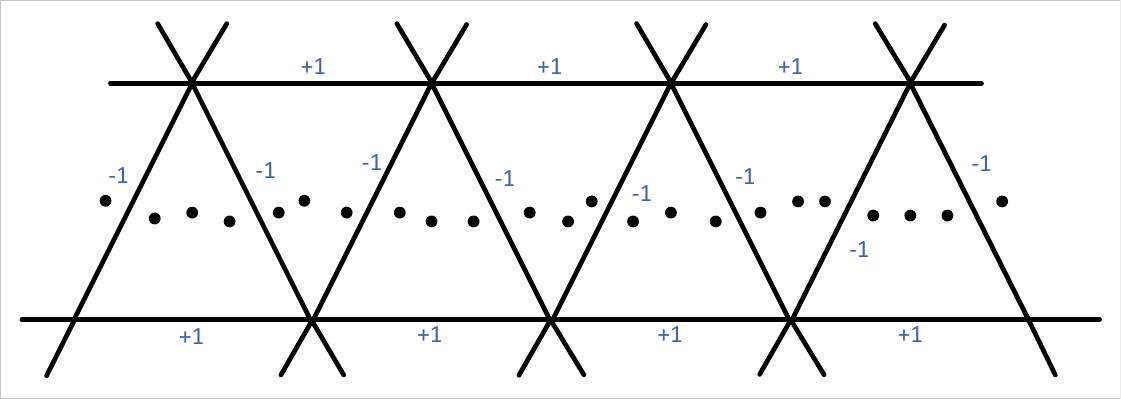}\vspace{0pt}  
}
\subfigure[]{
    \includegraphics[width=0.45\columnwidth,trim={20 20 20 20},clip]{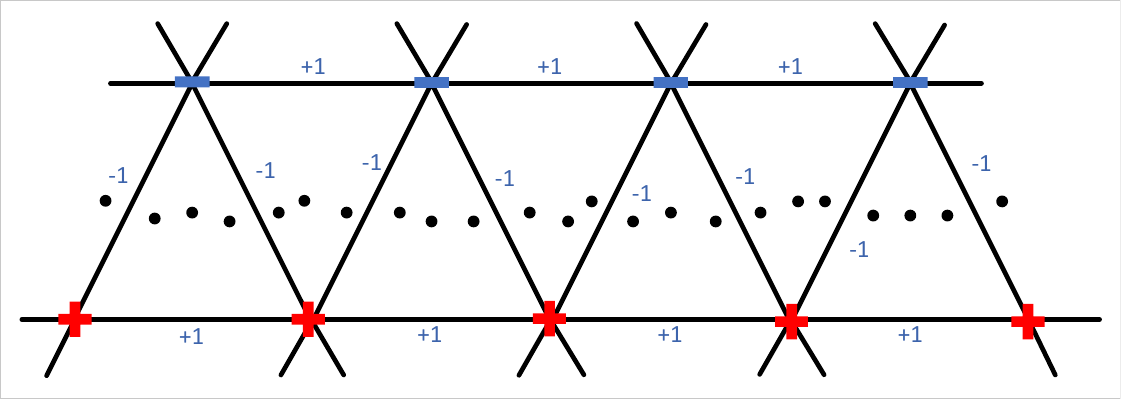}\vspace{0pt} 
}
\caption{Solving for a signed implicit function from the sign guesses for edges. (a) The sign guess for an edge is determined by detecting the crossings with sublevel sets of the unsigned distance function to the points. An edge is labeled $-1$ if it crosses the point clouds, and $+1$ otherwise. (b) A signed implicit function is then solved from the sign guesses of the edges. Red crosses denote the positive vertices, and blue minus signs denote the negative vertices. The signed implicit function is piecewise linear.}
\label{fig-sign-solver}
\end{figure}

To avoid the trivial solution, we enforce the sum of the signed implicit values to be a constant, i.e., $\sum_{\mathbf{v}} d_s(\mathbf{v}) = |\mathbf{V}|$, where $|\mathbf{V}|$ denotes the number of vertices. Fig.~\ref{fig-sign-solver} illustrates the signing process.

We can re-write the minimization of the objective function~\eqref{signing} in matrix form as
\begin{equation*}
\begin{aligned}
  \min_{\mathbf{x}} \quad & ||\mathbf{S}\mathbf{x}||^2 +\lambda||\mathbf{B}\mathbf{x}||^2\\
  \textrm{s.t.} \quad & \mathbf{C}\mathbf{x} = |\mathbf{V}|,
\end{aligned}
\end{equation*}
where $\mathbf{S}$ denotes the sign guess matrix. Each row corresponds to an edge of $\mathbf{Tr}$ and contains two elements, either set to $+1$ or $-1$. $\mathbf{B}$ are the barycentric coordinates of each input point, and $\mathbf{x}$ are the values for the signed implicit function $d_s(\cdot)$ that we need to solve for at each vertex. $\mathbf{C}$ is a one-dimensional constraint matrix of size $1 \times |\mathbf{V}|$ and filled with $1$.
Both $\mathbf{S}$ and $\mathbf{B}$ are sparse. By applying the Karush--Kuhn--Tucker (KKT) conditions, we have
\begin{equation*}
\begin{aligned}
  \min_{\mathbf{x},\mathbf{z}} \quad & (\mathbf{S}\mathbf{x})^T(\mathbf{S}\mathbf{x}) + \lambda(\mathbf{B}\mathbf{x})^T(\mathbf{B}\mathbf{x}) + \mathbf{z}^T(\mathbf{C}\mathbf{x} - |\mathbf{V}|) 
\end{aligned}
\end{equation*}

Finally, we solve for the signed implicit function values $\mathbf{x}$ and the Lagrange multiplier $\mathbf{z}$ in the least squares sense,
\begin{equation}
\begin{aligned}
\begin{bmatrix}
2\mathbf{S}^{T}\mathbf{S} + 2 \lambda \mathbf{B}^{T}\mathbf{B} & \mathbf{C}^T \\ 
\mathbf{C} & \mathbf{0}
\end{bmatrix} 
\begin{bmatrix}
\mathbf{x} \\ 
\mathbf{z}
\end{bmatrix} 
=
\begin{bmatrix}
\mathbf{0} \\ 
|\mathbf{V}|
\end{bmatrix}.
\end{aligned} 
\label{signing-matrix-form}
\end{equation}

\subsubsection{Signed Robust Distance Function}

The previous step solves for a signed implicit function defined on the 3D multi-domain. To obtain a robust implicit function that is more resilient to noise, we use the binary sign value from the signed implicit function $d_s(\cdot)$ to sign the unsigned robust distance function $\hat{d}_{u}(\cdot)$ to obtain a new implicit function $\hat{d}_s(\cdot)$ whose zero level set is the target surface
\begin{equation}
  \hat{d}_s(\cdot) = \text{sign}(d_s(\cdot)) \cdot \hat{d}_{u}(\cdot),
  \label{sign-rbf}
\end{equation}
where $d_s(\cdot)$ denotes the solved signed implicit function values after solving~\eqref{signing-matrix-form}.

\subsection{LFS-Aware Meshing}

A simple iso-surfacing method such as marching tetrahedra \cite{doi1991efficient} would not yield the final reconstructed surface mesh with the desired properties (sizing and well-shaped tetrahedra). We utilize instead a Delaunay refinement meshing \cite{jamin2015cgalmesh} to extract the final size-varying isotropic mesh. The Delaunay refinement meshing takes an implicit function with a mesh sizing function as input, optionally with sharp feature curves. It outputs an isotropic triangular mesh contouring the zero level set of the input implicit function with the mesh sizing function controlling the refinement termination. If sharp feature curves are provided, the final mesh can preserve sharp features. In our case, we feed both the signed robust distance function $\hat{d}_s(\cdot)$ together with mesh sizing function $\textrm{size}(\cdot)$, both of which are derived from the input point cloud directly, into the Delaunay refinement. Furthermore, we use NerVE \cite{zhu2023nerve} to detect the sharp feature curves for CAD shapes with sharp features.

We define the mesh sizing function, which is LFS-aware, as
\begin{equation}
\begin{aligned}
  \textrm{size}(\mathbf{x}) = \frac{\textrm{lfs}(\mathbf{x}) - I_R}{\textrm{lfs}_{\textrm{max}} - I_R}
  \cdot (\textrm{size}_{\textrm{max}} - \textrm{size}_{\textrm{min}}) + \textrm{size}_{\textrm{min}},
\end{aligned}
\end{equation}
where $\textrm{lfs}(\mathbf{x})$ denotes the local feature size, $I_R$ denotes the surface reach, $\textrm{size}_{\textrm{max}}$ denotes the user-specified maximal facet size, and $\textrm{size}_{\textrm{min}}$ denotes the minimal facet size set as a user-specified ratio to the thickness of the envelope. Finally, we apply smoothing to the facet sizing function.

\section{Experiments}

The CGAL library \cite{cgal} was used to implement the multi-domain discretization and Delaunay refinement steps. We use the conjugate gradient solver from the Eigen library \cite{eigenweb} to solve for the implicit function. Experiments are conducted on a Dell laptop with a 2.60GHz Intel i7-10750H CPU and 32GB memory.

\subsection{LFS estimation}

\subsubsection{Validation on canonical primitives} 
\begin{figure}[htbp!]
\centering
\subfigure[sphere]{
    \begin{minipage}[h]{0.29\columnwidth}
    \includegraphics[width=\linewidth,trim={300 50 300 50},clip]{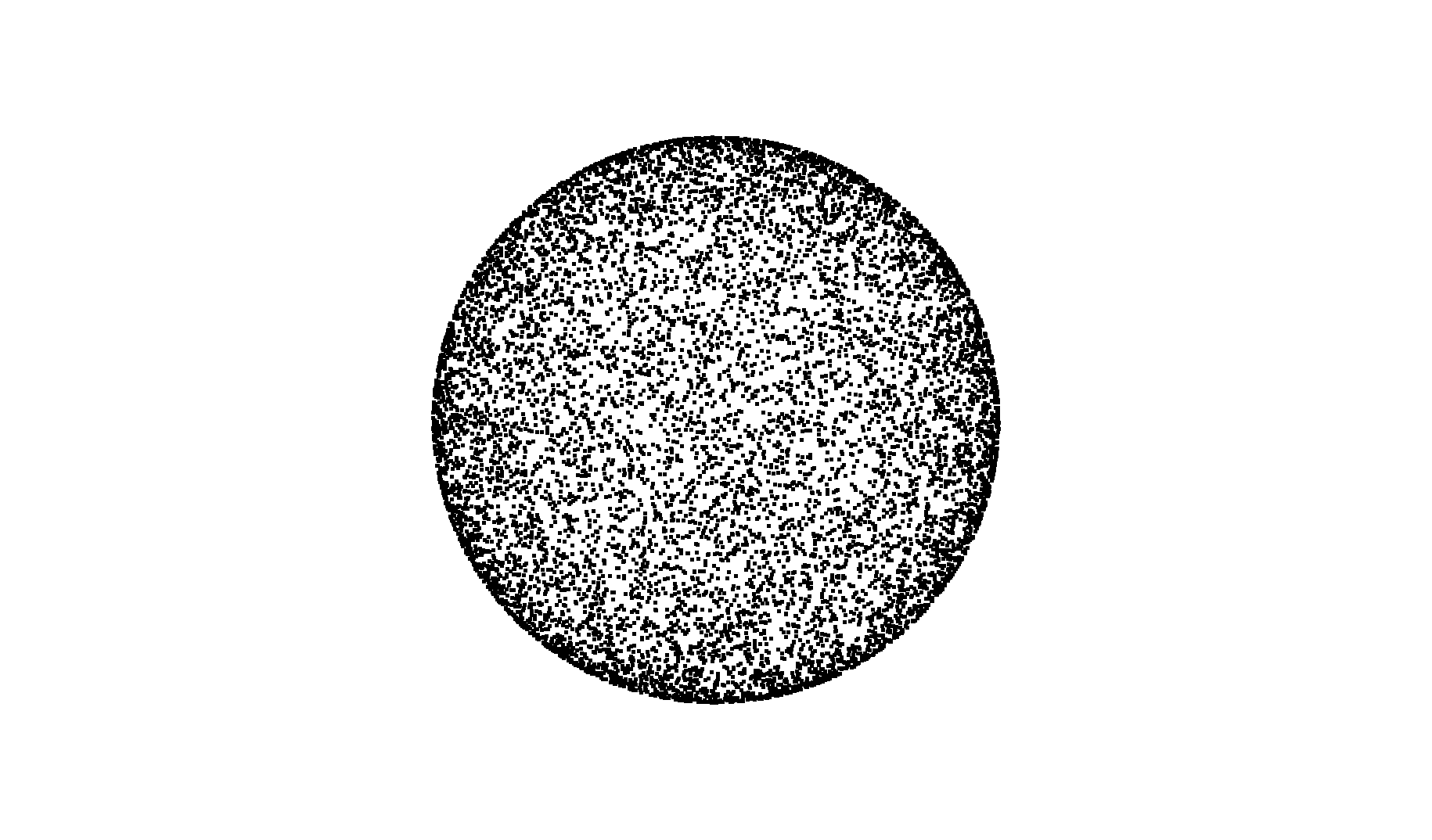}\vspace{0pt} 
    \includegraphics[width=\linewidth,trim={300 50 300 50},clip]{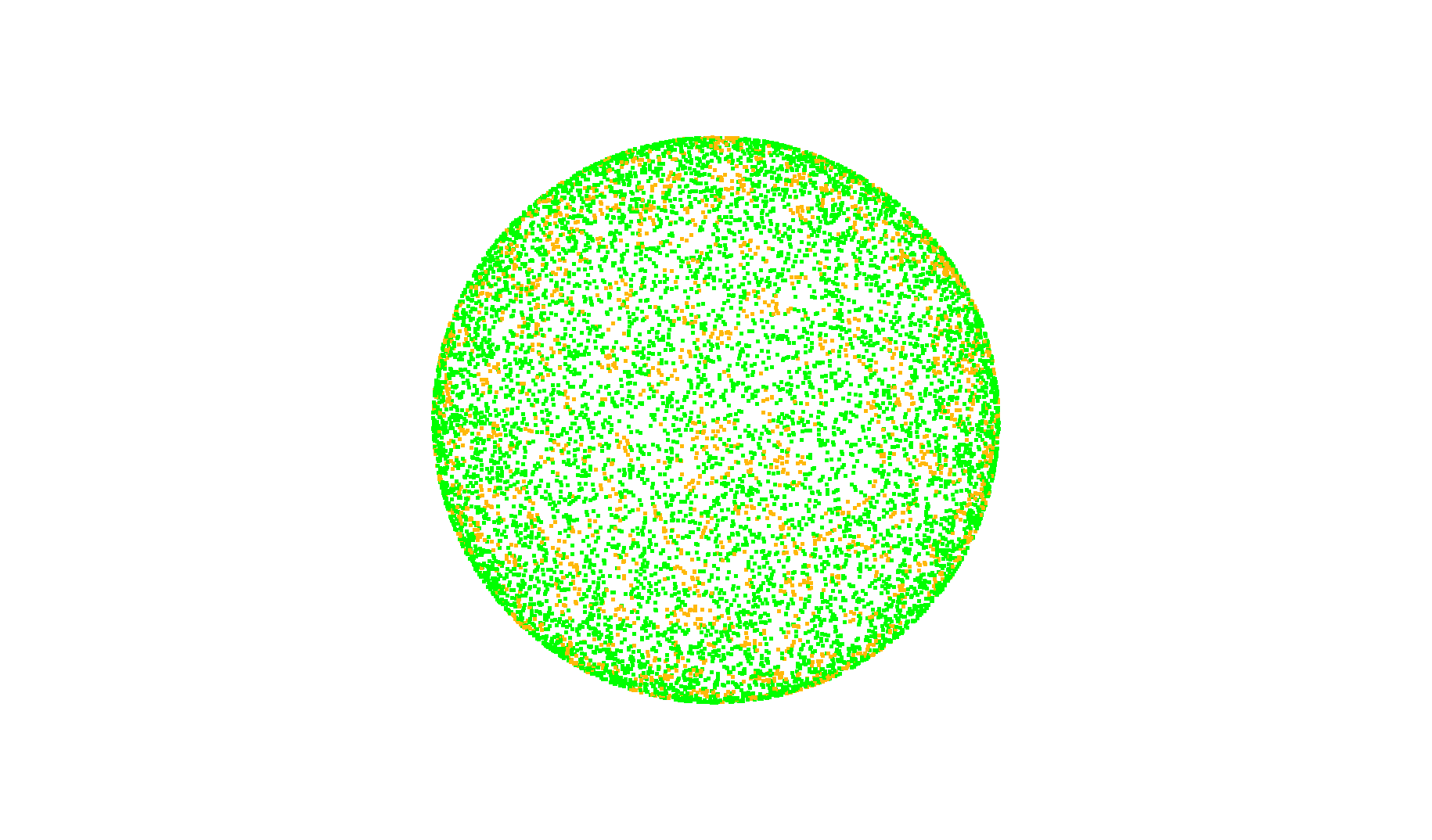}\vspace{0pt}
    \includegraphics[width=\linewidth,trim={300 50 300 50},clip]{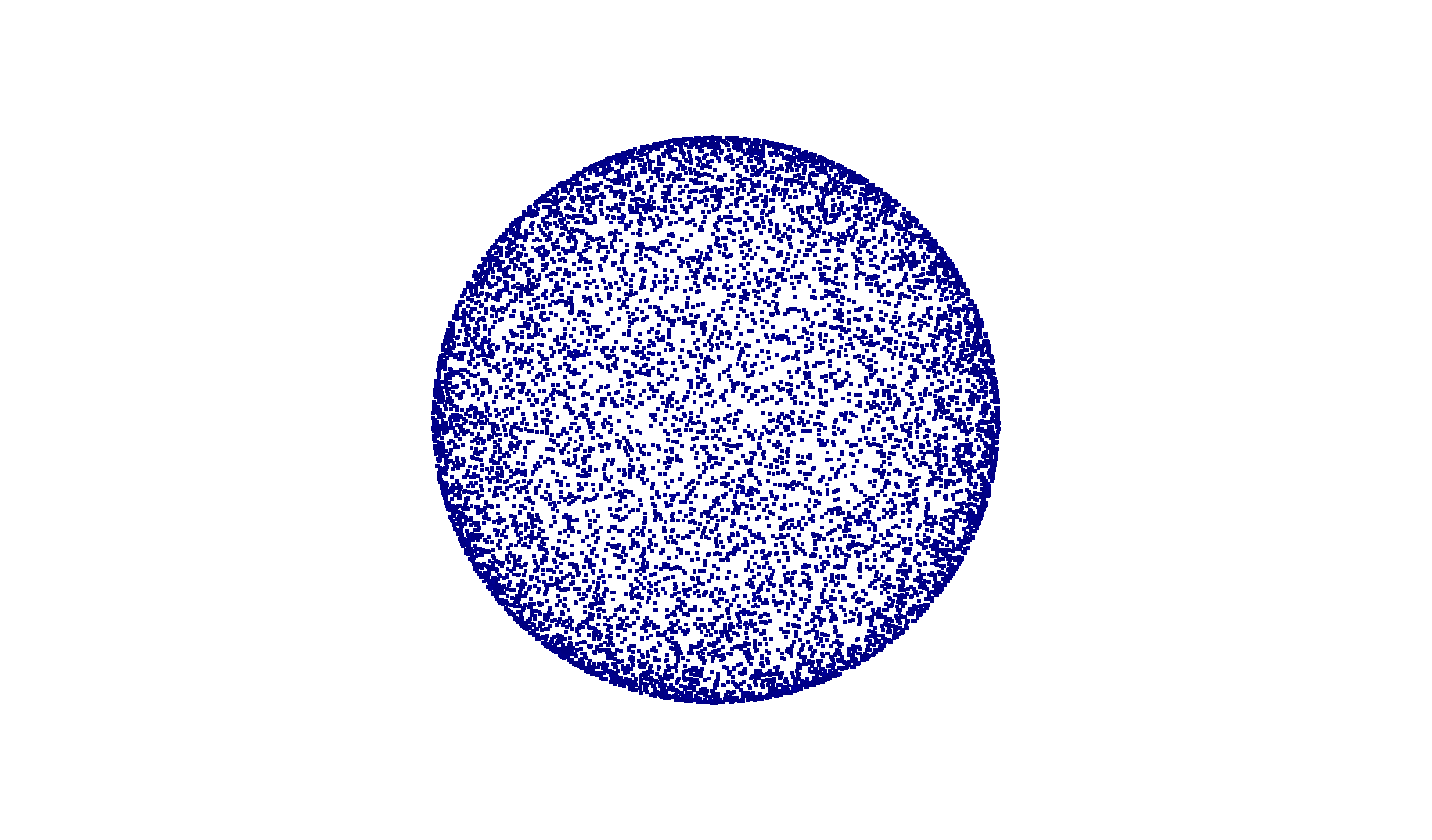}\vspace{0pt} 
    \includegraphics[width=\linewidth,trim={300 50 300 50},clip]{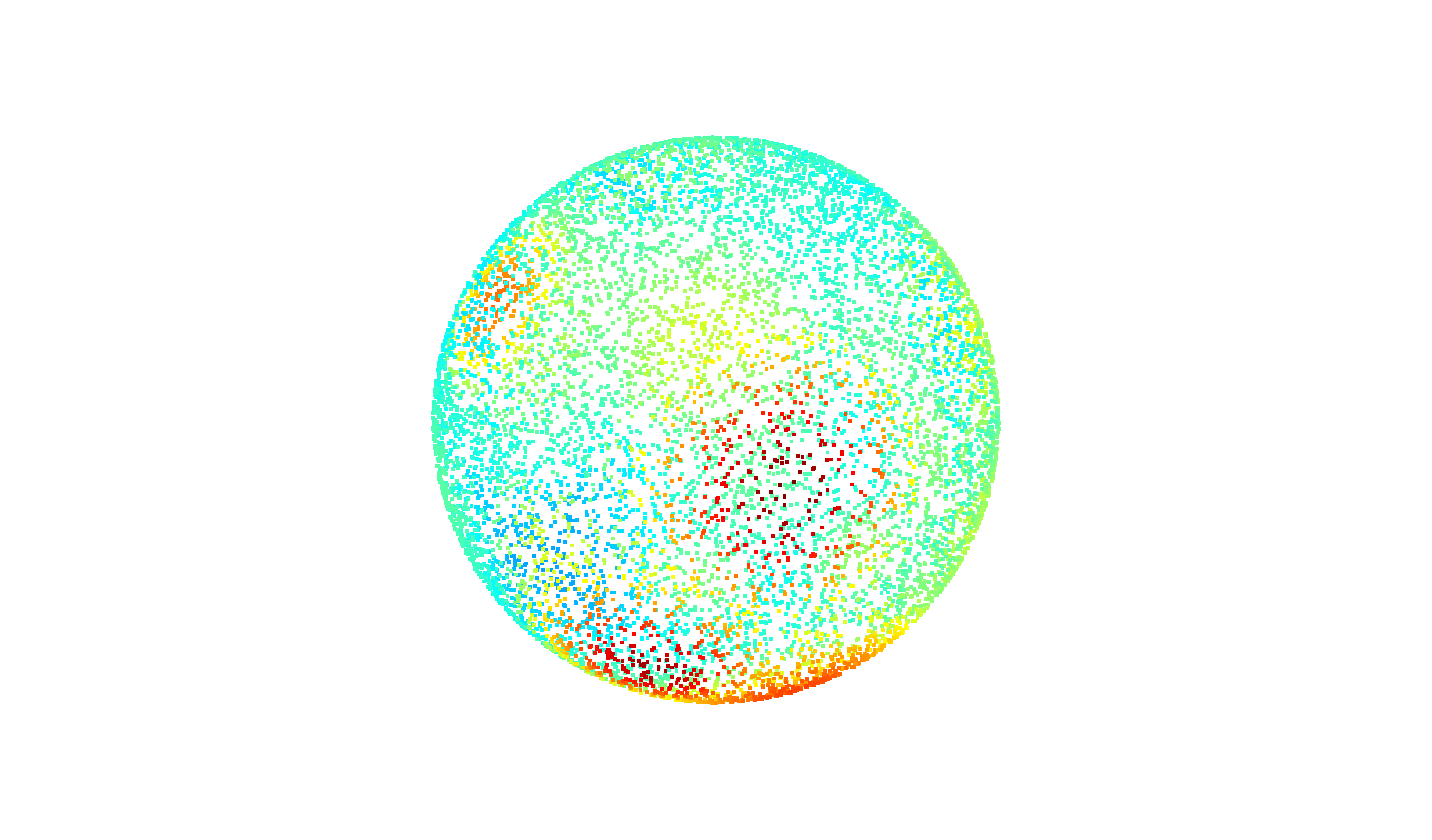}\vspace{0pt} 
    \includegraphics[width=\linewidth]{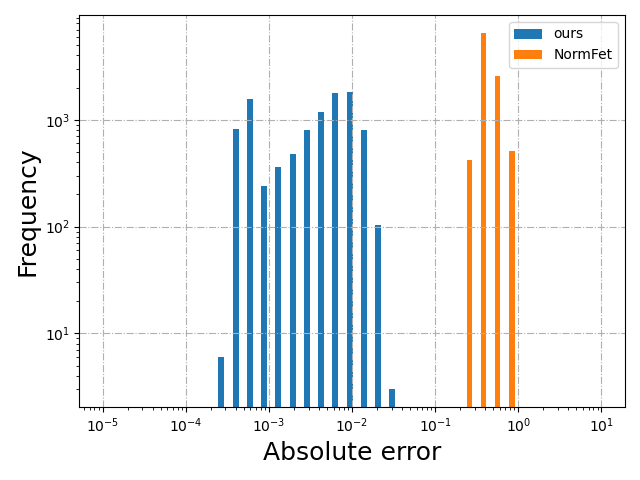}\vspace{0pt} 
    \end{minipage}
}
\subfigure[cone]{
    \begin{minipage}[h]{0.29\columnwidth}
    \includegraphics[width=\linewidth,trim={300 50 300 50},clip]{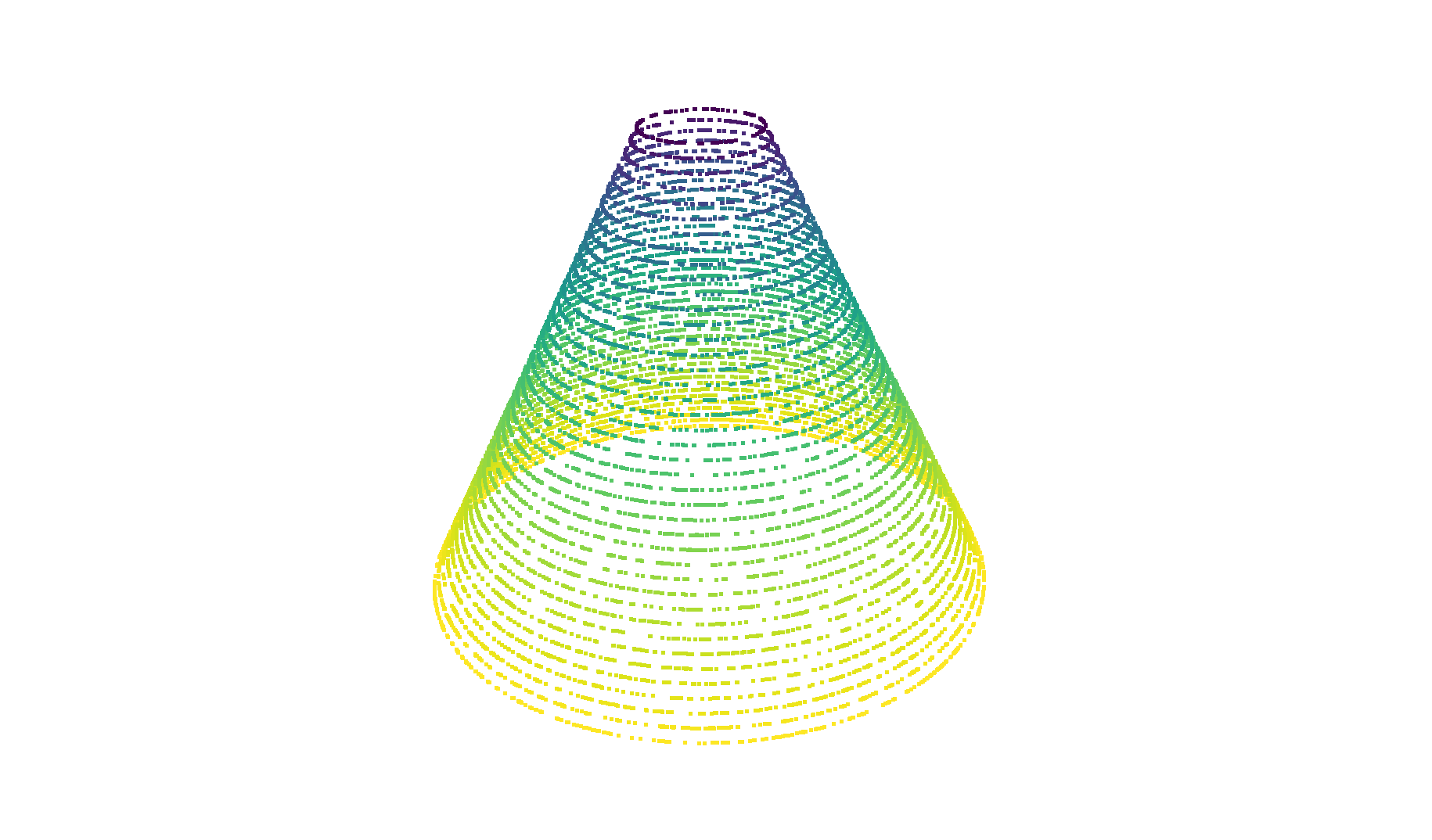}\vspace{0pt} 
    \includegraphics[width=\linewidth,trim={300 50 300 50},clip]{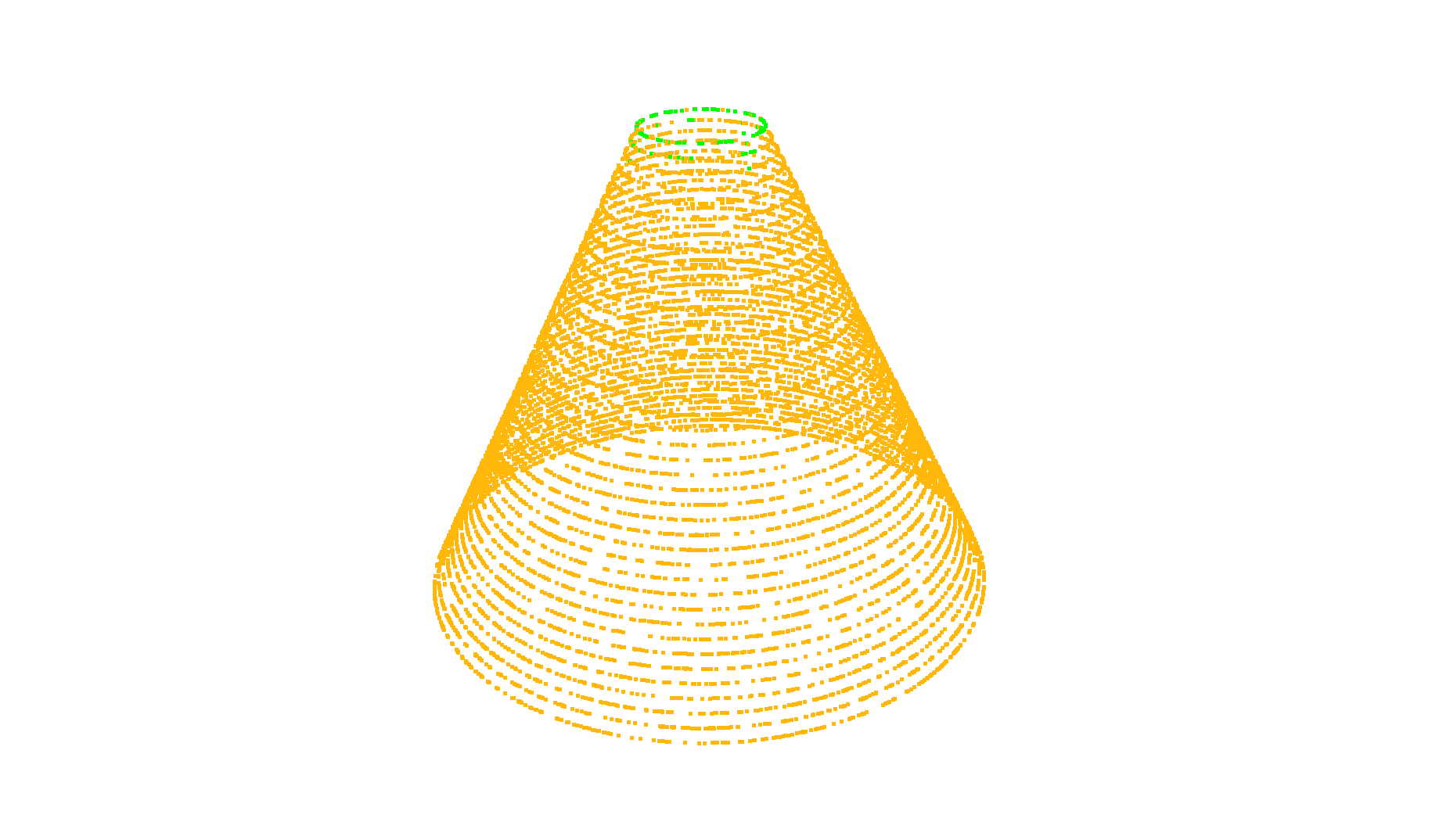}\vspace{0pt}
    \includegraphics[width=\linewidth,trim={300 50 300 50},clip]{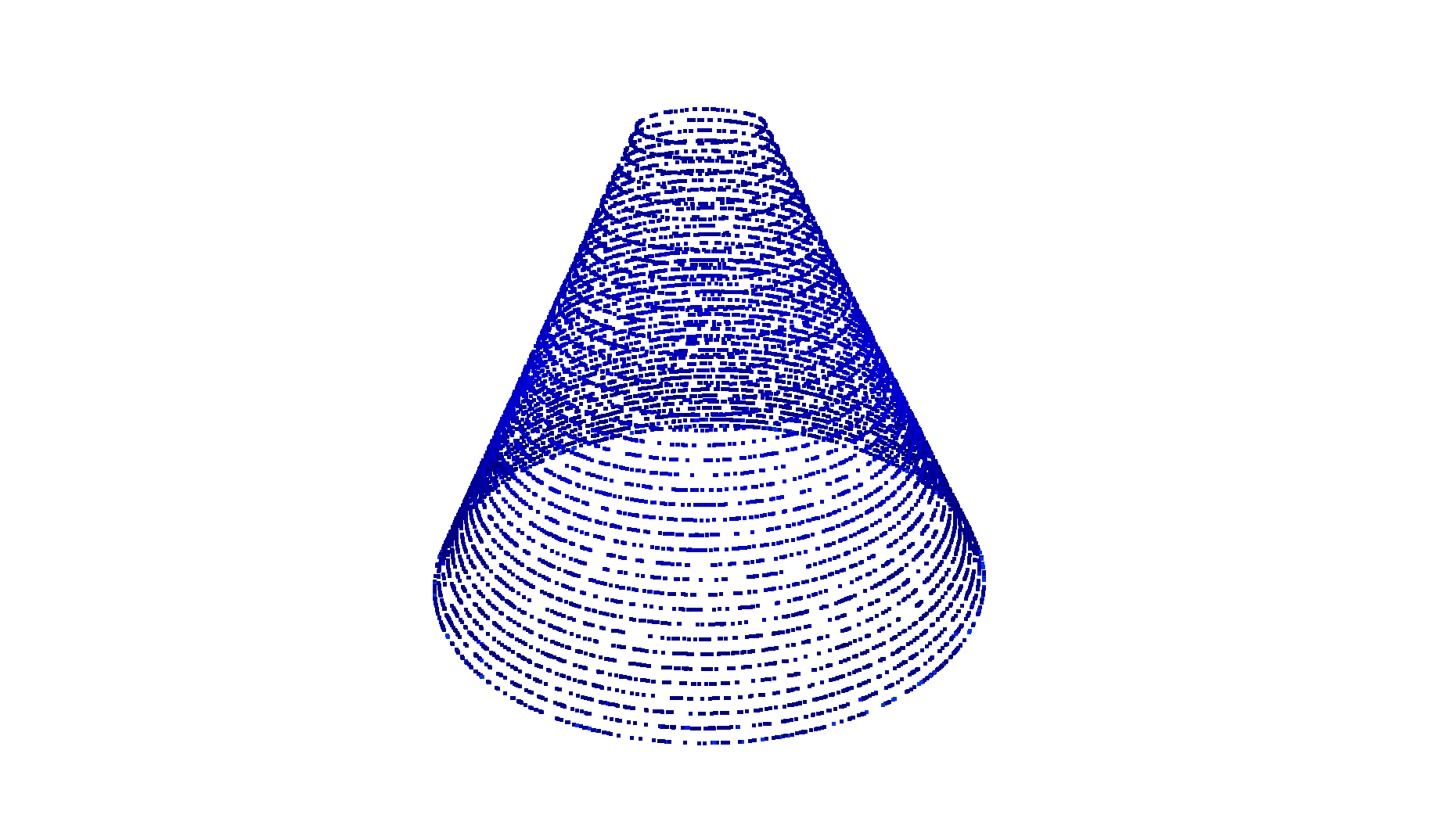}\vspace{0pt} 
    \includegraphics[width=\linewidth,trim={300 50 300 50},clip]{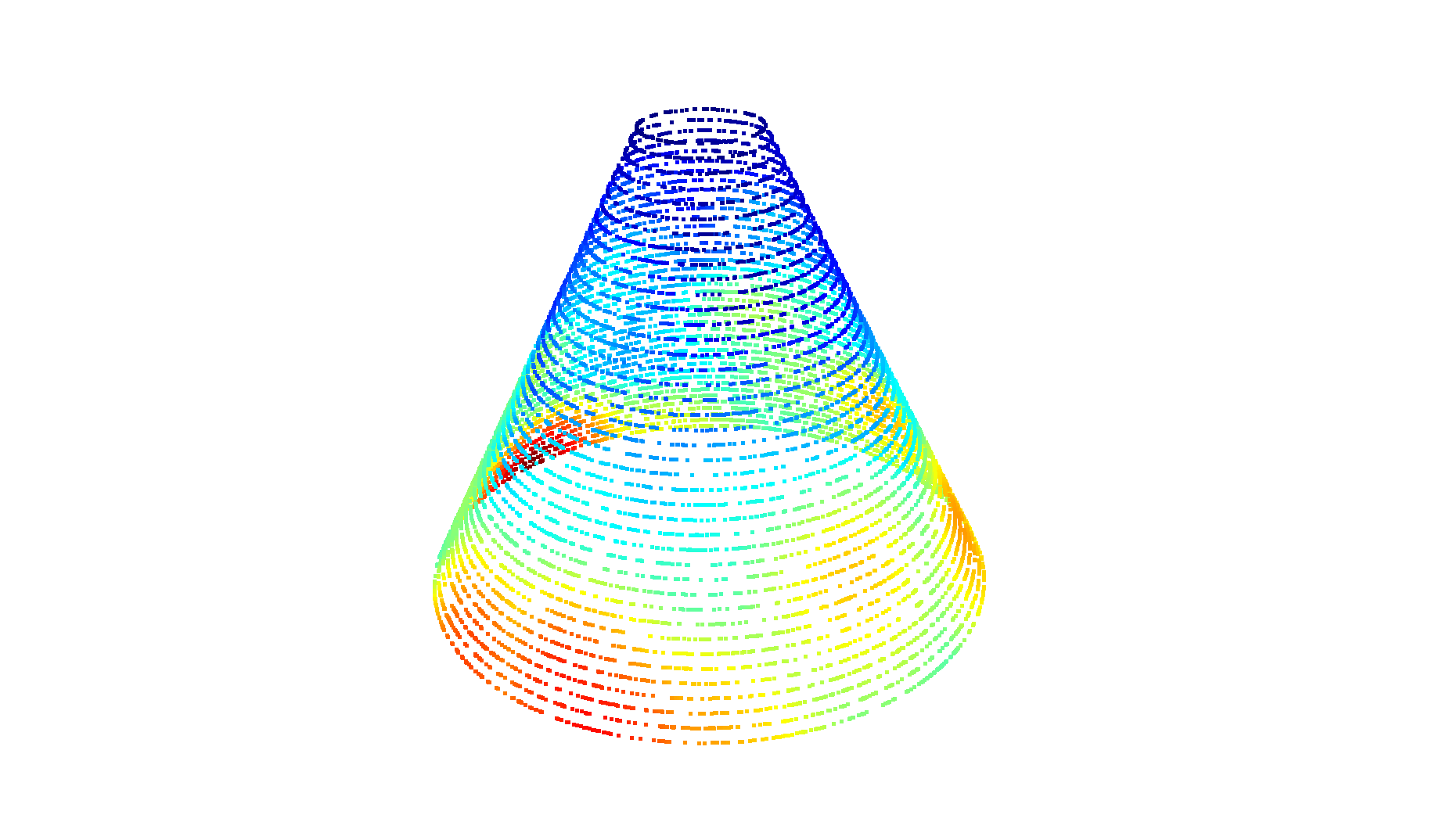}\vspace{0pt} 
    \includegraphics[width=\linewidth]{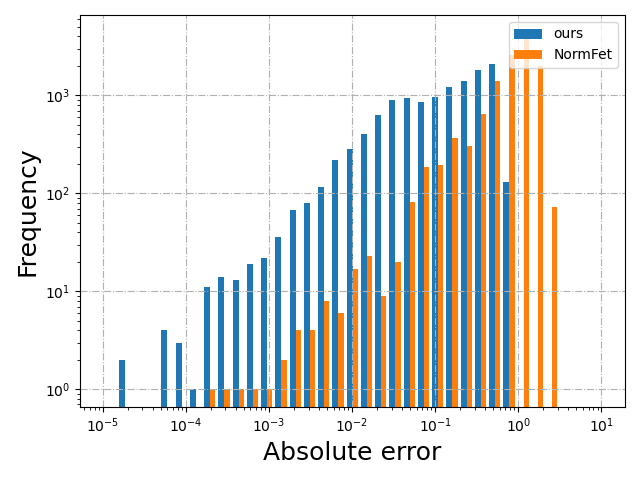}\vspace{0pt} 
    \end{minipage}
}
\subfigure[ellipsoid]{
    \begin{minipage}[h]{0.29\columnwidth}
    \includegraphics[width=\linewidth,trim={300 50 300 50},clip]{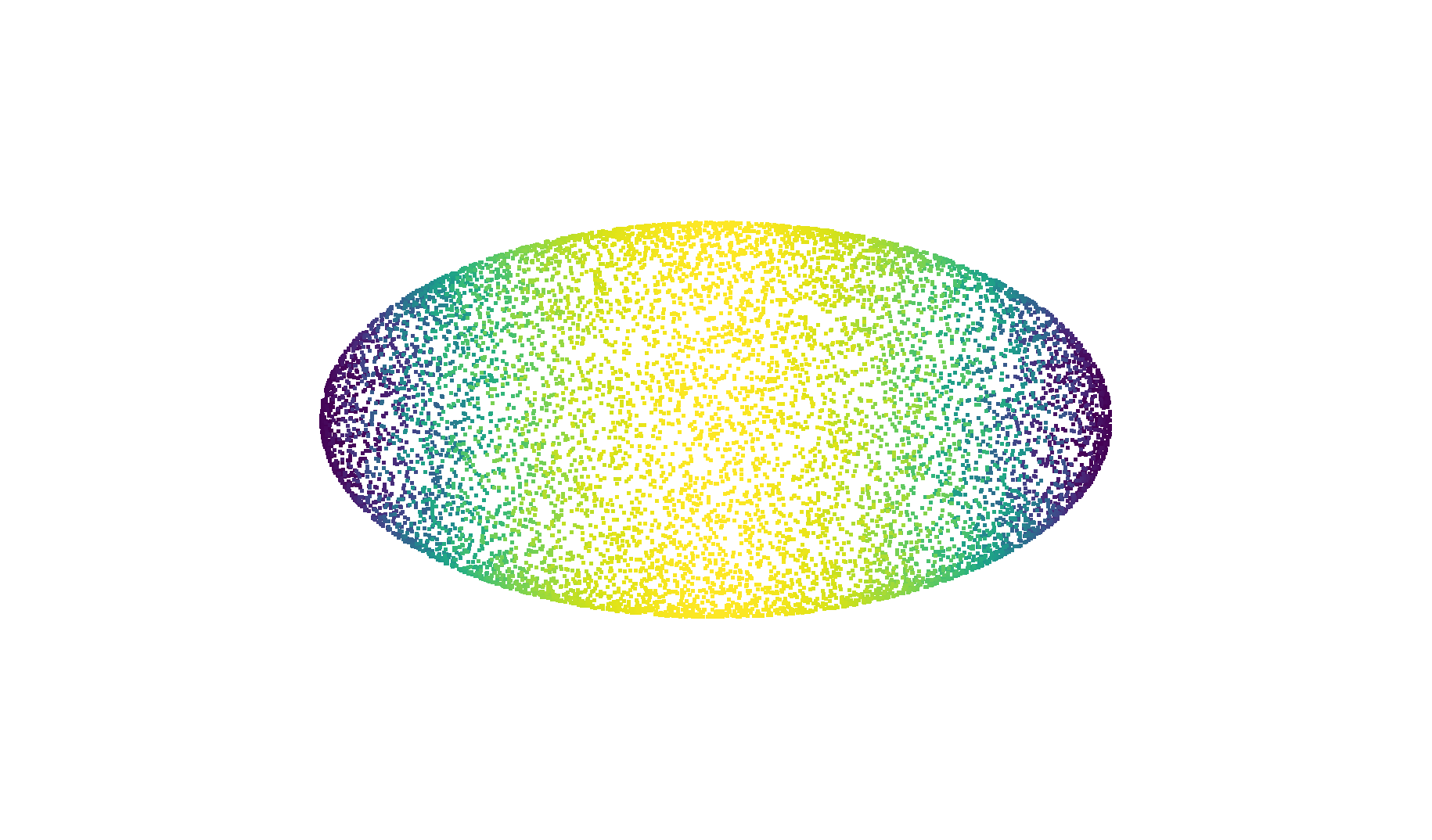}\vspace{0pt} 
    \includegraphics[width=\linewidth,trim={300 50 300 50},clip]{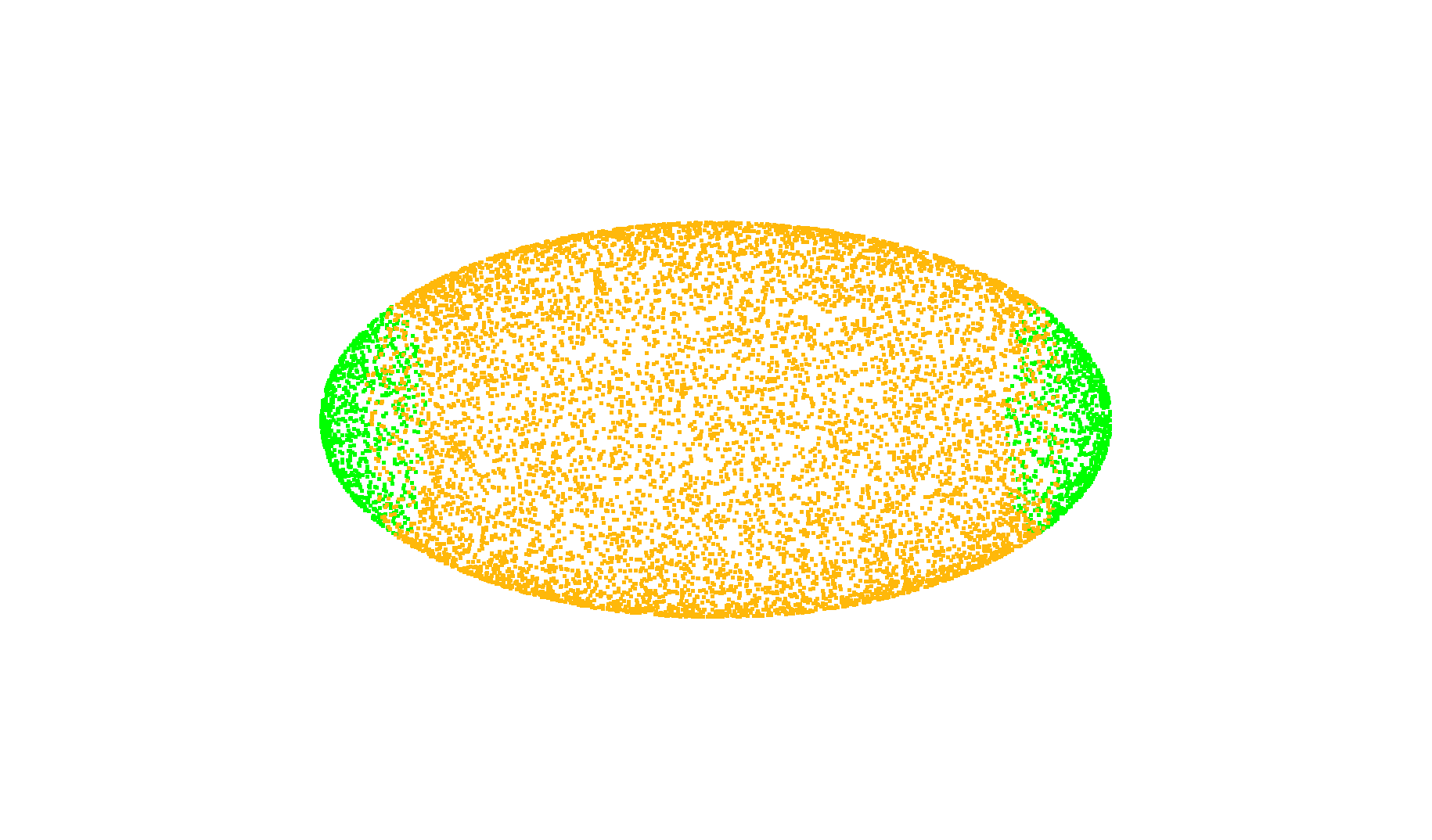}\vspace{0pt} 
    \includegraphics[width=\linewidth,trim={300 50 300 50},clip]{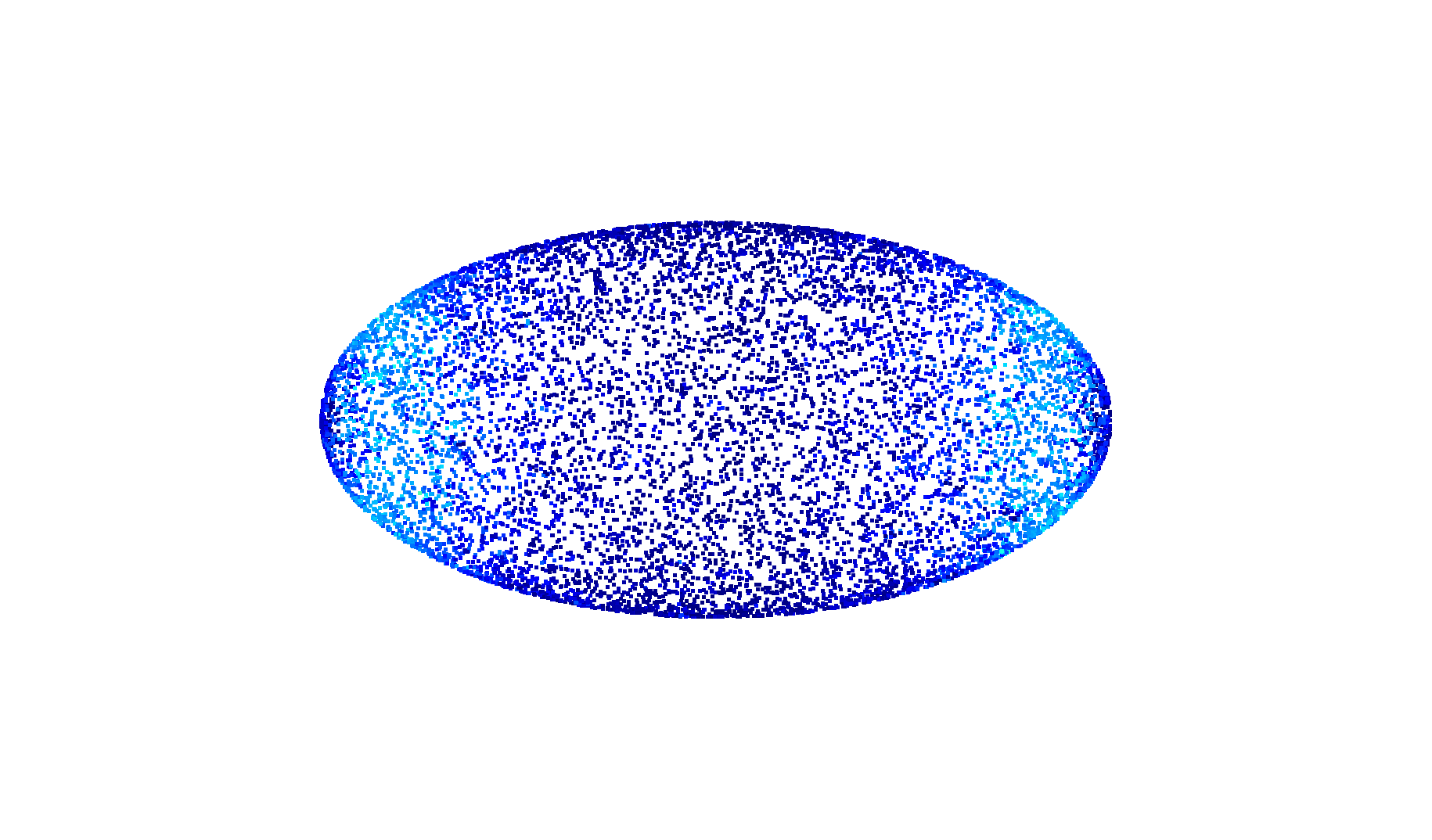}\vspace{0pt} 
    \includegraphics[width=\linewidth,trim={300 50 300 50},clip]{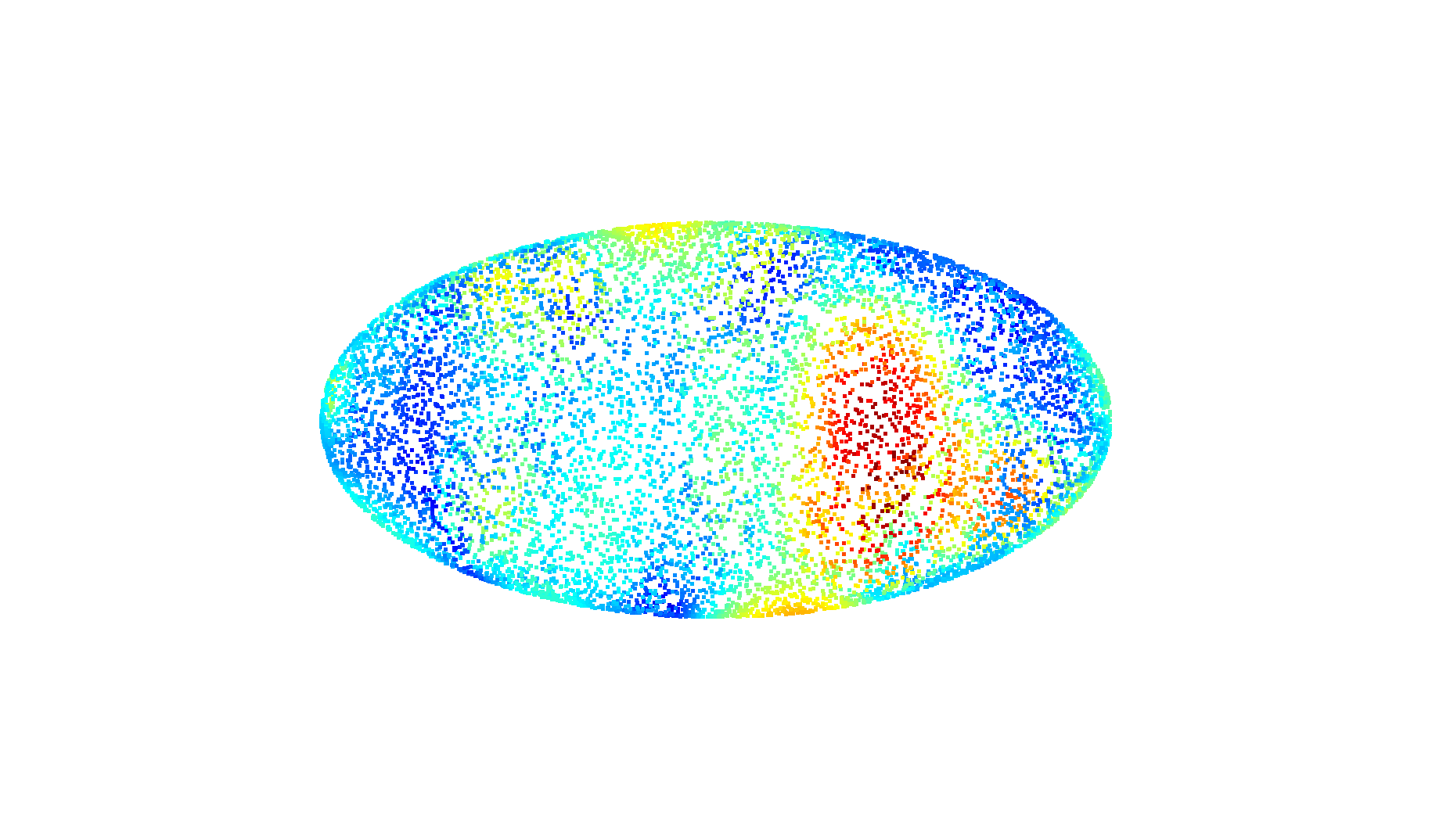}\vspace{0pt} 
    \includegraphics[width=\linewidth]{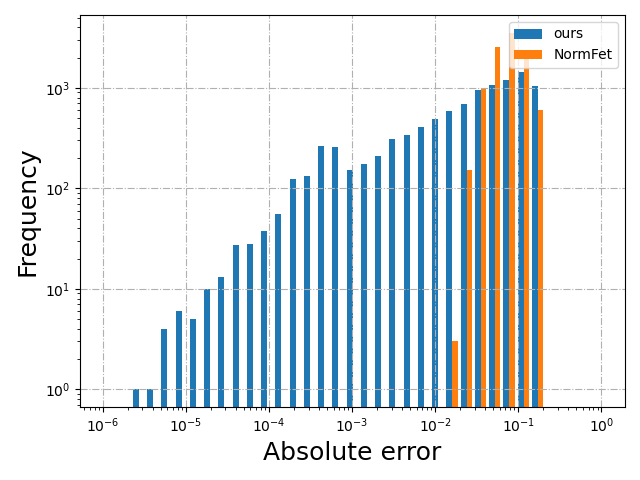}\vspace{0pt} 
    \end{minipage}
}
\\
\includegraphics[width=\linewidth,trim={30 0 30 0},clip]{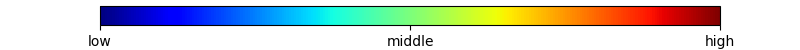}\vspace{0pt} 
\caption{Visual analysis of LFS estimation on canonical primitives (sphere, cone, and ellipsoid) sampled non-uniformly, with a little noise added. 
First row: ground truth LFS using the viridis color ramp. 
Second row: LFS type, where green denotes curvature-based and orange denotes diameter-based. Third row: estimated LFS error depicted for our algorithm using the jet color ramp.
Fourth row: estimated LFS error depicted for the NormFet algorithm. 
Fifth row: distribution of absolute estimation errors for all input points.}
\label{lfs-primitve}
\end{figure}

We first validate the proposed LFS estimation algorithm on 3D point clouds sampled from canonical primitives. We generate canonical primitives for which we know the ground-truth LFS analytically. We then measure the error made by our estimation algorithm (without smoothing) and compare it with NormFet \cite{dey2006normal}, which computes the medial axis via constructing the Voronoi diagram. More specifically, we sample three primitives (a sphere, a cone, and an ellipsoid) with non-uniform sampling and a small amount of added noise. We measure both the mean absolute error and max absolute error, as recorded in Table \ref{tab-primitive-lfs-cmp}. 
\begin{table}[htbp!]
\begin{center} 
\scalebox{0.82}{
\begin{tabular}{l| cc| cc| cc}
\toprule[1.2pt]
Capsule &\multicolumn{2}{c|}{Sphere (\#648)} &  \multicolumn{2}{c|}{Cone (\#2,610)} & \multicolumn{2}{c}{Ellipsoid (\#16,374)}  \\
 & mean($\downarrow$) & max($\downarrow$) & mean($\downarrow$) & max($\downarrow$) & mean($\downarrow$) & max($\downarrow$)  \\
\hline
NormFet & 4.079E-1 & 8.191E-1 & 9.413E-1 & 2.233 & 7.835E-2 & 1.917E-1 \\
Ours & \textbf{5.511E-3} & \textbf{3.190E-2} & \textbf{2.135E-1} & \textbf{7.751E-1} & \textbf{5.471E-2} & \textbf{1.786E-1} \\
\bottomrule[1.2pt]
\end{tabular}}
\end{center}
\caption{LFS comparisons with NormFet on canonical primitives. 
``mean'' denotes the mean absolute error. 
``max'' denotes the maximal absolute error. }
\label{tab-primitive-lfs-cmp}
\end{table}
Our experimental results show that our LFS estimation algorithm outperforms NormFet in terms of the mean and maximal absolute errors. The experiments highlight the robustness of our algorithm to non-uniform sampling and small amounts of noise. NormFet estimates LFS by first constructing the medial axis from the Voronoi diagram, but a good estimation of the medial axis requires a point cloud that is dense and noise-free. Recall that we define LFS as the minimum of the curvature radius and the shape diameter halved. We can thus depict the origin of LFS, as shown in Fig.~\ref{lfs-primitve} (second row). Green points depict the points with a curvature radius smaller than half of the shape diameter, and orange points depict the other points. As expected, the sphere, whose curvature radius equates to half of the shape diameter, yields a mixture of green and orange points. On the ellipsoid, the two tips correspond to a curvature-based LFS, while the rest are diameter-based. Fig.~\ref{lfs-primitve} offers a comprehensive visual depiction.

\subsubsection{Free-form point clouds} 
\begin{figure}[t!]
\centering
\subfigure[\scriptsize{Smooth cone (\#31,368)}]{
   \includegraphics[width=0.29\linewidth,trim={430 200 430 200},clip]{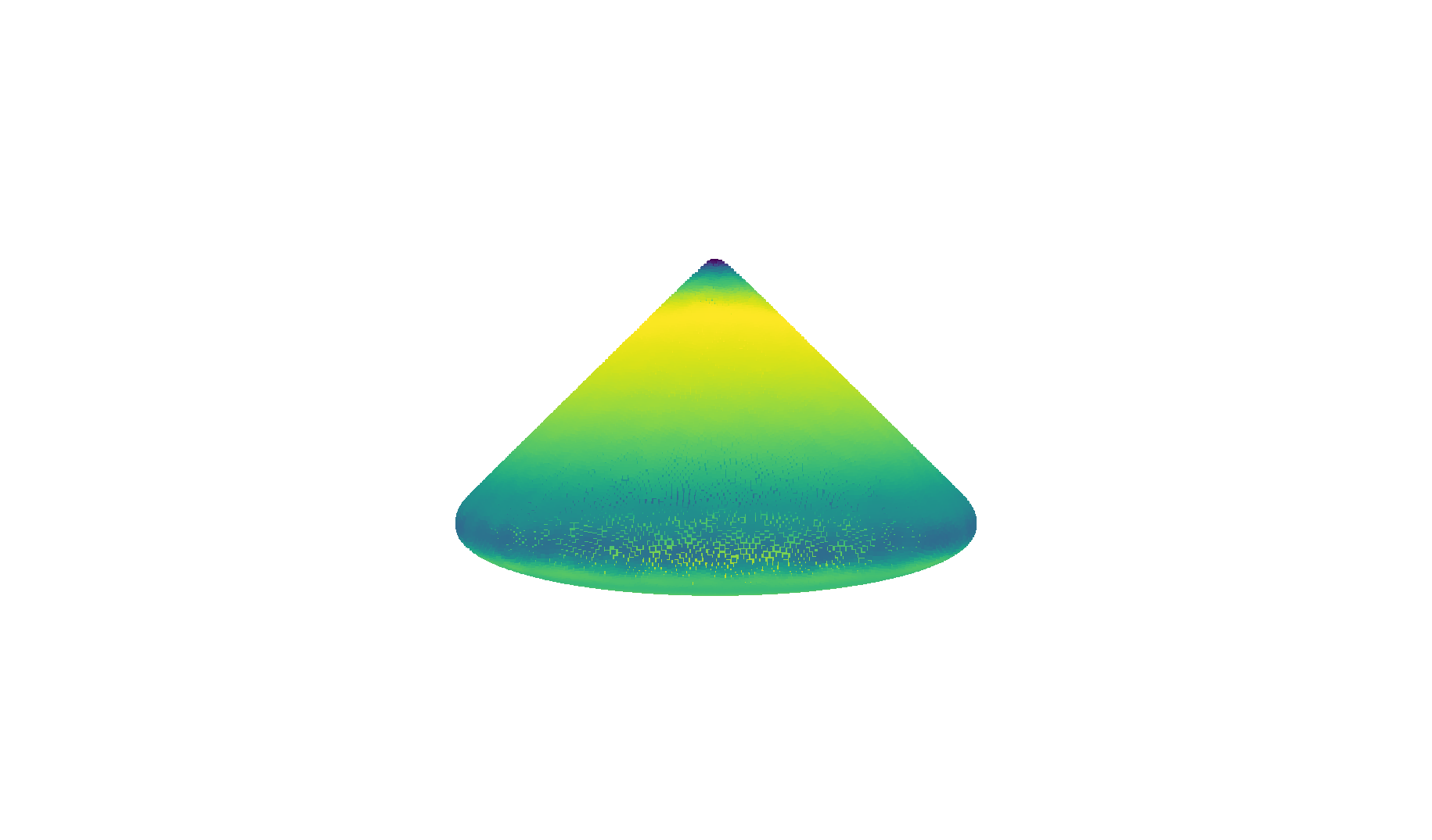}\vspace{0pt}  
}
\subfigure[\scriptsize{Knot (\#115,200)}]{
   \includegraphics[width=0.29\linewidth,trim={400 100 400 100},clip]{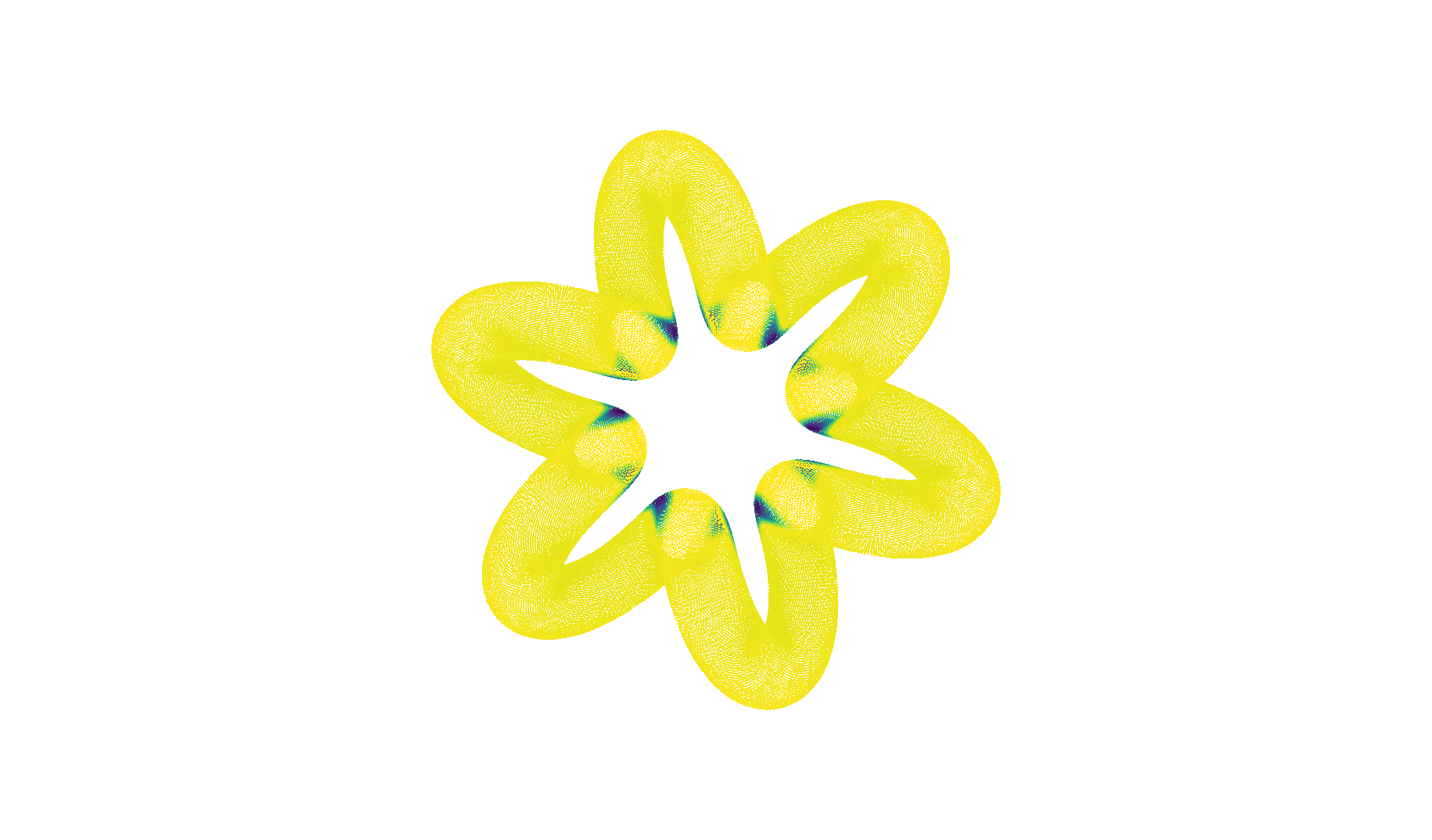}\vspace{0pt}  
}
\subfigure[\scriptsize{Tangle cube (\#8,545)}]{
   \includegraphics[width=0.29\linewidth,trim={550 270 550 270},clip]{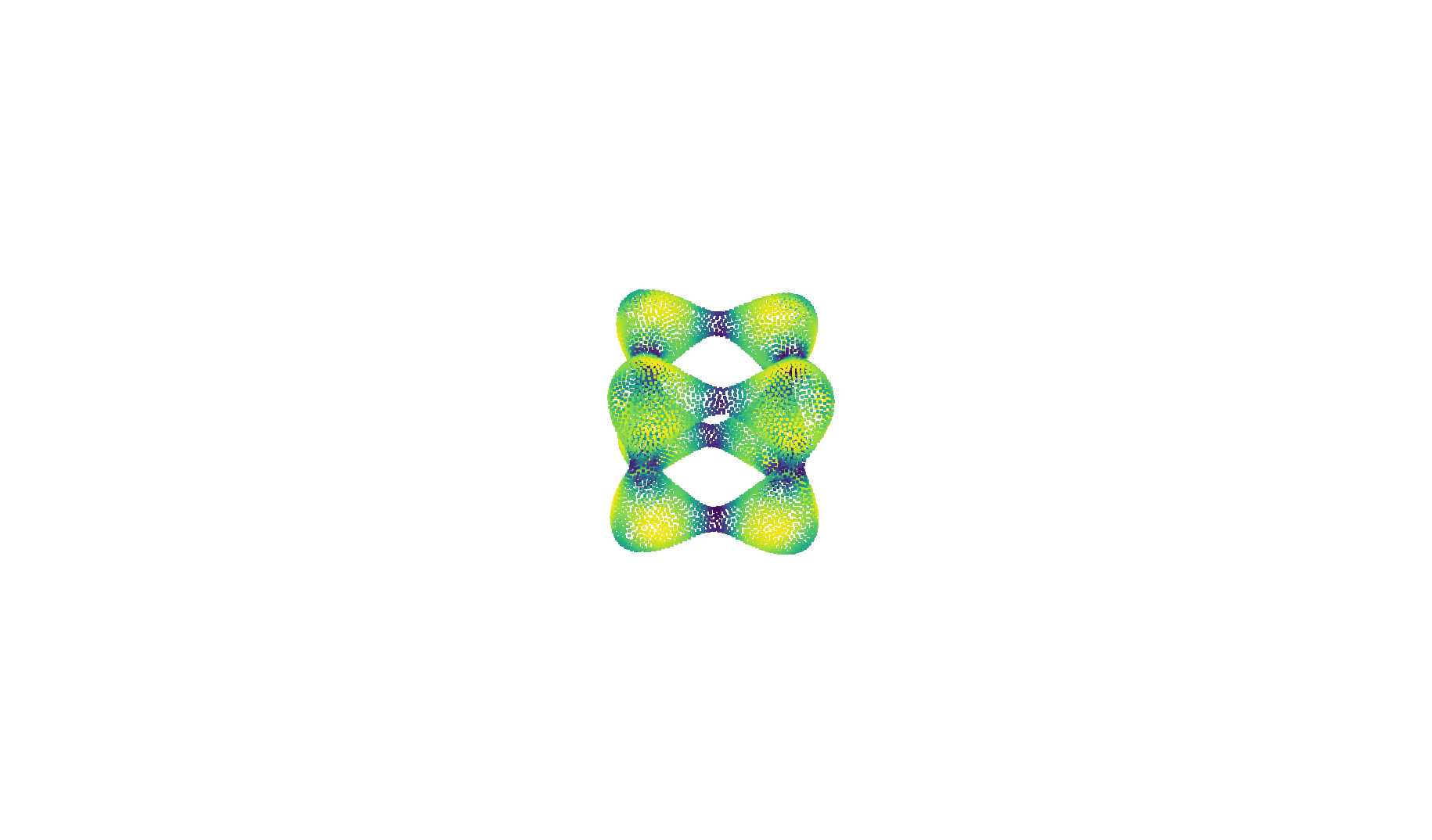}\vspace{0pt}
}\\[-2ex]
\subfigure[\scriptsize{Kitten (\#20,380)}]{
   \includegraphics[width=0.29\columnwidth,trim={350 10 350 10},clip]{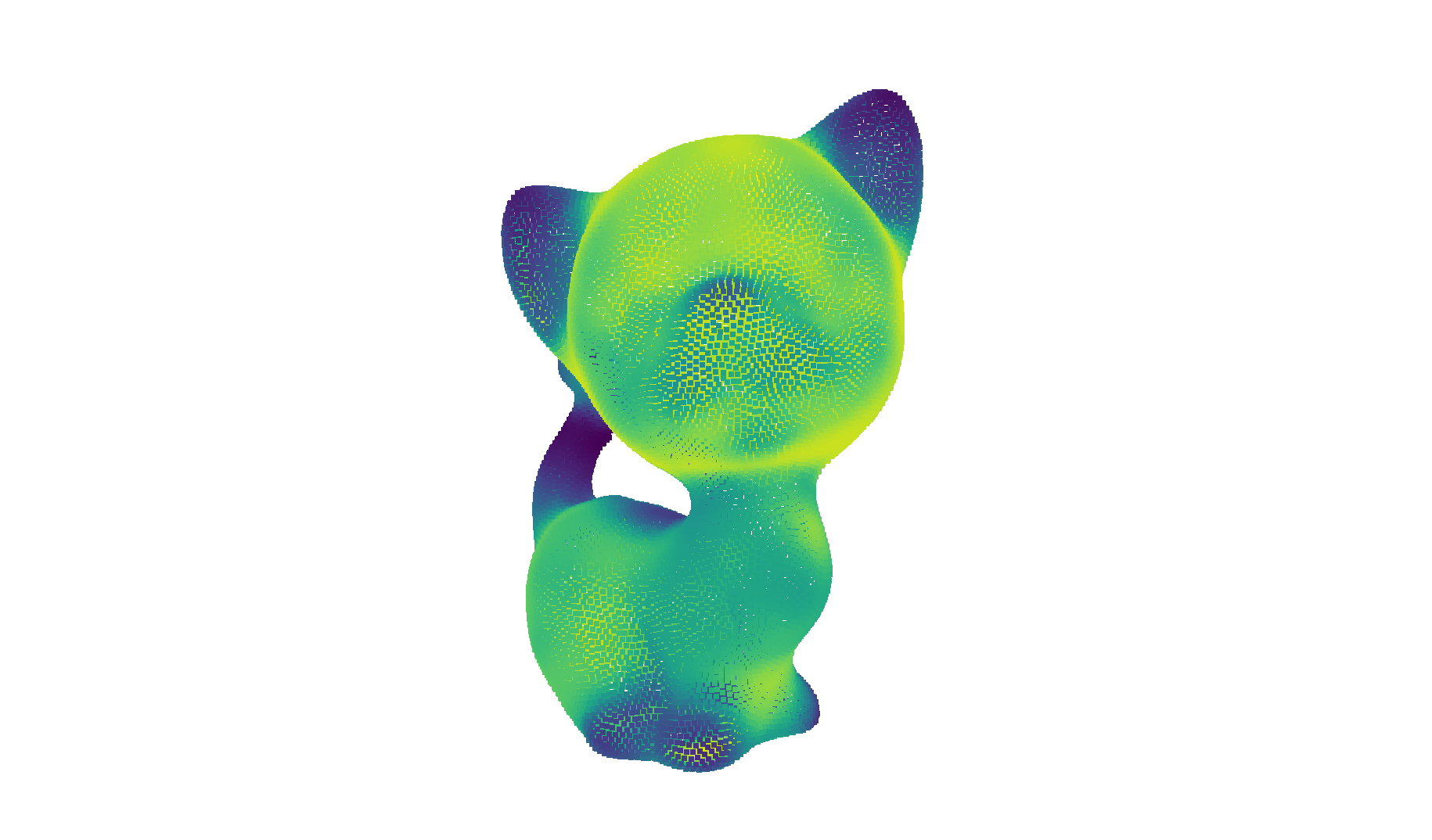}\vspace{-3pt} 
}
\subfigure[\scriptsize{Bust (\#75,002)}]{
   \includegraphics[width=0.29\columnwidth,trim={350 10 350 10},clip]{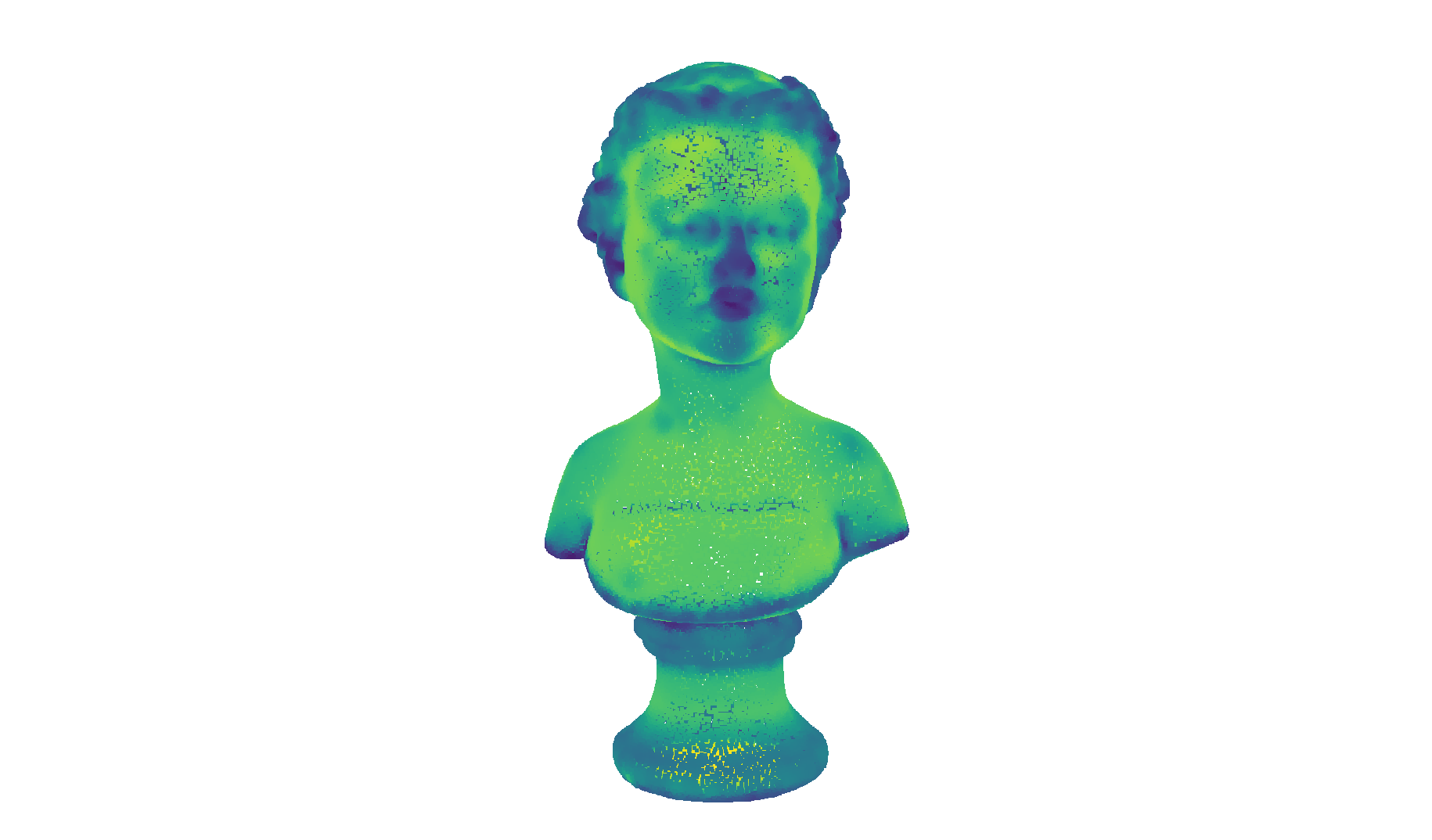}\vspace{-3pt}
}
\subfigure[\scriptsize{Elephant (\#92,964)}]{
   \includegraphics[width=0.29\columnwidth,trim={360 10 360 10},clip]{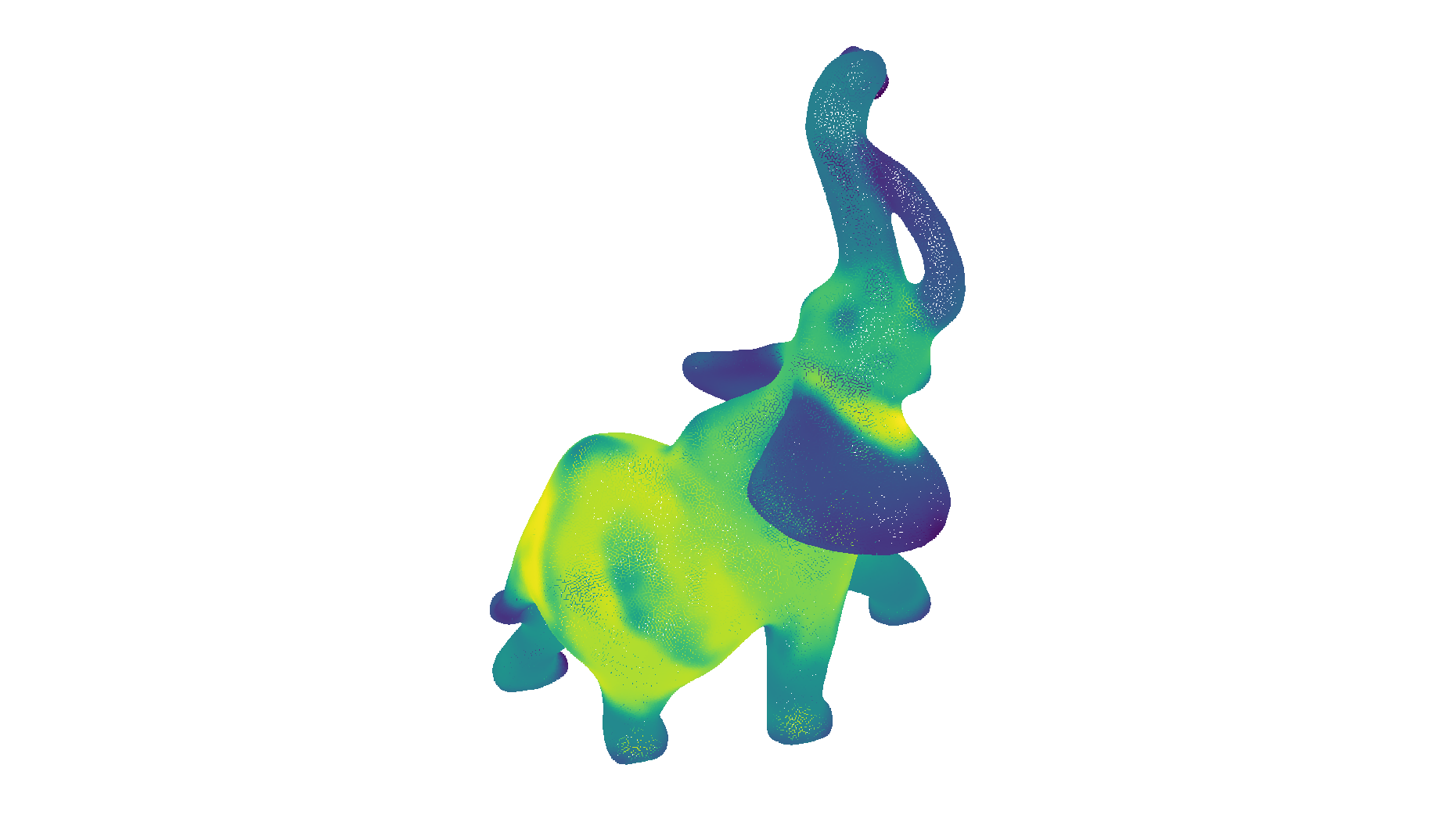}\vspace{-3pt}
}\\[-2ex]
\subfigure[\scriptsize{Hippo (\#58,188)}]{
   \includegraphics[width=0.29\columnwidth,trim={200 10 250 10},clip]{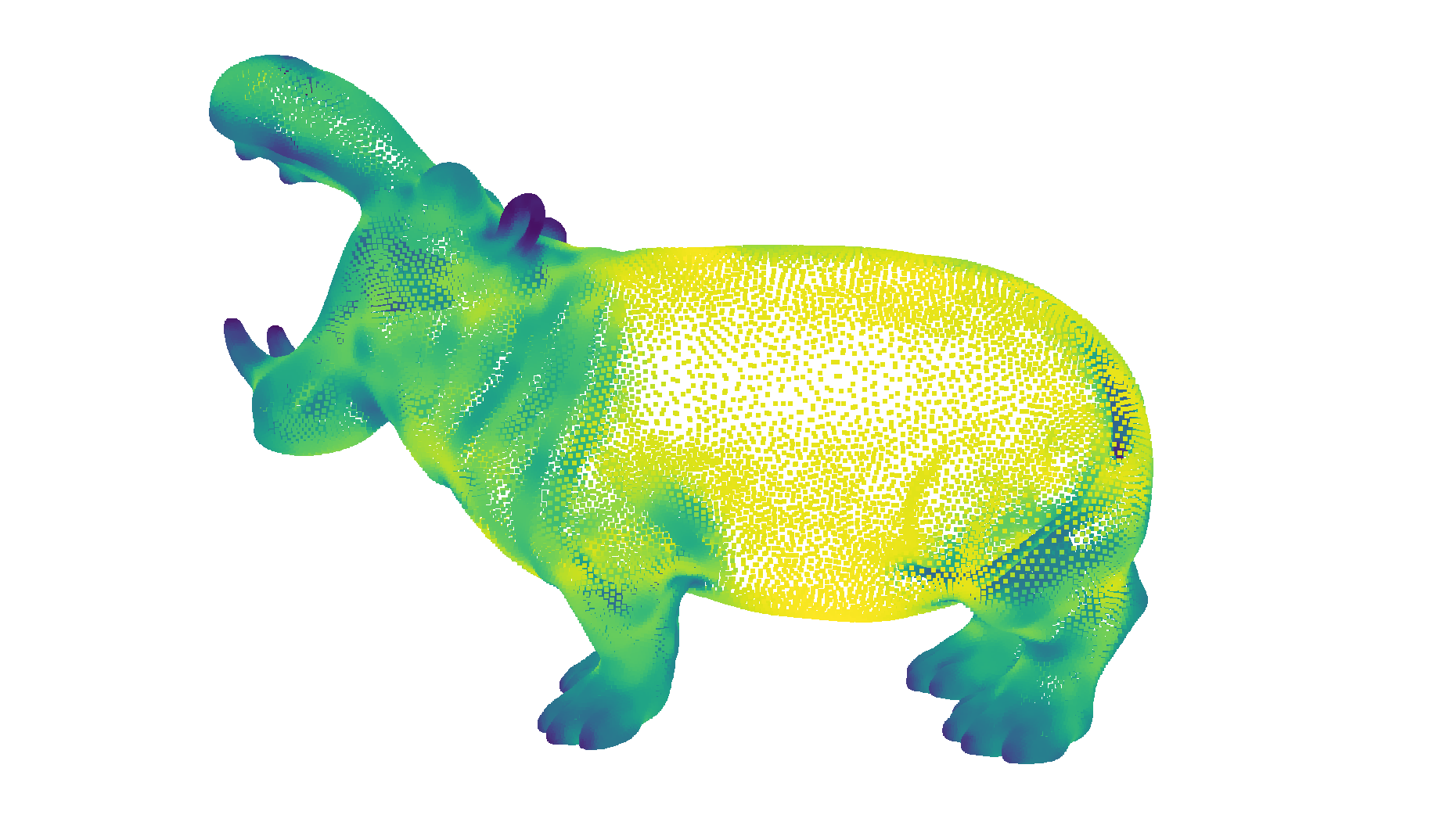}\vspace{-3pt} 
}
\subfigure[\scriptsize{Buddha (\#719,560)}]{
   \includegraphics[width=0.29\columnwidth,trim={320 10 320 10},clip]{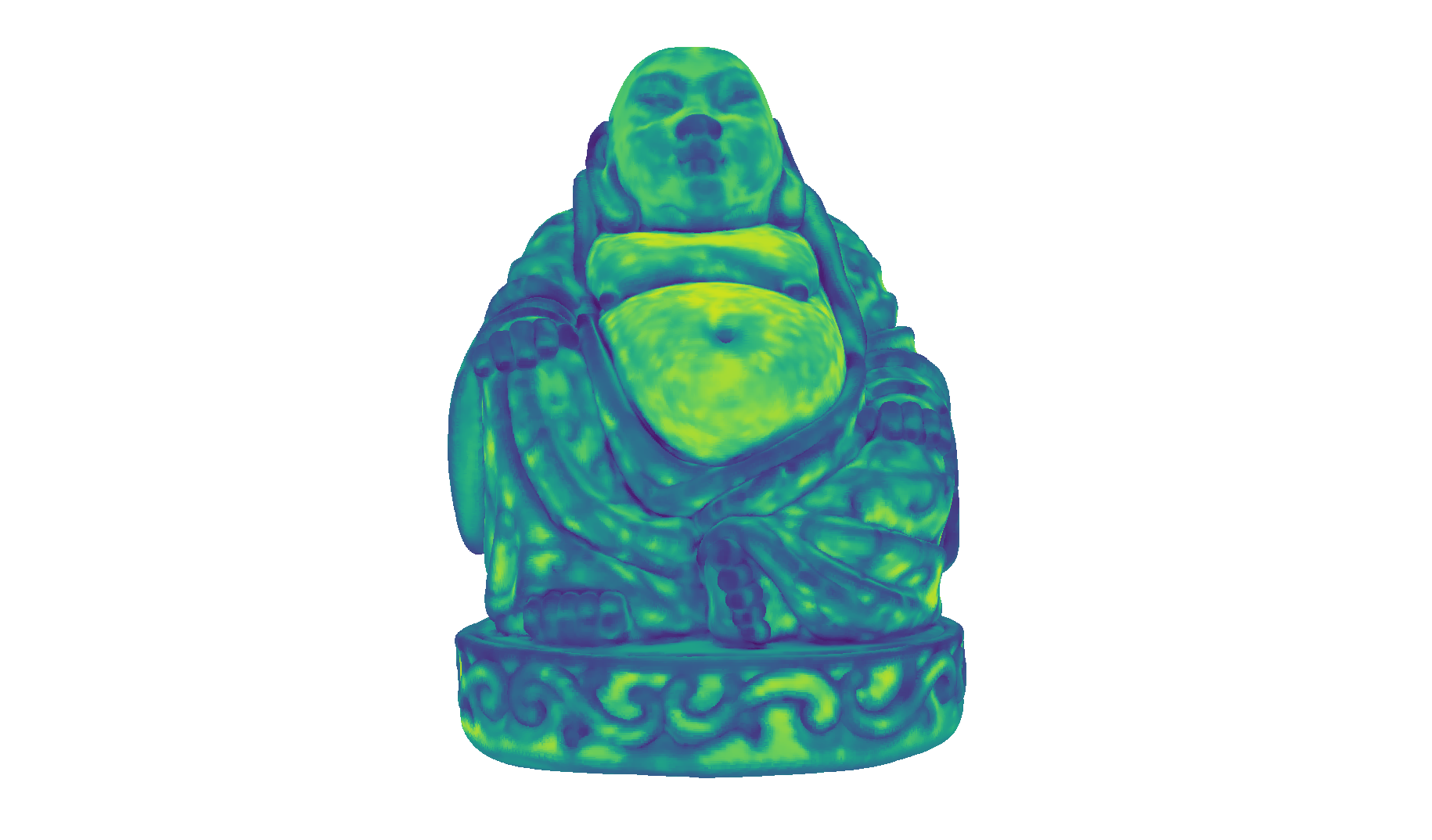}\vspace{-3pt} 
}
\subfigure[\scriptsize{Hand (\#369,045)}]{
   \includegraphics[width=0.29\columnwidth,trim={350 10 350 10},clip]{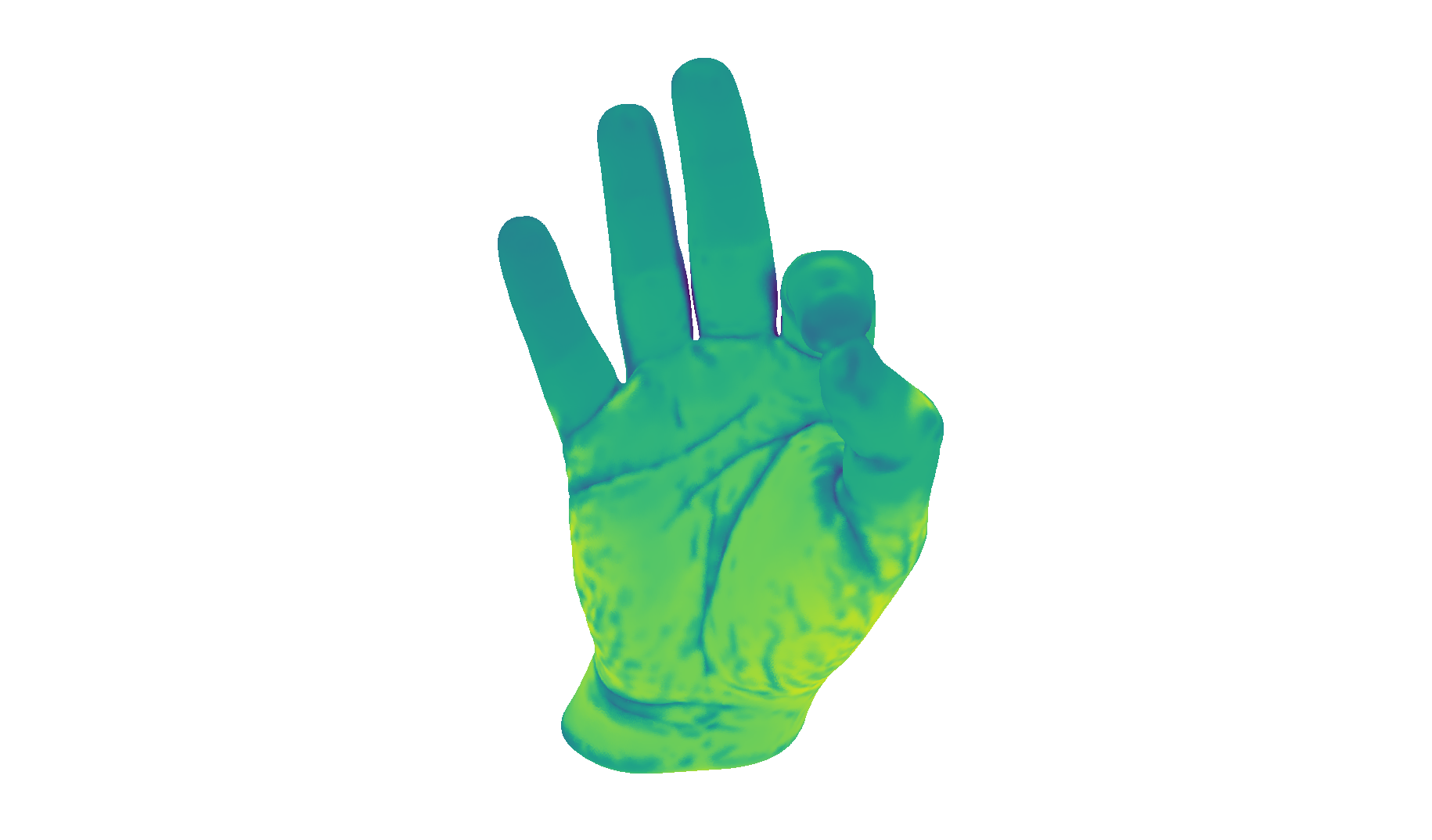}\vspace{-3pt} 
}\\[-2ex]
\subfigure[\scriptsize{Feline (\#49,864)}]{
   \includegraphics[width=0.29\columnwidth,trim={250 10 250 10},clip]{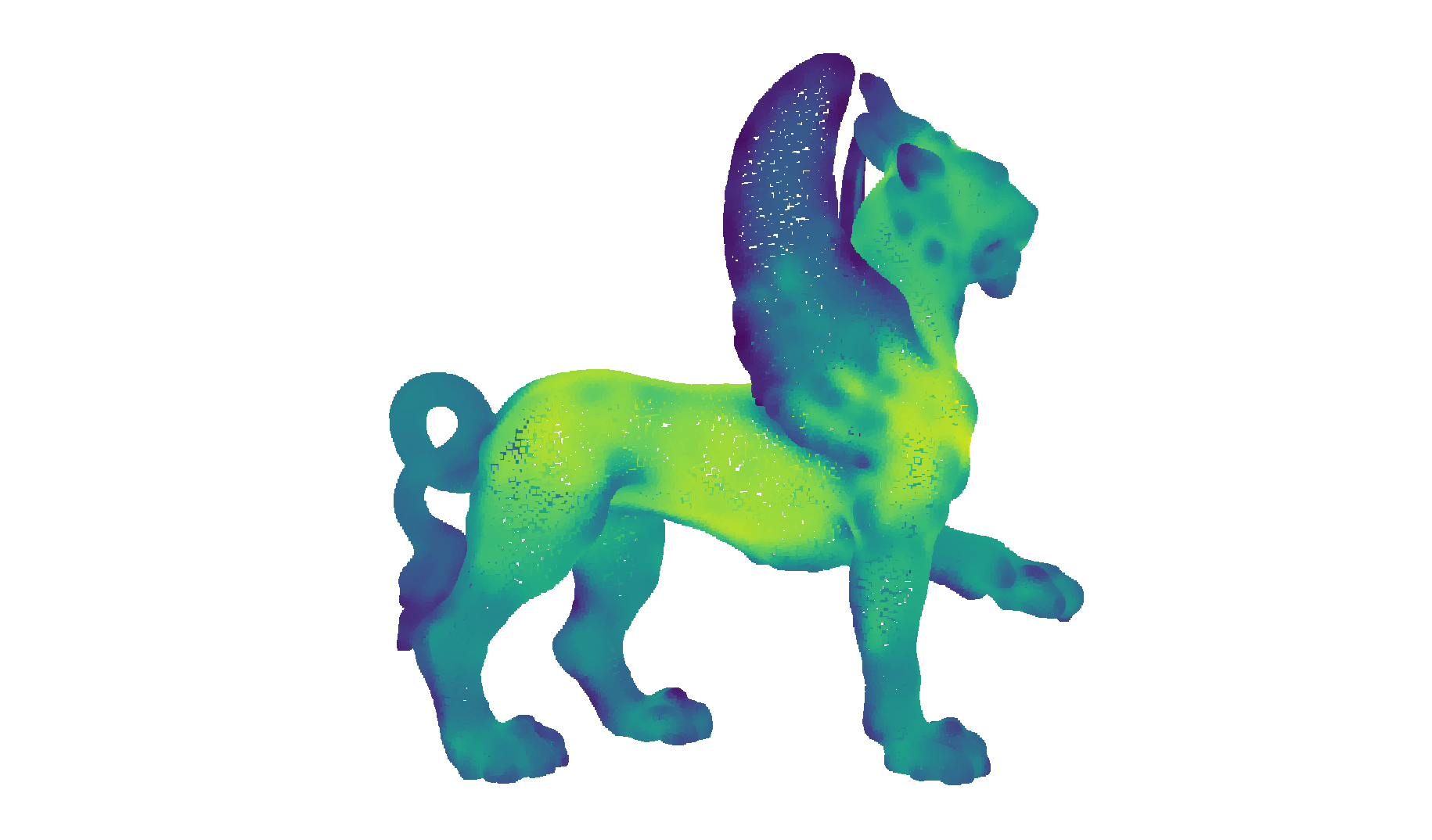}\vspace{-3pt} 
}
\subfigure[\scriptsize{Isidore (\#235,379)}]{
   \includegraphics[width=0.29\columnwidth,trim={250 10 250 10},clip]{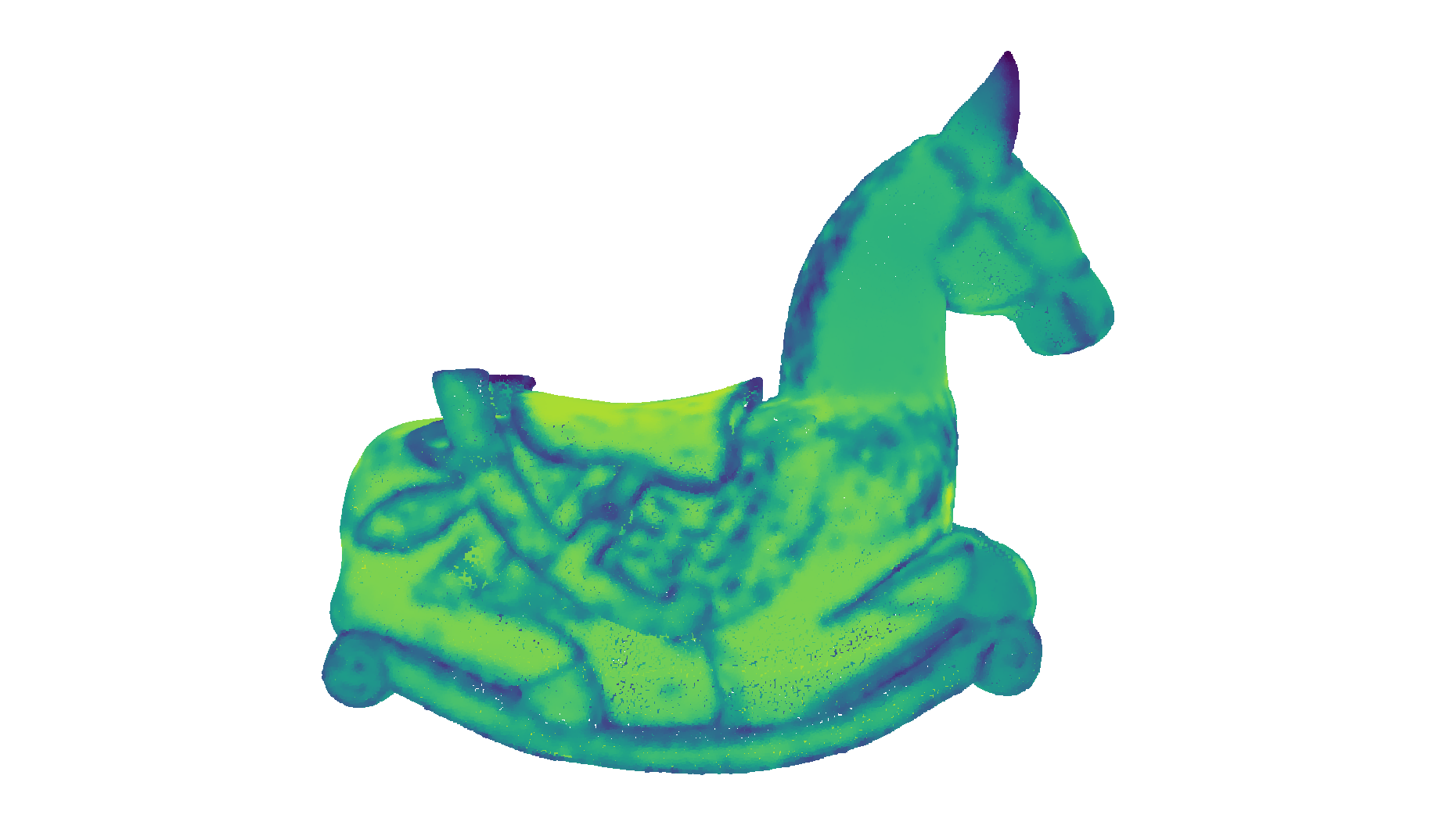}\vspace{-3pt} 
}
\subfigure[\scriptsize{Dragon (\#296,200)}]{
   \includegraphics[width=0.29\columnwidth,trim={180 50 250 50},clip]{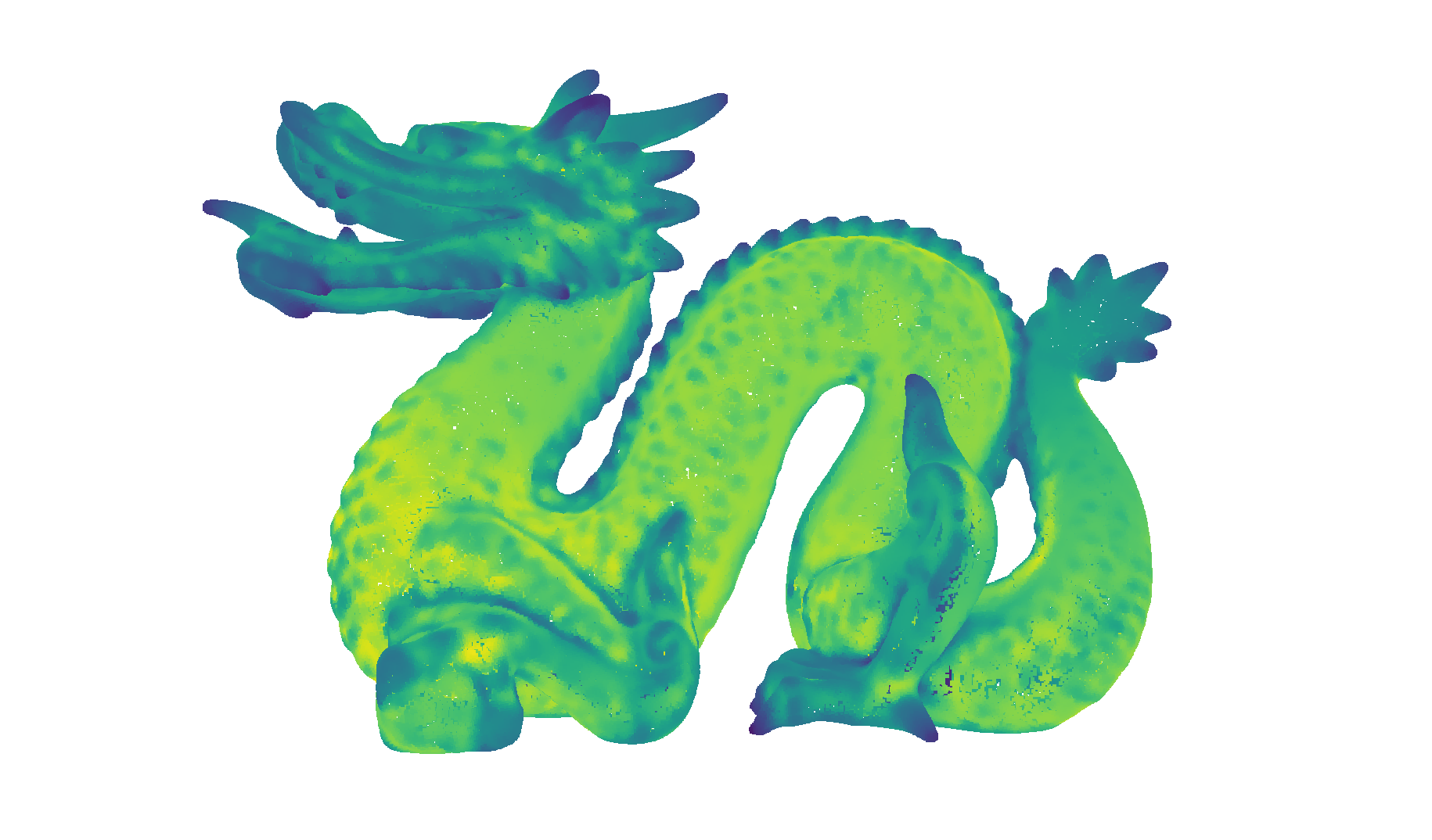}\vspace{-3pt} 
}\par
\includegraphics[width=\linewidth,trim={30 0 30 0},clip]{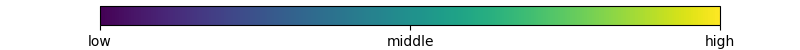}\vspace{0pt} 
\caption{LFS estimation for 3D point clouds sampled on free-form shapes. The viridis color ramp is used to map the smoothed LFS values for each point cloud.}
\label{lfs-more}
\end{figure}

We now evaluate our LFS estimation approach for point clouds sampled on free-form shapes that lack ground-truth data. Smoothing is performed on the raw estimated LFS. Refer to Fig.~\ref{lfs-more} for visual results.

\subsection{Implicit Function}
\begin{figure}[ht]
\centering
\subfigure[Kitten]{
    \begin{minipage}[h]{0.29\columnwidth}
    \includegraphics[width=\linewidth,trim={290 0 290 0},clip]{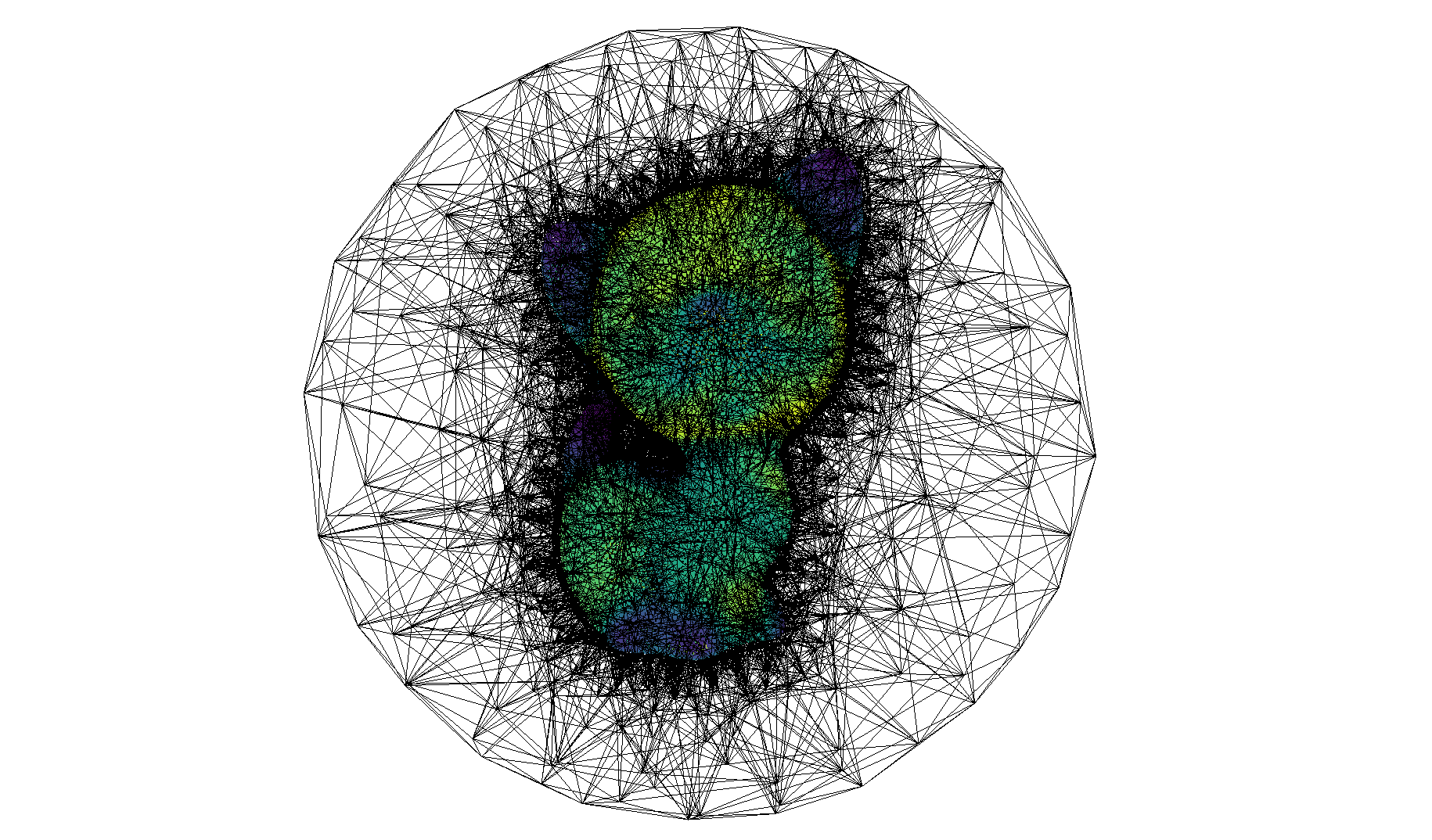}\vspace{0pt} 
    \includegraphics[width=\linewidth,trim={290 0 290 0},clip]{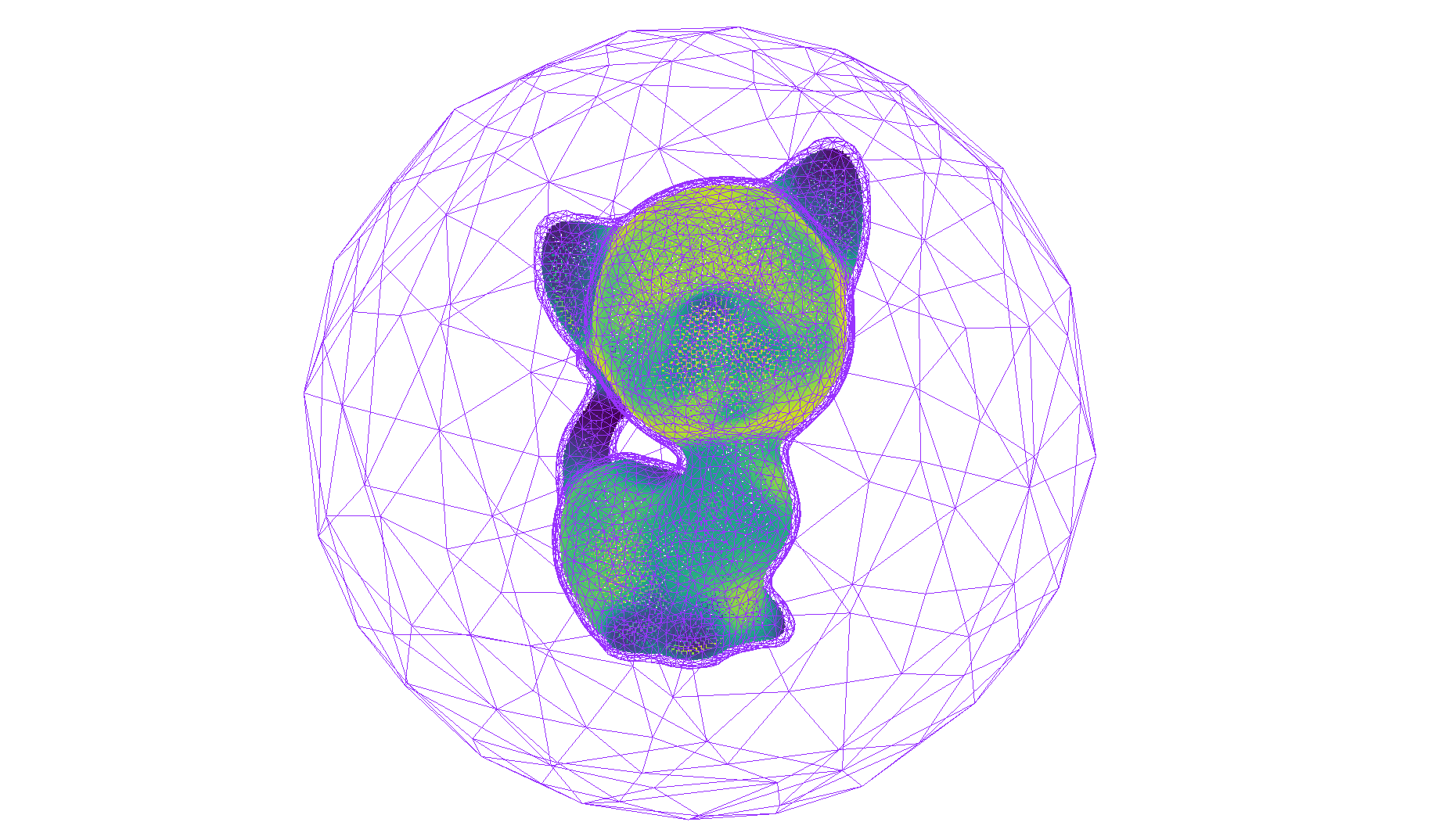}\vspace{0pt}  
    \includegraphics[width=\linewidth,trim={420 80 420 80},clip]{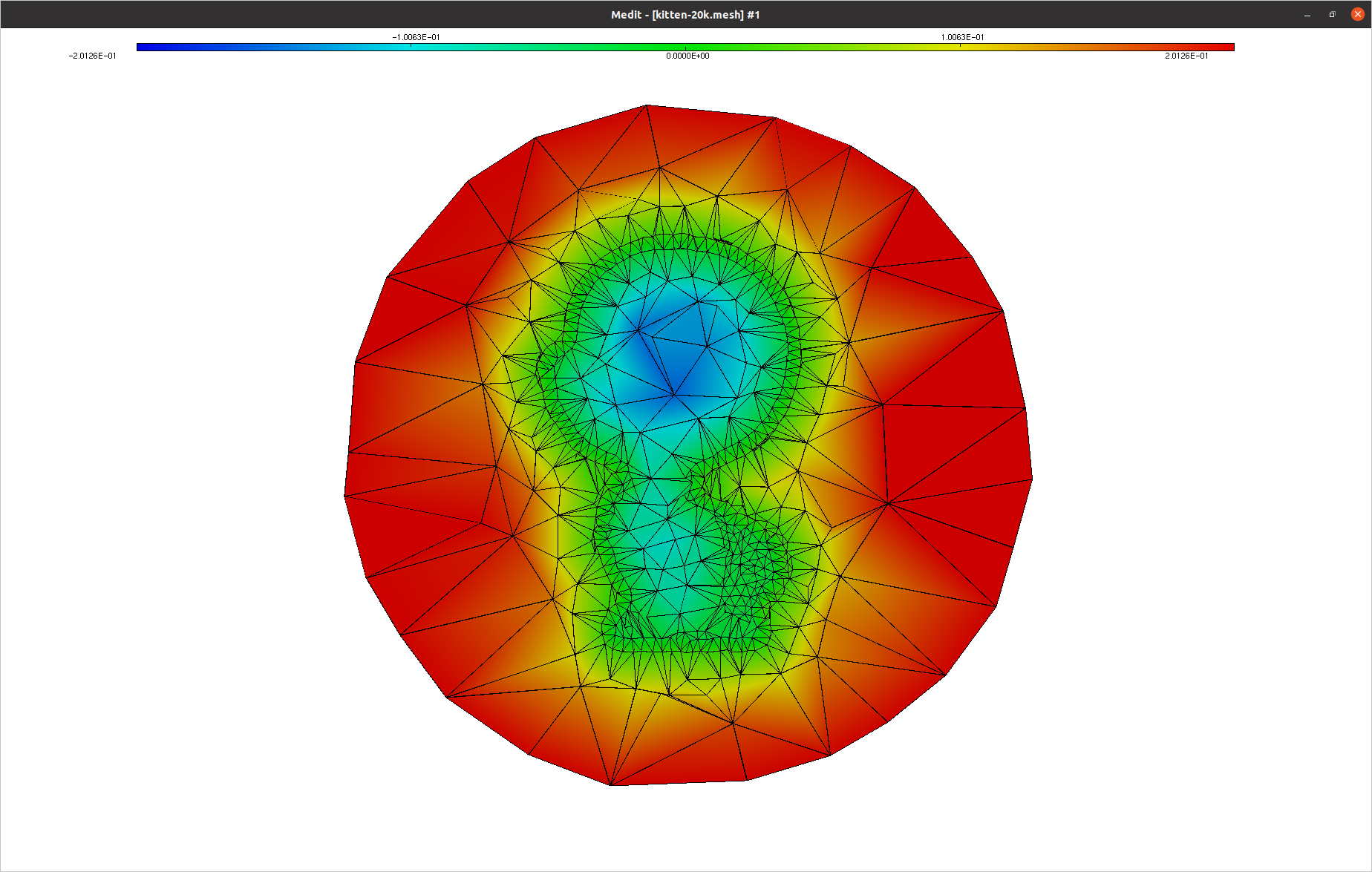}\vspace{0pt} 
    \includegraphics[width=\linewidth]{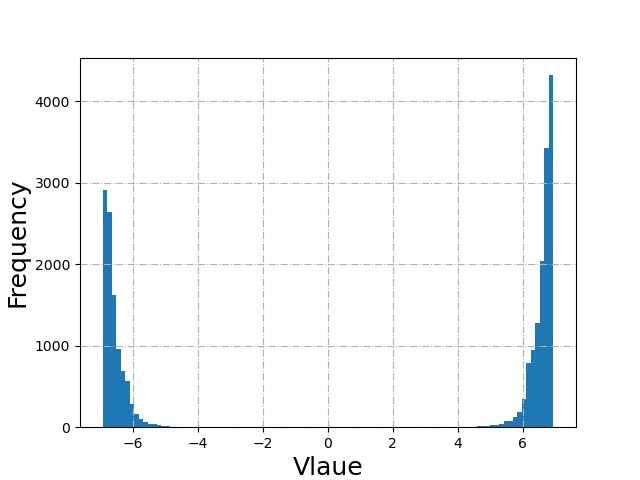}\vspace{5pt}
    \end{minipage}
}\hfill
\subfigure[Bust]{
    \begin{minipage}[h]{0.29\columnwidth}
    \includegraphics[width=\linewidth,trim={290 0 290 0},clip]{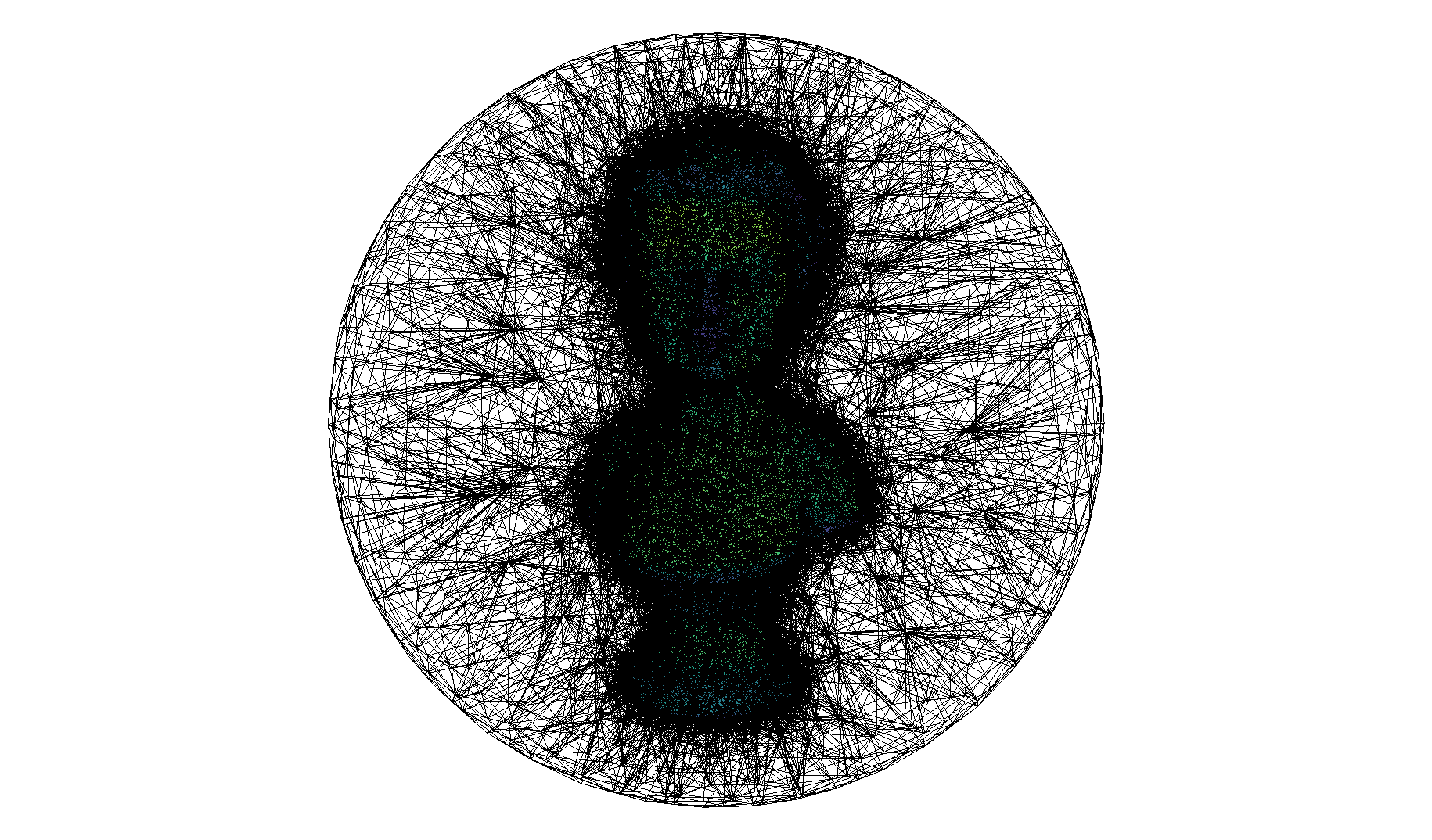}\vspace{0pt} 
    \includegraphics[width=\linewidth,trim={290 0 290 0},clip]{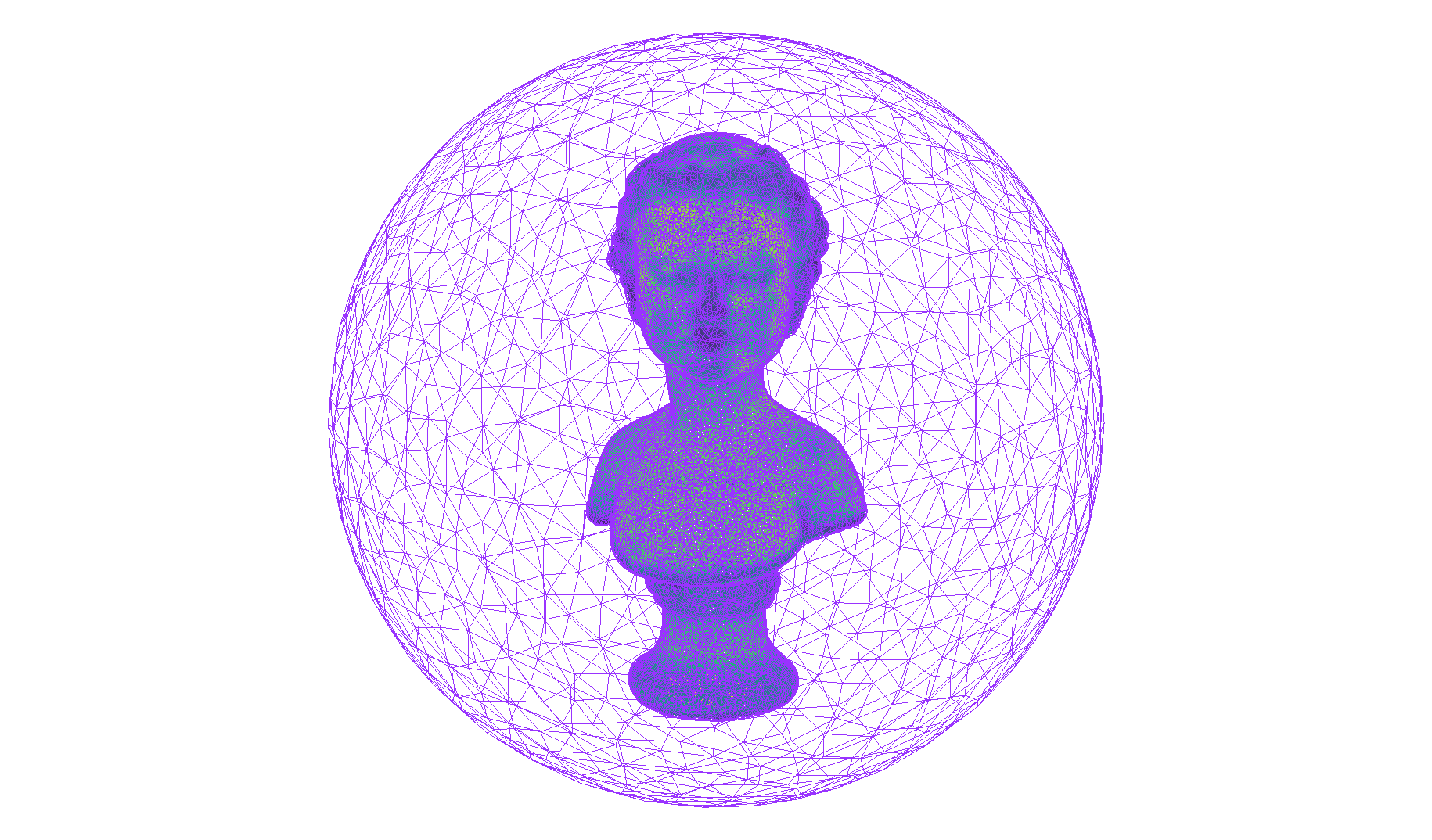}\vspace{0pt} 
    \includegraphics[width=\linewidth,trim={420 80 420 80},clip]{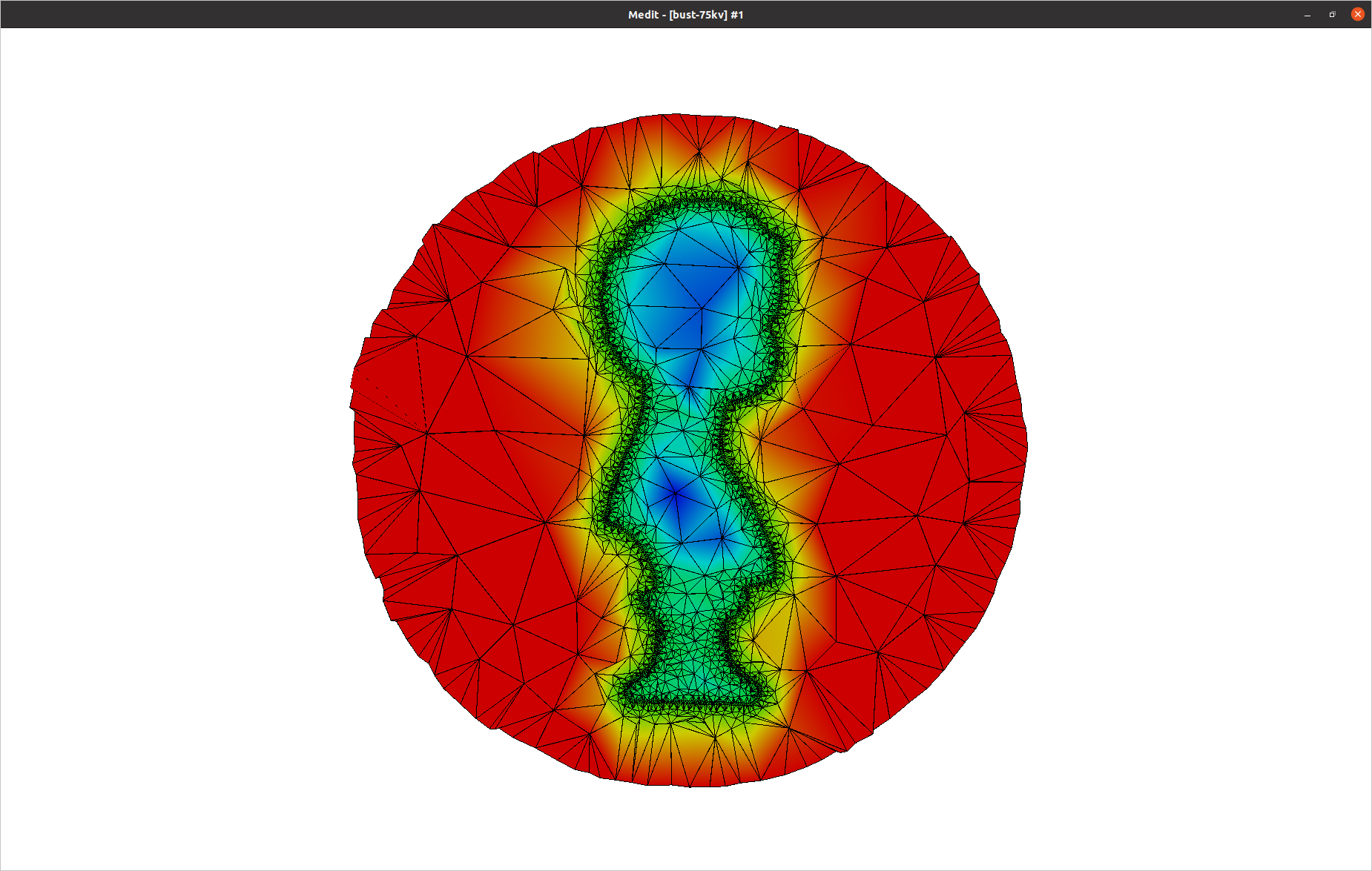}\vspace{0pt} 
    \includegraphics[width=\linewidth]{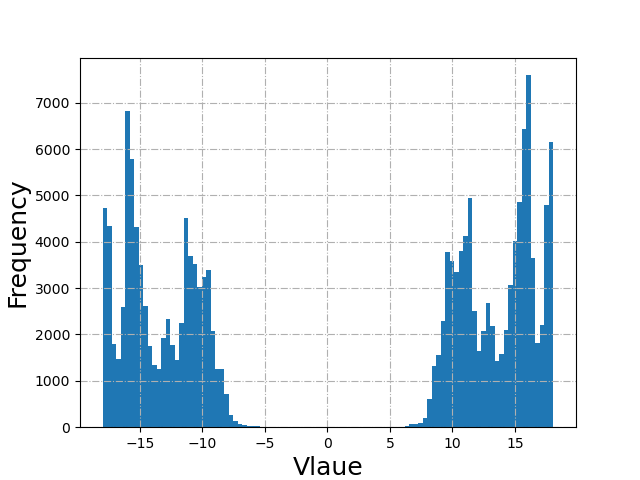}\vspace{5pt} 
    \end{minipage}
}\hfill
\subfigure[Elephant]{
    \begin{minipage}[h]{0.29\columnwidth}
    \includegraphics[width=\linewidth,trim={290 0 290 0},clip]{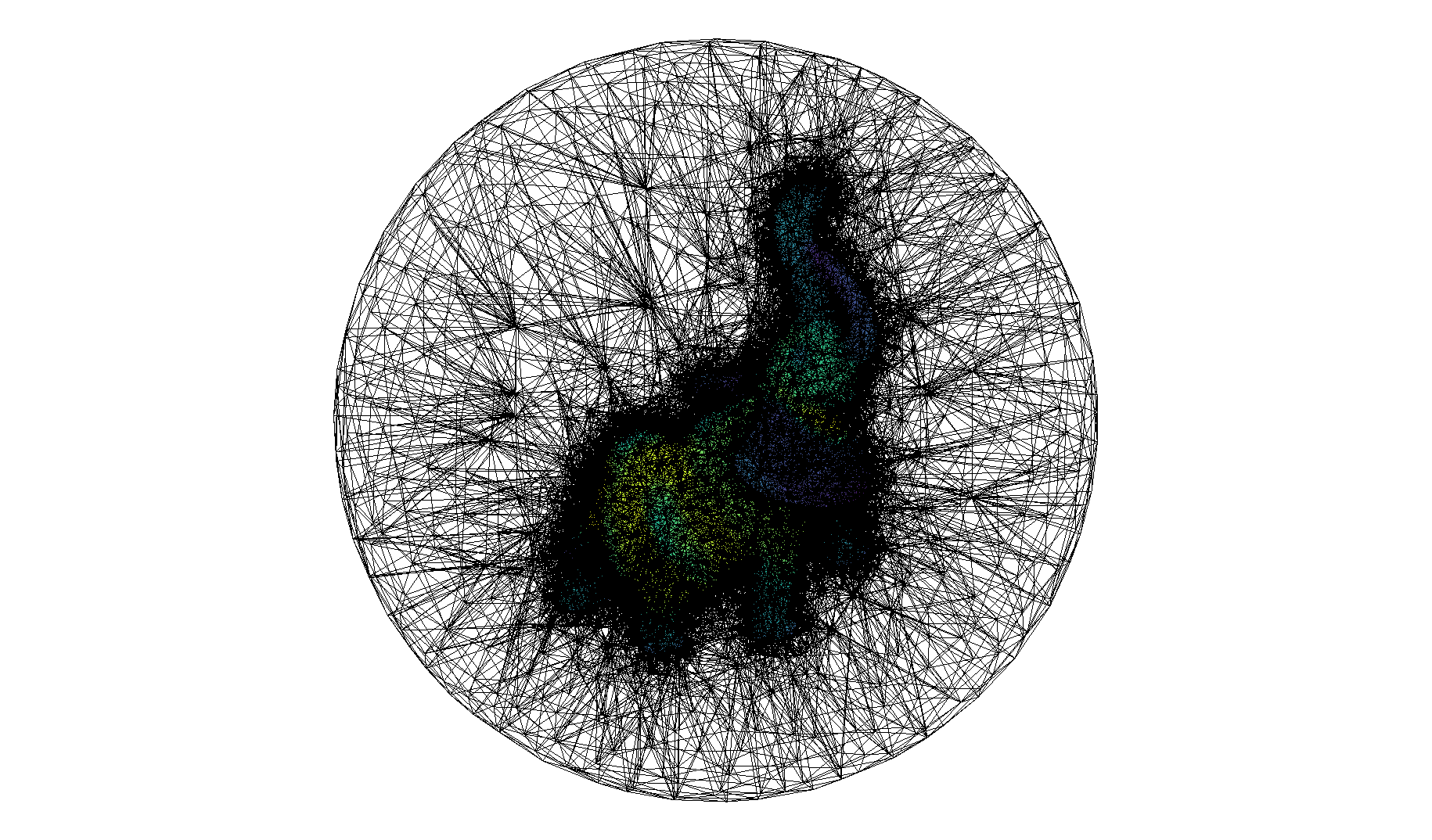}\vspace{0pt} 
    \includegraphics[width=\linewidth,trim={290 0 290 0},clip]{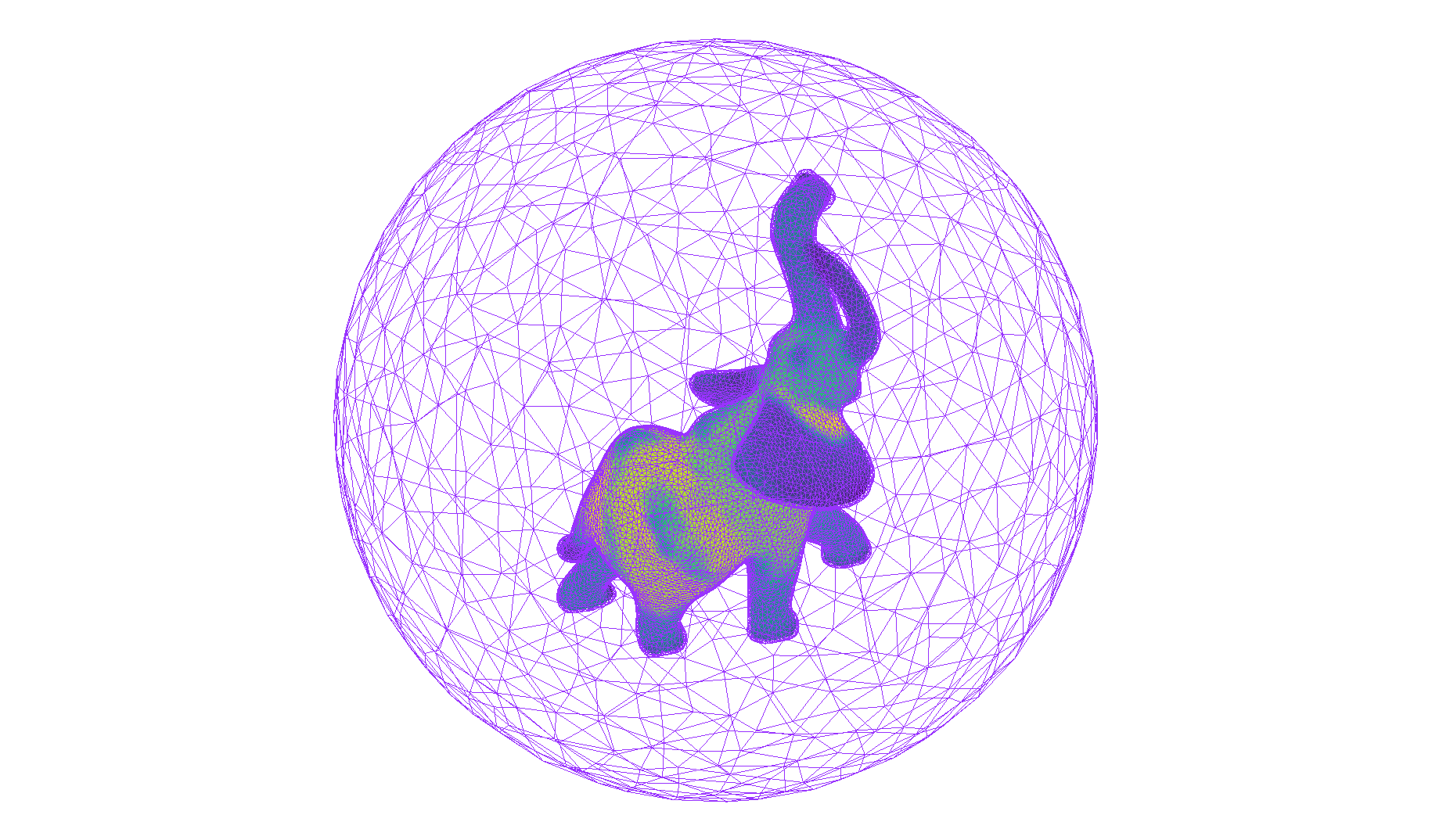}\vspace{0pt} 
    \includegraphics[width=\linewidth,trim={420 80 420 80},clip]{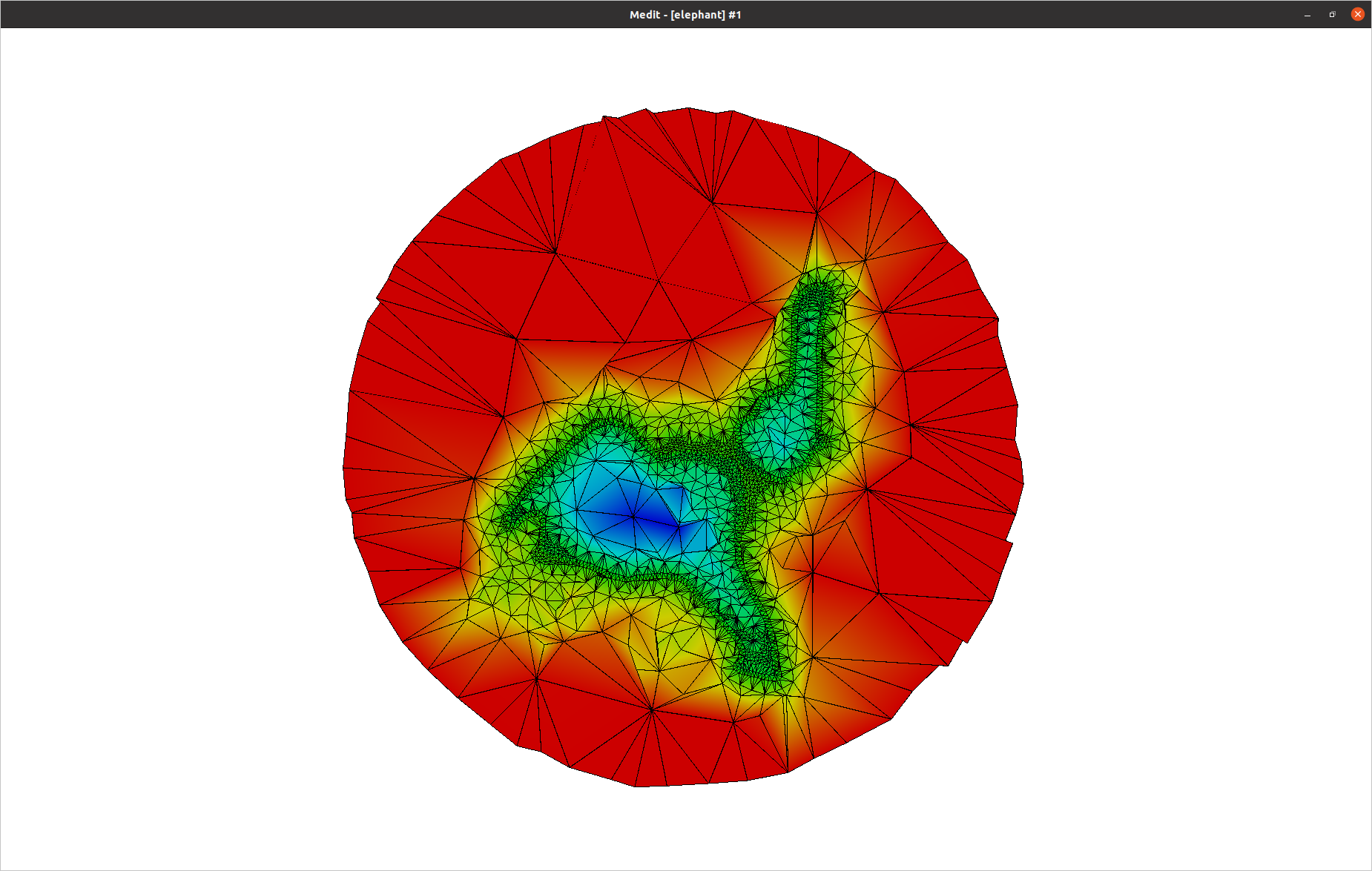}\vspace{0pt} 
    \includegraphics[width=\linewidth]{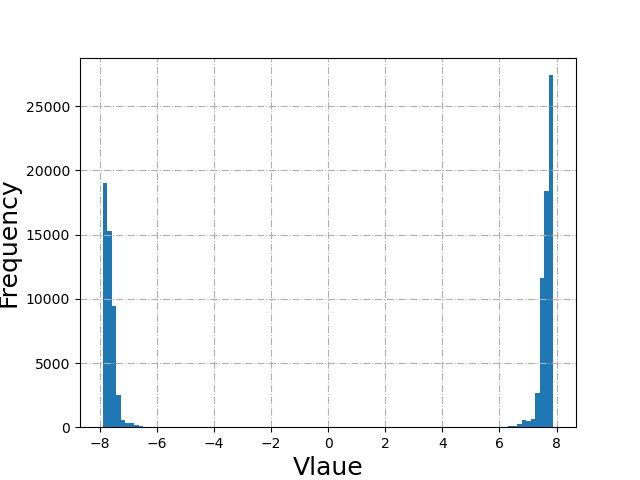}\vspace{5pt} 
    \end{minipage}
}\hfill
\caption{Implicit functions for the kitten, bust, and elephant 3D point clouds. 
First row: global depiction of the multi-domain. 
Second row: boundary of the multi-domain. The reach-aware envelope encloses the input points embedded in a bounded sphere. 
Third row: cut view of the piecewise linear implicit function. The tetrahedra are well-shaped by Delaunay refinement.
Fourth row: distribution of solved signed implicit values at the triangulation vertices. The two main peaks correspond to two sets of vertices located inside and outside the inferred surface. Again, we find the thin envelope layer sandwiching the inputs, and the tetrahedra are isotropic from the clip view.}
\label{domain-imf}
\end{figure}

The implicit function is defined on the reach-aware 3D multi-domain and represented as a piecewise linear function on a 3D Delaunay triangulation. The multi-domain is used to ensure that the triangulation is dense inside the envelope where the input point set is located and sparser outside the envelope and inside the loose bounding sphere. The reach-aware property ensures that the piecewise linear function closely approximates the input point cloud. Meshing the entire bounding sphere increases the numerical stability of our solver. Fig.~\ref{domain-imf} depicts the implicit function defined on the reach-aware multi-domain.   

\subsection{LFS-aware Meshing}
\begin{figure}[t!]
\centering
\subfigure[Kitten]{
\begin{minipage}[t]{0.29\columnwidth}
   \includegraphics[width=\linewidth,trim={400 0 400 0},clip]{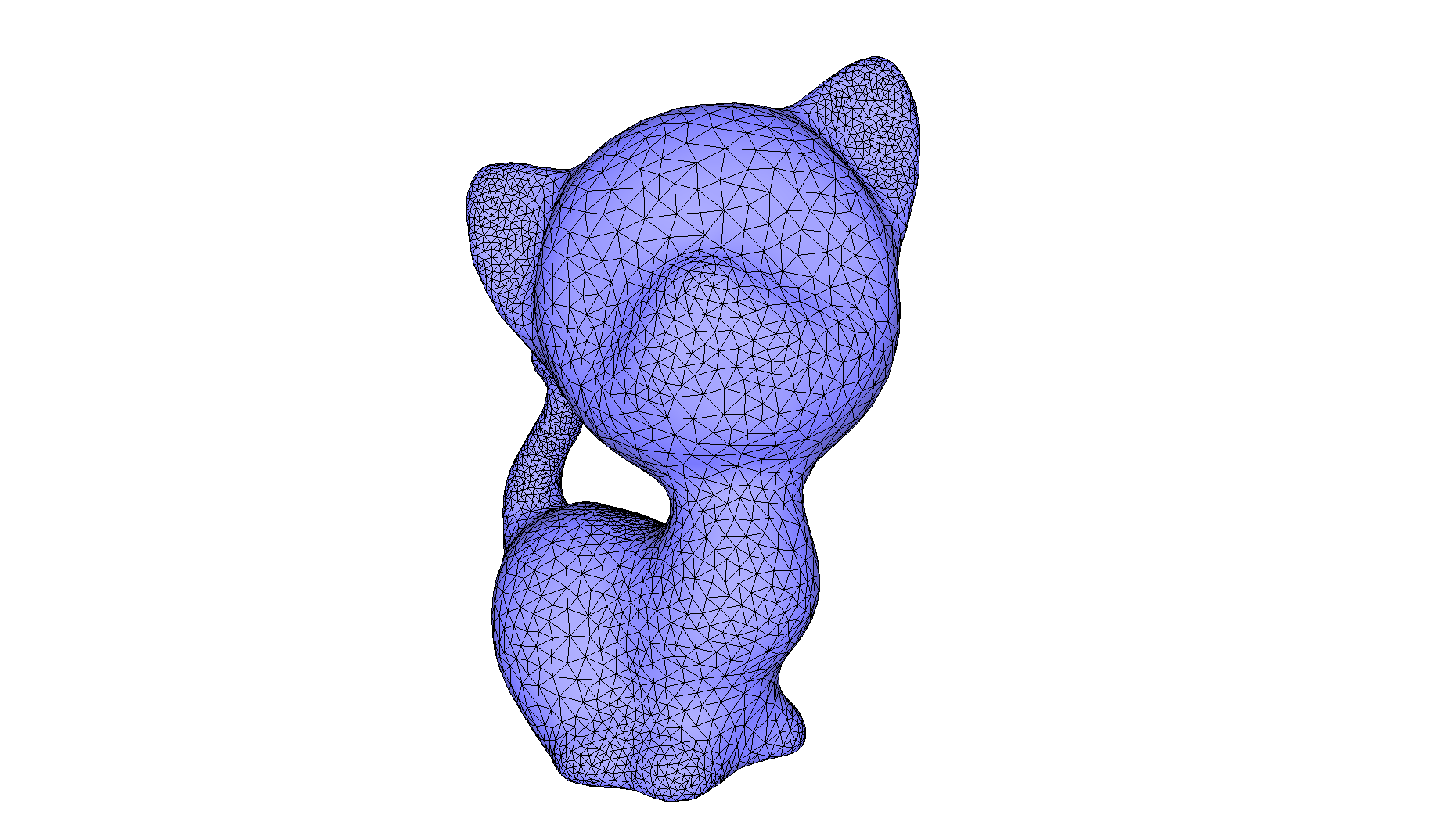}\vspace{0pt} 
   \includegraphics[width=\linewidth]{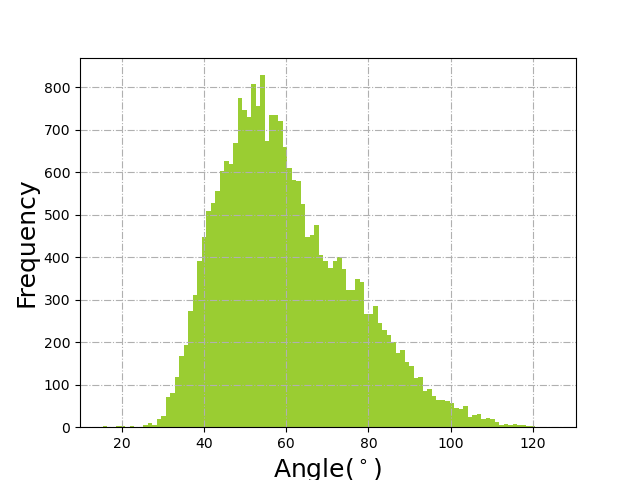}\vspace{5pt} 
\end{minipage}
}
\subfigure[Bust]{
\begin{minipage}[t]{0.29\columnwidth}
   \includegraphics[width=\linewidth,trim={400 0 400 0},clip]{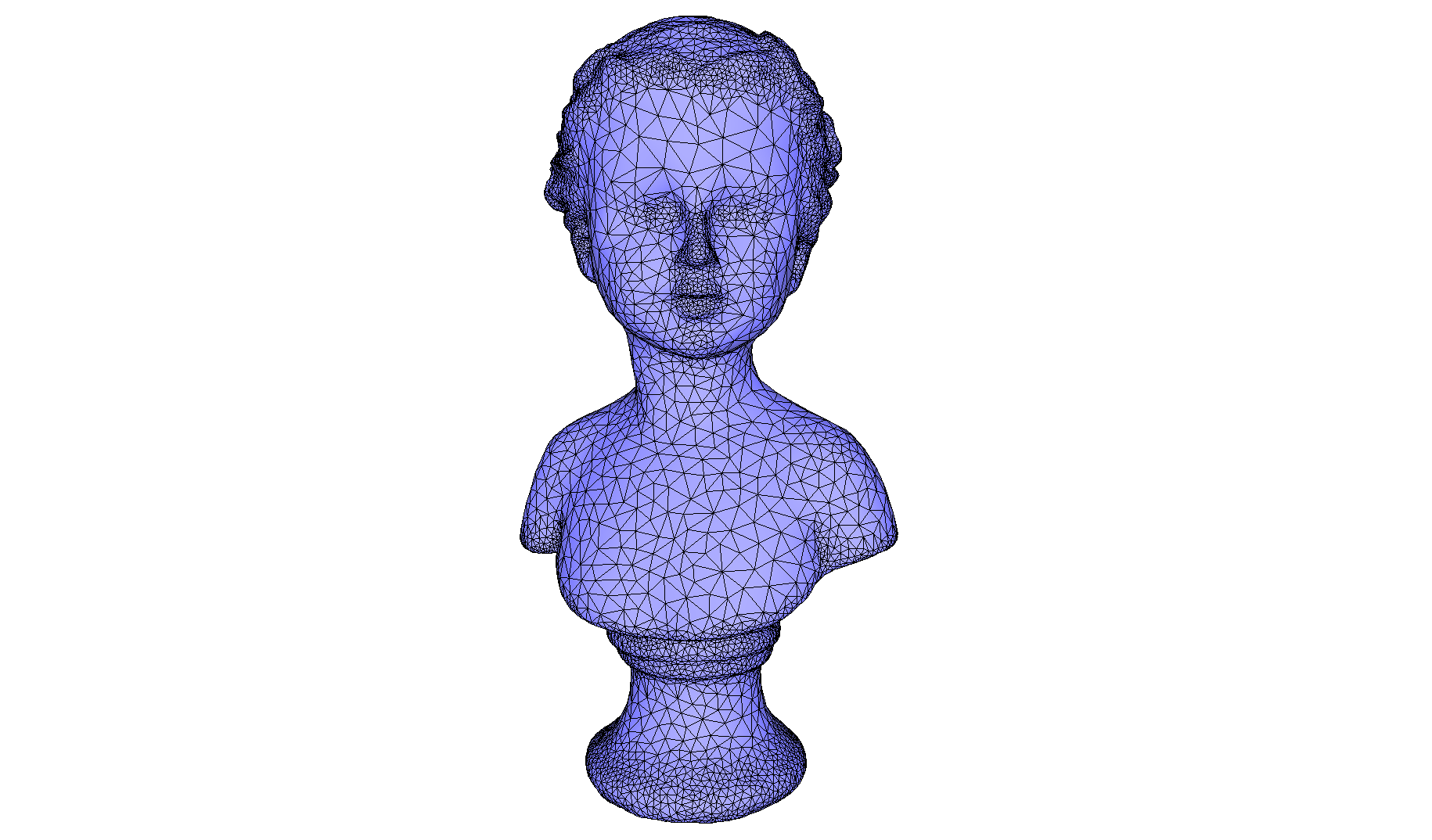}\vspace{0pt}
   \includegraphics[width=\linewidth]{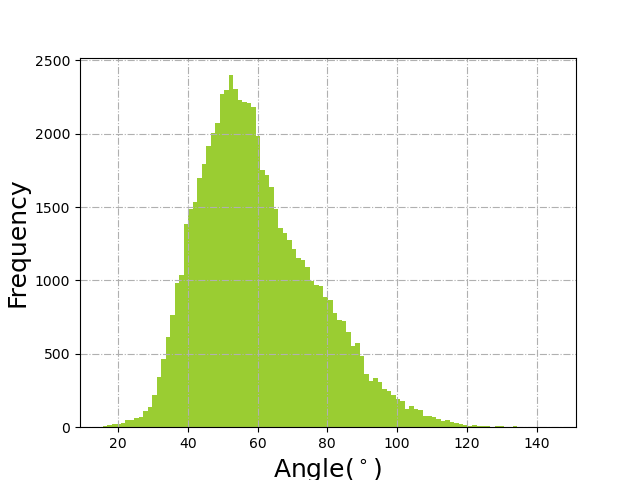}\vspace{5pt} 
\end{minipage}
}
\subfigure[Elephant]{
\begin{minipage}[t]{0.29\columnwidth}
   \includegraphics[width=\linewidth,trim={400 0 400 0},clip]{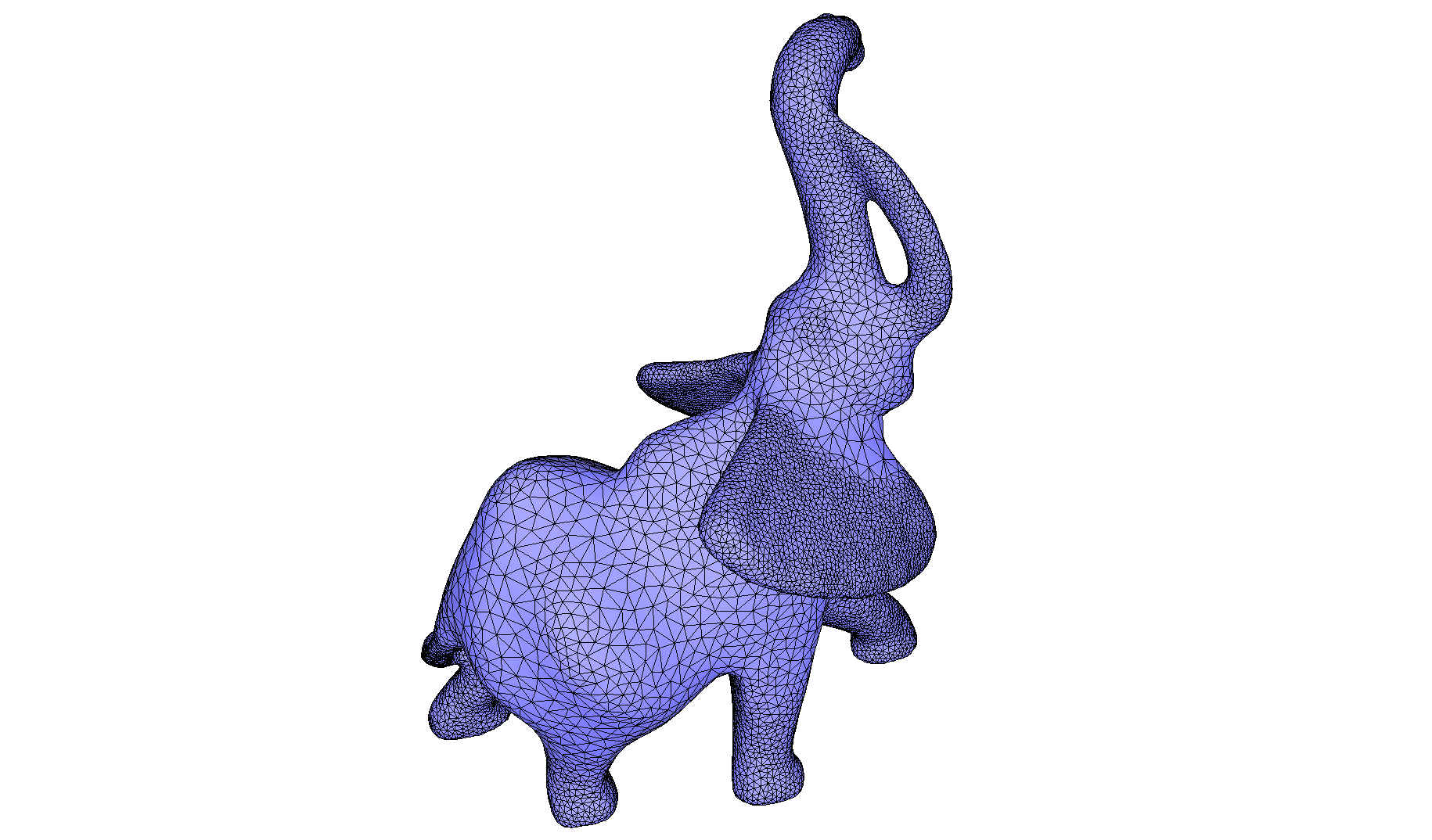}\vspace{0pt}
   \includegraphics[width=\linewidth]{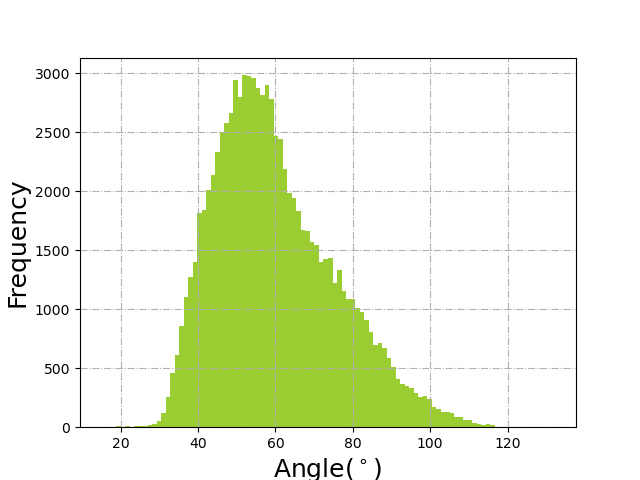}\vspace{5pt} 
\end{minipage}
}
\caption{Reconstructed LFS-aware meshes with distribution of facet angles. The facet sizing varies in accordance to the estimated LFS, and the output meshes are isotropic.}
\label{lfs-mesh-6}
\end{figure}

The final output of our algorithm is an isotropic surface triangle mesh with a sizing function that adapts to the estimated LFS. The output mesh is generated by Delaunay refinement~\cite{jamin2015cgalmesh} on the zero level set of the signed robust distance function. An LFS-aware facet sizing function is designed to ensure that facets are larger where the LFS is larger and vice versa. Fig.~\ref{lfs-mesh-6} illustrates output LFS-aware meshes, where the facet sizing varies significantly, while preserving the details of the input 3D point cloud. The distribution of facet angles is reported to validate the meshes' isotropy.

\begin{figure}[t!]
\centering
\includegraphics[width=\columnwidth,trim={0 0 0 0},clip]{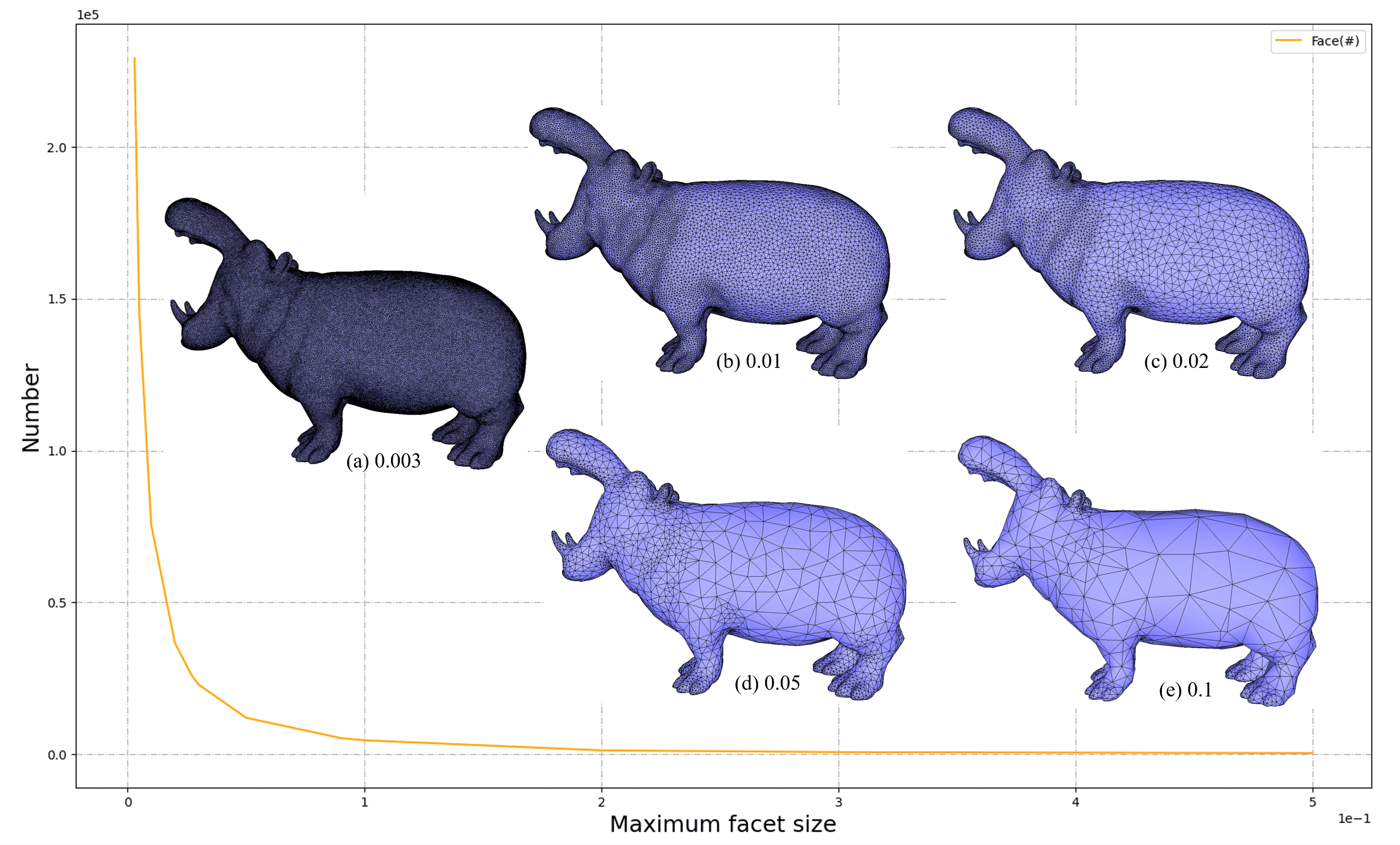}\vspace{0pt}
\caption{Output of reconstructed meshes with different $\textrm{size}_\textrm{max}$ parameters on the hippo 3D point cloud. The numbers under each hippo denote $\textrm{size}_\textrm{max}$, which determines the maximum sizing of the facets in the output mesh. 
The curve plots the number of output facets against $\textrm{size}_\textrm{max}$.} 
\label{adaptive-mesh-ablation}
\end{figure}

We also conduct experiments to verify the impact of the user-defined parameter $\textrm{size}_\textrm{max}$ on the output meshes.  $\textrm{size}_\textrm{max}$ controls the maximum sizing of the facets, while $\textrm{size}_\textrm{min}$ is a fixed ratio to the thickness of the envelope. If $\textrm{size}_\textrm{max}$ is set to equate to $\textrm{size}_\textrm{min}$, then the output mesh is uniform and isotropic. When increasing $\textrm{size}_\textrm{max}$, the variance in facet sizing increases. Fig.~\ref{adaptive-mesh-ablation} illustrates the impact of $\textrm{size}_\textrm{max}$ on the hippo 3D point cloud.

\subsection{Robustness}

\subsubsection{Noise, outliers, and missing data}  
\begin{figure}[htb!]
\centering
\subfigure[(0.5\%, 5)]{
    \begin{minipage}[h]{0.29\columnwidth} 
    \includegraphics[width=0.95\linewidth,trim={450 80 450 80},clip]{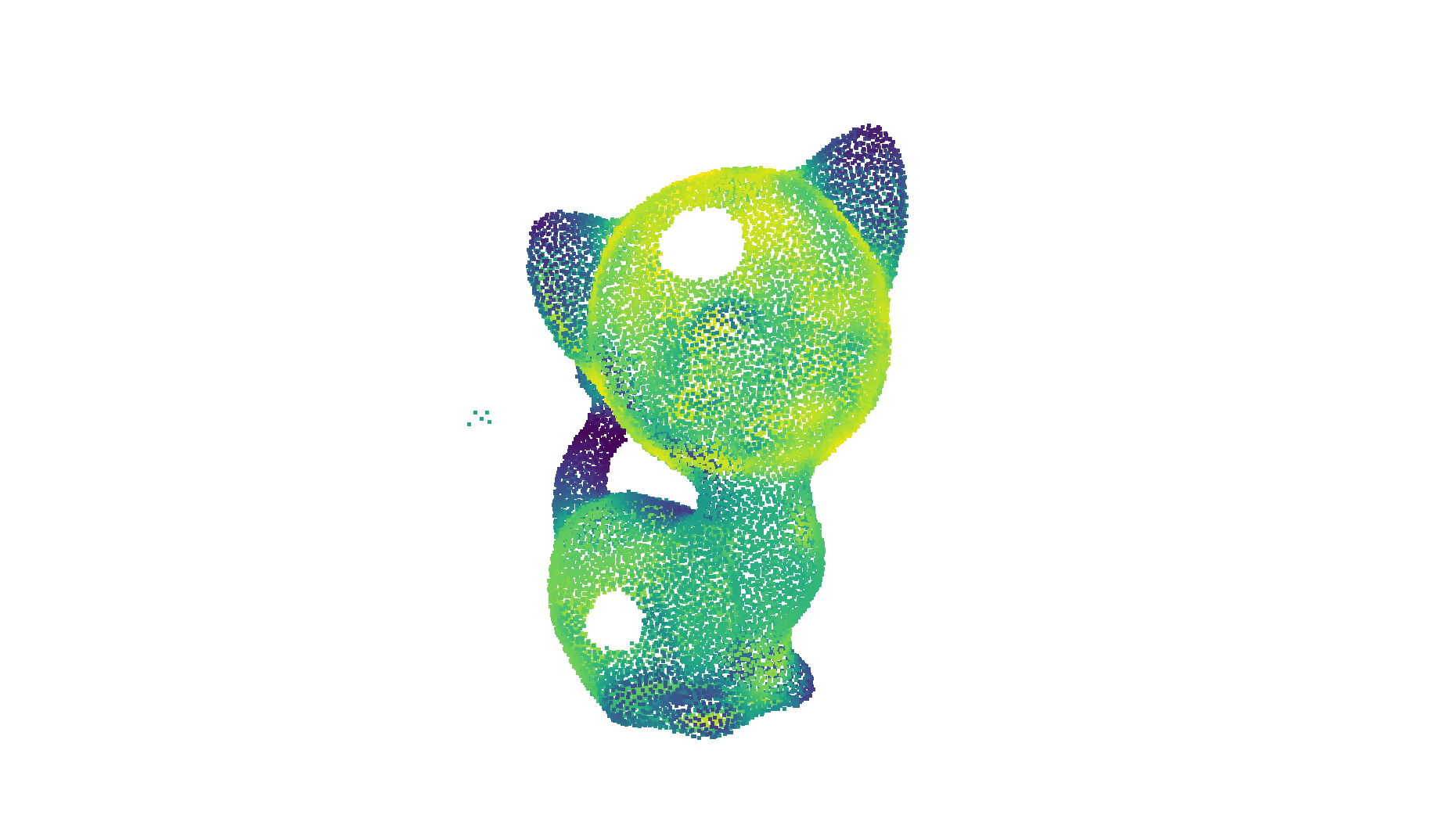}\vspace{0pt}
    \includegraphics[width=\linewidth,trim={340 0 340 0},clip]{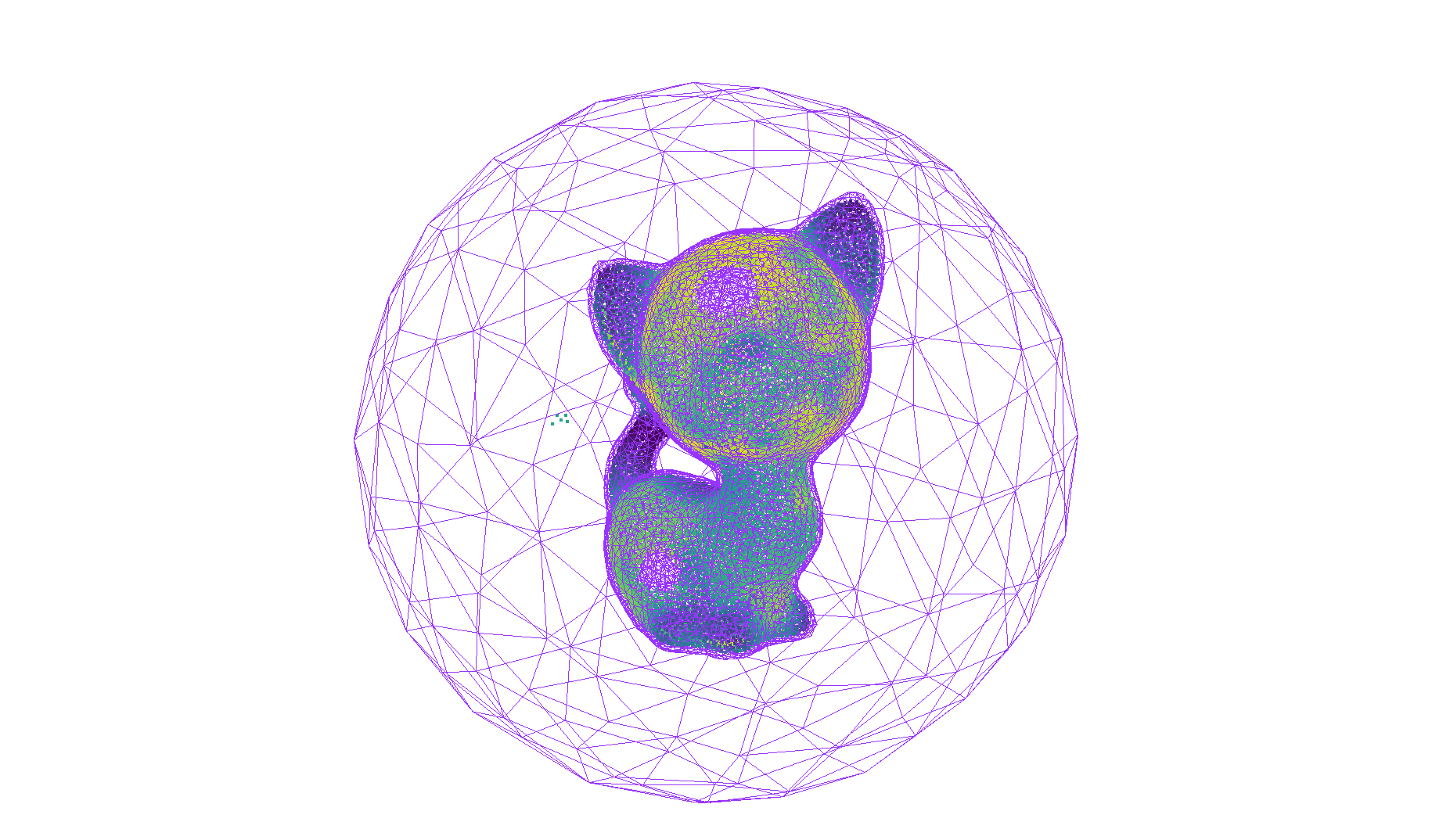}\vspace{0pt}
    \includegraphics[width=\linewidth,trim={100 50 300 20},clip]{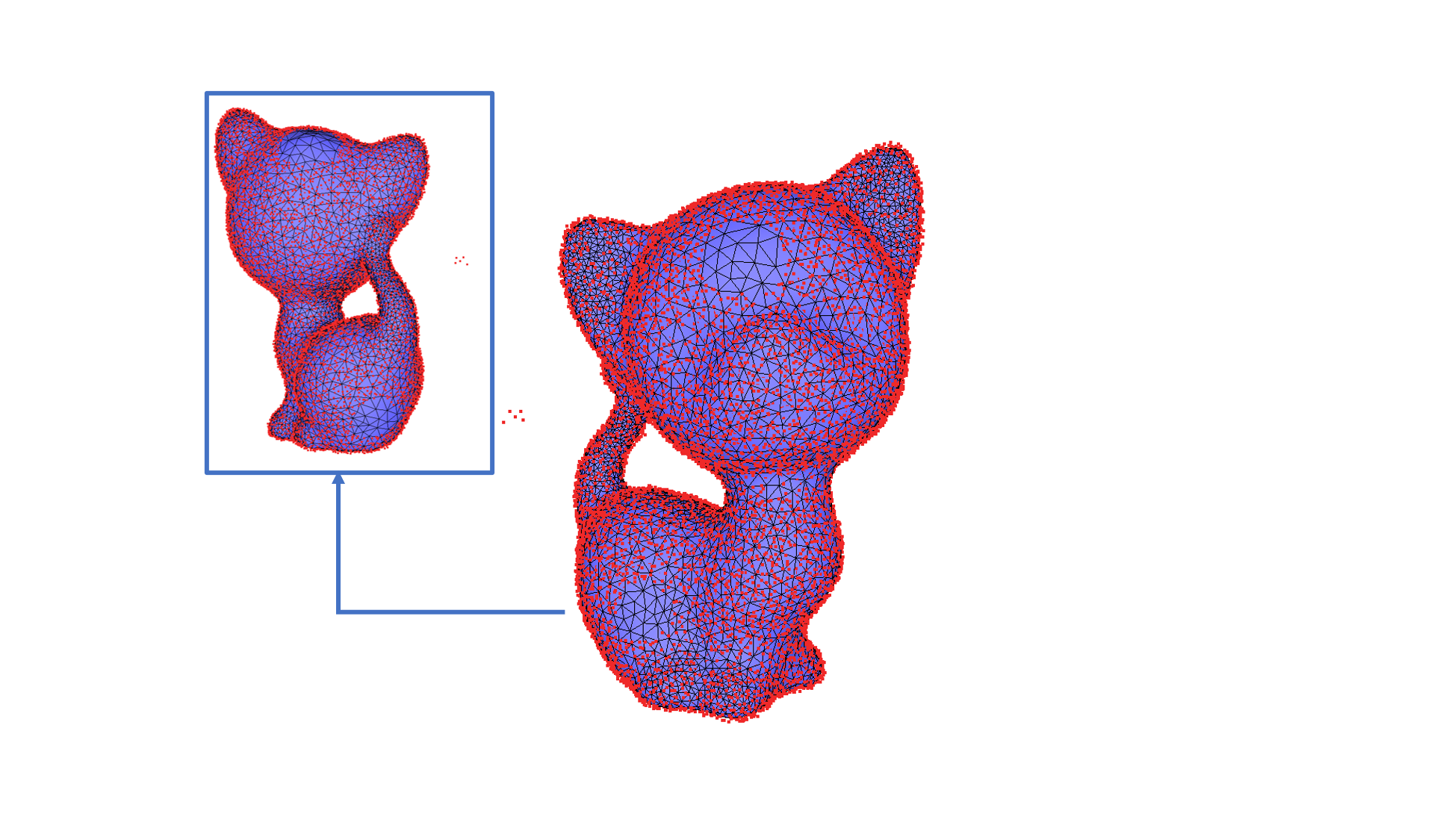}\vspace{0pt}
    \includegraphics[width=\linewidth,trim={450 80 450 100},clip]{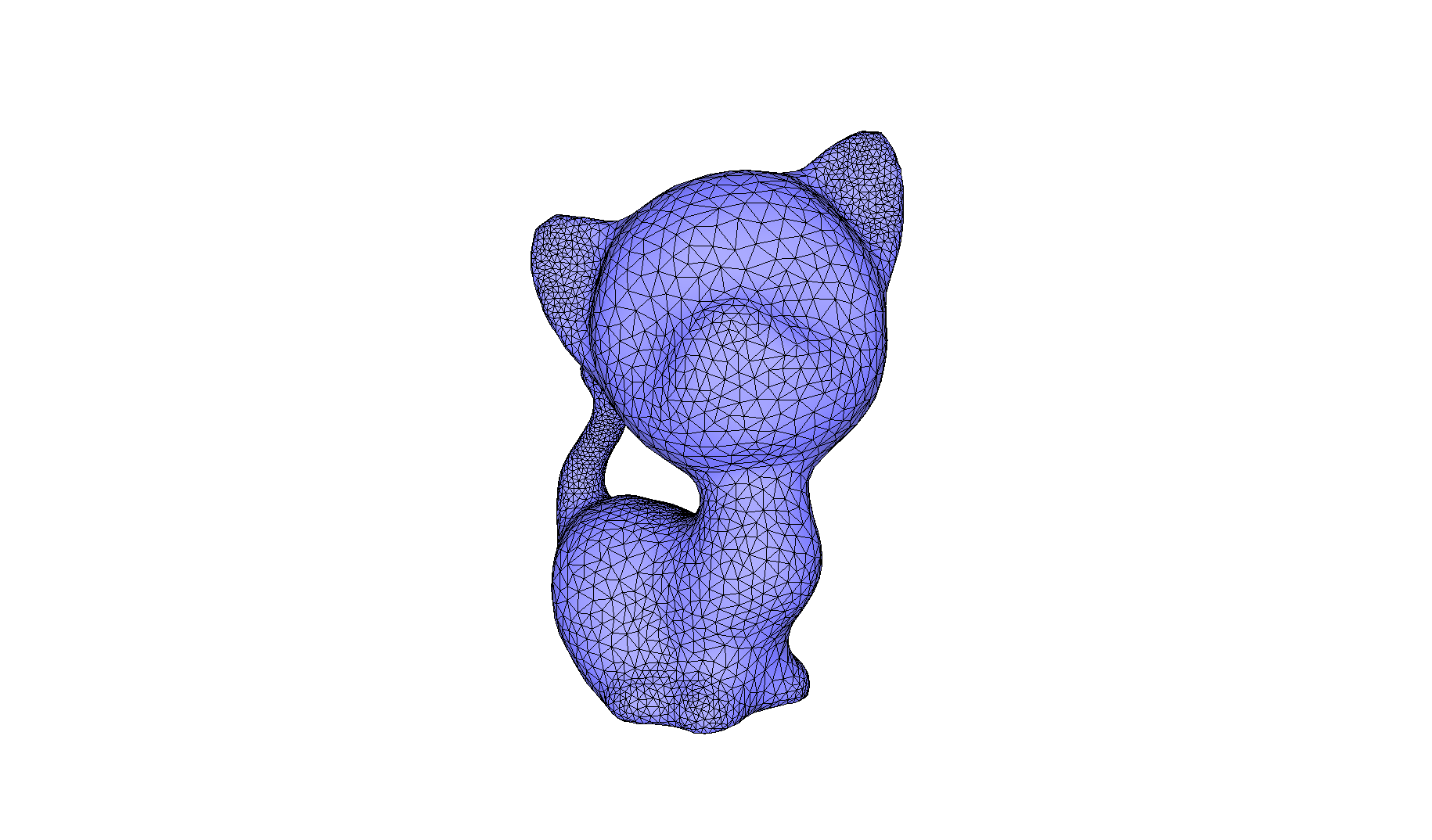}\vspace{3pt}
    \end{minipage}
}
\subfigure[(1.0\%, 10)]{
    \begin{minipage}[h]{0.29\columnwidth}
    \includegraphics[width=0.95\linewidth,trim={450 80 450 80},clip]{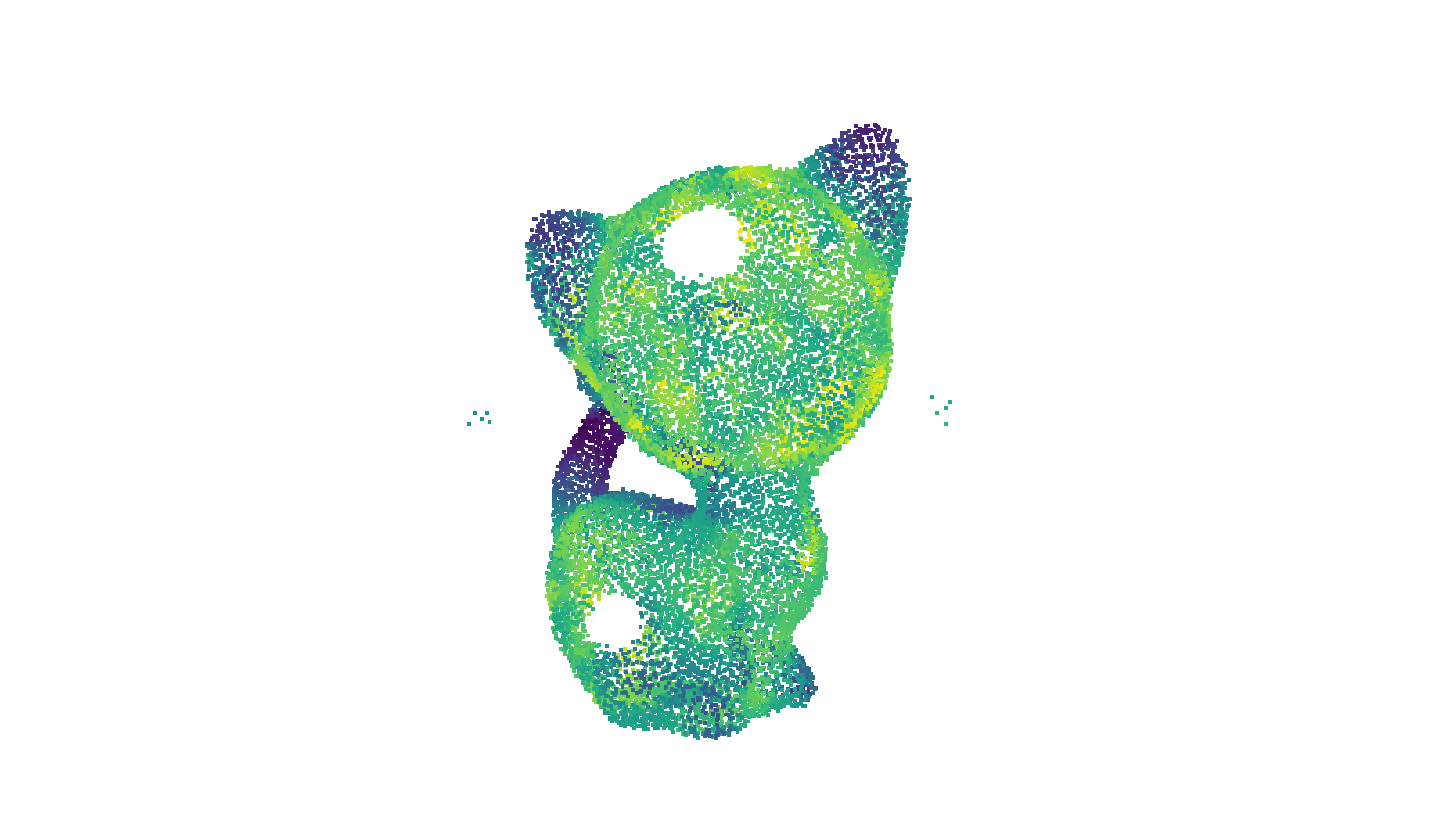}\vspace{0pt} 
    \includegraphics[width=\linewidth,trim={340 0 340 0},clip]{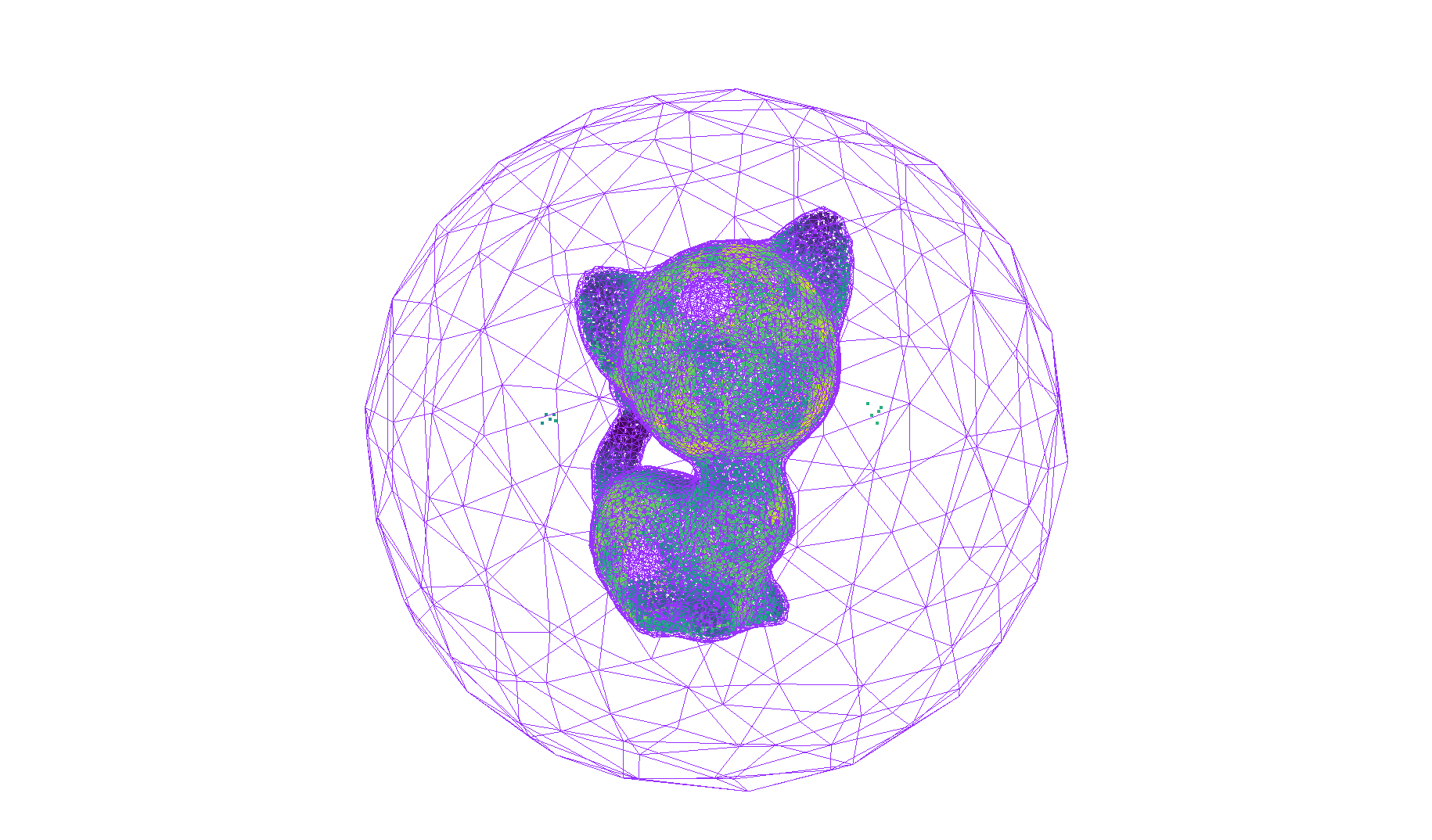}\vspace{0pt}
    \includegraphics[width=\linewidth,trim={100 50 300 20},clip]{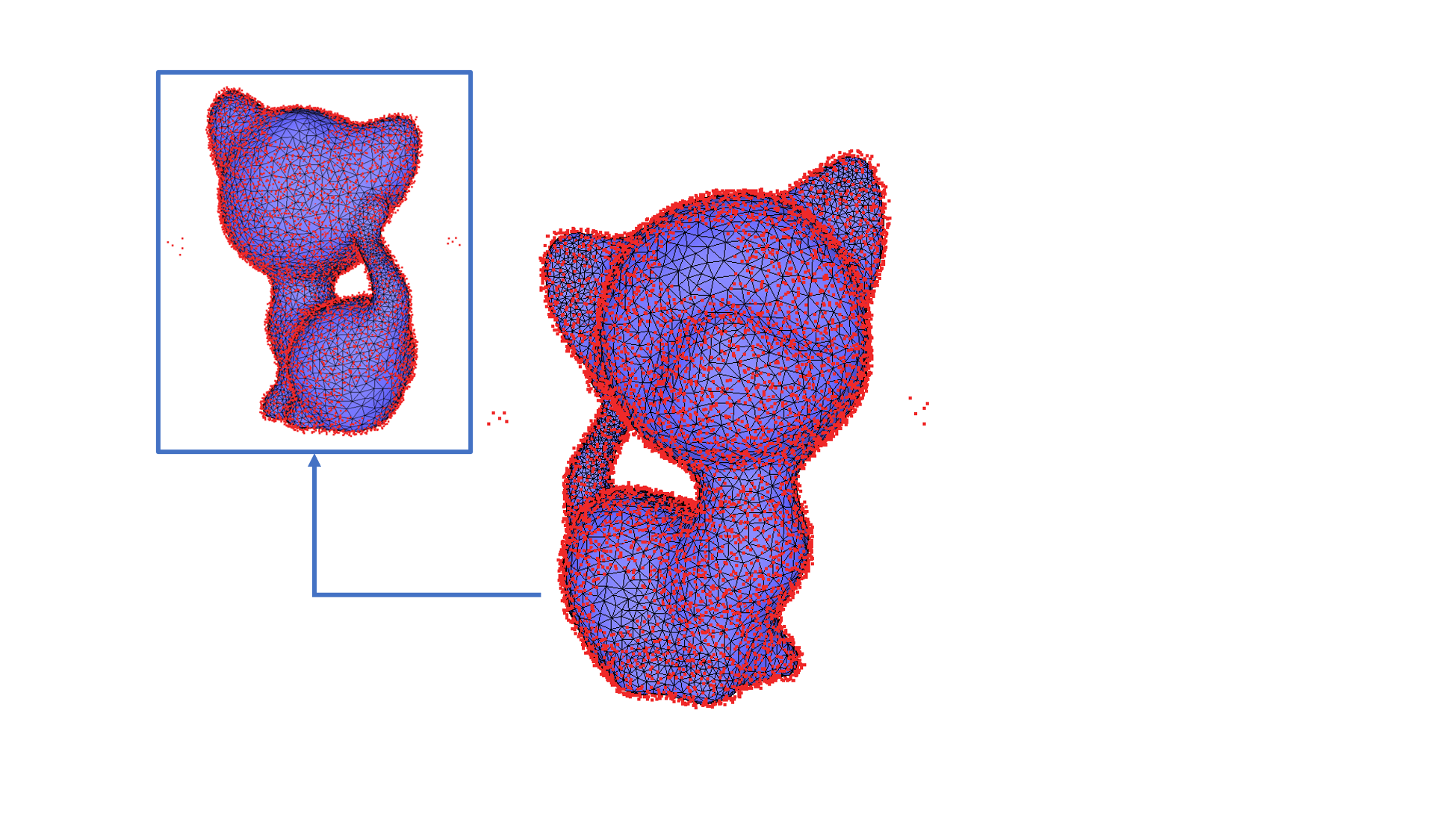}\vspace{0pt}
    \includegraphics[width=\linewidth,trim={450 80 450 100},clip]{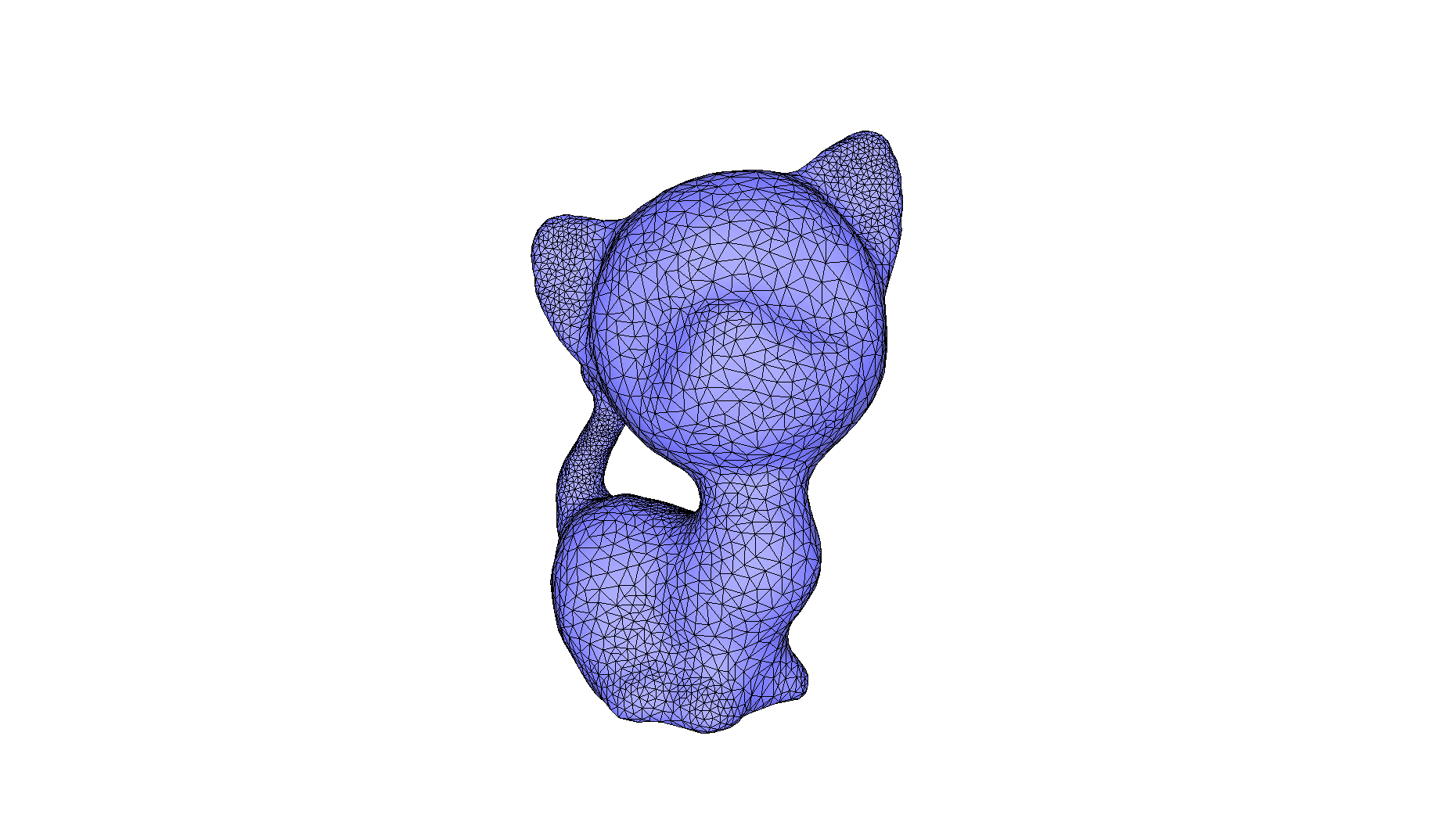}\vspace{3pt} 
    \end{minipage}
}
\subfigure[(1.5\%, 15)]{
    \begin{minipage}[h]{0.29\columnwidth}
    \includegraphics[width=0.95\linewidth,trim={450 80 450 80},clip]{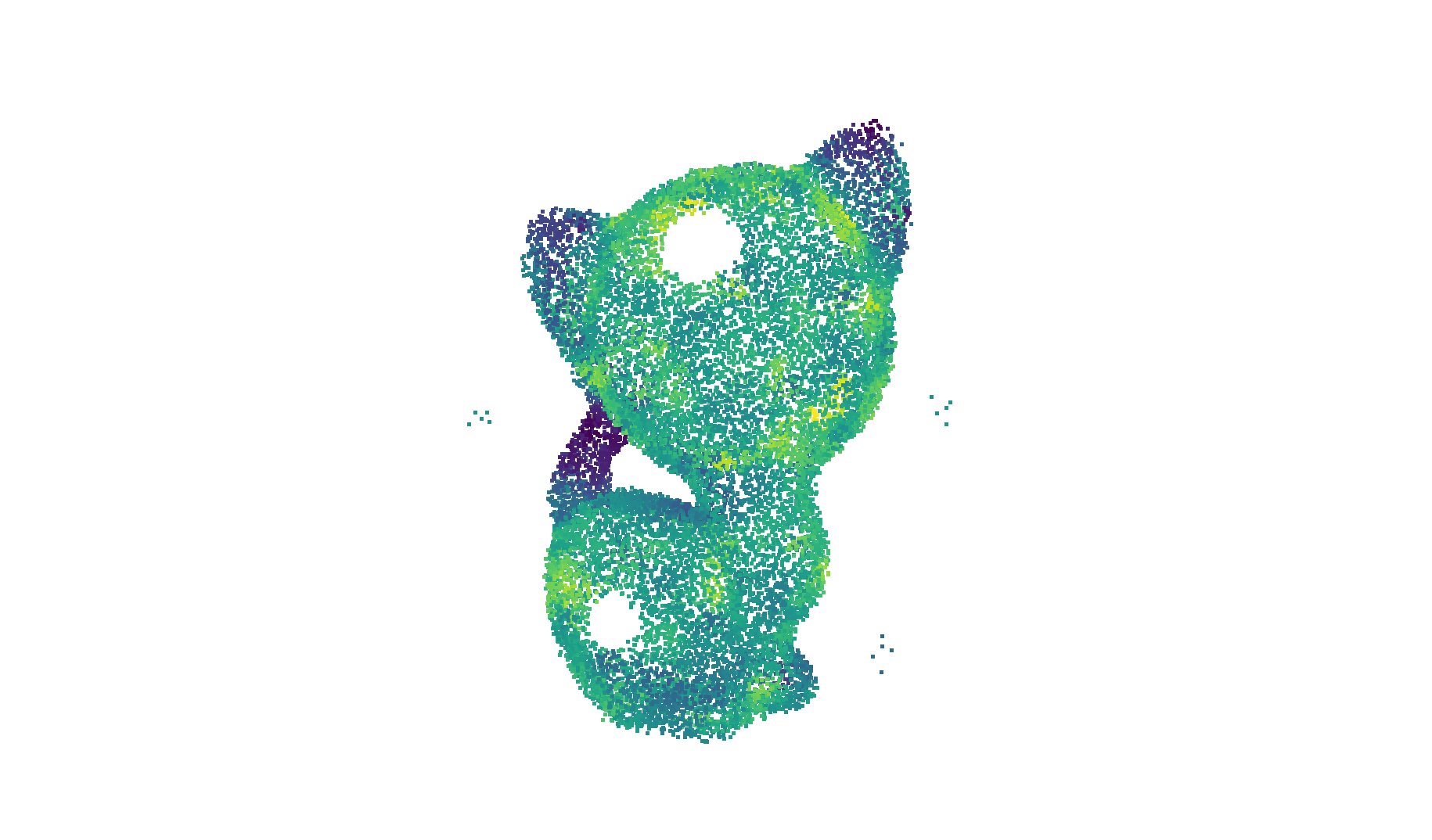}\vspace{0pt} 
    \includegraphics[width=\linewidth,trim={340 0 340 0},clip]{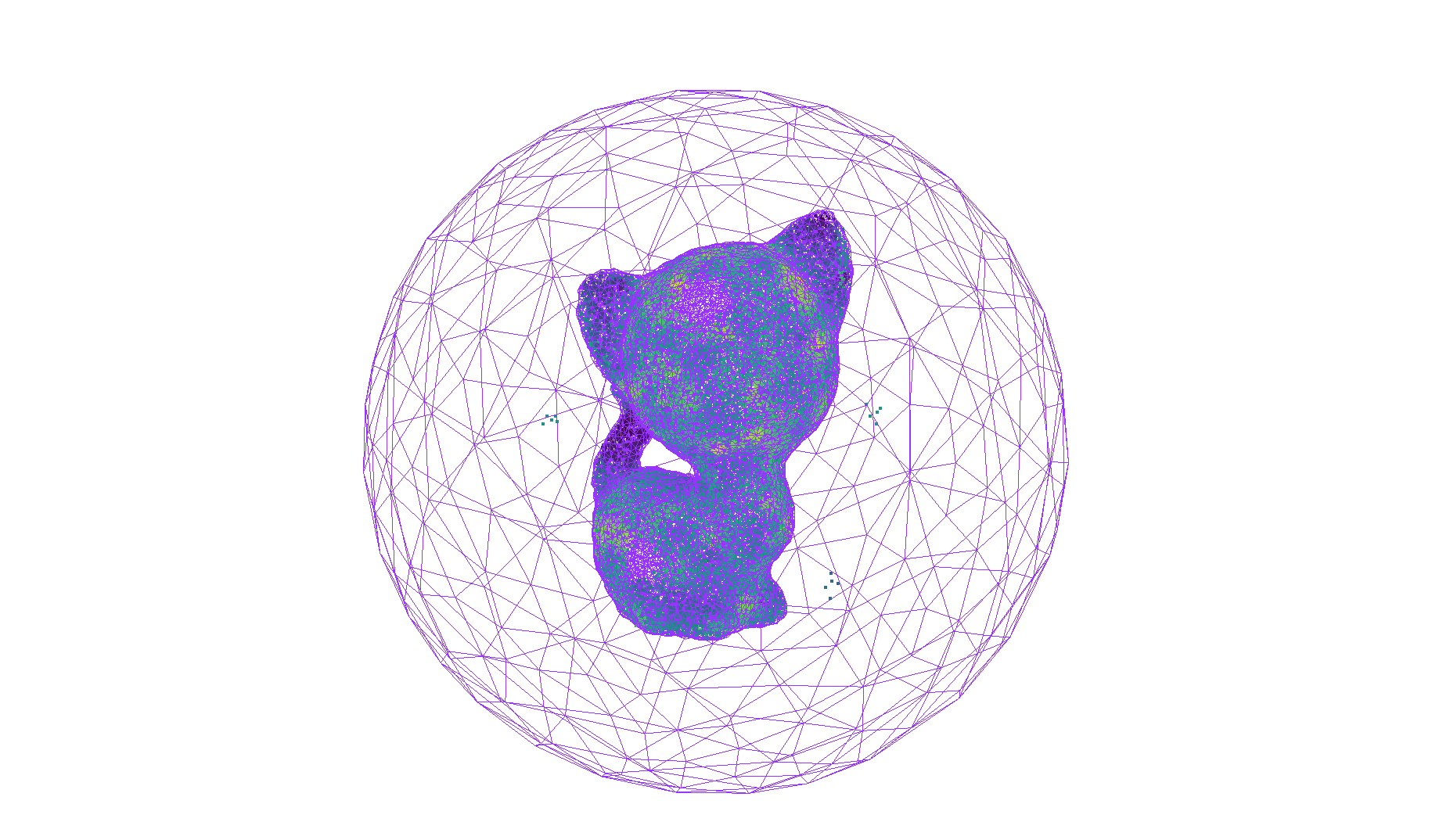}\vspace{0pt}
    \includegraphics[width=\linewidth,trim={100 50 300 20},clip]{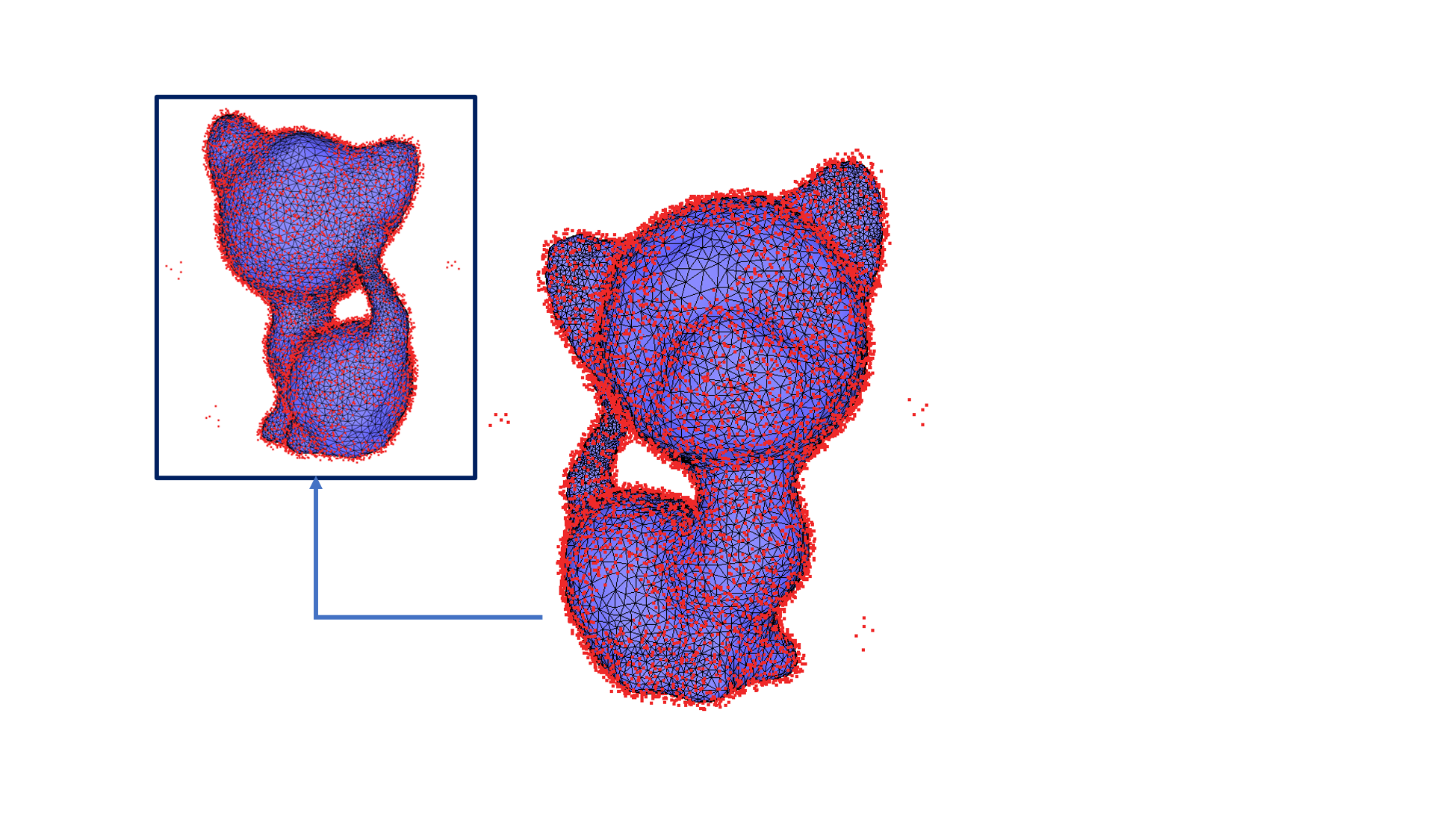}\vspace{0pt}
    \includegraphics[width=\linewidth,trim={450 80 450 100},clip]{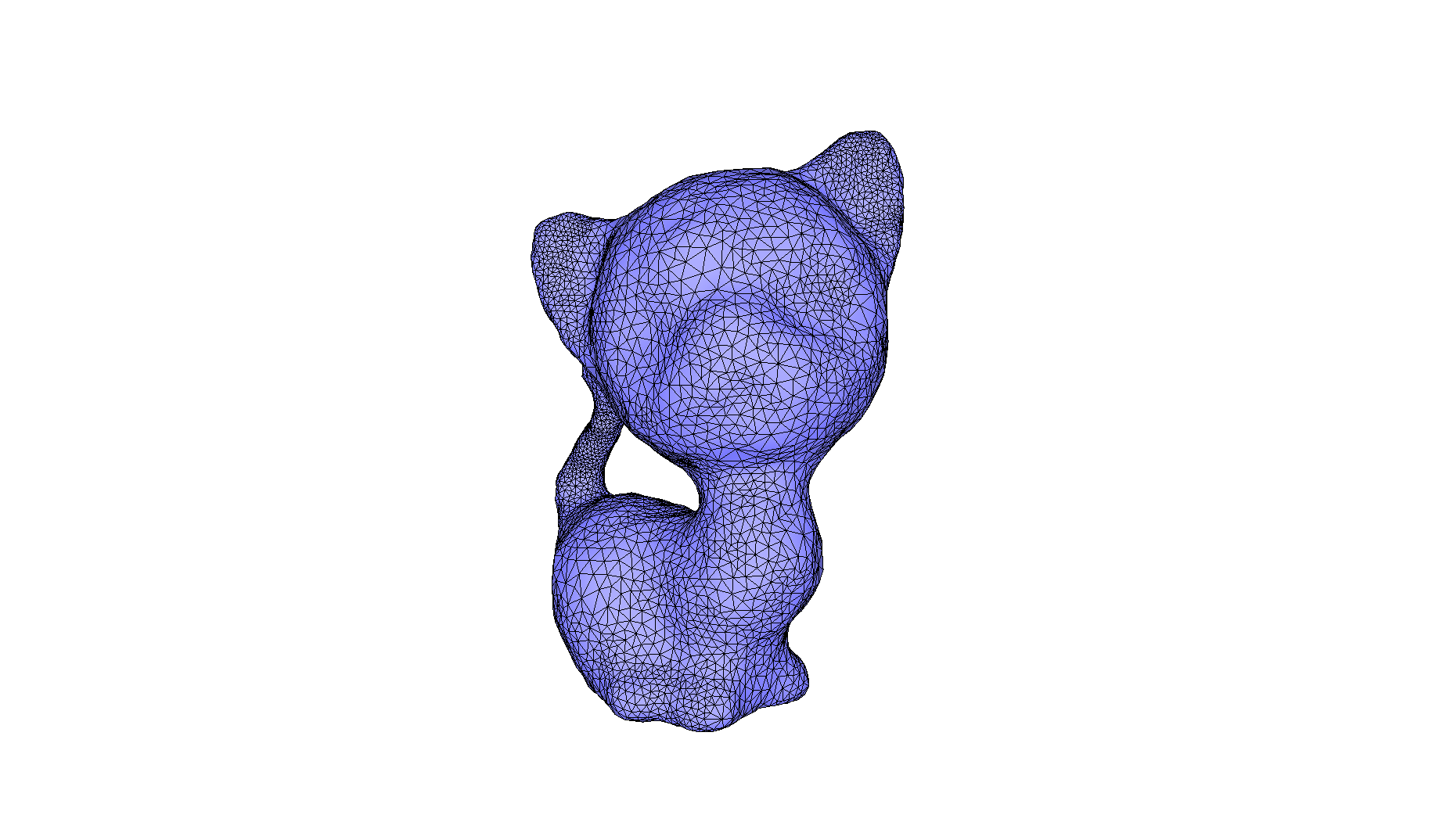}\vspace{3pt} 
    \end{minipage}
}
\caption{Evaluation of robustness on the kitten point cloud (5k points) with increasing levels of noise and number of outliers, as well as missing data. The noise level increases from 0.5\% to 1.5\%, and the numbers of outliers increases from 5 to 15. First row: estimated LFS.
Second row: the boundary of the multi-domain ignores the outliers. 
Third and fourth rows: output LFS-aware meshes.}
\label{kitten-robust-tests}
\end{figure}

We now evaluate the robustness of our surface reconstruction approach to noise, outliers, and missing data. We add Gaussian white noise at varying levels, ranging from 0.5\% to 1.5\% of the maximum edge length of the bounding box of the point cloud. 
We also introduce a moderate amount of random structured outliers at each noise level by generating one to  three small clusters in the loose bounding sphere of the input, each containing five random points. Finally, we simulate missing data by creating two holes in the head and two holes in the body of the kitten point cloud. 
Our approach is robust to noise thanks to the property of the robust distance function. 
Our approach is also resilient to moderate amounts of random structured outliers thanks to the Delaunay refinement process~\cite{jamin2015cgalmesh}. The Delaunay refinement can eliminate the outliers outside the envelope. Thus, there is no impact on the implicit function solver, as all edges outside the envelope are considered to connect two vertices with similar signs. 
Fig.~\ref{kitten-robust-tests} illustrates the robustness of our approach on the kitten point cloud.

\begin{figure}[t!]
\centering
\subfigure[Multi-domain]{
\includegraphics[height=0.5\columnwidth,trim={3 3 3 3},clip]{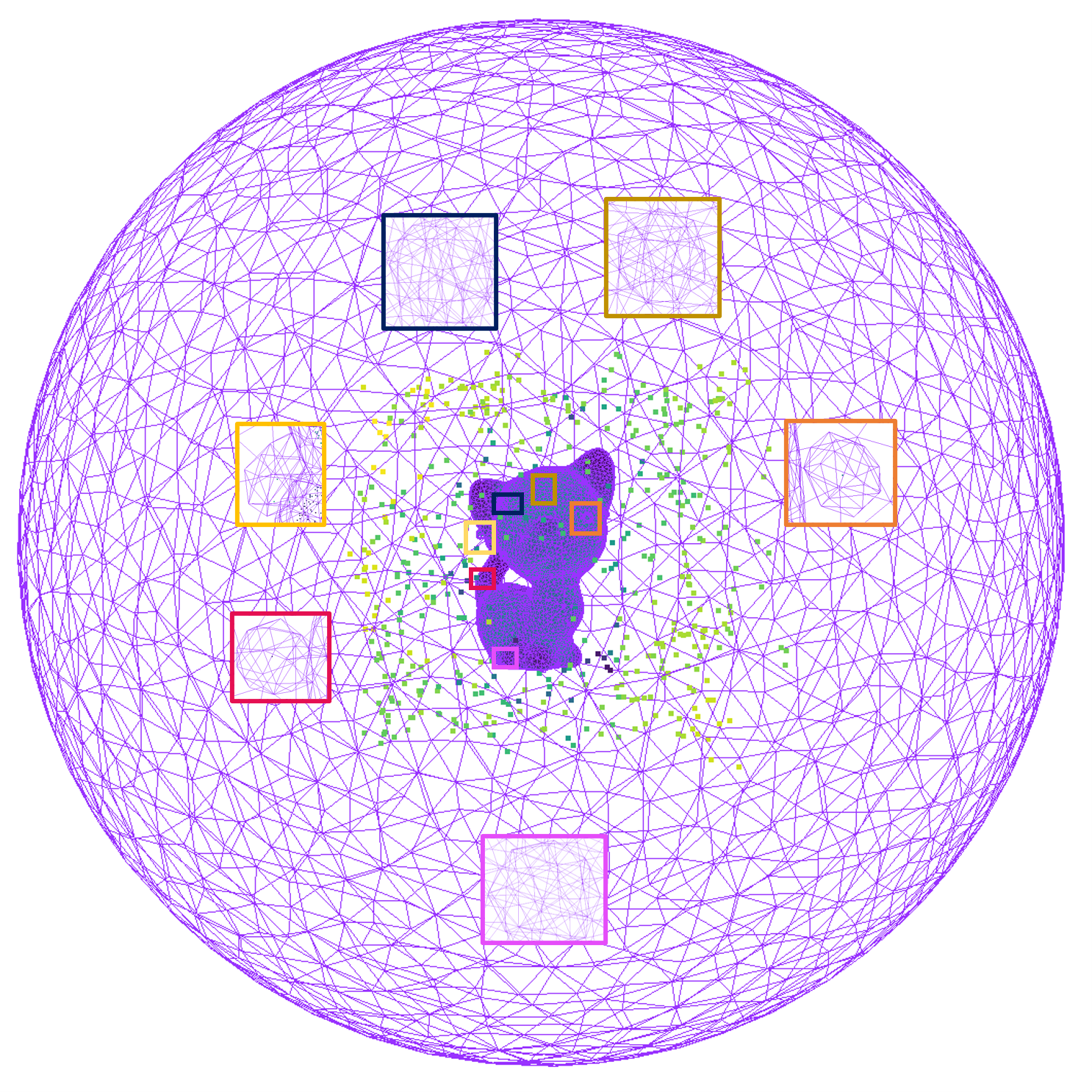}
}
\subfigure[Mesh]{
\includegraphics[height=0.4\columnwidth,trim={0 0 0 0},clip]{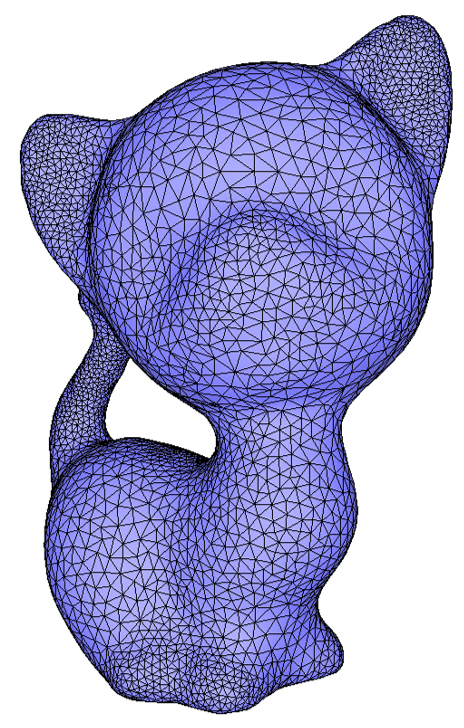}
}
\caption{Resilience to outliers evaluated on the kitten point cloud (5k points) with 500 random background outliers. A handful of bubbles are generated around some outliers inside the multi-domain discretization. A close-up of selected bubbles is delineated with a colored rectangle. Note that some bubbles are inside the point cloud, thus blocked by the front views. The bubbles do not affect the final result because the implicit function solver erases them.}
\label{kitten-robust-tests-backgroud-outlers}
\end{figure}

We now evaluate the resilience to random dense outliers of our implicit function solver by uniformly adding dense background outliers to the kitten point cloud. We generate 500 outliers from a random uniform distribution, bounded by a loose bounding box containing the kitten point cloud. The dense outliers affect the Delaunay refinement, resulting in a handful of bubbles in the multi-domain discretization. Nevertheless, thanks to the stability of our implicit function solver, the bubbles do not affect the signing with the data fitting step. The final mesh has no artifacts, as shown in Fig.~\ref{kitten-robust-tests-backgroud-outlers}.

\subsubsection{Large hole filling on real-scan data} 
\begin{figure}[t!]
\centering
\subfigure[\scriptsize{Perfume (\#56,411)}]{
    \begin{minipage}[h]{0.29\columnwidth}
    \includegraphics[width=\linewidth,trim={450 100 450 100},clip]{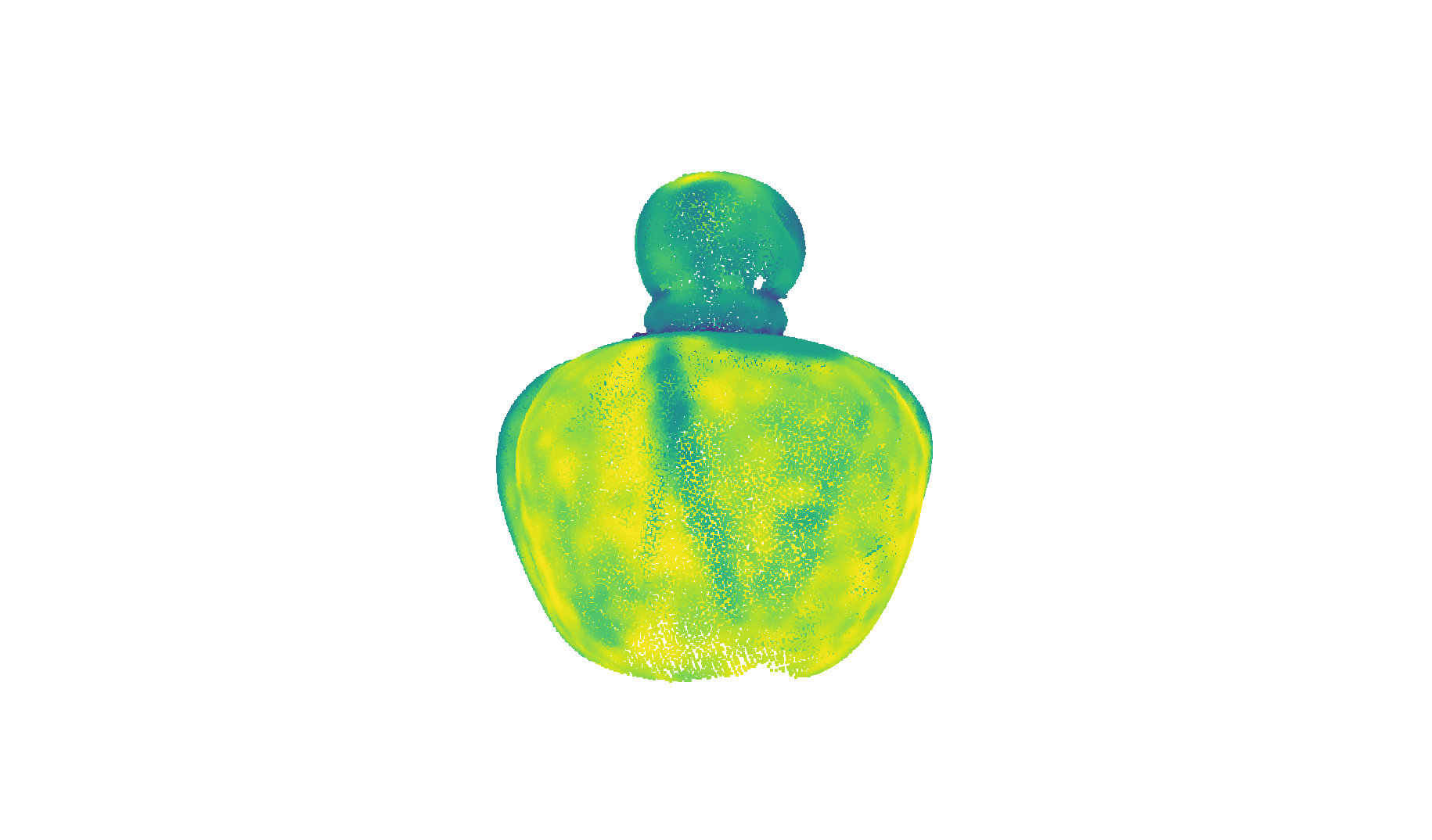}\vspace{0pt} 
    \includegraphics[width=\linewidth,trim={450 100 450 100},clip]{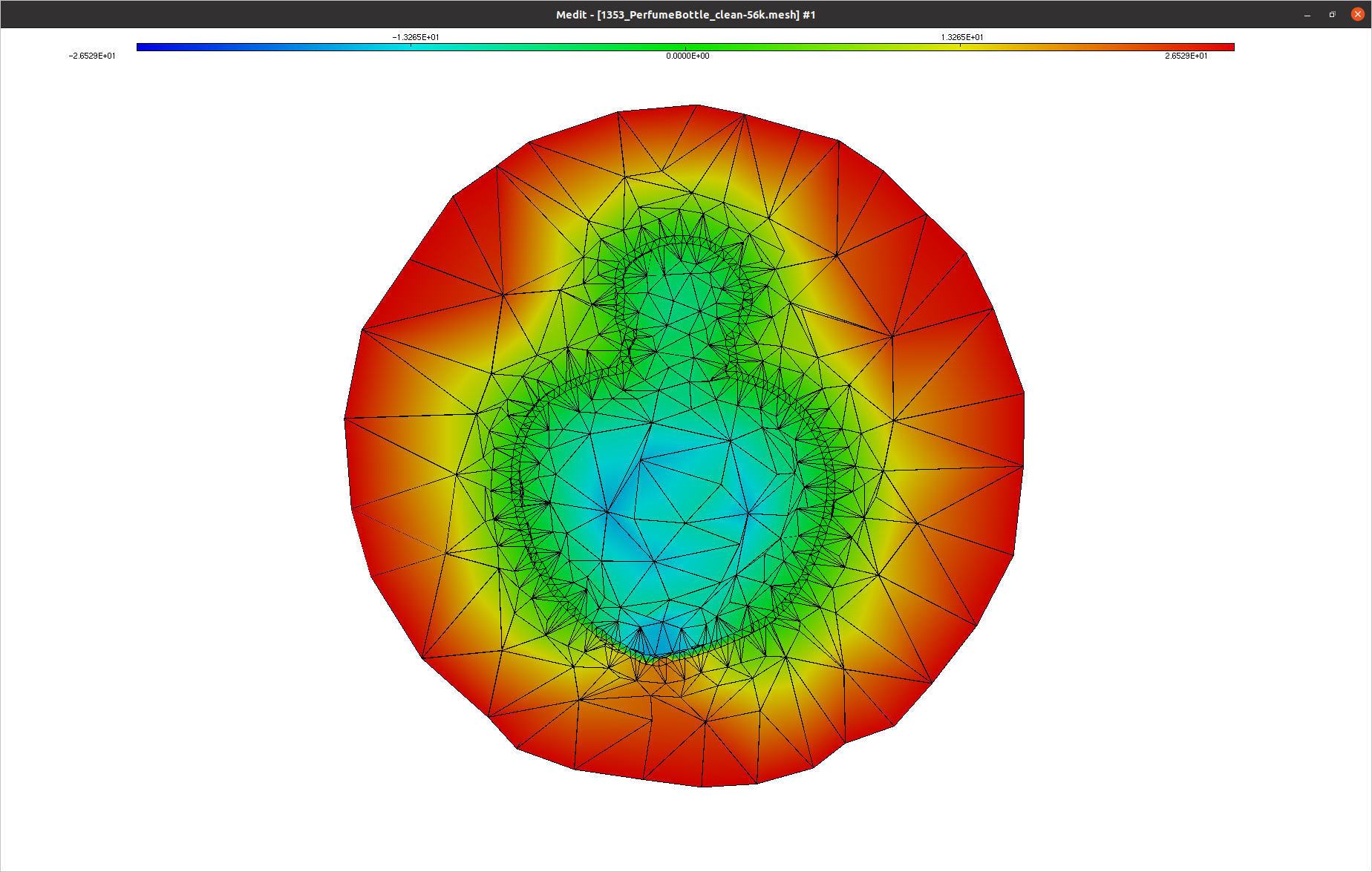}\vspace{0pt} 
    \includegraphics[width=1.2\linewidth,trim={100 0 80 0},clip]{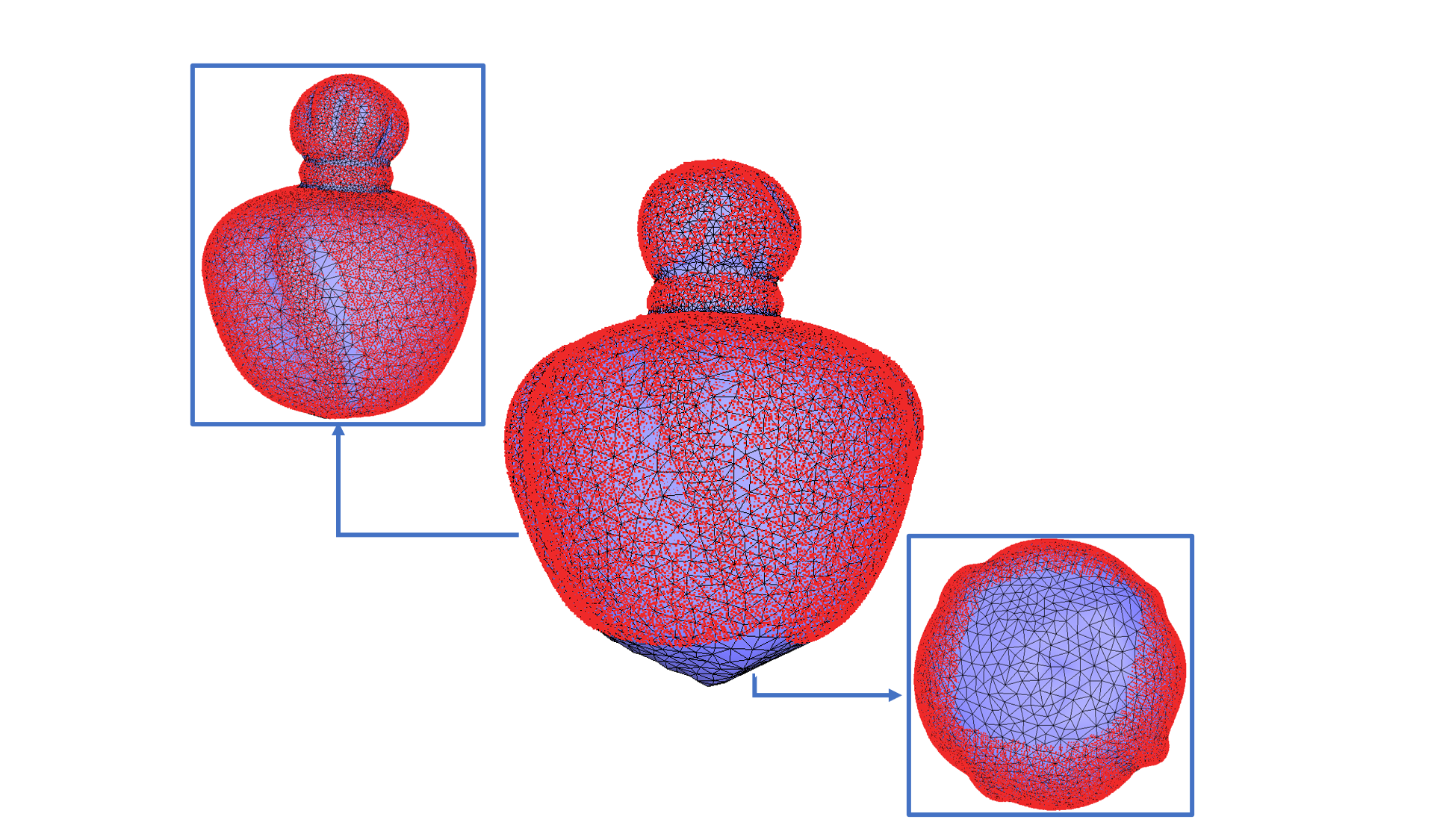}\vspace{0pt} 
    \includegraphics[width=\linewidth,trim={450 80 450 100},clip]{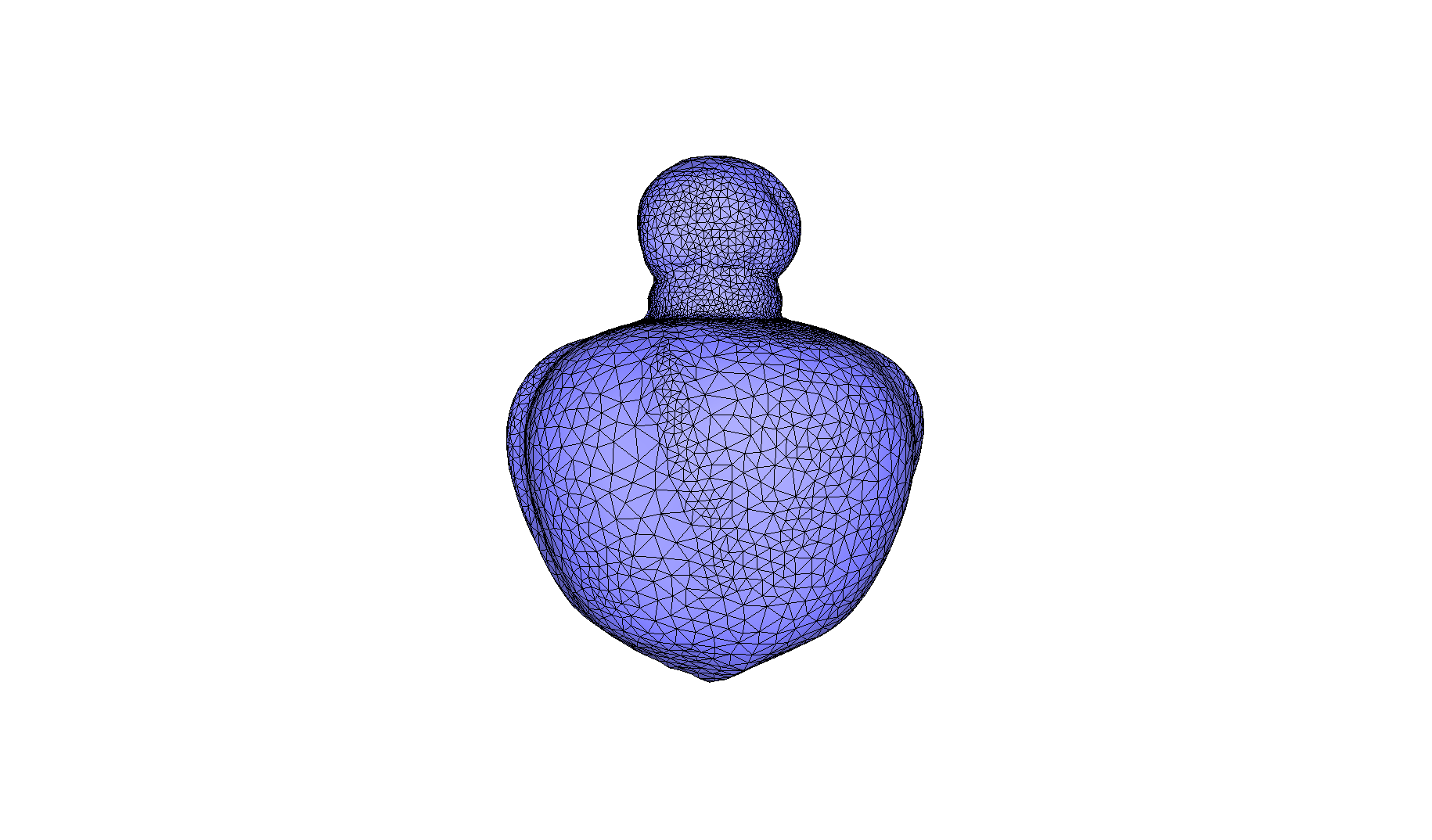}\vspace{0pt}    
    \end{minipage}
}
\subfigure[\scriptsize{Hand kreon (\#206,621)}]{  
    \begin{minipage}[h]{0.29\columnwidth}
    \includegraphics[width=\linewidth,trim={450 100 450 100},clip]{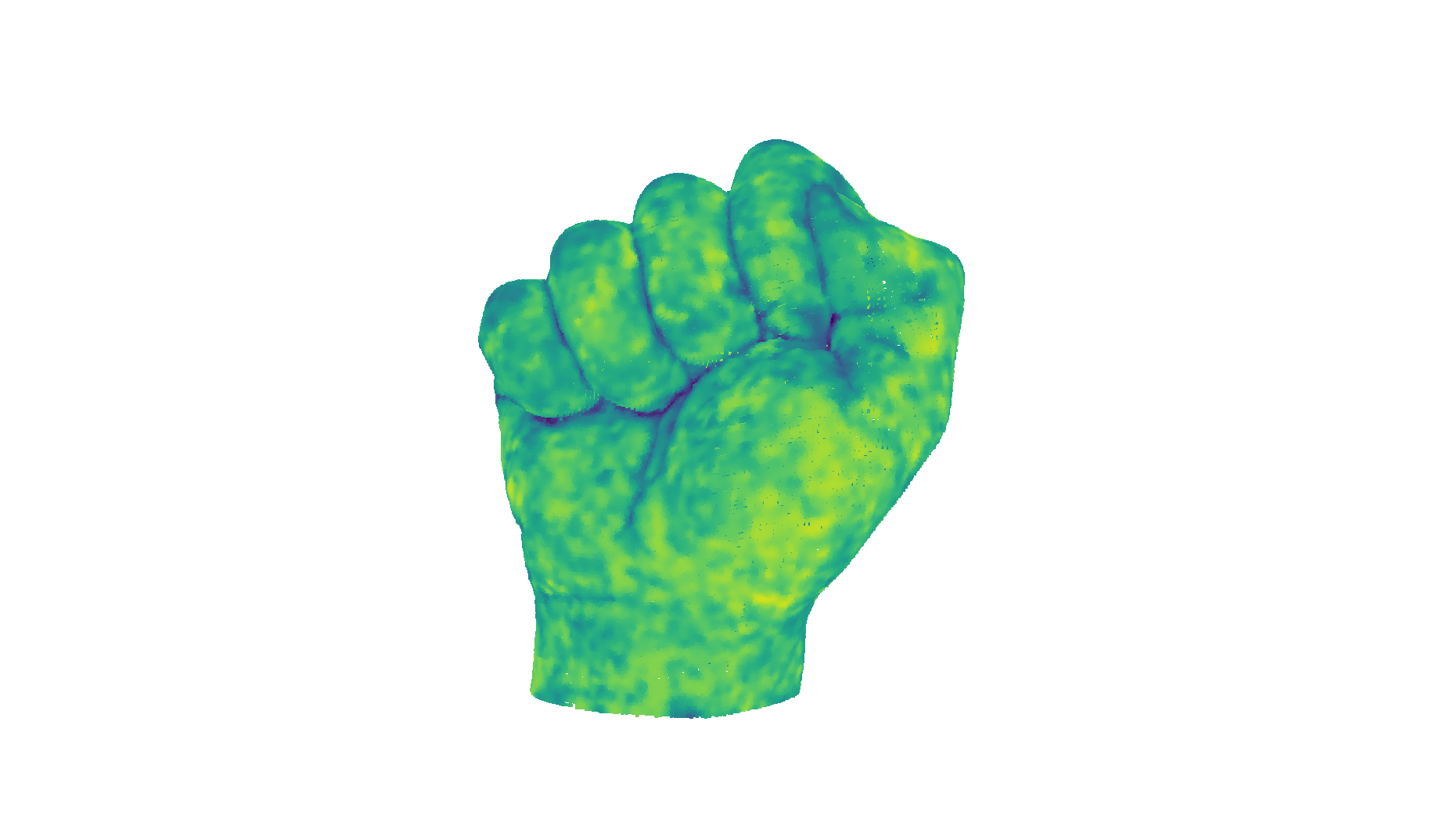}\vspace{0pt}
    \includegraphics[width=\linewidth,trim={450 100 450 100},clip]{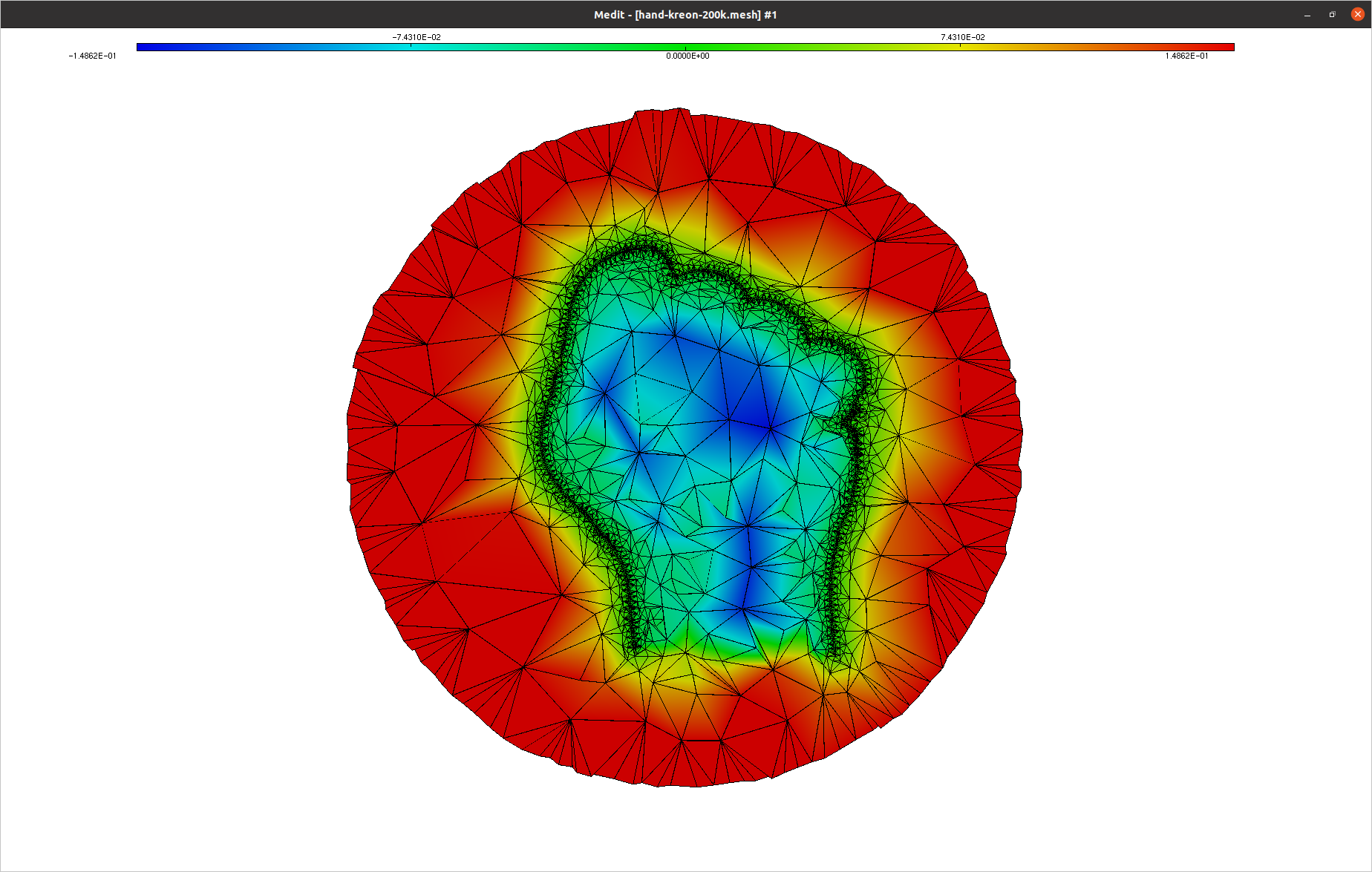}\vspace{0pt}
    \includegraphics[width=1.2\linewidth,trim={100 0 80 0},clip]{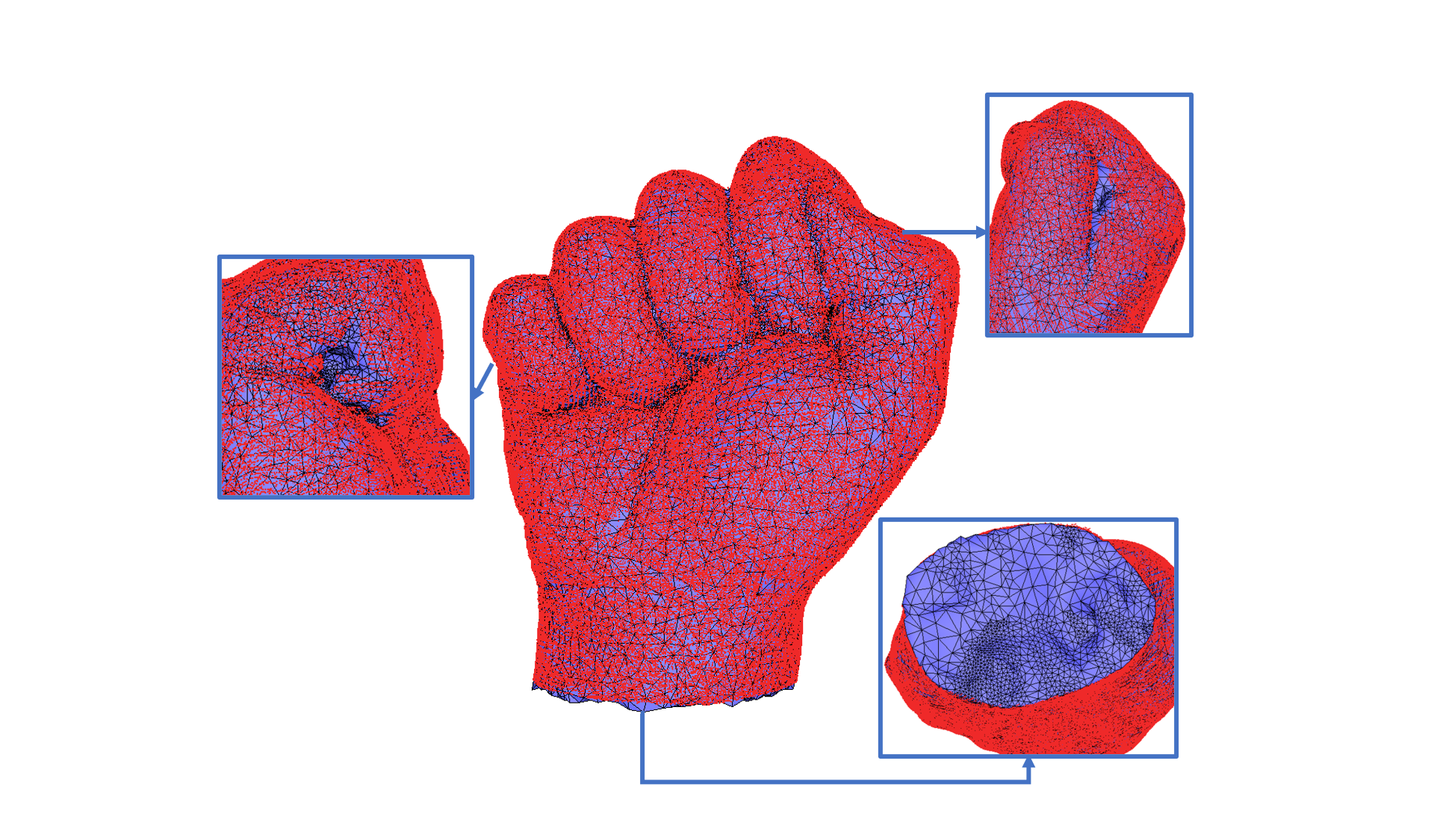}\vspace{0pt} 
    \includegraphics[width=\linewidth,trim={450 80 450 100},clip]{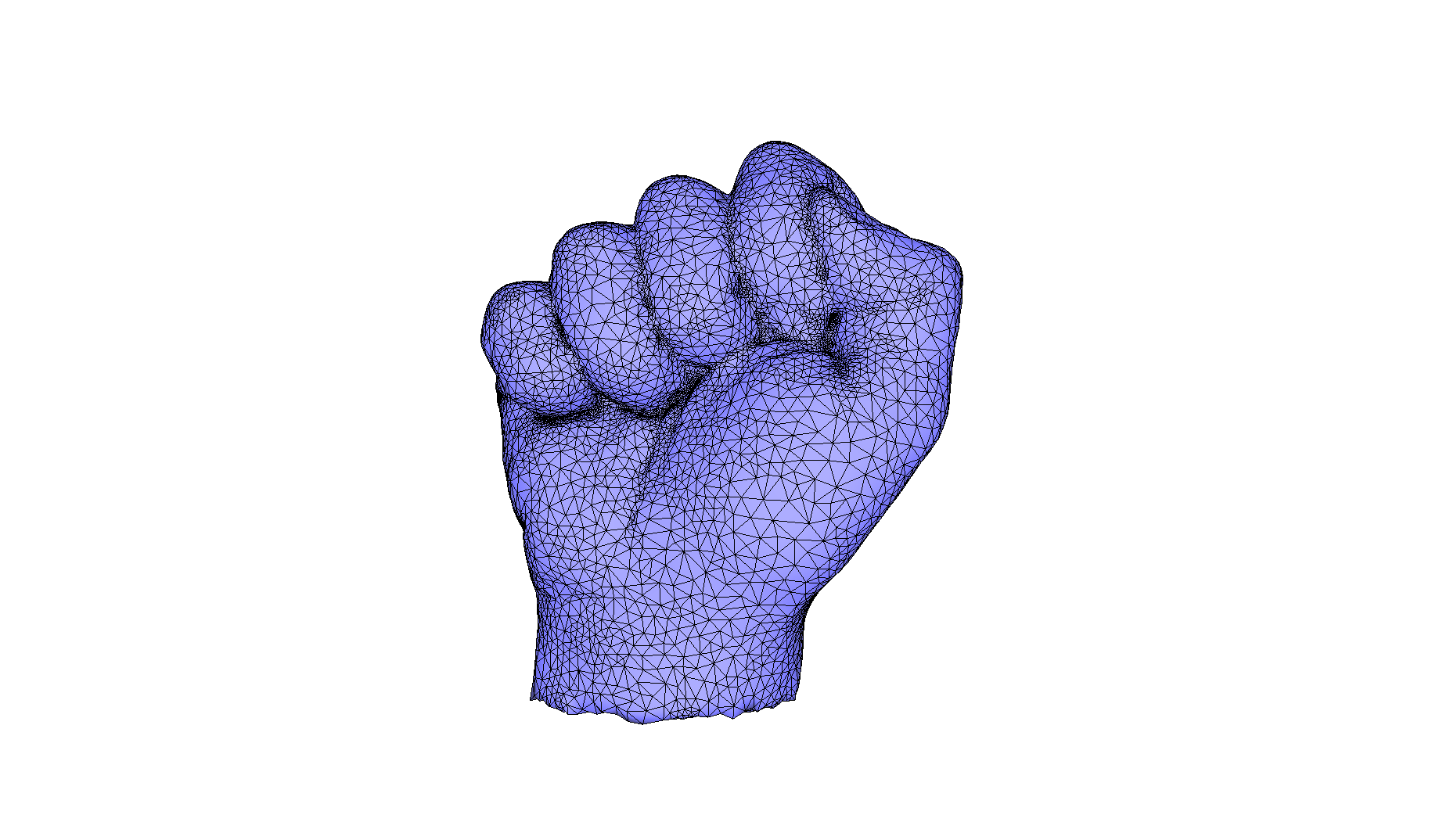}\vspace{0pt} 
    \end{minipage}
}
\subfigure[\scriptsize{Squirrel (\#43,535)}]{   
    \begin{minipage}[h]{0.29\columnwidth}
    \includegraphics[width=\linewidth,trim={450 100 450 100},clip]{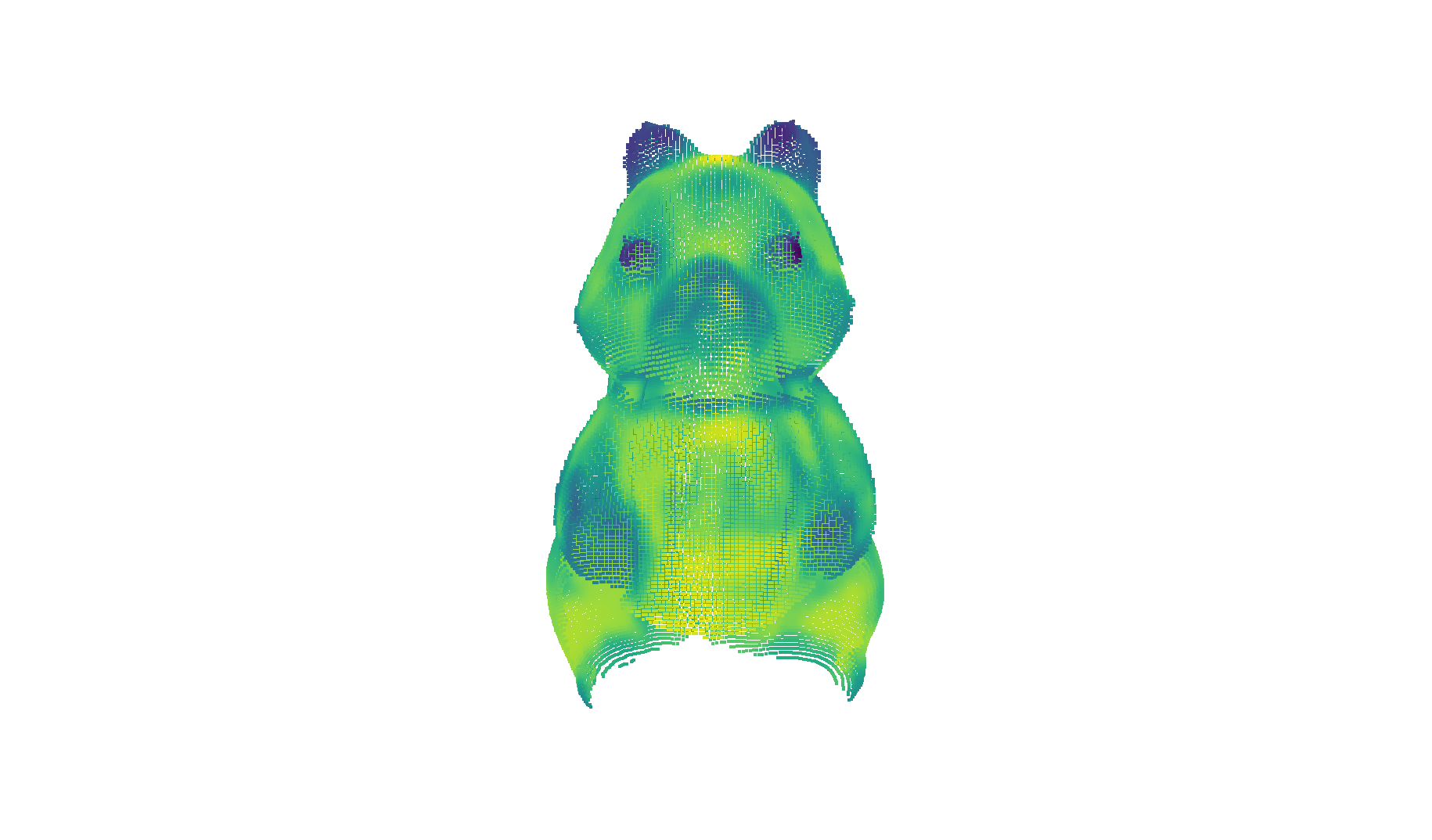}\vspace{0pt}
    \includegraphics[width=\linewidth,trim={450 100 450 100},clip]{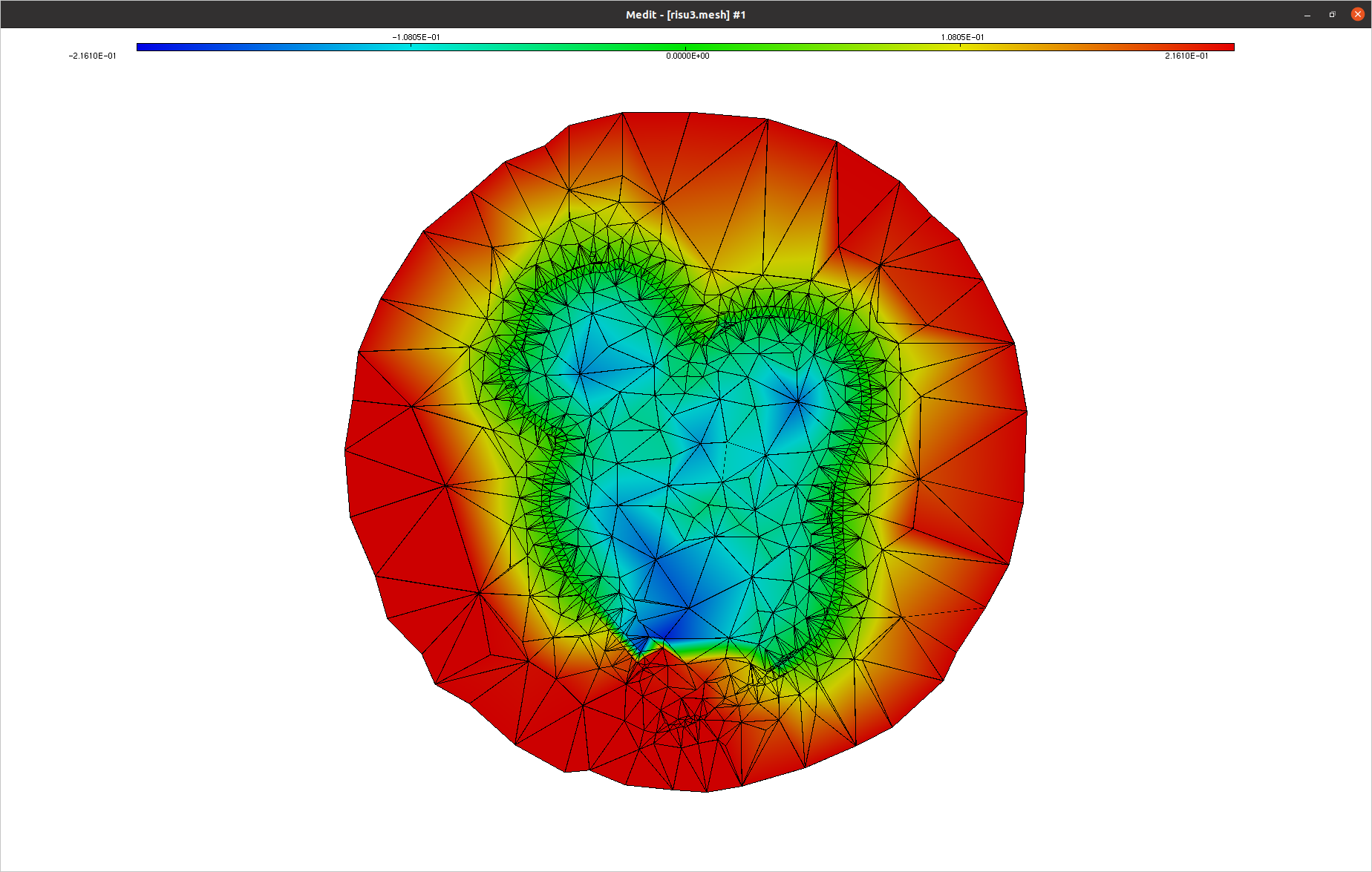}\vspace{0pt}
    \includegraphics[width=1.2\linewidth,trim={100 0 80 0},clip]{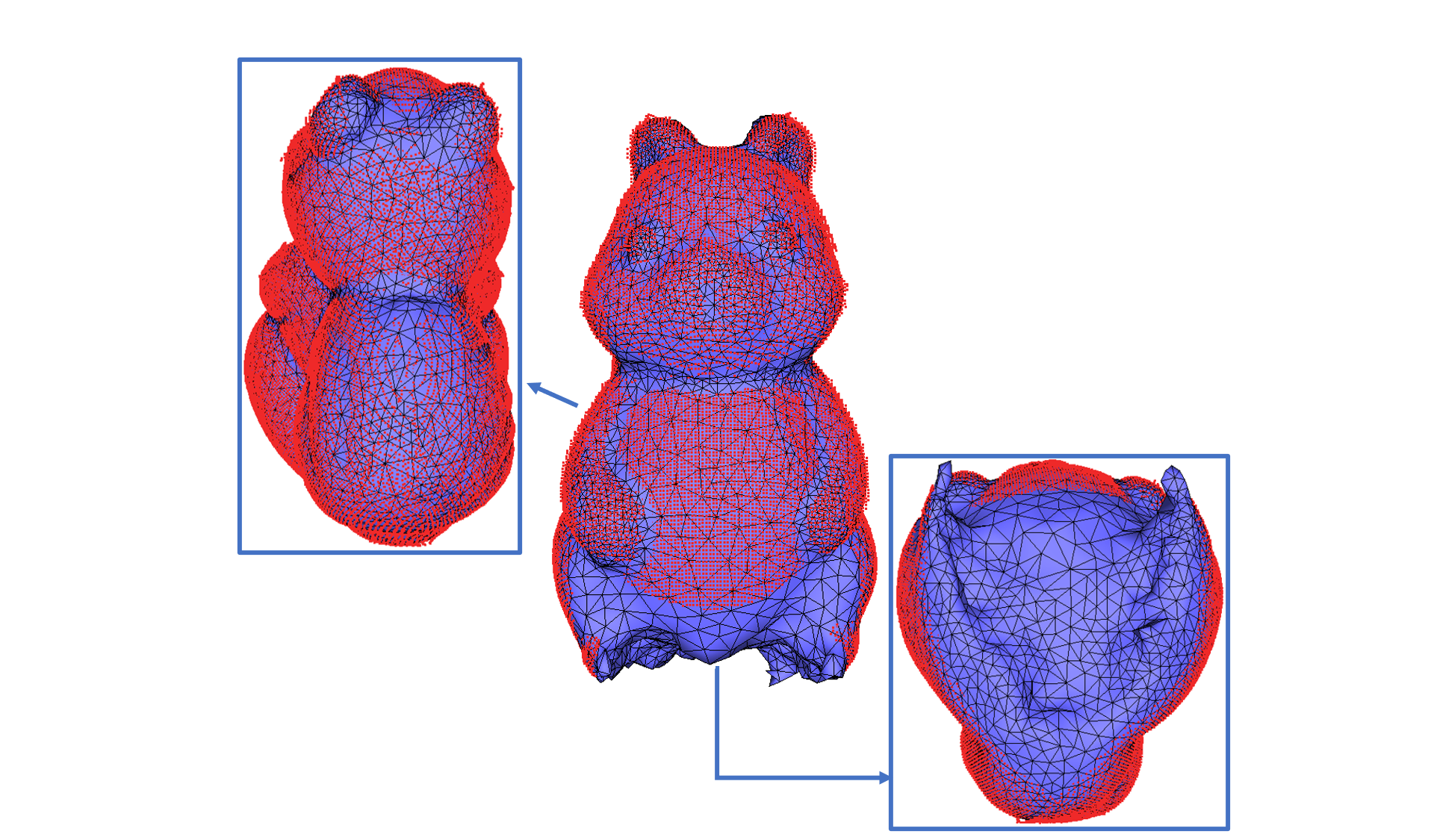}\vspace{0pt} 
    \includegraphics[width=\linewidth,trim={450 80 450 100},clip]{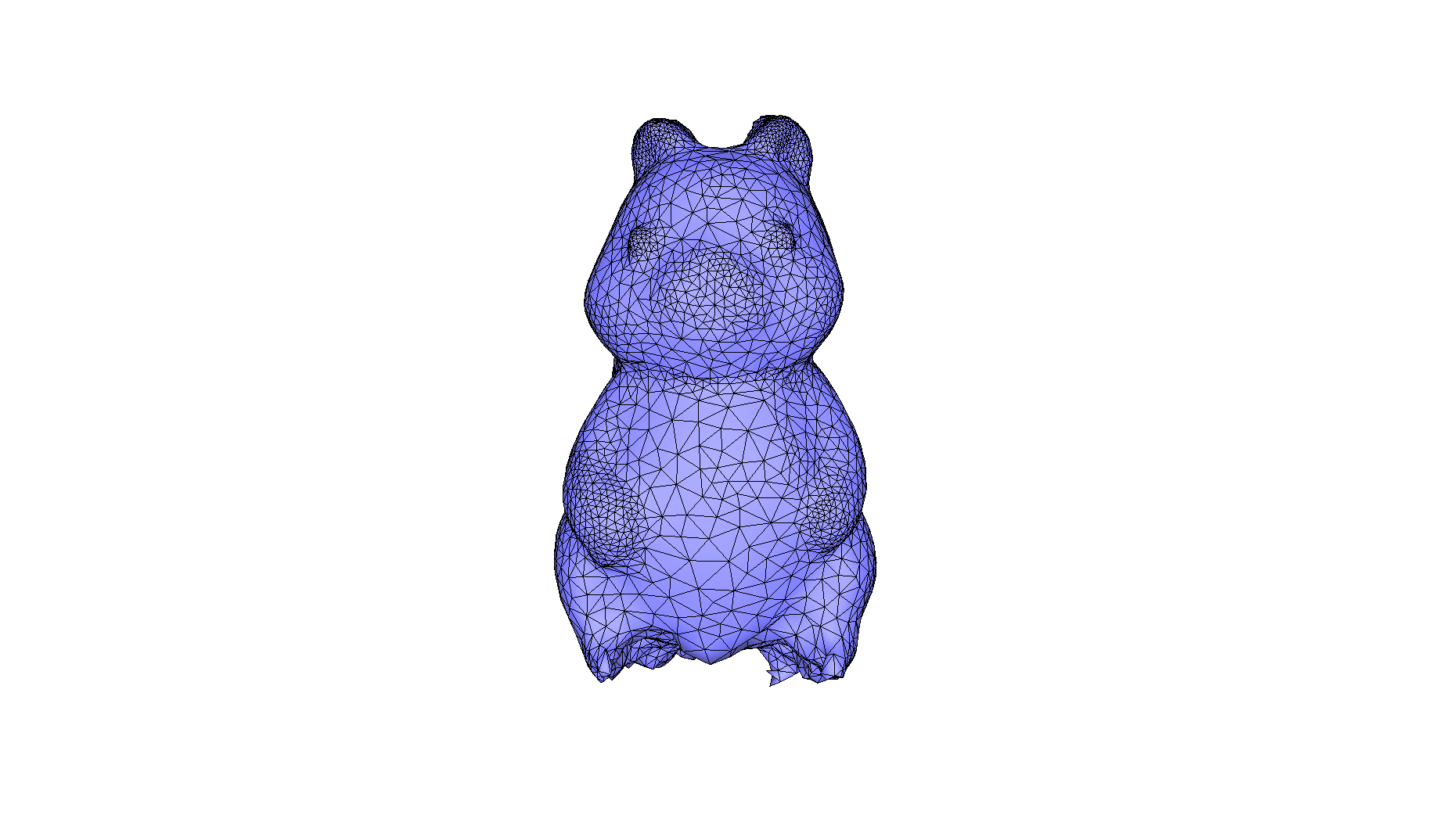}\vspace{0pt}
    \end{minipage}
}
\caption{Reconstructions on real-world data with large holes. 
First row: estimated LFS. 
Second row: clipped view of the signed robust distance function. 
Third and fourth rows: LFS-aware meshes.}
\label{hole-filling-real-scan}
\end{figure}

We now conduct experiments to verify the ability of our algorithm to fill large holes in real-world data acquired by laser scanners provided by the AIM@Shape dataset (see Fig.~\ref{hole-filling-real-scan}). Such data commonly feature non-uniform sampling and measurement noise, making hole filling challenging.
On areas where the point set is dense, we observe a high density of edges with opposite sign guesses. Areas near holes are more ambiguous, with a mixture of edges with opposite or similar signs. Intuitively, the signing step is robust to large holes (with many missing data) thanks to the least-squares solver that propagates the sign on the triangulation vertices via consolidating sign guess hypotheses emitted for all triangulation edges.

\subsubsection{Sharp features} 
\begin{figure}[t!]
\centering
\includegraphics[width=\columnwidth,trim={5 5 5 5},clip]{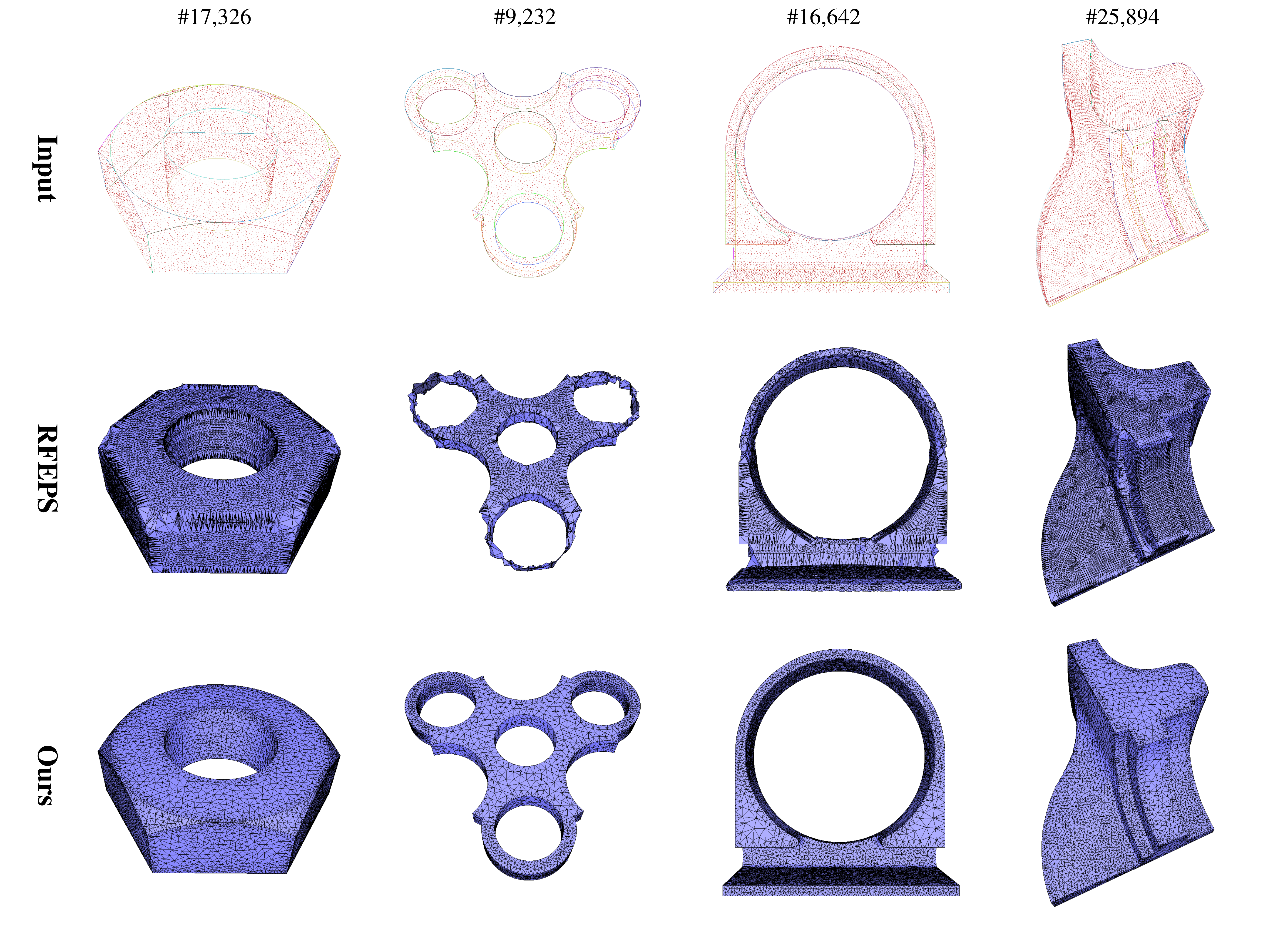}\vspace{0pt}
\caption{Experiments on CAD models with sharp features. The top row shows the input point clouds with detected sharp features using NerVE \cite{zhu2023nerve}. The middle row is the results of RFEPS~\cite{xu2022rfeps}. The bottom row shows the results of our algorithm.} 
\label{sharp-features}
\end{figure}

Our algorithm also excels in preserving sharp features. To achieve this, we initially employ the NerVE method \cite{zhu2023nerve} to detect these sharp feature curves from the input point cloud, and then we convert the sharp feature curves into piecewise linear feature polylines. As we utilize the Delaunay refinement method \cite{jamin2015cgalmesh} to extract the mesh from the solved implicit function, if feature polylines are provided, the Delaunay refinement method will generate protecting balls centered at these feature lines to avoid refinement on the sharp feature lines. Thus, the final mesh will preserve the sharp features. We verified the ability of sharp feature preservation on the ABC dataset \cite{Koch_2019_CVPR} and compared our algorithm with RFEPS \cite{xu2022rfeps}. Compared to RFEPS, our method can split sharp features that are close to each other and maintain the mesh's isotropy.
Please refer to Fig.~\ref{sharp-features}.

\subsection{Ablation Study}

\subsubsection{LFS vs. curvature} 
\begin{figure}[t!]
\centering
\subfigure[LFS]{
\includegraphics[width=0.29\columnwidth,trim={350 200 350 200},clip]{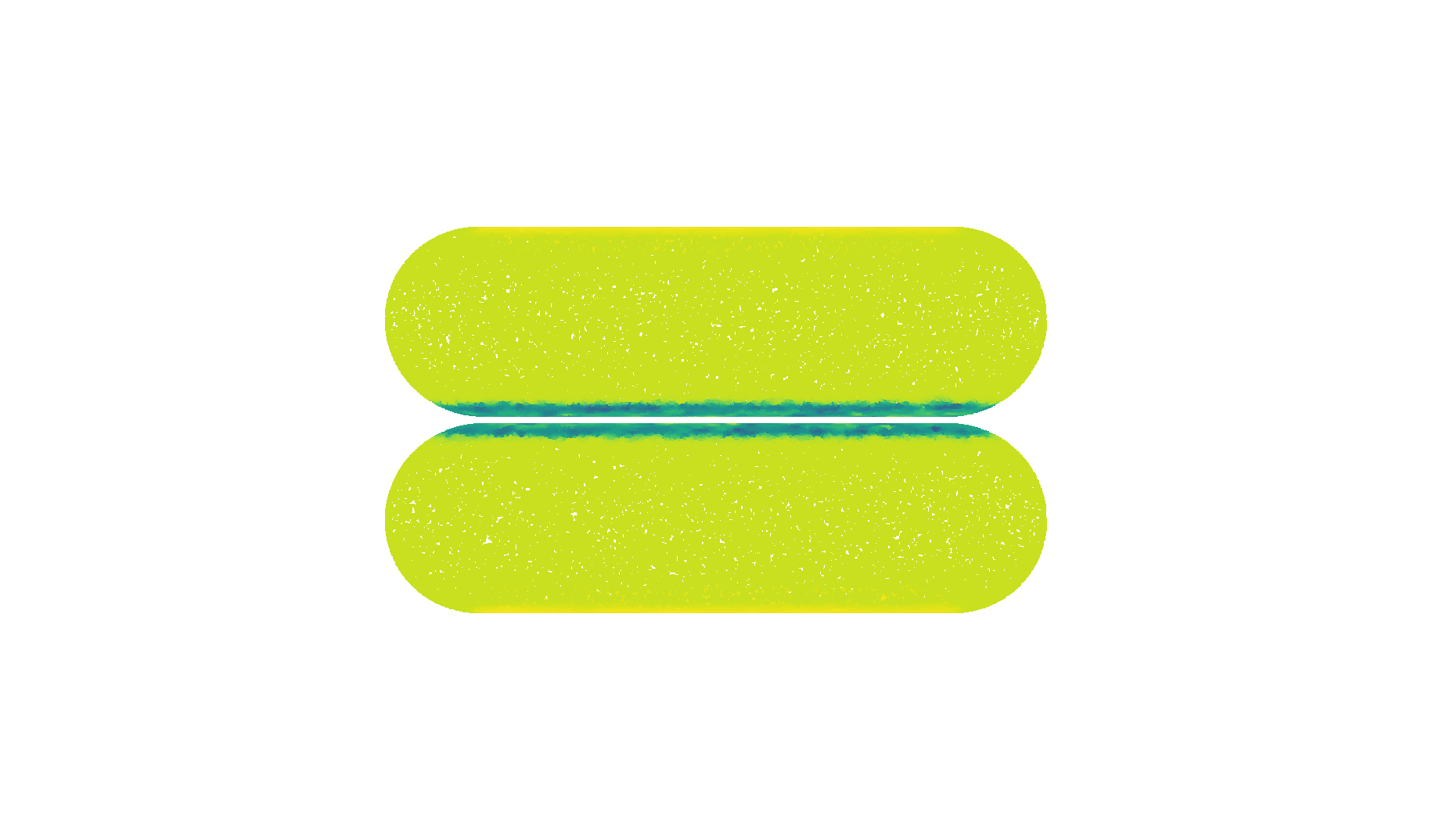}\vspace{0pt}
}
\subfigure[curvature]{
\includegraphics[width=0.29\columnwidth,trim={350 200 350 200},clip]{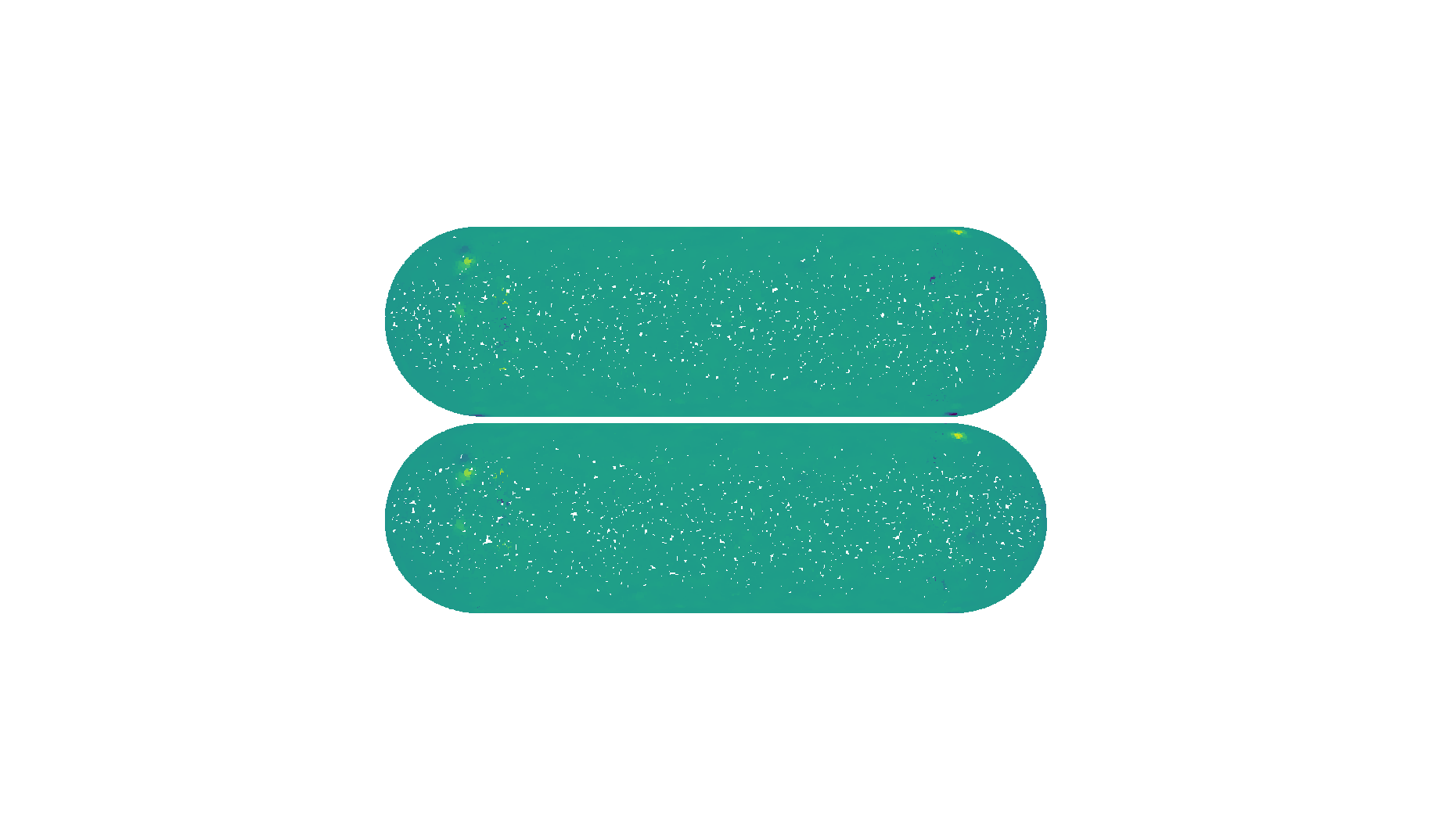}\vspace{0pt}
}
\\
\subfigure[reach-aware]{
\includegraphics[width=0.29\columnwidth,trim={300 0 300 0},clip]{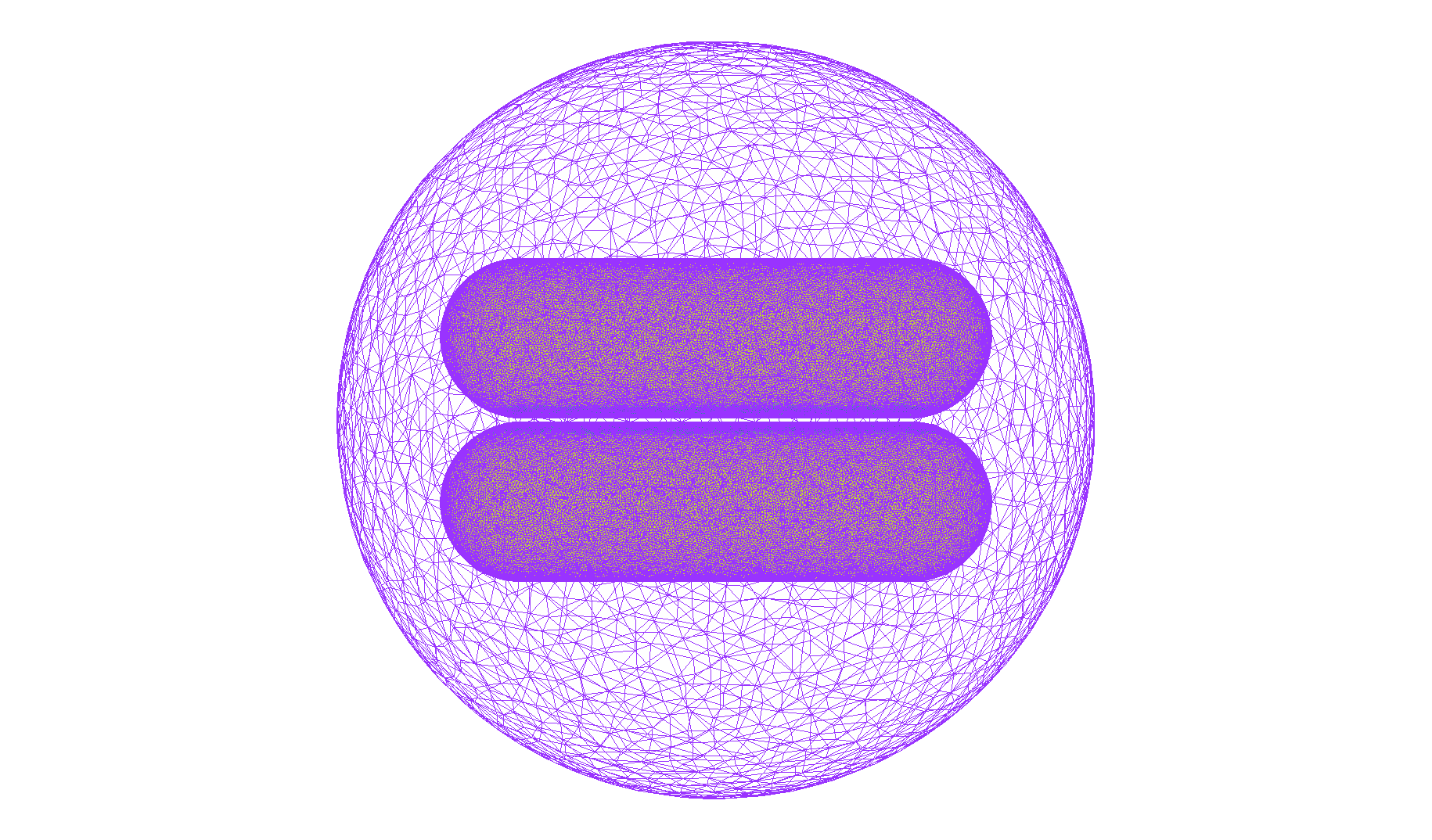}\vspace{0pt}
}
\subfigure[min-curvature-aware]{
\includegraphics[width=0.29\columnwidth,trim={300 0 300 0},clip]{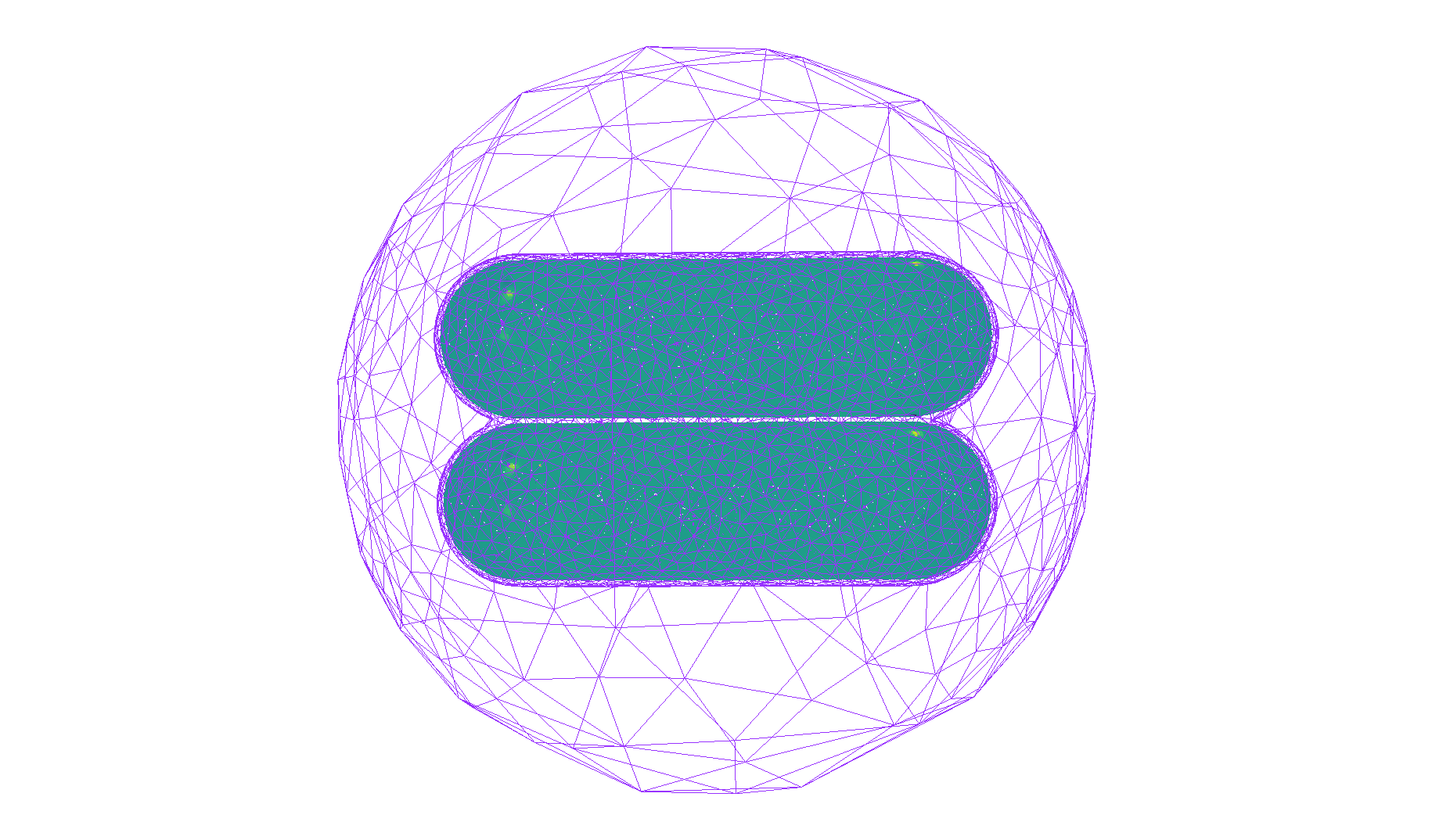}\vspace{0pt}
}
\\
\subfigure[LFS-aware mesh]{
\includegraphics[width=0.29\columnwidth,trim={350 200 350 200},clip]{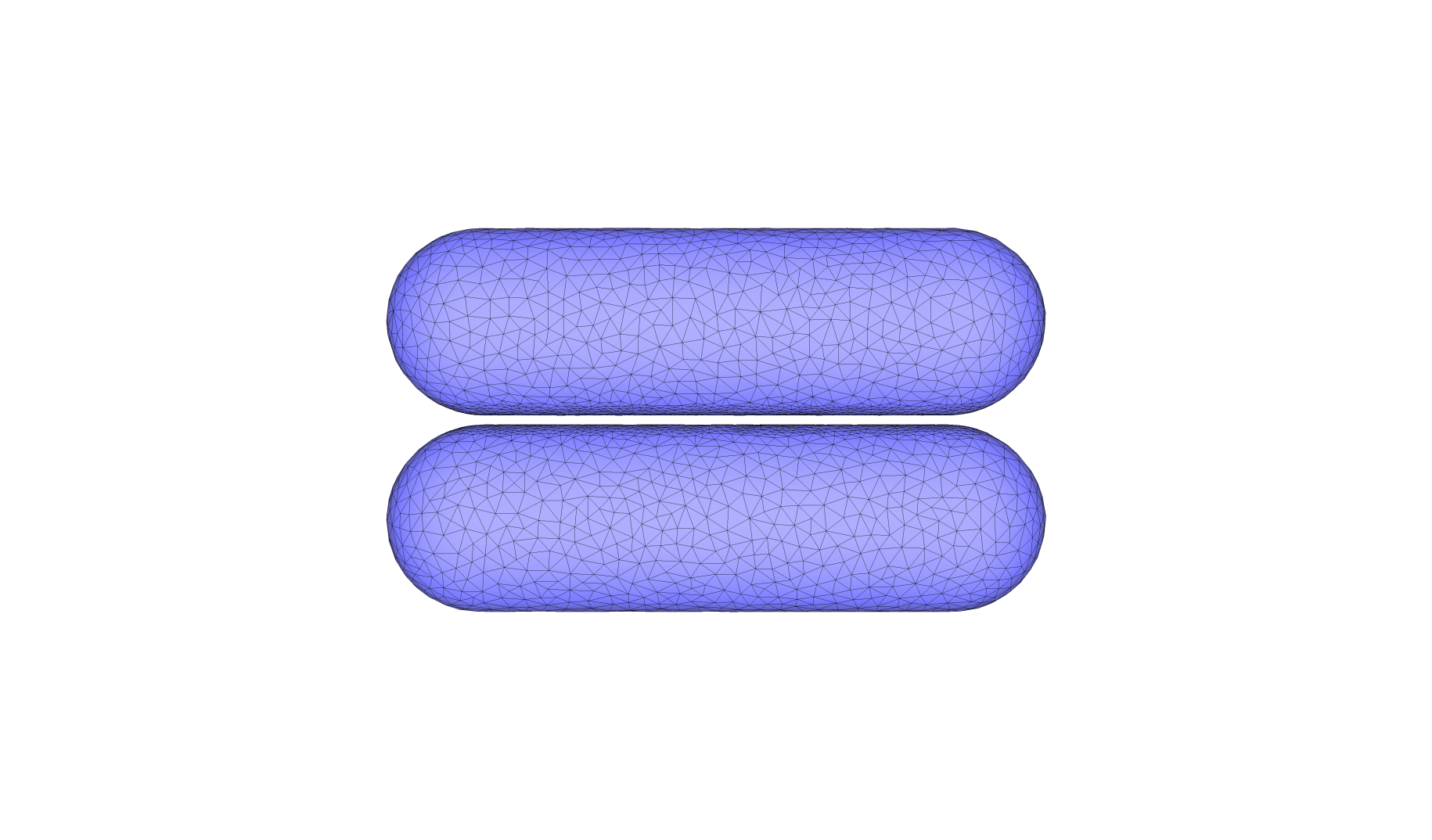}\vspace{0pt}
}
\subfigure[curvature-mesh of (c)]{
\includegraphics[width=0.29\columnwidth,trim={350 200 350 200},clip]{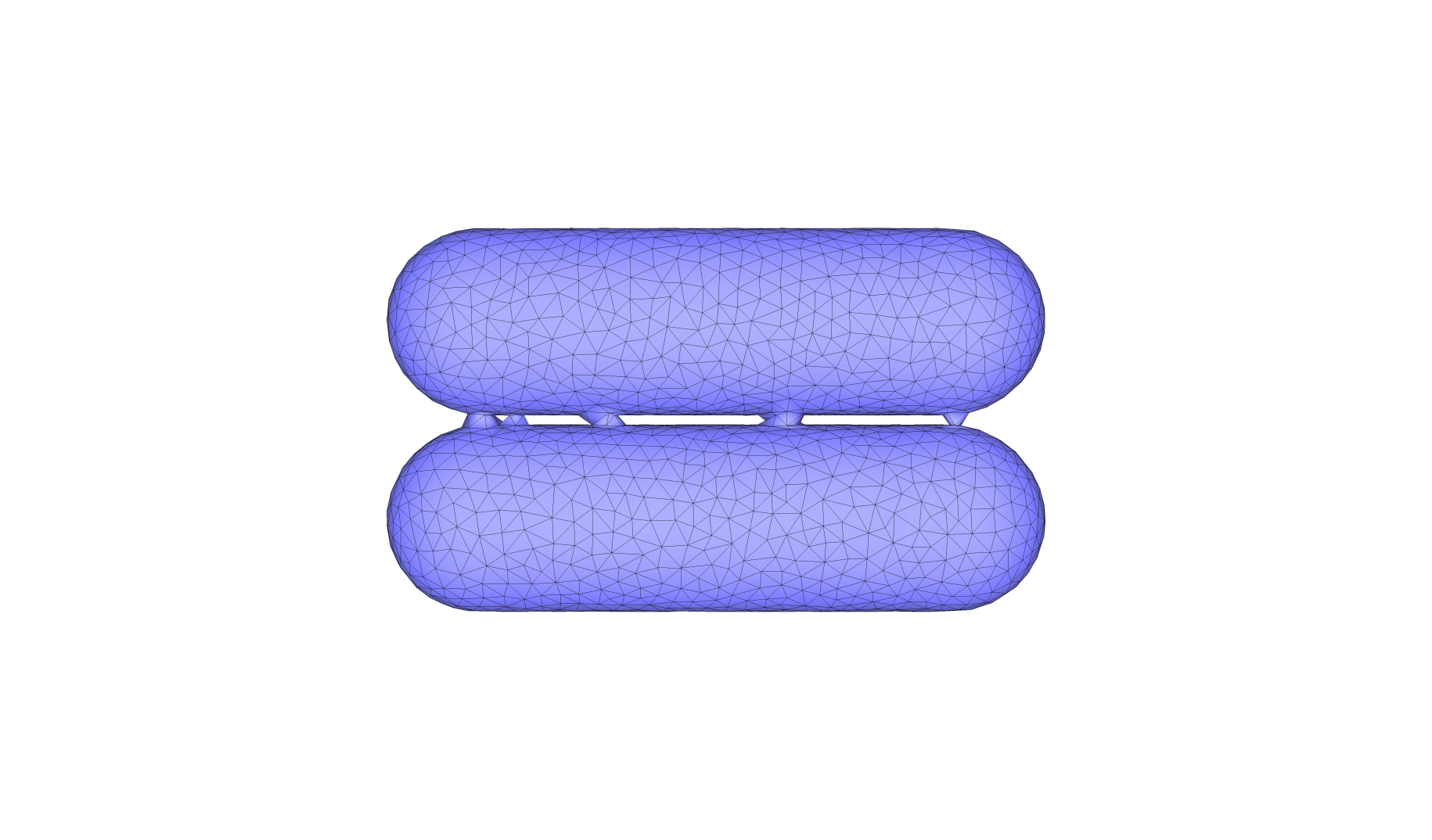}\vspace{0pt}
\label{fig-lfs-vs-curvature-c}
}
\subfigure[curvature-mesh of (d)]{
\includegraphics[width=0.29\columnwidth,trim={350 200 350 200},clip]{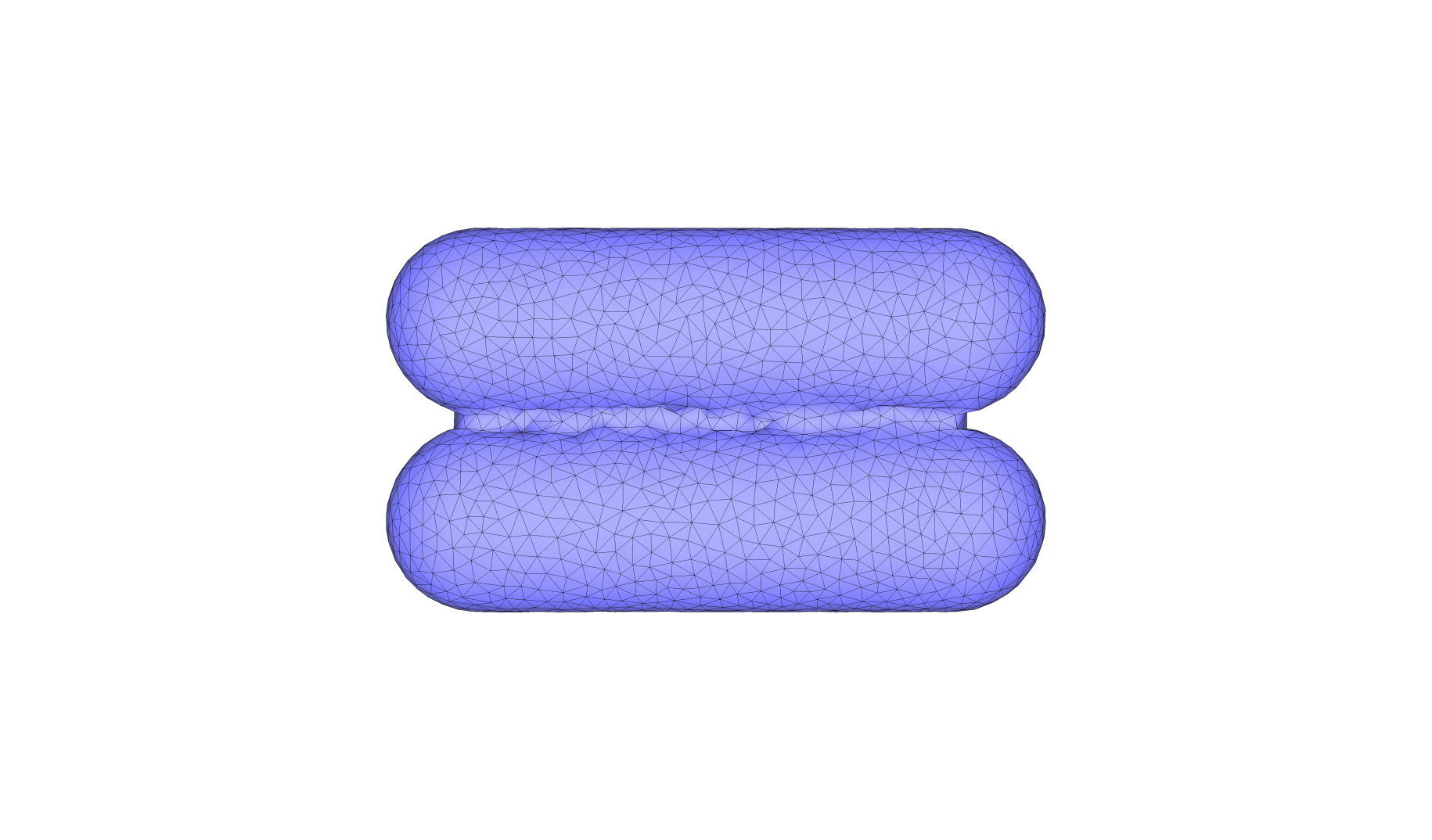}\vspace{0pt}
}
\caption{LFS vs. curvature. (a) The estimated LFS. (b) The estimated curvature. (c) The reach-aware multi-domain. (d) The minimum-curvature-aware multi-domain. (e) The LFS-aware mesh extracted from the implicit function solved on (c). (f) The curvature-aware mesh extracted from the implicit function solved on (c). (g) The curvature-aware mesh extracted from the implicit function solved on (d).}
\label{fig-lfs-vs-curvature}
\end{figure}

We now conduct experiments to verify the added value of using LFS instead of curvature. A first experiment is conducted on two adjacent capsules with a smaller separation distance than the curvature radius, see Fig.~\ref{fig-lfs-vs-curvature}. We use three settings: (1) the estimated minimum LFS is used to construct the reach-aware envelope, and LFS is used for sizing the facets; (2) we also construct a reach-aware envelope but use only the curvature for sizing the facets; (3) we use only the minimum curvature radius to construct the envelope and the curvature to size the facets. Setting (1) yields the best results, where the curvature and topology are captured, while settings (2) and (3) fail to reconstruct the topology as the two capsules merge. More specifically, using only the minimum curvature considers neither the thickness nor the separation, thus leading to a thicker envelope and insufficient resolution for the implicit function. In addition, using only the curvature for sizing the output mesh is insufficient even when the implicit function is solved on the reach-aware domain, as the Delaunay refinement process fails to separate nearby surface sheets (see Fig.~\ref{fig-lfs-vs-curvature-c}). The middle of the two capsules is connected as the size of the facets derived from the curvature is larger than the separation distance. 

\subsection{Comparisons}

\begin{figure*}[htb!]
\centering
\includegraphics[width=1.0\linewidth,trim={5 5 5 5},clip]{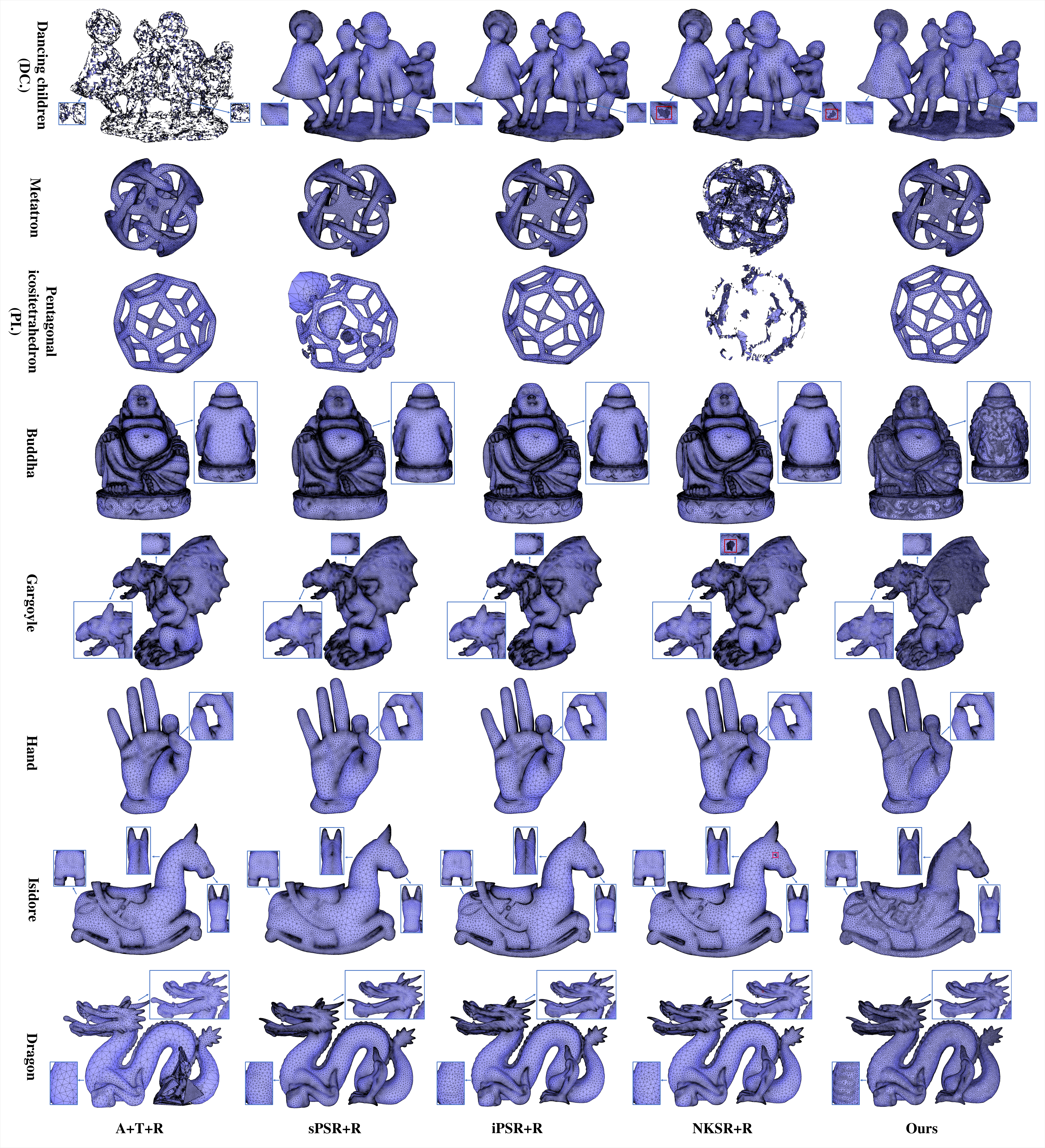}\vspace{0pt} 
\caption{Visual comparison with different algorithms: (A+T+R) AMLS reconstruction, followed by Tight Cocone and remeshing; (sPSR+R) screened Poisson surface reconstruction, followed by remeshing; (iPSR+R) iterative Poisson surface reconstruction, followed by remeshing; (NKSR+R) NKSR, followed by remeshing.}
\label{lfs-mesh-cmp}
\end{figure*}

\begin{table*}[t!]
\begin{center} 
\scalebox{0.79}{
\begin{tabular}{l| ccccc| ccccc| ccccc}
\toprule[1.2pt]
Model &\multicolumn{5}{c|}{Chamfer($\downarrow$)} &  \multicolumn{5}{c|}{ Hausdorff($\downarrow$)} & \multicolumn{5}{c}{ Face(\#)}  \\
 & A+T+R & sPSR+R & iPSR+R & NKSR+R & Ours & A+T+R & sPSR+R & iPSR+R & NKSR+R & Ours & A+T+R & sPSR+R & iPSR+R & NKSR+R & Ours \\
\hline
DC.(\#724k) & 4.88E-1 & 1.37E-2 & 1.19E-2 & 1.07E-2 & \textbf{6.10E-3} & 3.99 & 1.72E-1 & 1.65E-1 & 1.48 & \textbf{1.63E-1} & 95k & 92k & 92k & 92k & 92k \\
Metatron(\#69k) & 4.73E-1 & 6.35E-2 & 5.72E-2 & 6.32E-1 & \textbf{4.76E-2} & 4.01 & 6.11E-1 & 5.49E-1 & 5.98 & \textbf{5.40E-1} & 61k & 61k & 61k & 64k & 61k \\
PI.(\#9k) & 3.03E-1 & 1.79E-1 & 1.48E-1 & 1.53 & \textbf{9.69E-2} & 9.31E-1 & 1.92 & 7.20E-1 & 7.24 & \textbf{4.02E-1} & 15k & 15k & 15k & 15k & 15k \\
Buddha(\#719k) & 4.60E-4 & 3.50E-4 & 3.05E-4 & 3.13E-4 & \textbf{2.49E-4} & 3.81E-3 & 4.30E-3 & 3.78E-3 & 3.71E-3 & \textbf{2.73E-3} & 128k & 128k & 128k & 128k & 128k \\
Gargoyle(\#963k) & 6.12E-2 & 1.34E-2 & 1.03E-2 & 1.31E-2 & \textbf{4.82E-3} & 6.49E-1 & 1.89E-1 & 1.23E-1 & 2.17 & \textbf{1.21E-1} & 180k & 180k & 180k & 180k & 180k \\
Hand(\#369k) & 1.12E-3 & 2.09E-4 & 2.14E-4 & 2.28E-4 & \textbf{2.06E-4} & 7.08E-3 & 5.31E-3 & 5.60E-3 & 6.48E-3 & \textbf{5.28E-3} & 35k & 35k & 35k & 35k & 35k \\
Isidore(\#235k) & 4.12E-4 & 3.78E-4 & 3.62E-4 & 3.99E-4 & \textbf{3.15E-4} & 6.93E-3 & 7.09E-3 & 6.95E-3 & 7.18E-3 & \textbf{6.91E-3} & 72k & 72k & 72k & 72k & 72k \\
Dragon(\#296k) & 1.02E-3 & 3.05E-4 & 1.85E-4 & 3.10E-4 & \textbf{1.60E-4} & 3.29E-2 & 1.44E-2 & 1.37E-2 & 9.98E-3 & \textbf{6.99E-3} & 164k & 158k &  158k & 158k & 158k \\
\bottomrule[1.2pt]
\end{tabular}}
\end{center}
\caption{Comparison with different algorithms: (A+T+R) AMLS reconstruction, followed by Tight Cocone and remeshing; (sPSR+R) screened Poisson surface reconstruction, followed by remeshing; (iPSR+R) iterative Poisson surface reconstruction, followed by remeshing; (NKSR+R) NKSR, followed by remeshing.}
\label{tab-mesh-cmp}
\end{table*}

We now conduct more experiments to compare our algorithm with other baseline methods. Comparisons with RFEPS \cite{xu2022rfeps} on CAD models with sharp features are already presented in Fig.~\ref{sharp-features}. We compare  against AMLS \cite{dey2005adaptive}, screened Poisson surface reconstruction (sPSR) \cite{kazhdan2013screened}, iPSR \cite{hou2022iterative}, and NKSR \cite{huang2023neural}.  AMLS falls into the class of explicit reconstruction methods that also use LFS as a prior. In contrast, sPSR and iPSR fall into the category of implicit surface reconstruction, and NKSR is a learning-based method. In addition, to obtain isotropic meshes with LFS-aware sizing, we perform LFS-based remeshing using the Geogram library \cite{levy2015geogram}. This involves sampling points on the reconstructed raw mesh and constructing the Voronoi diagram to estimate the LFS, which is then used to size the facets. To make a fair comparison, we ensure that all algorithms output roughly the same mesh complexity in terms of the number of faces. We then evaluate the reconstruction fidelity using the point-to-mesh Chamfer distance and the Hausdorff distance \cite{zhao2021progressive, wang2023neural}. Fig.~\ref{lfs-mesh-cmp} provides a visual comparison and Table~\ref{tab-mesh-cmp} reports the evaluation metrics. In Fig.~\ref{lfs-mesh-cmp}, our method may appear to produce more triangles for each model visually compared to the baselines. But this is not the case. The baselines generate extremely dense and tiny triangles in detailed regions, such as the creases of the hand model, which gives the impression of fewer triangles overall in Fig.~\ref{lfs-mesh-cmp}. From the experiment, our approach achieves the smallest reconstruction errors for both the Chamfer and Hausdorff distance, and a visual inspection shows that it preserves more details, such as the back of the Buddha and the tail of the Isidore horse point cloud. Our approach better captures the topology by leveraging the LFS estimation, while AMLS and NKSR fail to reconstruct from the Metatron and PI point clouds. Additionally, the output of our method is water-tight, while NKSR can leave holes. We mark those holes in the red rectangles in Fig.~\ref{lfs-mesh-cmp}.

\section{Conclusion}

This paper introduces a novel approach for estimating the local feature size (LFS) and leveraging it for LFS-aware surface reconstruction from unorganized 3D point cloud data. The primary output of our method is a valid and intersection-free isotropic triangle surface mesh with facet sizing that captures the estimated LFS. One of the key contributions of our approach is the incorporation of LFS into the surface reconstruction process. By doing so, we generate a mesh with, ideally, ``just-enough'' complexity, seeking a balance between capturing fine details in the data, while maintaining robustness to various data defects such as noise, outliers, and missing data. This results in a more accurate and visually pleasing reconstructed surface. Our experiments demonstrate the robustness of our LFS estimation method and show that it can handle moderate noise and non-uniform sampling effectively. Additionally, our surface reconstruction algorithm outperforms some existing baselines in terms of preserving details and accurately representing thin topological features.

\appendix
\label{appendix}

\section{Lipschitz-guided recursive dichotomic search}
\label{lipschitz-search}

We propose a Lipschitz-guided recursive dichotomic search algorithm to detect the sublevels of a distance function. More specifically, we  utilize the said dichotomic search to find the antipodal points when estimating the local shape diameter. We also use it to determine a sign guess for an edge inside the envelope enclosed in the multi-domain discretization. Unlike an exhaustive search that requires a dense point sampling on a ray to ensure that all intersections are detected, such a dichotomic search accelerates computations by avoiding unnecessary point sampling.

\begin{figure}[htb]
  \centering
\includegraphics[width=\columnwidth,trim={20 20 20 20},clip]{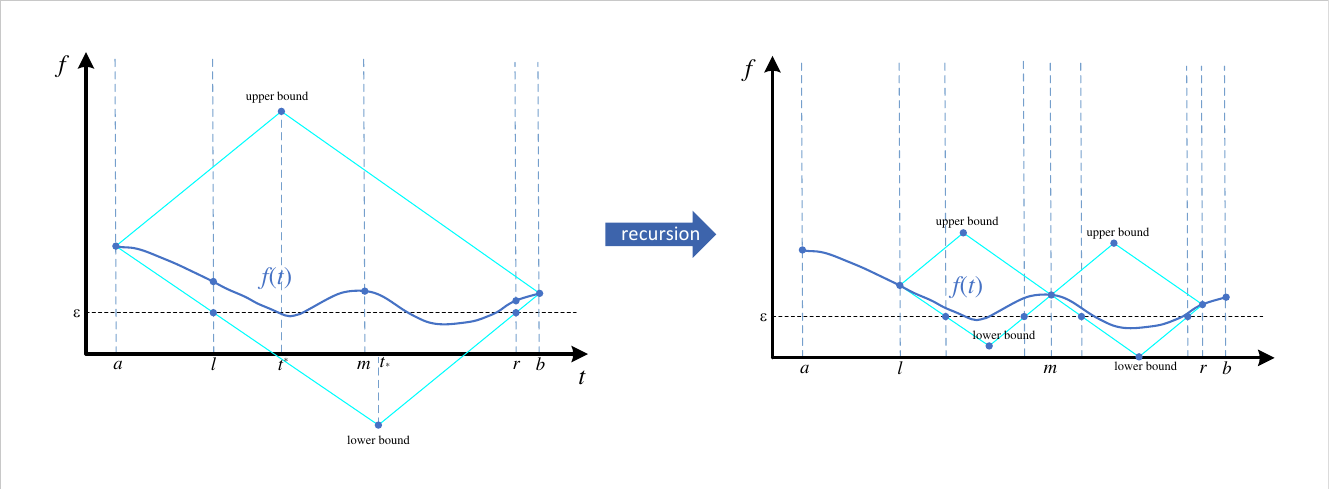}
  \caption{Lipschitz-guided recursive dichotomic search. The algorithm first searches in $[a, b]$, then the following recursion searches in $[l, m]$ and $[m, r]$ after bisection.}
  \label{fig-recursive}
\end{figure}

Let $\mathbf{\Omega} \subset \mathbb{R}^3$ denote a 3D domain and $\partial\mathbf{\Omega}$ denote the boundary of $\mathbf{\Omega}$. The distance function $f(\mathbf{x}) \in \mathbb{R}$ is defined as $f(\mathbf{x})=\mathrm{inf}_{\mathbf{y} \in \partial\mathbf{\Omega}}{|\mathbf{x}-\mathbf{y}|}$, where $\mathbf{x} \in \mathbb{R}^3$ denotes a query point. From the definition, $f(\mathbf{x})$ is $1$-Lipschitz continuous. Since we search along a ray, the query point $\mathbf{x}$ can be parameterized by a scalar parameter $t \in \mathbb{R}$. For simplicity, we use this parameter $t$ to refer to a query point $\mathbf{x}$ on a given ray. For a parameterized query point $t_1, t_2 \in \mathbb{R}$ on a given ray, we have $|f(t_1) - f(t_2)| \leq |t_1 - t_2|$, i.e., the distance function satisfies the $1$-Lipschitz continuity condition. Therefore, given a ray trimmed by a search interval $[a, b]$, we leverage the Lipschitz continuity to shrink the search interval. More specifically, we obtain the inequalities
\begin{equation}\label{lipschitz-square-bound}
\begin{aligned}
|f(t) - f(a)| &\le |t - a|, \\
|f(t) - f(b)| &\le |t - b|.
\end{aligned}
\end{equation}
Equation (\ref{lipschitz-square-bound}) reflects that the plot $f(t)$ is bounded by a parallelogram-shaped region. By choosing a small value $\epsilon$, we find two points $l$ and $r$ satisfying $|l - a| = \epsilon$ and $|r - b| = \epsilon$, as depicted in Fig.~\ref{fig-recursive}. This means that the points that satisfy $f(t)=\epsilon$ lie in the interval $[l, r]$. Algorithm~\ref{alg-recursive} details the steps of the searching algorithm, which recursively bisects the search interval.

\begin{algorithm}[htb]
\caption{Recursive\_dichotomic\_search($a, b$)}
\label{alg-recursive}
\KwIn{the search interval $[a, b]$, a $1$-Lipschitz continuous function $f(t)$, a small value $\epsilon$, a global empty container $R_t$}
\KwOut{approximated points $x_r$ will be stored in $R_t$ such that $R_t=\{x_r:|f(x_r)|\approx\epsilon\}$}

\tcc{calculate the lower bound} 
$t_{*} \gets \frac{1}{2}[a + b - f(b) + f(a)]$, $y_{*} \gets f(a) - (t_{*} - a)$ 

\tcc{calculate the upper bound}
$t^{*} \gets \frac{1}{2}[a + b + f(b) - f(a)]$, $y^{*} \gets f(a) + (t^{*} - a)$ 

 \If{$y_{*} \ge \epsilon$ \textrm{\textbf{or}} $y^{*} \le \epsilon$ }
 {
    \Return
 }

\lIf{$f(a) > \epsilon$}{  $k_a  \gets -1$ \textbf{else} $k_a  \gets 1$}
\lIf{$f(b) > \epsilon$}{  $k_b  \gets 1$ \textbf{else} $k_b  \gets -1$}

$l \gets a + \frac{\epsilon-f(a)}{k_a}$, $r \gets b + \frac{\epsilon-y_b}{k_b}$, $m \gets \frac{l+r}{2}$ \;

\If{ $|f(m) - \epsilon| \le 10^{-7}$ {\bfseries\emph{and}} $|r - l| \le \epsilon$ }
{
 \tcc{compare the binary signs}
 \eIf {$\mathrm{sign}(f(a)-\epsilon) == \mathrm{sign}(f(b)-\epsilon)$}  
 {
  $R_t.\mathrm{push}(m)$ \;
  $R_t.\mathrm{push}(m)$ \;
 } {
   $R_t.push(m)$;
 }
 \Return;
} 
Recursive\_dichotomic\_search($l$, $m$)\;
Recursive\_dichotomic\_search($m$, $r$);
\end{algorithm}

\begin{figure}[htbp!]
  \centering
\includegraphics[width=\columnwidth,trim={2 2 2 2},clip]{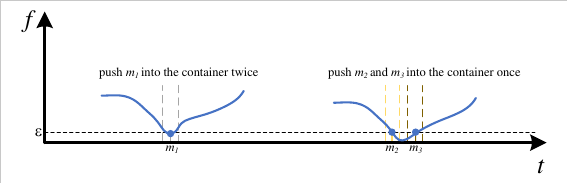}
  \caption{Points stored in the container $R_t$. Algorithm \ref{alg-recursive} pushes $m_1$ twice into container $R_t$, and pushes $m_2$ and $m_3$ once. }
  \label{fig-recursive-roots}
\end{figure}

In Algorithm~\ref{alg-recursive}, $m$ will be the approximated point if $f(m)$ approaches $\epsilon$ and the length $|r-l|$ is smaller than $\epsilon$. Note that the container $R_t$ stores the pair values approximating $\epsilon$. When Algorithm~\ref{alg-recursive} finds an approximated point $m$, if both $f(a)$ and $f(b)$ are greater or smaller than $\epsilon$, we push $m$ twice into the container $R_t$; otherwise, $m$ is pushed once (see Fig.~\ref{fig-recursive-roots}). The container stores the points such that \mbox{$R_t=\{x_r:f(x_r)\approx\epsilon\}$}. Even though $\epsilon$ is a small value, the approximated points do not directly estimate the zero sublevel. To make a better approximation, we take the average values of $m_2$ and $m_3$ (Fig.~\ref{fig-recursive-roots}) as the final points if $f(a)$ and $f(b)$ on the opposite side of $\epsilon$. That explains why we push $m_1$ twice into the container and push $m_2$ and $m_3$ into the container only once. Finally, we regard the final point as the average of the adjacent two values stored in $R_t$.

\section{Smoothing LFS}
In theory, LFS is a 1-Lipschitz continuous function. However, our LFS estimation procedure may lead to a noisy function when applied to point clouds with variable sampling density and noise. For example, two nearby query points may have a very large curvature estimated by jet fitting due to noise, or a super large shape diameter equating to the radius of the loose bounding sphere due to sparse sampling. This results in salt-and-pepper noise in the estimated LFS function. We thus apply a two-step filtering process to denoise and smooth out the estimated LFS function. We first apply a median filter to remove the salt-and-pepper noise and then apply Laplacian smoothing to further smooth out the denoised LFS function. 

\section{Delaunay refinement}
\label{delaunay_refinement}

We utilize the Delaunay refinement paradigm ([$4$] in the main paper) to generate 3D  triangulations with well-shaped triangles and tetrahedra, both for discretizing a multi-domain and for  LFS-aware isosurfacing of an implicit function.
Starting from an initial triangulation generated by sampling random points on the domain boundary $\partial\mathbf{\Omega}$, Delaunay refinement adds Steiner points defined either as intersection points between Voronoi edges and $\partial\mathbf{\Omega}$, or as circumsphere centers of tetrahedra. 
In our context, $\mathbf{\Omega}$ is defined as a sublevel of the solved implicit function, and we design an LFS-aware sizing function for governing the local size of the mesh elements.

\section{Smoothing facet sizing function}

The facet sizing function $\textrm{size}(\mathbf{x})$ is used to control the final mesh sizing. The function is also Lipschitz-continuous as it derives from a linear transformation of LFS. To avoid too large facet sizes and make the sizing function $1$-Lipschitz, we use a breadth-first propagation to further smooth the sizing function. Starting from the point with a local minimum of the facet size values, we check if the size values from its neighbor satisfy the Lipschitz continuity. We refer to Algorithm 2 for details.

\begin{algorithm}[htb]
\caption{BFS\_lipschitz\_continuity\_propagation}
\label{alg-bfs}
\KwIn{$k$-NN graph of the input points, facet sizing function $\textrm{size}(\mathbf{x})$ }
\KwOut{ smoothed facet sizing function $\textrm{size}(\mathbf{x})$}
establish a priority queue $Q$ whose top stores the minimum value\;
establish a map $\mathit{visited}$ to mark if the point has been visited\;
\For{$\mathbf{p}_i \in \mathbf{P}$}{
    $Q$.push($\textrm{size}(\mathbf{p}_i)$) ;
}
\While{$Q$ \rm{is not empty}}{
    $\textrm{size}(\mathbf{p}_{\textrm{cur}})$ $\gets$ $Q$.top() \;
    $Q$.pop() \;
    \For{ $\mathbf{p}_{\rm nei} \in N_k(\mathbf{p}_{\rm cur})$}{
       \If{$\mathit{visited}[\mathbf{p}_{\rm nei}]$ \rm{has been marked}} {
         \textbf{continue} \;
       }

       \eIf{${\rm size}(\mathbf{p}_{\rm nei}) > {\rm size}(\mathbf{p}_{\rm cur})$} {
          \If{${\rm size}(\mathbf{p}_{\rm nei}) - {\rm size}(\mathbf{p}_{\rm cur}) > |\mathbf{p}_{\rm nei} - \mathbf{p}_{\rm cur}| $ } {
           ${\rm size}(\mathbf{p}_{\rm nei}) = {\rm size}(\mathbf{p}_{\rm cur}) + |\mathbf{p}_{\rm nei} - \mathbf{p}_{\rm cur}|$
          }
       } {
         $Q$.push($\textrm{size}(\mathbf{p}_{cur}$) ;
       }
    }    
}
\end{algorithm}

\section{More experiments}
We attach more experiments to verify our algorithm.

\subsection{LFS estimation}

\subsubsection{Sampling density} 
\begin{figure}[t!]
\centering
\subfigure[low density]{
    \begin{minipage}[h]{0.29\columnwidth}
    \includegraphics[width=\linewidth,trim={250 180 250 180},clip]{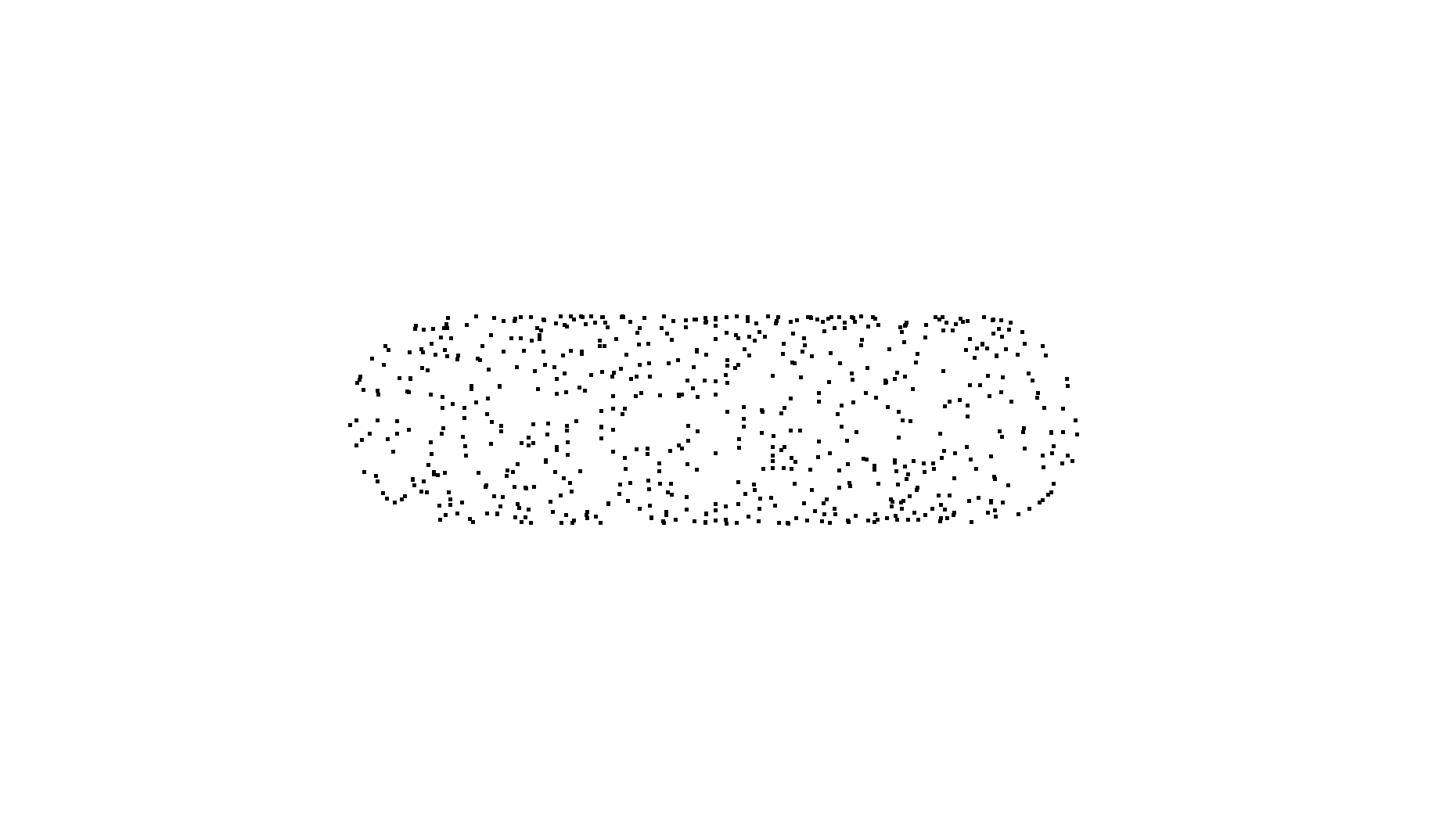}\vspace{0pt} 
    \includegraphics[width=\linewidth,trim={250 180 250 180},clip]{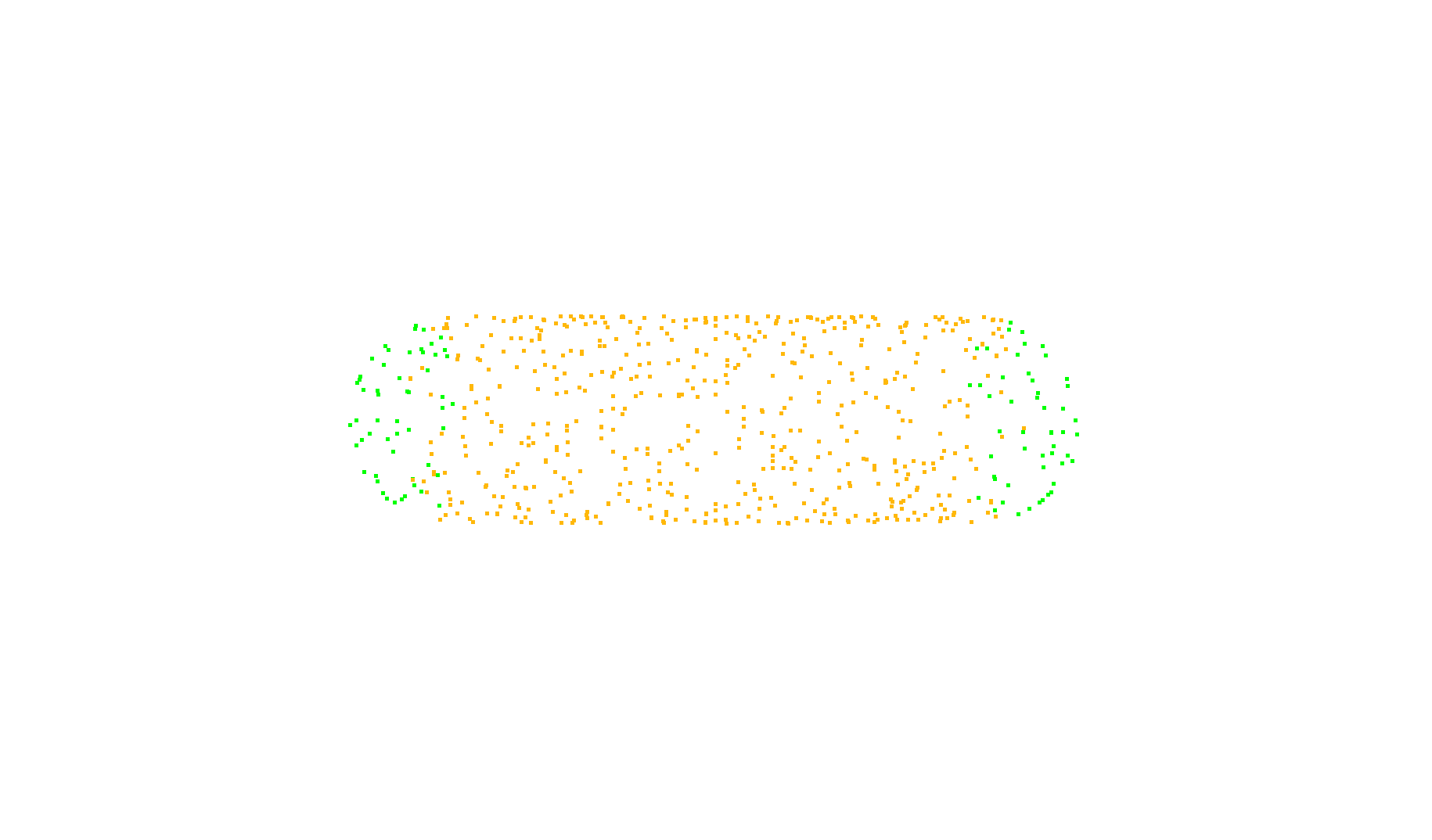}\vspace{0pt} 
    \includegraphics[width=\linewidth,trim={250 180 250 180},clip]{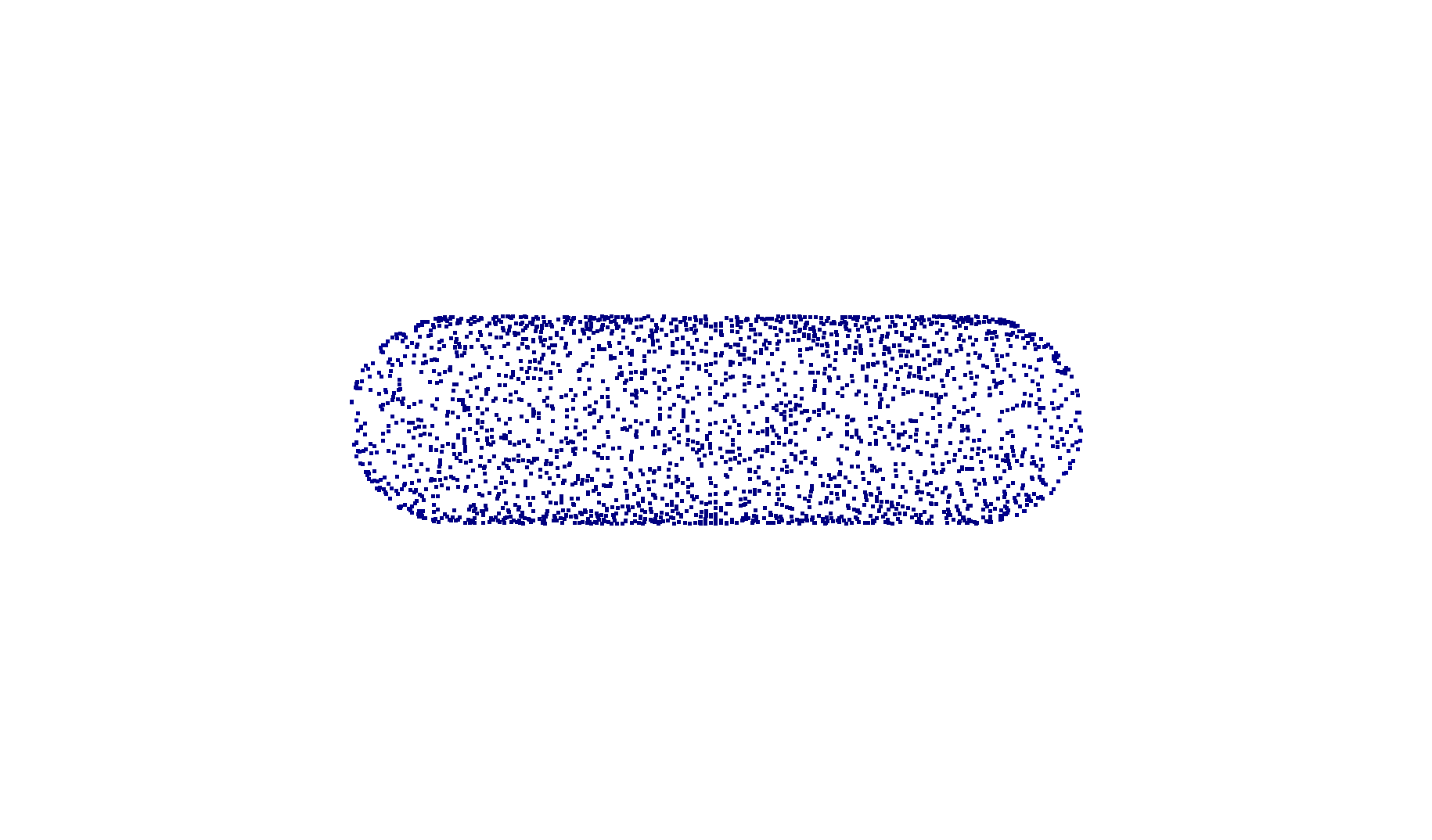}\vspace{0pt} 
    \includegraphics[width=\linewidth,trim={250 180 250 180},clip]{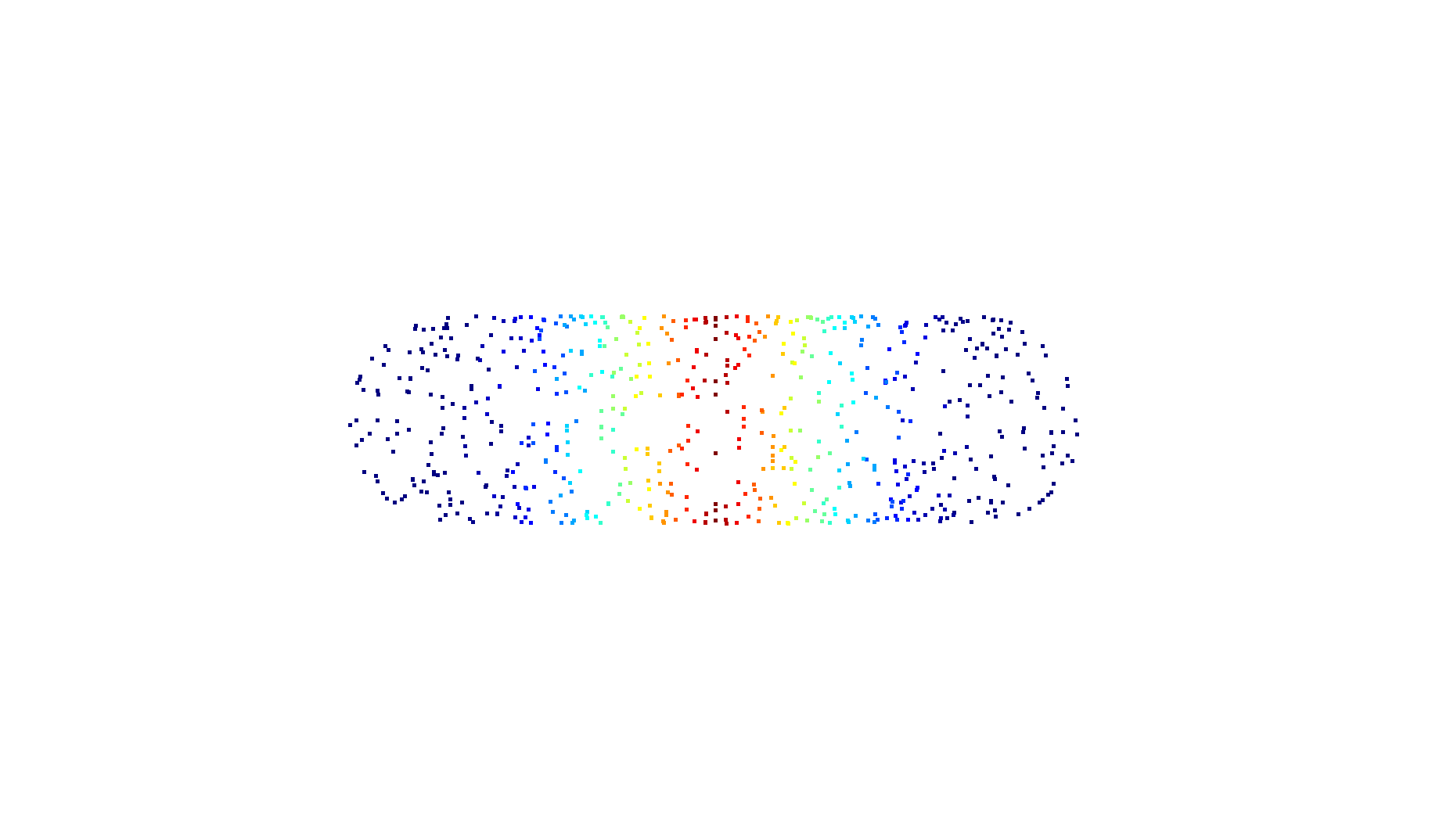}\vspace{0pt} 
    \includegraphics[width=\linewidth]{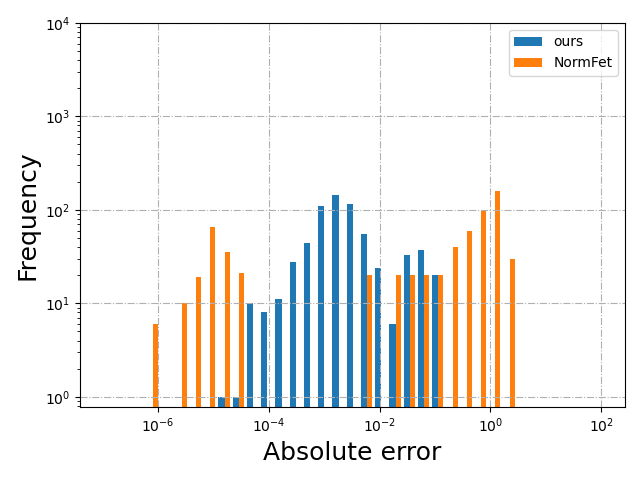}\vspace{0pt} 
    \end{minipage}
}
\subfigure[moderate density]{
    \begin{minipage}[h]{0.29\columnwidth}
    \includegraphics[width=\linewidth,trim={250 180 250 180},clip]{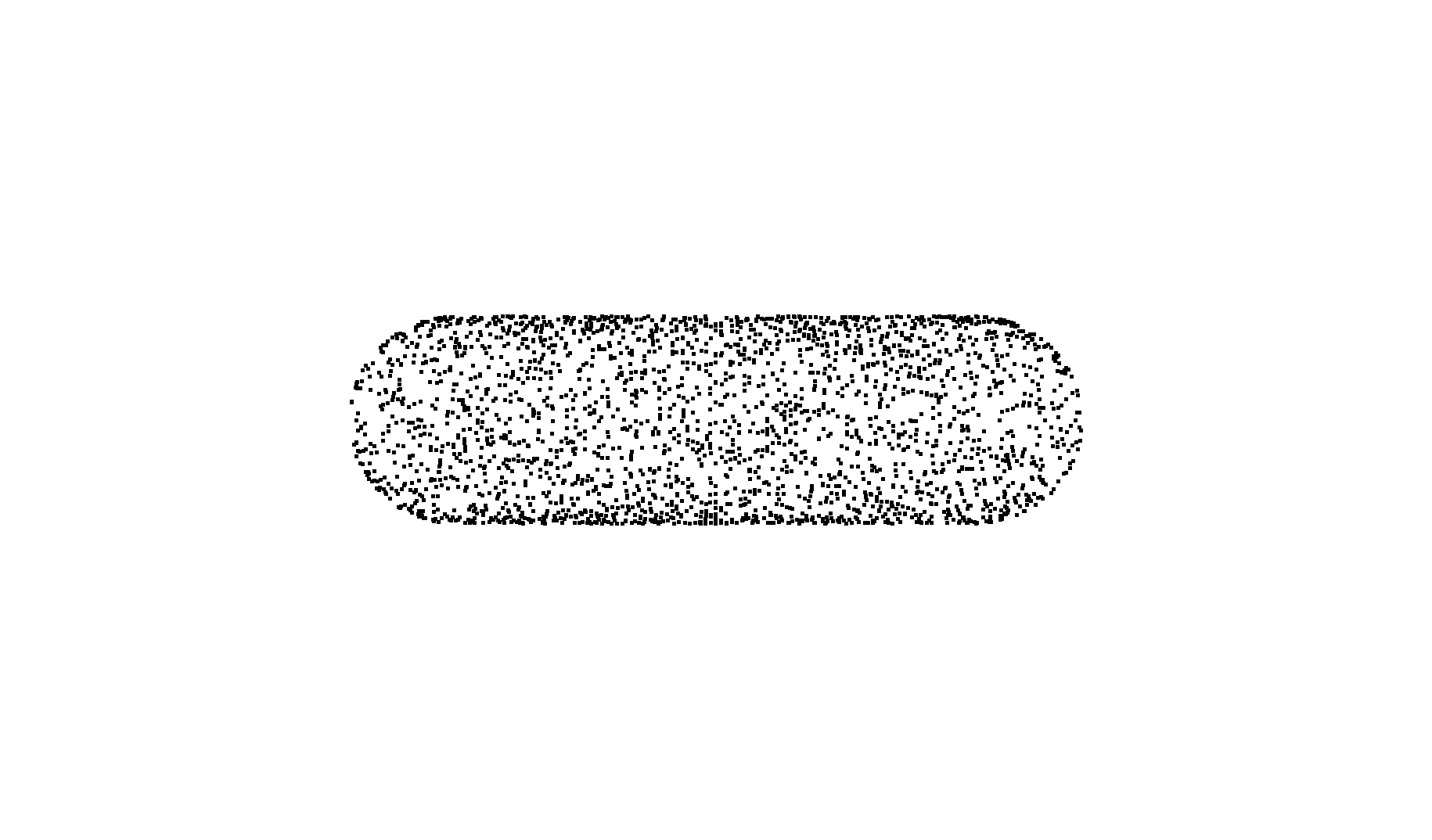}\vspace{0pt} 
    \includegraphics[width=\linewidth,trim={250 180 250 180},clip]{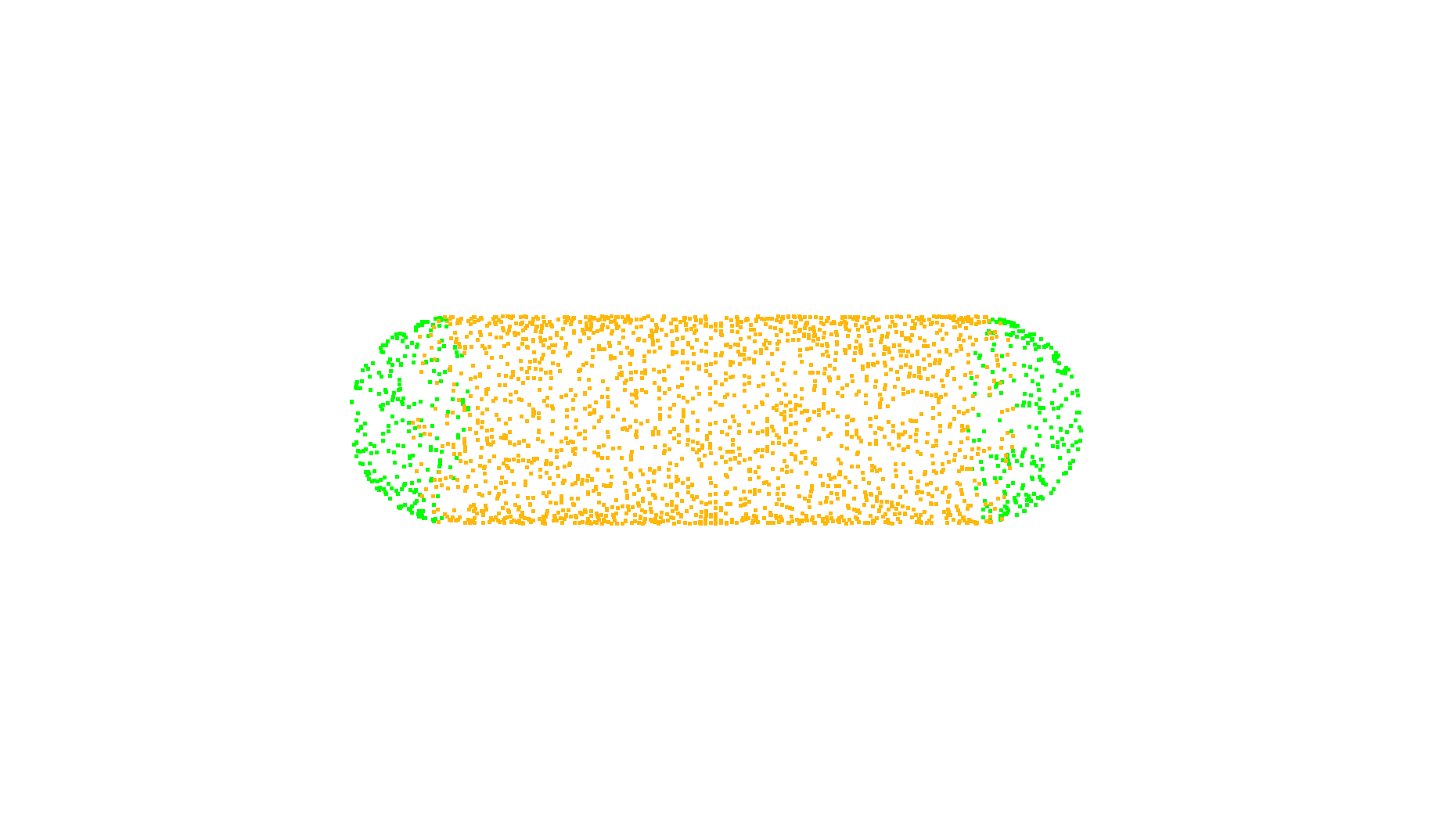}\vspace{0pt} 
    \includegraphics[width=\linewidth,trim={250 180 250 180},clip]{fig/lfs-capsule-cmp/capsule-2610-lfs-abs-err.png}\vspace{0pt} 
    \includegraphics[width=\linewidth,trim={250 180 250 180},clip]{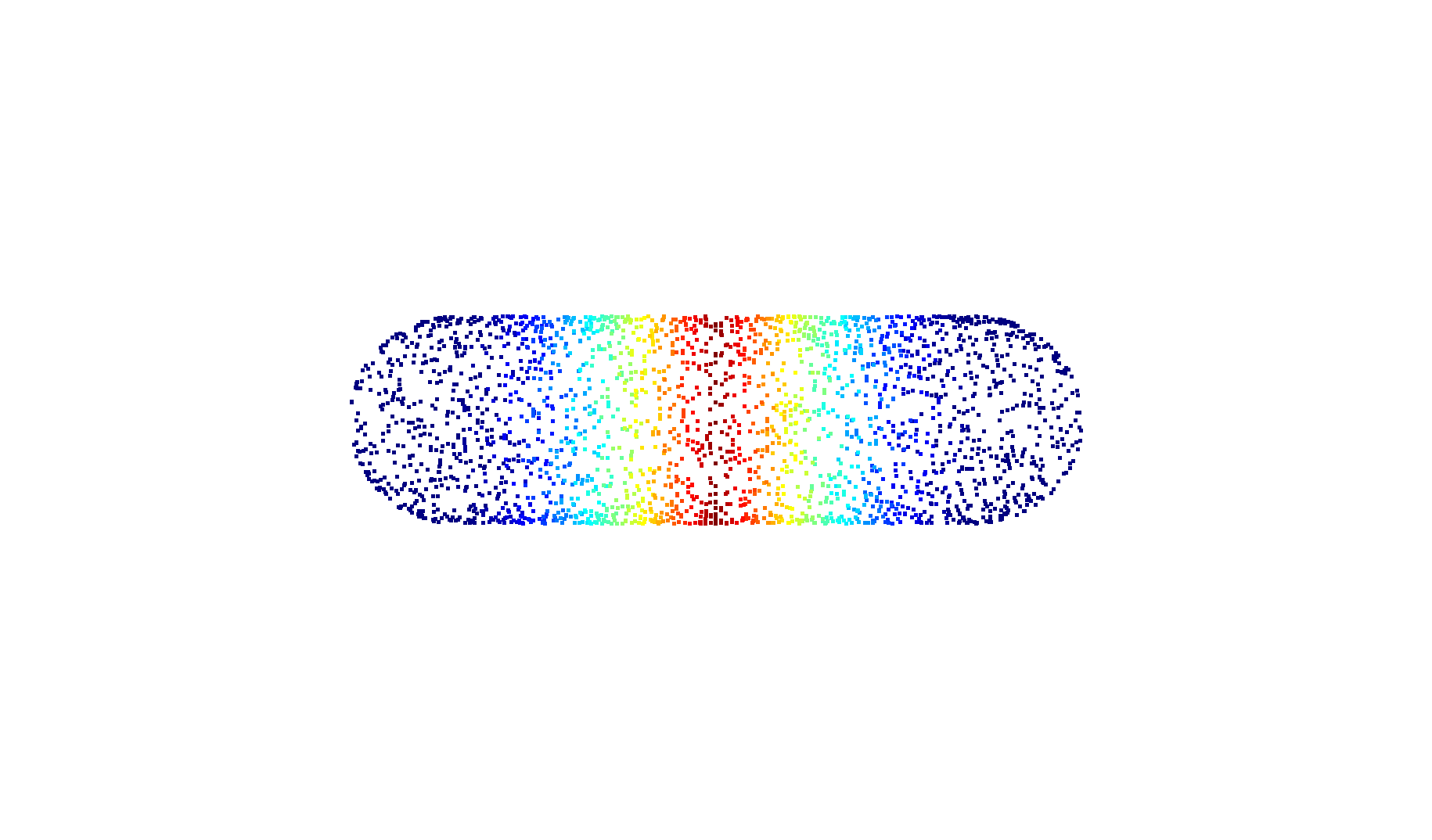}\vspace{0pt} 
    \includegraphics[width=\linewidth]{fig/lfs-capsule-cmp/capsule-648-abs-err-hist-cmp.png}\vspace{0pt} 
    \end{minipage}
}
\subfigure[high density]{
    \begin{minipage}[h]{0.29\columnwidth}
    \includegraphics[width=\linewidth,trim={250 180 250 180},clip]{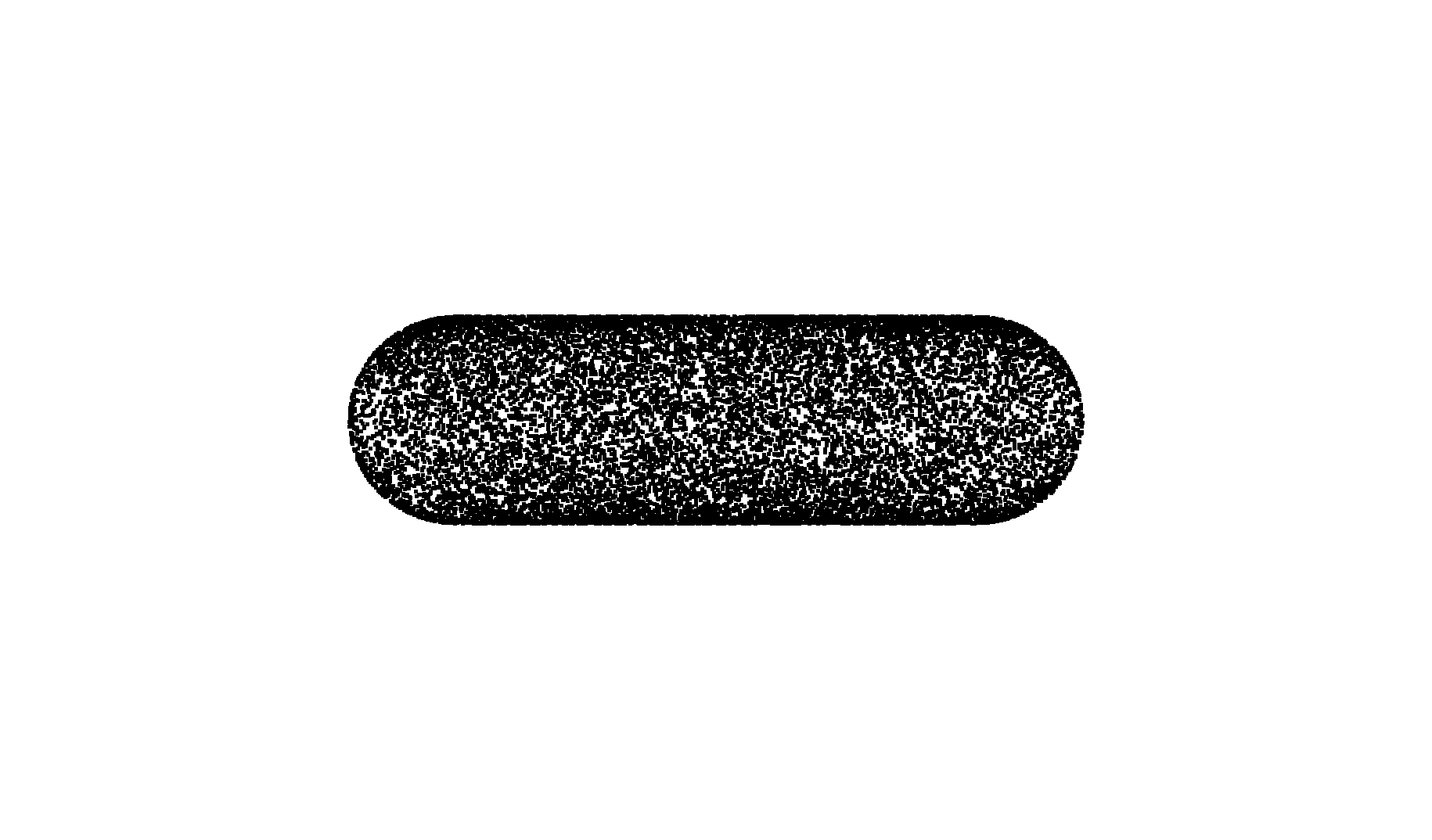}\vspace{0pt} 
    \includegraphics[width=\linewidth,trim={250 180 250 180},clip]{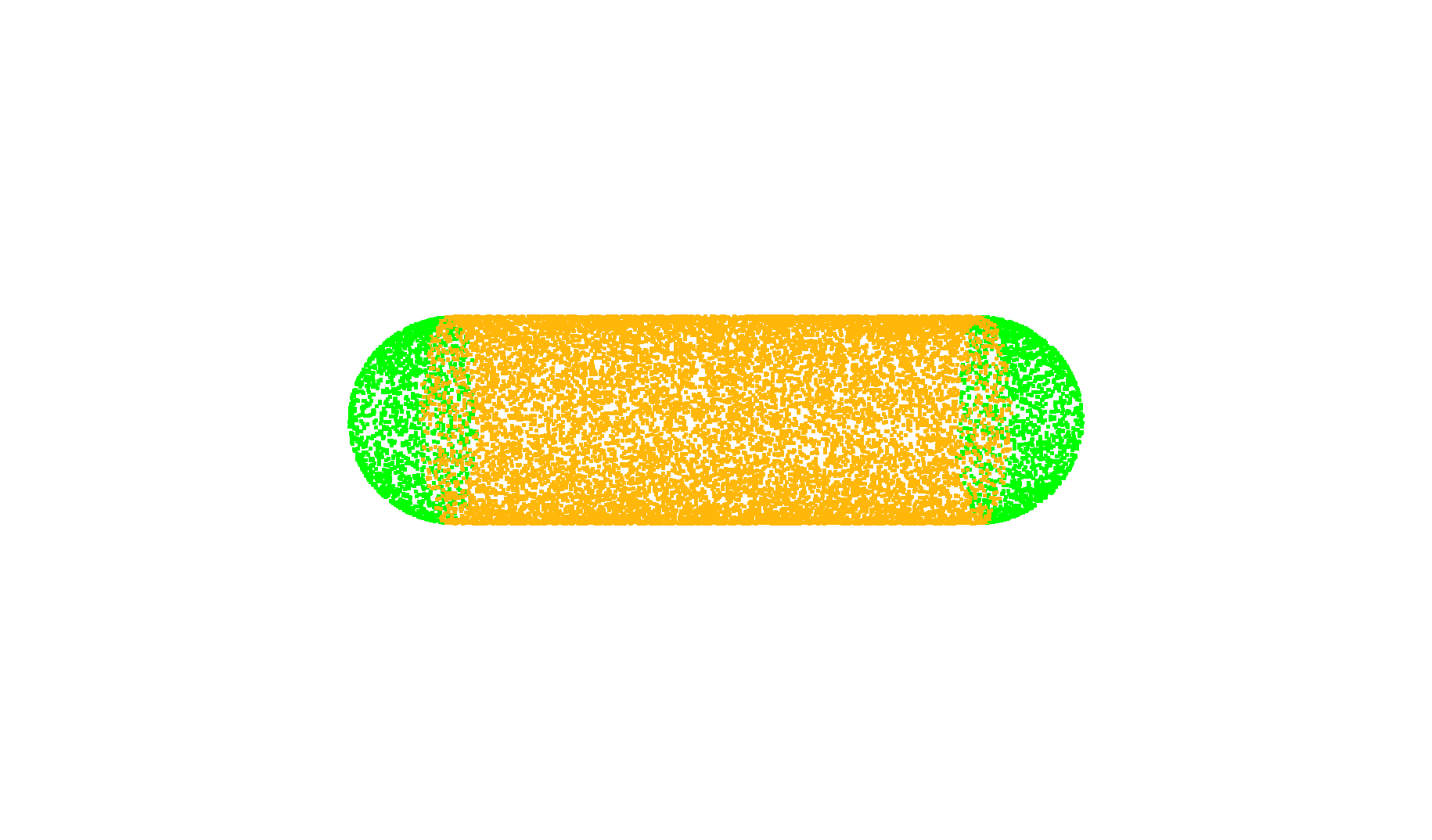}\vspace{0pt} 
    \includegraphics[width=\linewidth,trim={250 180 250 180},clip]{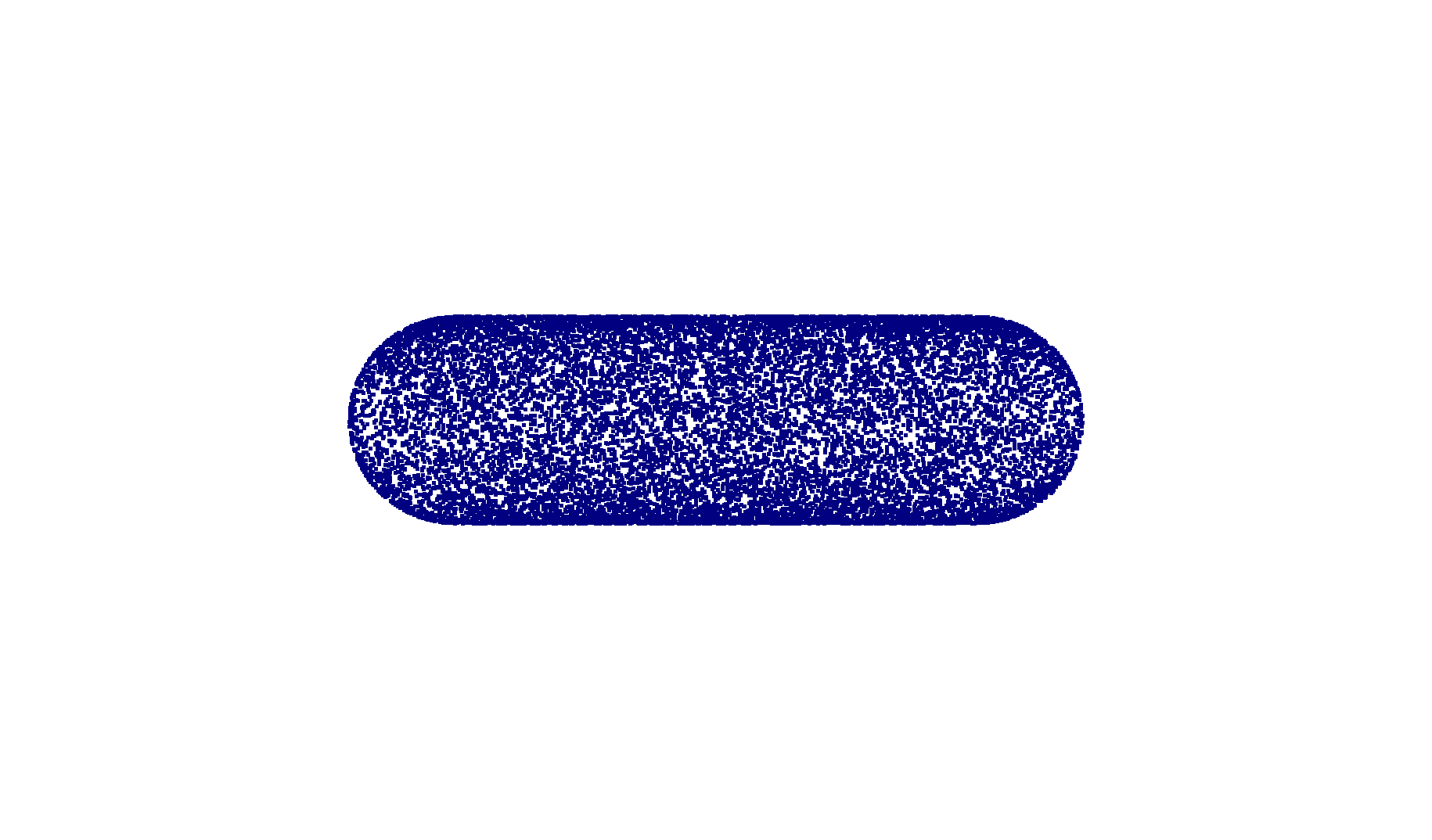}\vspace{0pt} 
    \includegraphics[width=\linewidth,trim={250 180 250 180},clip]{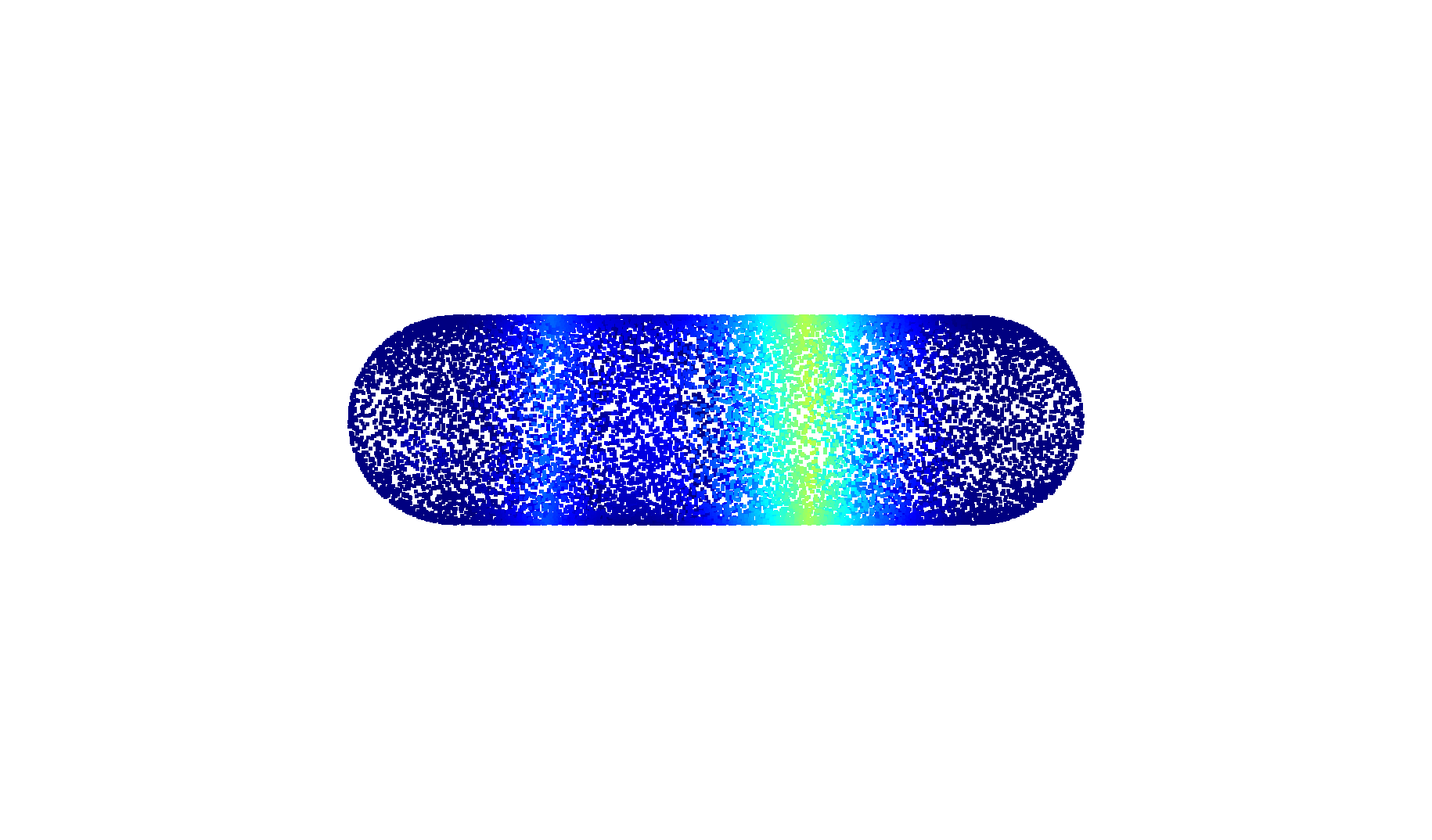}\vspace{0pt} 
    \includegraphics[width=\linewidth]{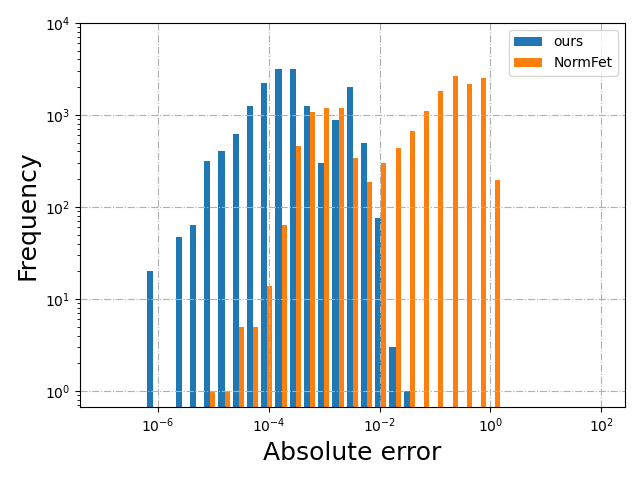}\vspace{0pt} 
    \end{minipage}
}
\\
\includegraphics[width=\linewidth,trim={30 0 30 0},clip]{fig/jet-colorbar.png}\vspace{0pt} 
\caption{LFS estimation with varying sampling density. 
From left to right: 648, 2,610, and 16,374 sample points. 
First row: input point clouds with increasing sampling density. 
Second row: LFS types: orange for diameter-based and green for curvature-based.
Third row: estimated LFS error map for our algorithm, using the jet color ramp.
Fourth row: estimated LFS error map for NormFet. 
Fifth row: distribution of estimated LFS errors for all sample points. }
\label{lfs-capusle-cmp}
\end{figure}

We now evaluate the performance of our LFS estimation method (without smoothing) under different sampling densities and compare it with NormFet. To conduct this experiment, we construct a capsule model with ground-truth LFS, with a cylinder and two half-spheres whose radii are known a priori. We then randomly sample points on the capsule with three different densities (see Table~\ref{tab-capsule-cmp} and Fig.~\ref{lfs-capusle-cmp}). The table indicates that our algorithm provides a reasonable LFS estimation in all three sampling conditions, while NormFet is more sensitive to the density of the point cloud.

\begin{table}[t!]
\begin{center} 
\scalebox{0.82}{
\begin{tabular}{l| cc| cc| cc}
\toprule[1.2pt]
Capsule &\multicolumn{2}{c|}{Low (\#648)} &  \multicolumn{2}{c|}{Moderate (\#2610)} & \multicolumn{2}{c}{High (\#16374)}  \\
 & mean($\downarrow$) & max($\downarrow$) & mean($\downarrow$) & max($\downarrow$) & mean($\downarrow$) & max($\downarrow$)  \\
\hline
NormFet & 5.216E-1 & 1.693 & 5.178E-1 & 1.692 & 2.136E-1 & 9.550E-1 \\
\hdashline
$(0\degree,1)$ & 7.130E-2 & 1.030 & 1.698E-1 & 1.030 & 1.848E-1 & 1.004 \\
$(5\degree,10)$ & 2.239E-2 & 9.743E-1 & 9.808E-3 & 9.915E-1 & 1.190E-3 & 9.978E-1 \\
$(5\degree,100)$ & 1.630E-2 & 9.743E-1 & 5.272E-3 & 9.779E-1 & \textbf{7.781E-4} & \textbf{2.523E-2} \\
$(10\degree,10)$ & 1.477E-2 & 9.743E-1 & 4.311E-3 & \textbf{6.510E-2} & 1.068E-3 & \textbf{2.523E-2} \\
$(10\degree,100)$ & \textbf{1.023E-2} & \textbf{1.229E-1} & \textbf{4.168E-3} & \textbf{6.510E-2} & 8.655E-4 & \textbf{2.523E-2} \\
\bottomrule[1.2pt]
\end{tabular}}
\end{center}
\caption{LFS estimation on the capsule point cloud with increasing sampling density. ($D\degree$, N) denotes the parameters of the our dual cone search approach. $D$ denotes the apex angle and $N$ denotes the number of rays.}
\label{tab-capsule-cmp}
\end{table}

\subsubsection{Dual cone search}  
\begin{figure}[t!]
\centering
\subfigure[$(0\degree,1)$]{
    \begin{minipage}[h]{0.29\columnwidth}
    \includegraphics[width=\linewidth,trim={380 180 380 180},clip]{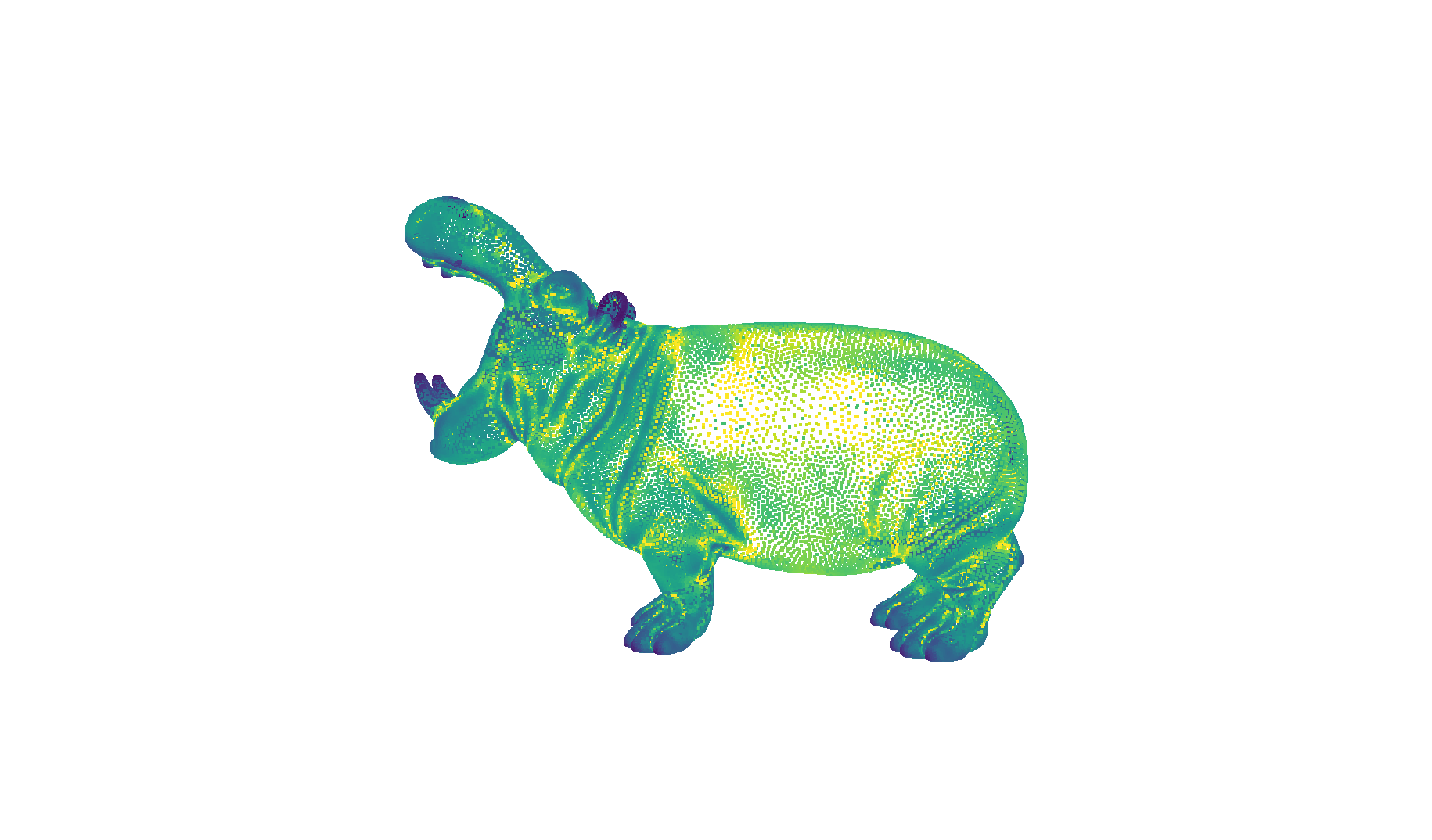}\vspace{5pt} 
    \end{minipage}
}
\subfigure[$(15\degree,15)$]{
    \begin{minipage}[h]{0.29\columnwidth}
    \includegraphics[width=\linewidth,trim={380 180 380 180},clip]{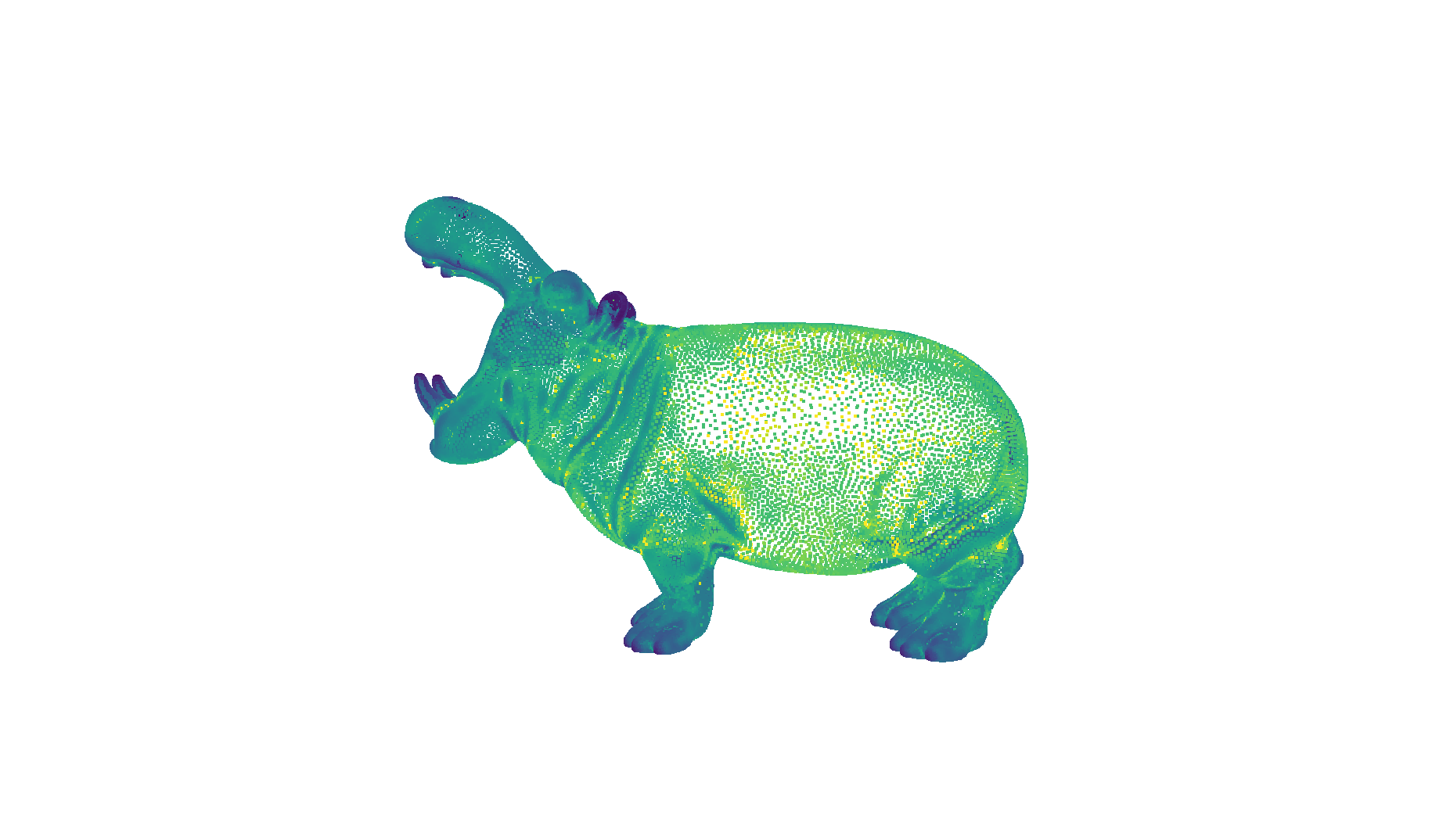}\vspace{5pt} 
    \end{minipage}
}
\subfigure[$(30\degree,30)$]{
    \begin{minipage}[h]{0.29\columnwidth}
    \includegraphics[width=\linewidth,trim={380 180 380 180},clip]{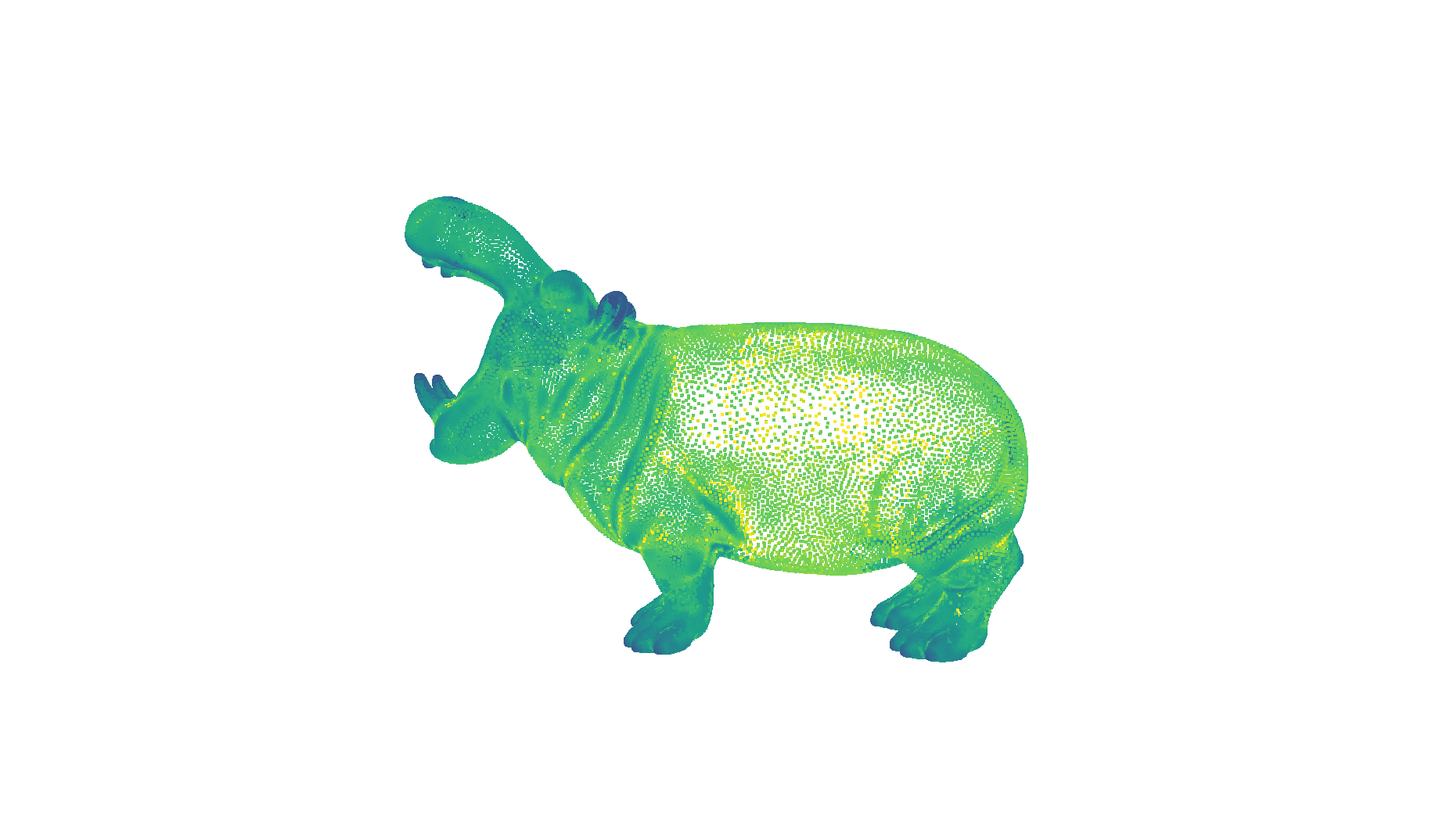}\vspace{5pt} 
    \end{minipage}
}
\\
\subfigure[$(30\degree,1)$]{
    \begin{minipage}[h]{0.29\columnwidth}
    \includegraphics[width=\linewidth,trim={380 180 380 180},clip]{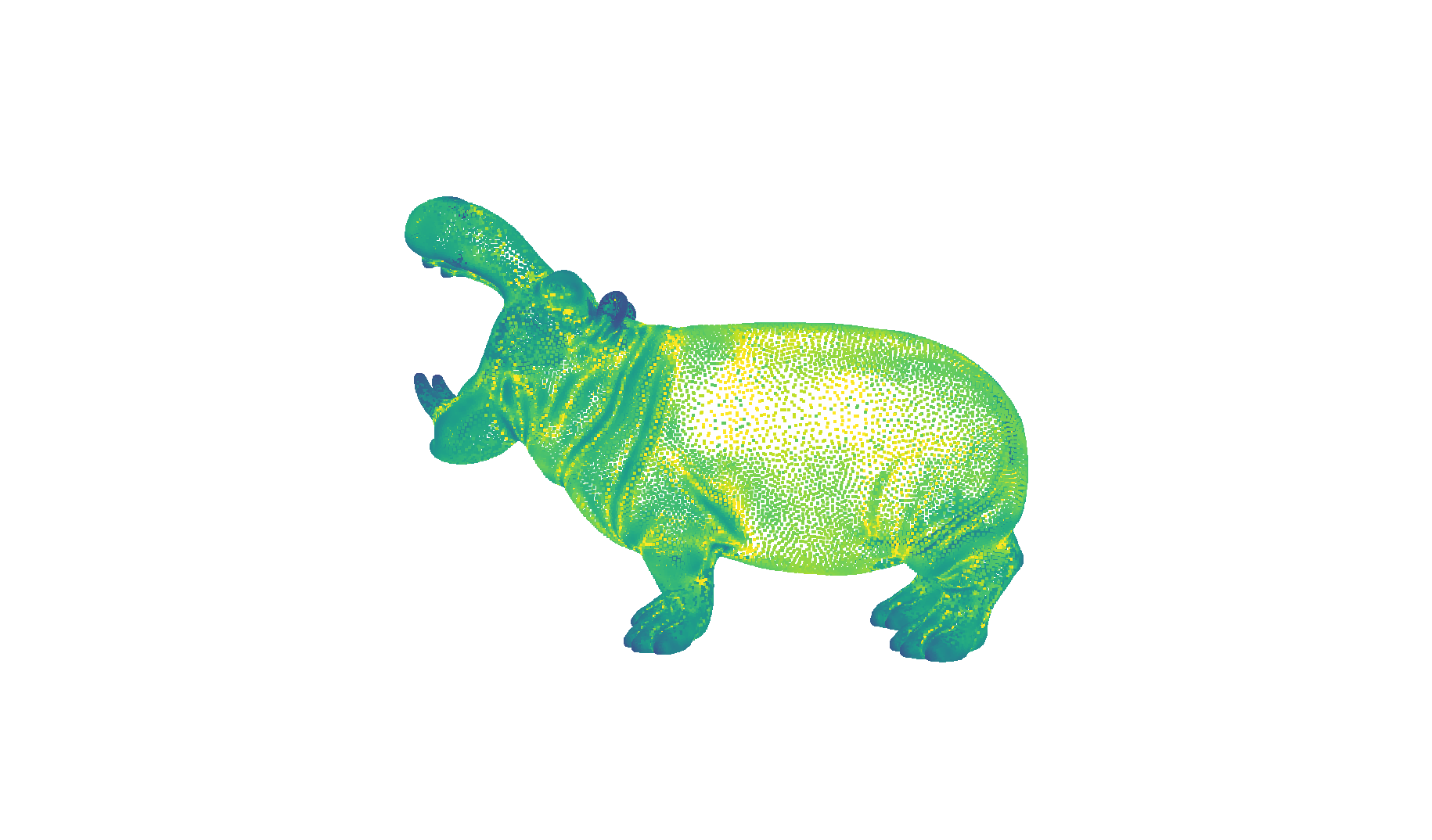}\vspace{5pt} 
    \end{minipage}
}
\subfigure[($30^{\circ}$, 15)]{
    \begin{minipage}[h]{0.29\columnwidth}
    \includegraphics[width=\linewidth,trim={380 180 380 180},clip]{fig/hippo-lfs-ablation/hippo-lfs-raw-apex30ray15.png}\vspace{5pt} 
    \end{minipage}
}
\subfigure[$(30\degree,60)$]{
    \begin{minipage}[h]{0.29\columnwidth}
    \includegraphics[width=\linewidth,trim={380 180 380 180},clip]{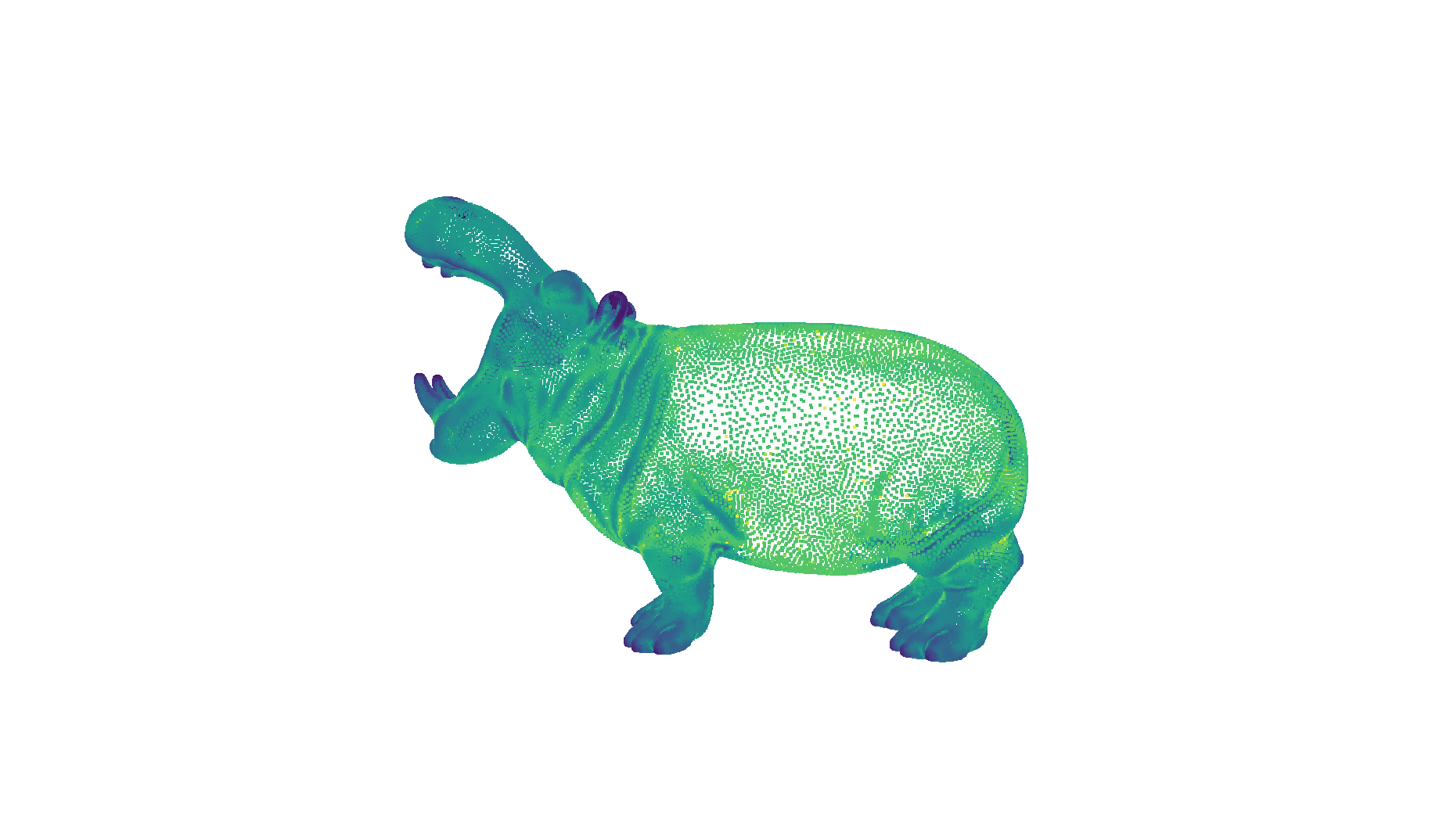}\vspace{5pt} 
    \end{minipage}
}
\\
\includegraphics[width=\linewidth,trim={30 0 30 0},clip]{fig/viridis-colorbar.png}\vspace{0pt} 
\caption{Visual ablation study on the hippo point cloud (\#58,188) using different settings of the dual cone search. We vary the apex angle ($D^{\circ}$) and the number of rays ($N$) to investigate their impact on the LFS estimation. The LFS values are visualized using the viridis color ramp without any smoothing applied. We perform min-max normalization in the log scale space to map the raw estimated LFS values for each hippo separately into the range $[0,1]$. Note that we lack ground truth LFS for the hippo. }
\label{hippo-lfs-ablation}
\end{figure}

In order to evaluate the impact of parameters of the dual cone search, we conduct experiments with different settings for the apex angle $\theta$ and the number of rays cast $N$ on the capsule and the hippo point cloud. The dual cone search devised to estimate the local shape diameter at a sample point may not find the antipodal points when the parameters are not set properly, resulting in LFS estimation errors. Table \ref{tab-capsule-cmp} records estimation errors for the capsule point cloud with increasing density, apex angles, and numbers of cast rays. Shooting a single ray along the normal direction $(0\degree,1)$ always results in large errors for all sampling conditions. Increasing angle $\theta$ and the number of rays $N$ helps reduce the errors significantly, but using small $\theta$ and $N$ can still find the correct antipodal points. As the maximum absolute error remains the same, we do not need to set large $\theta$ and $N$. For example, for the capsule with high density (16,374 points), $(5\degree,100)$ and $(10\degree,100)$ have the same absolute errors, even if $\theta$ differs. Similarly, $(10\degree,10)$ and $(10^\degree,100)$ also have the same maximum absolute error, even though the number of rays differs significantly. Fig.~\ref{hippo-lfs-ablation} shows the LFS estimation on the hippo point cloud. The hippo point cloud is sparse in the body and dense in other parts, making it challenging to find the correct antipodal point for the body part if only shooting a single ray. Increasing $\theta$ and $N$ helps obtain a smoother color map. 

\subsection{Complex topology}  
\begin{figure}[t!]
\centering
 \includegraphics[width=\linewidth,trim={0 10 0 0},clip]{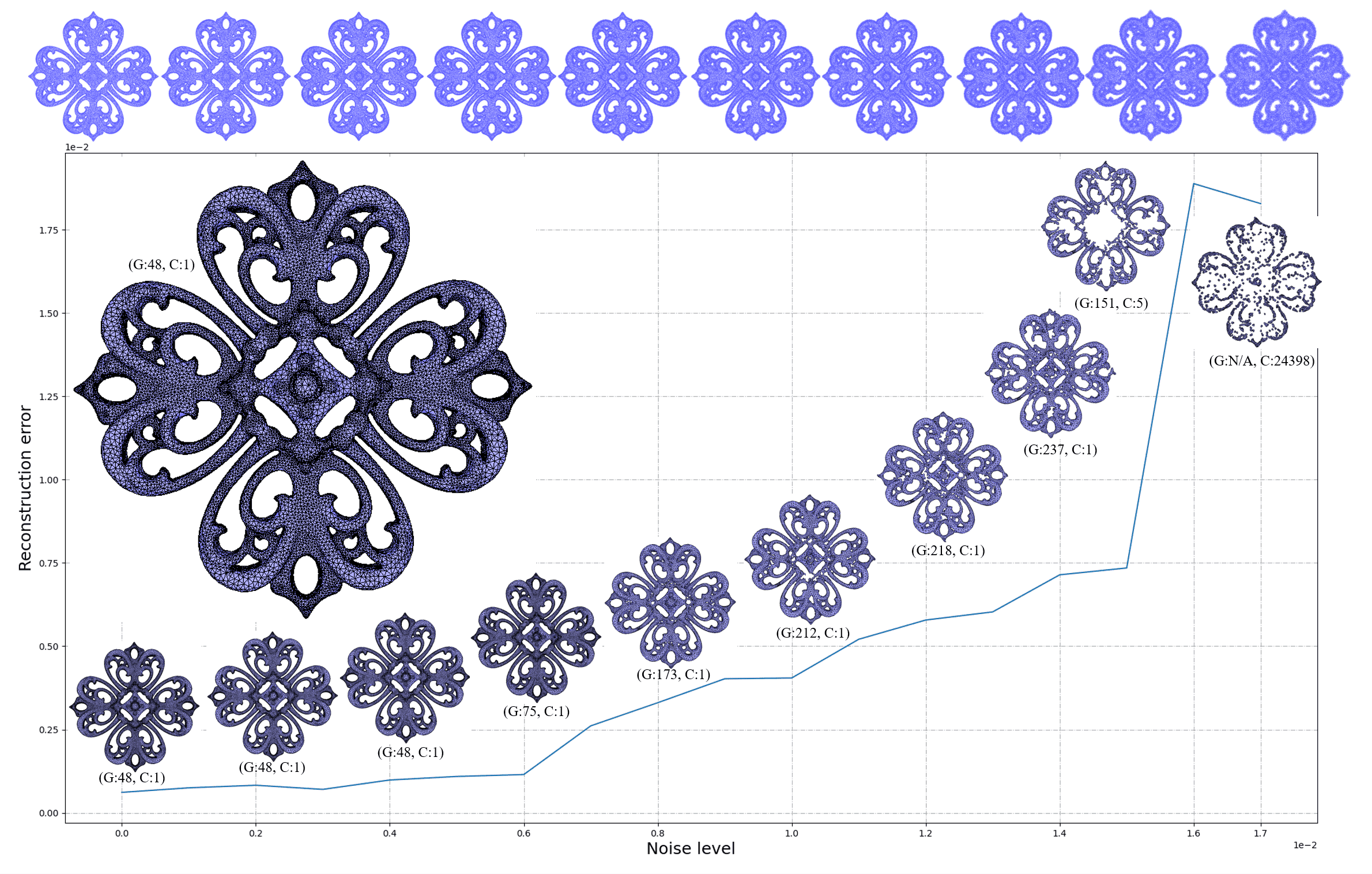}\vspace{0pt}
\caption{Reconstructing the filigree point cloud with increasing levels of noise. The largest mesh depicts the output for the noise-free case. The other series of meshes depicts the evolution of the reconstructed genus when noise increases. We record the genus (``G'') and number of connected components (``C'').}
\label{filigree-evolve}
\end{figure}

We evaluate the capability of our algorithm to deal with shapes with non-trivial topology, by reconstructing from the filigree point cloud. Adding increasing levels of noise translates into reconstructions that are robust to low levels of noise and degrade gracefully with noise. The high genus of the filigree is captured even when the surface becomes distorted. For high levels of noise, the output mesh breaks into multiple connected components, demonstrating the limitations of the algorithm under extreme noise conditions. Refer to Fig.~\ref{filigree-evolve} for the visual depiction.

\subsection{Sparsity and non-uniformity} We verify the robustness of our method by downsampling the Bimba model. The original Bimba model contained 49,445 points, resulting in a mesh with fine details. We employ random downsampling to generate the \emph{non-uniform} Bimba point cloud, and as the point count decreases, the level of detail reduces accordingly. Additionally, we generate a Bimba model with a dense point cloud for the head and a sparse point cloud for the body. Our method consistently produces faithful results across these variations.

\begin{figure}[t!]
\centering
 \includegraphics[width=\linewidth,trim={5 5 5 5},clip]{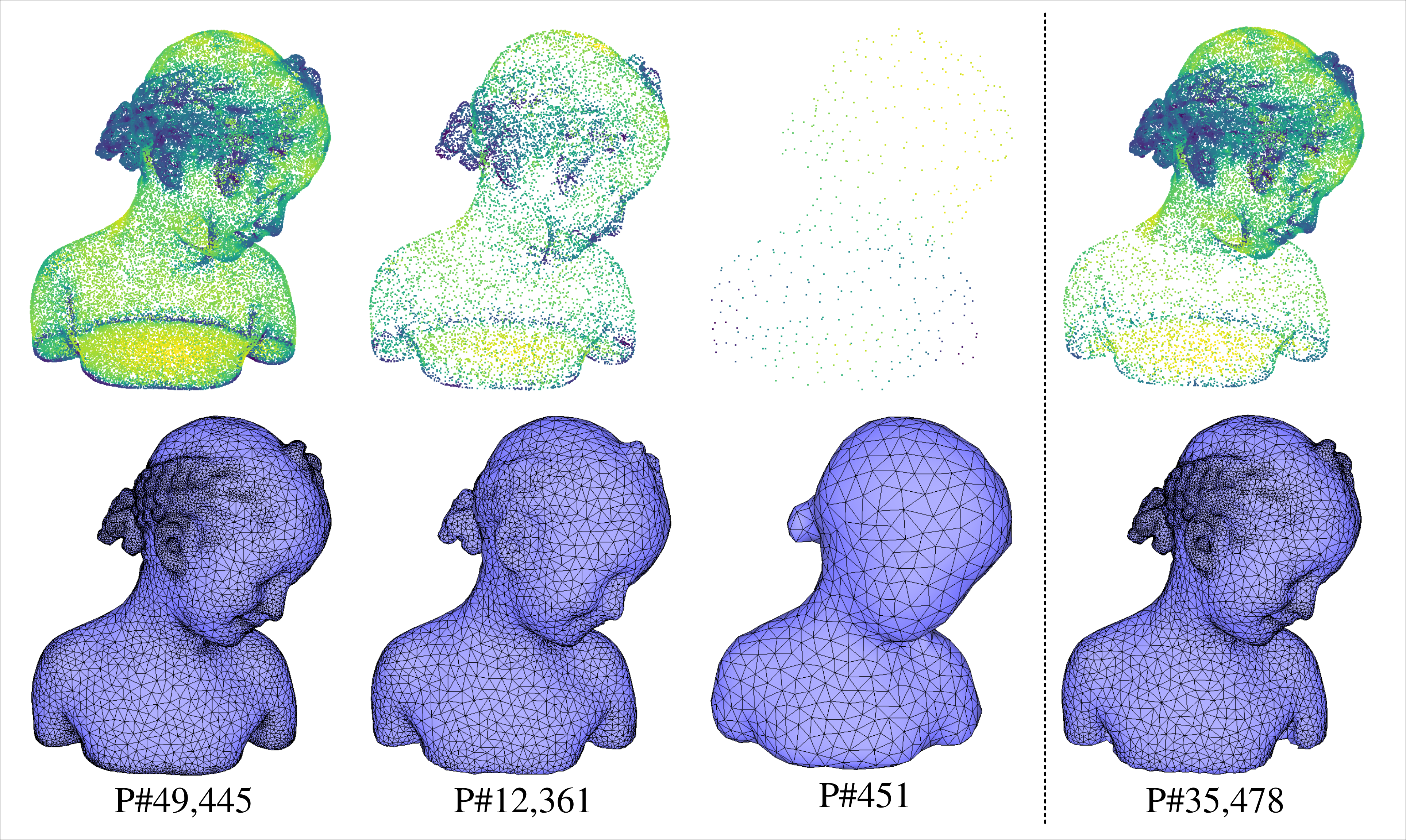}\vspace{0pt}
\caption{Reconstruction of the Bimba point clouds with decreasing number of input points. The number below each model indicates the number of points. The rightmost is the point cloud, where the head is dense and the body is sparse.}
\label{bimba-downsample}
\end{figure}

\subsection{Reconstruction Error Analysis} 
\label{error-bound}

We evaluate the reconstruction errors using the one-sided distance from the input points to the reconstructed surface mesh. For each facet in the 2D configuration, we use $\epsilon$ to denote the reconstruction error from the zero level set of the implicit function to the reconstructed surface triangle mesh. 
\begin{wrapfigure}{r}{0.25\columnwidth}
  \vskip-5pt%
  \includegraphics[width=0.25\columnwidth,trim={15 15 15 15},clip]{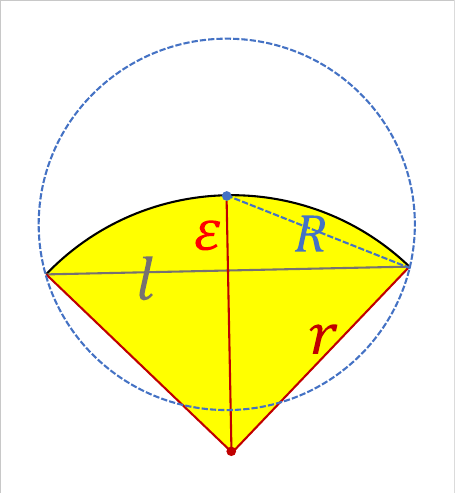}%
  \vskip-10pt
\end{wrapfigure}
We use $l$ to represent the local facet size and $R$ to represent the local surface Delaunay ball. The local surface Delaunay ball, as described in [$4$] in the main paper,
has its center on the surface and circumscribes the local facet, with no other vertices located inside the ball. By taking into consideration the local curvature radius $r$ and applying the Pythagorean theorem to this configuration, we derive an estimation for the reconstruction error $\epsilon$ in terms of $r$ and the surface Delaunay ball radius $R$. The relationship is given as $\epsilon = R^2/(2r)$.

\begin{figure}[htb!]
\centering
\subfigure[A+T+R]{
   \begin{minipage}[h]{0.29\columnwidth}
   \includegraphics[width=\linewidth,trim={320 10 320 10},clip]{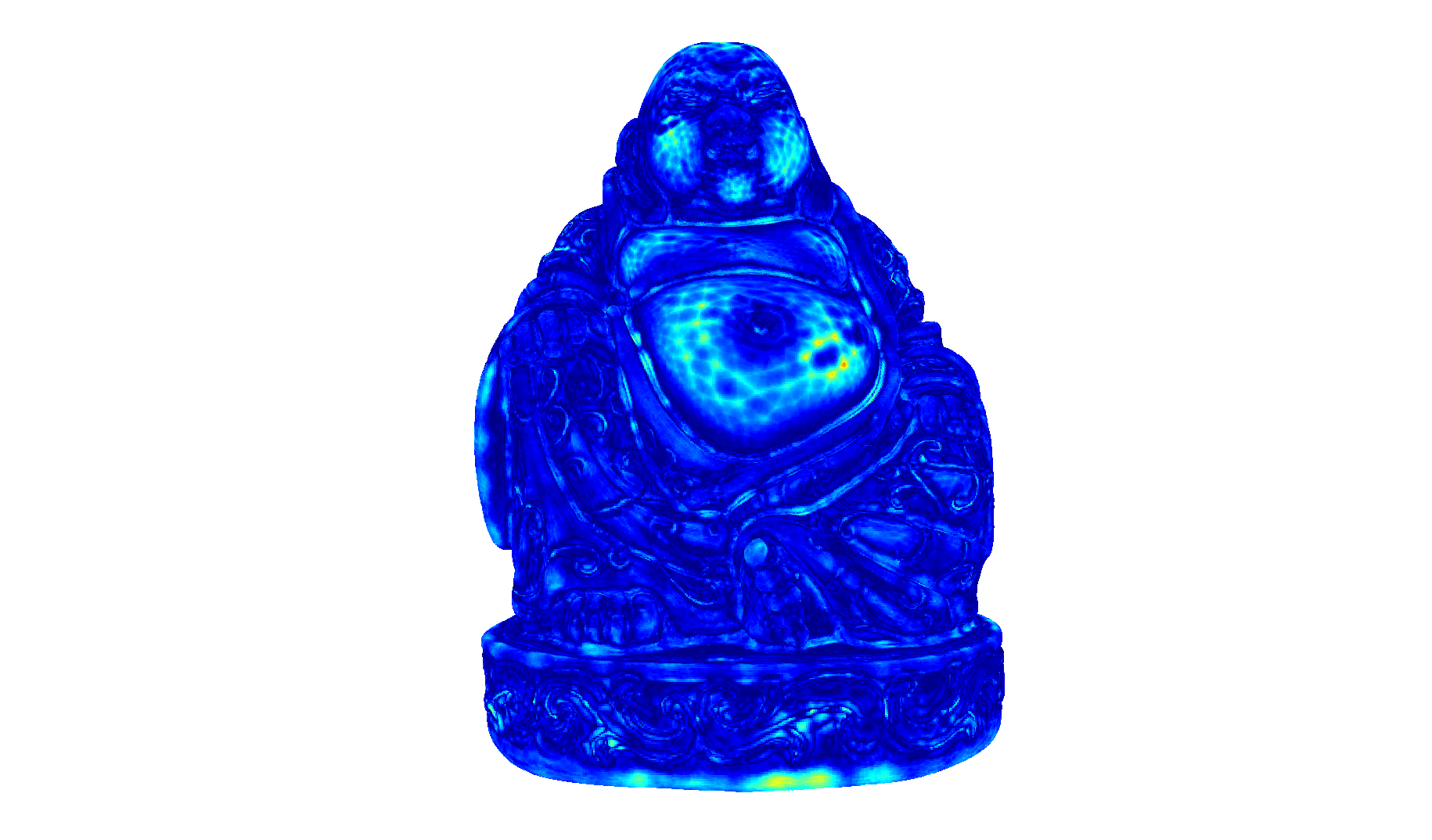}\vspace{0pt}
   \includegraphics[width=\linewidth]{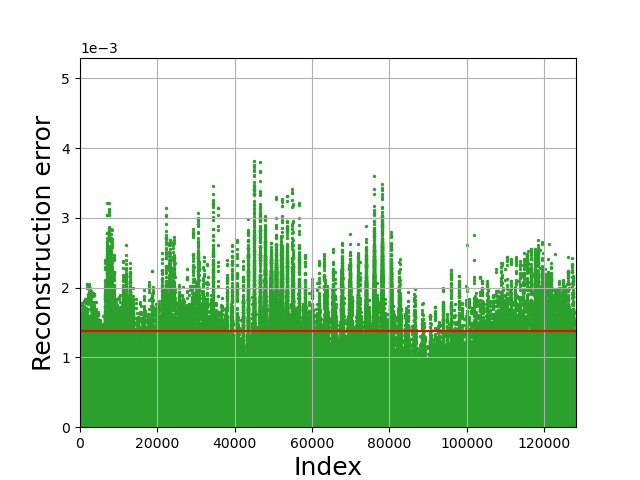}\vspace{0pt} 
   \includegraphics[width=\linewidth,trim={180 40 180 40},clip]{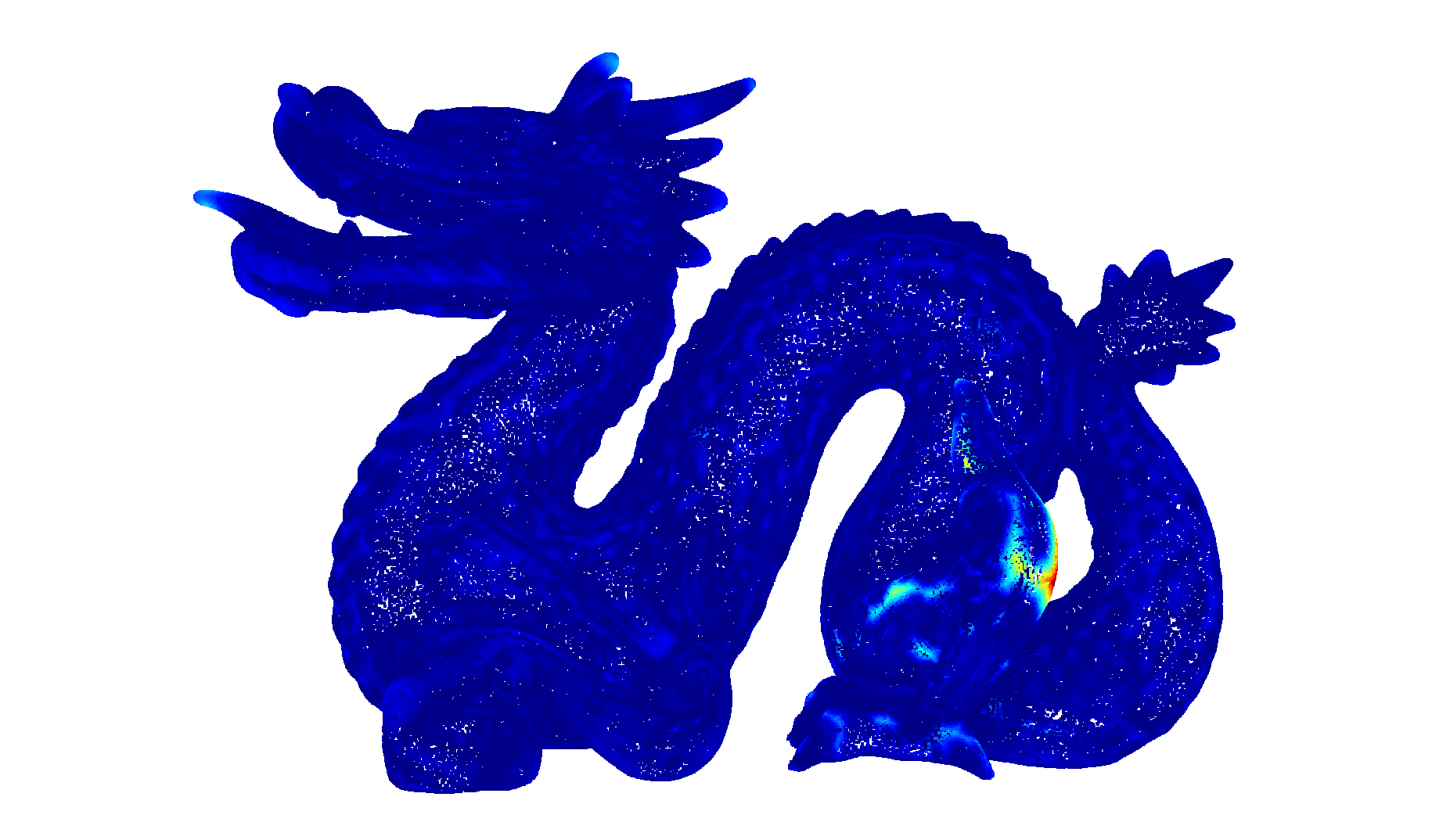}\vspace{0pt} 
   \includegraphics[width=\linewidth]{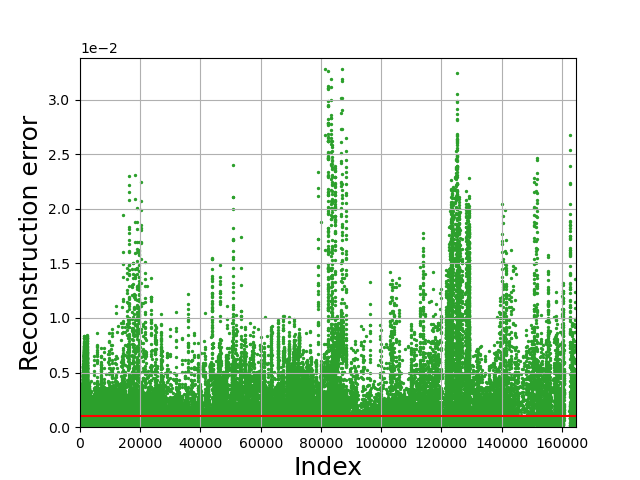}\vspace{0pt}
   \includegraphics[width=\linewidth,trim={250 10 250 10},clip]{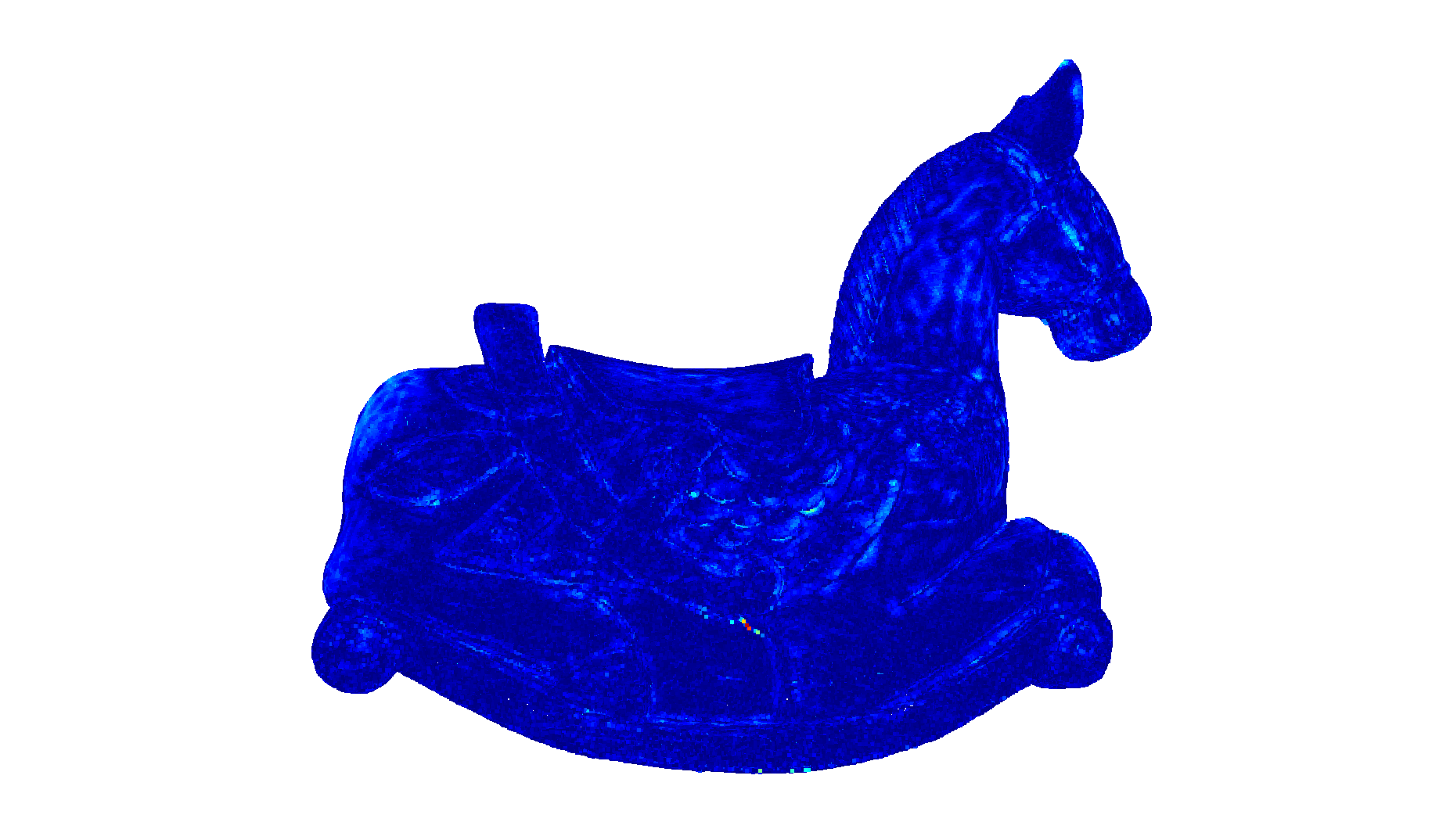}\vspace{0pt}
   \includegraphics[width=\linewidth]{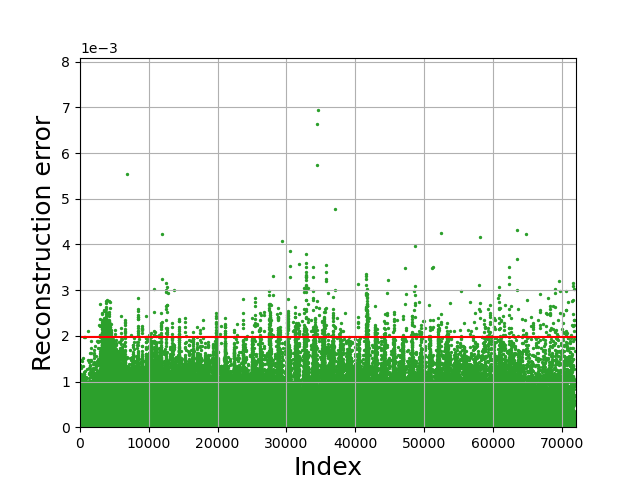}\vspace{0pt} 
   \end{minipage}
}
\subfigure[sPSR+R]{
   \begin{minipage}[h]{0.29\columnwidth}
   \includegraphics[width=\linewidth,trim={320 10 320 10},clip]{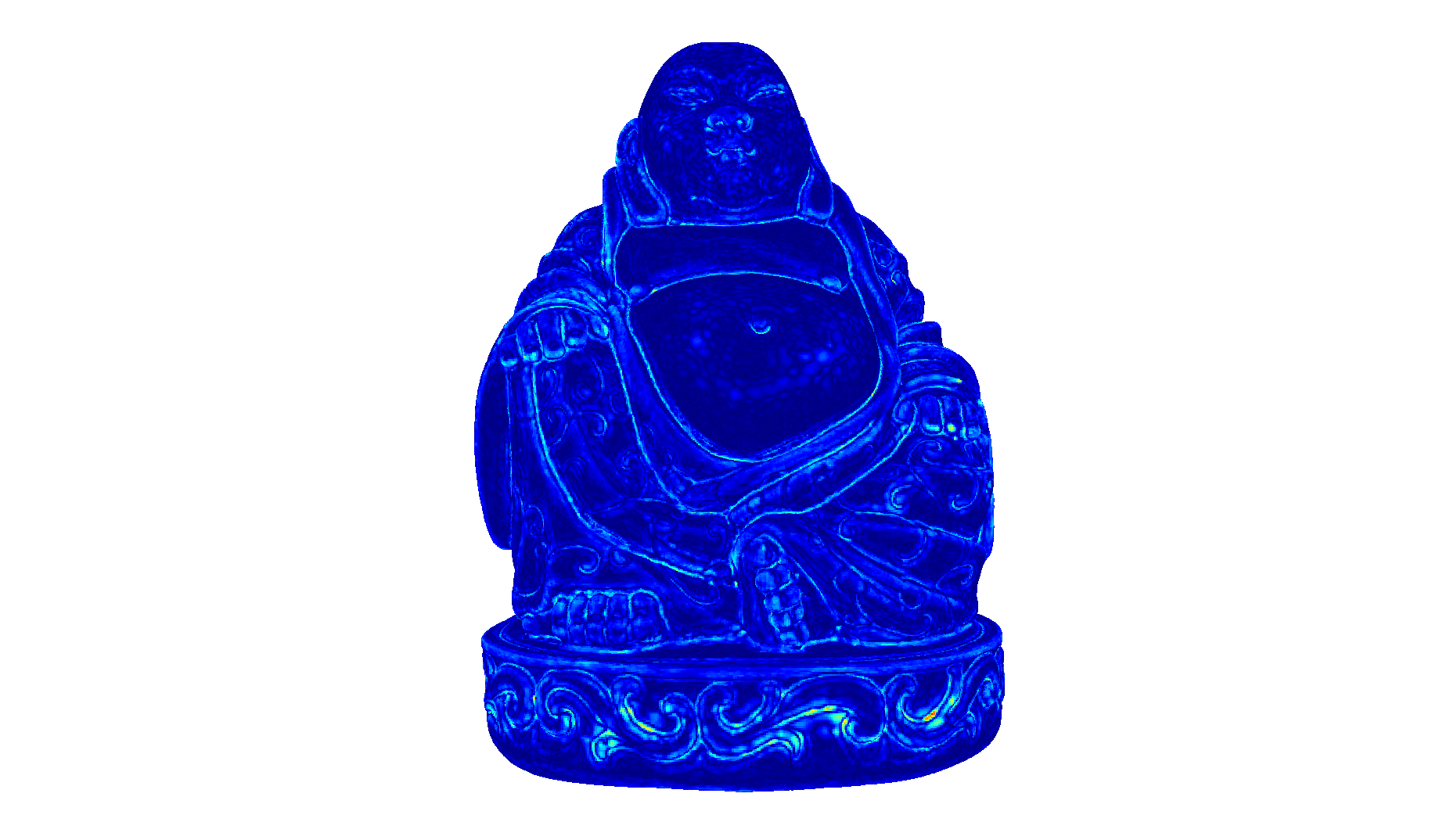}\vspace{0pt}
   \includegraphics[width=\linewidth]{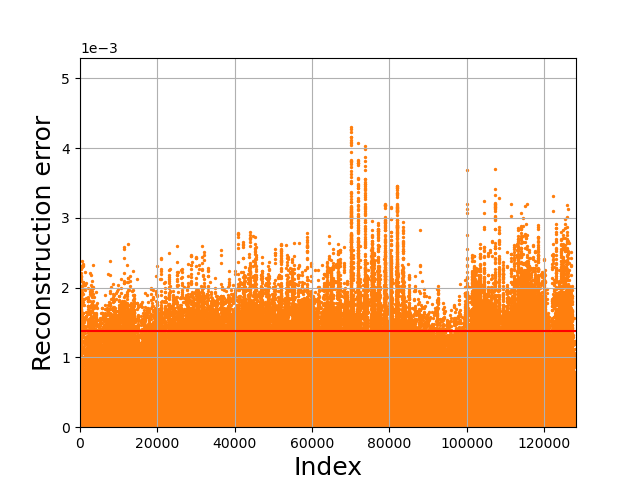}\vspace{0pt} 
   \includegraphics[width=\linewidth,trim={180 40 180 40},clip]{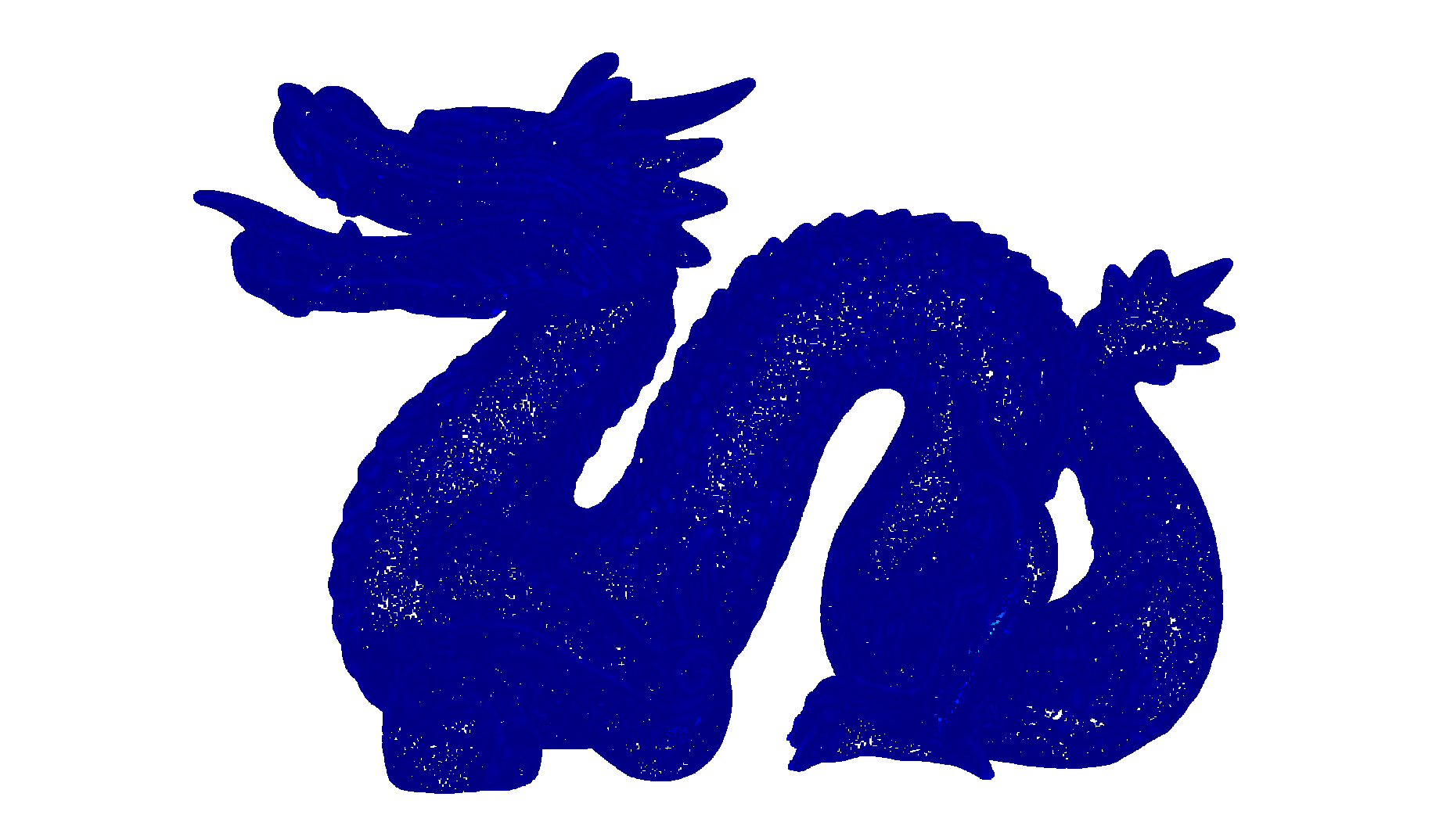}\vspace{0pt} 
   \includegraphics[width=\linewidth]{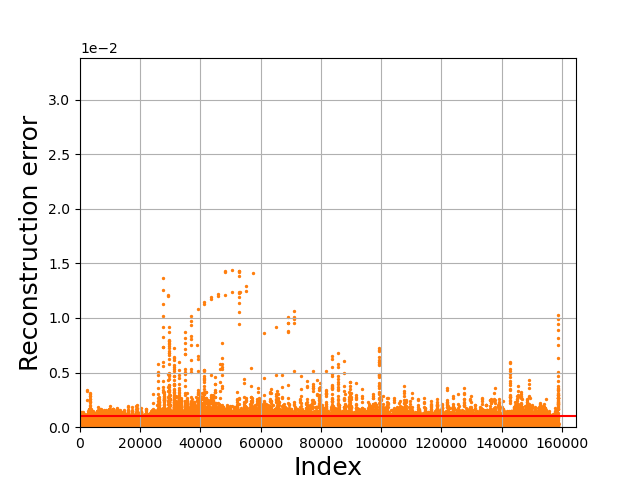}\vspace{0pt}
   \includegraphics[width=\linewidth,trim={250 10 250 10},clip]{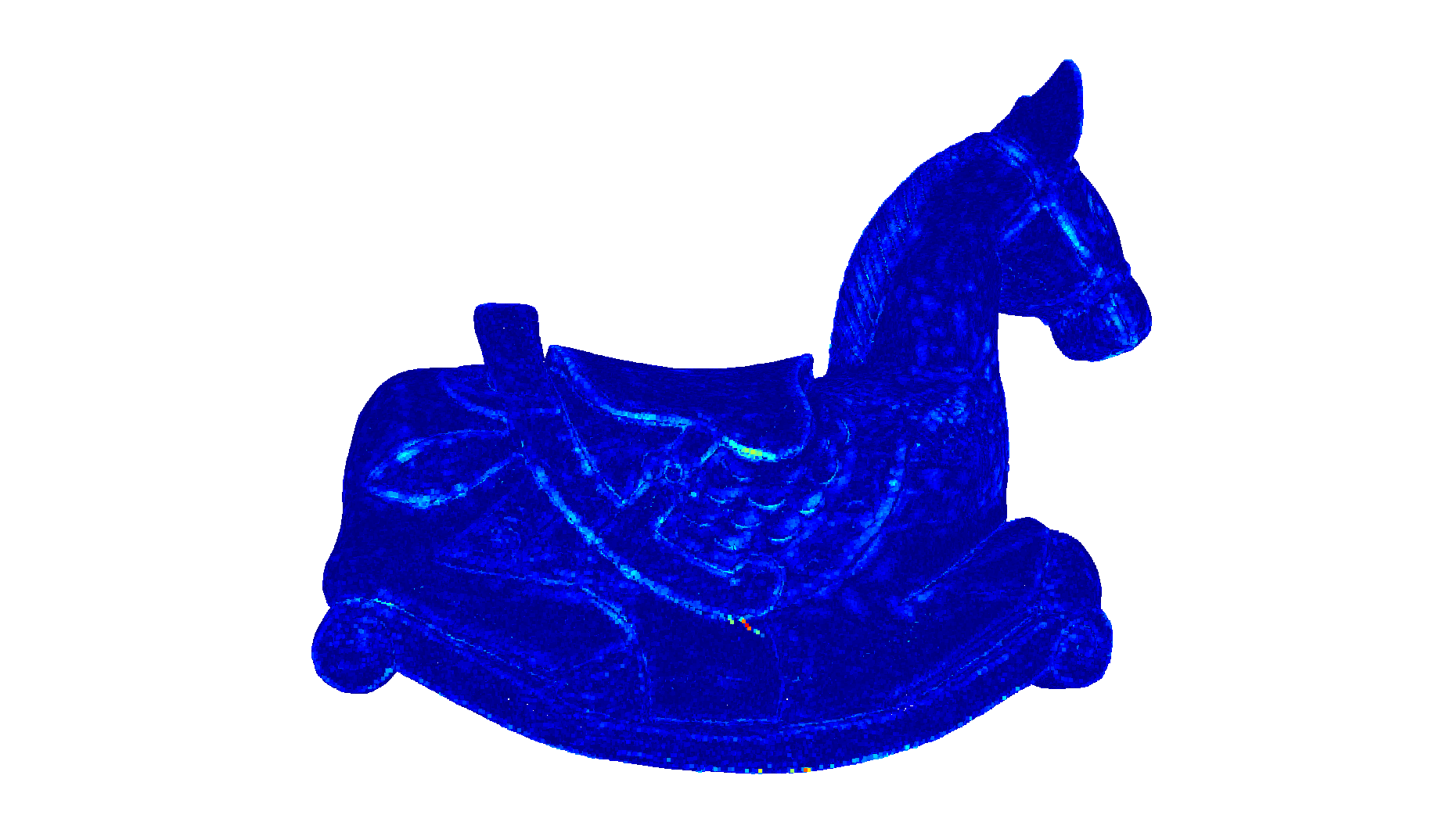}\vspace{0pt}
   \includegraphics[width=\linewidth]{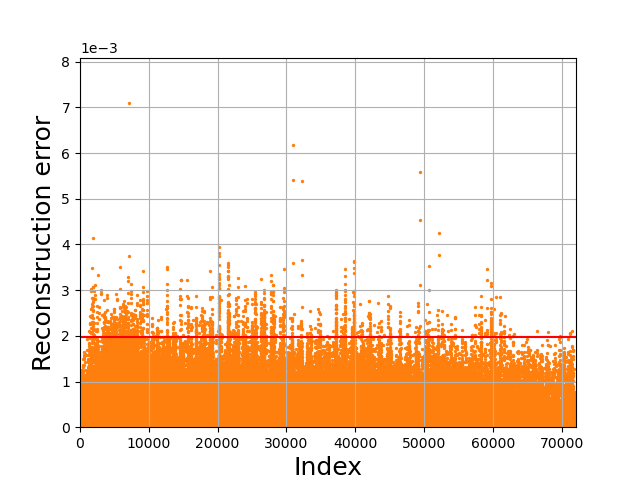}\vspace{0pt} 
   \end{minipage}
}
\subfigure[Ours]{
    \begin{minipage}[h]{0.29\columnwidth}
    \includegraphics[width=\linewidth,trim={320 10 320 10},clip]{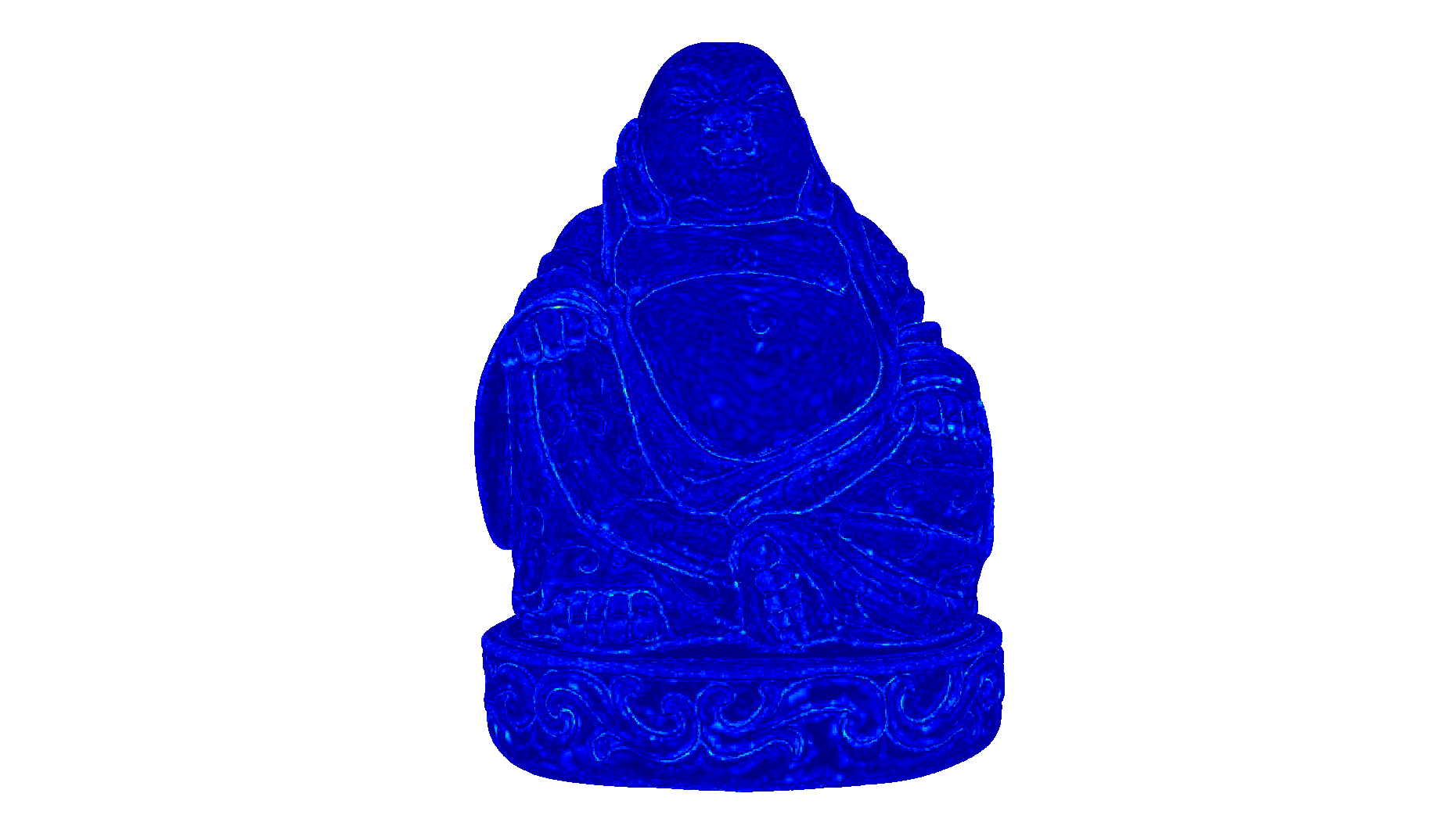}\vspace{0pt} 
    \includegraphics[width=\linewidth]{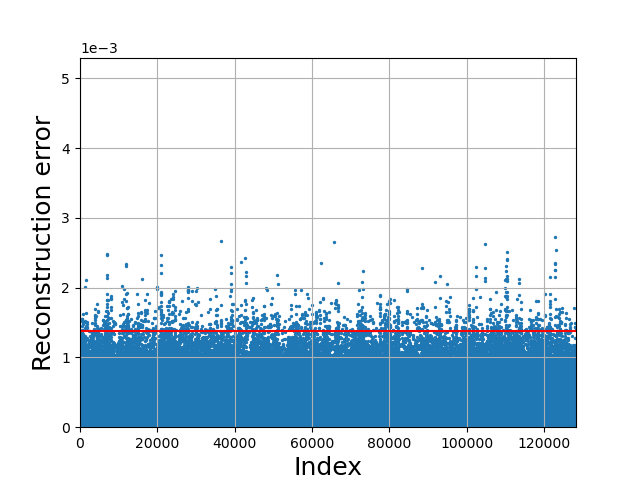}\vspace{0pt} 
    \includegraphics[width=\linewidth,trim={180 40 180 40},clip]{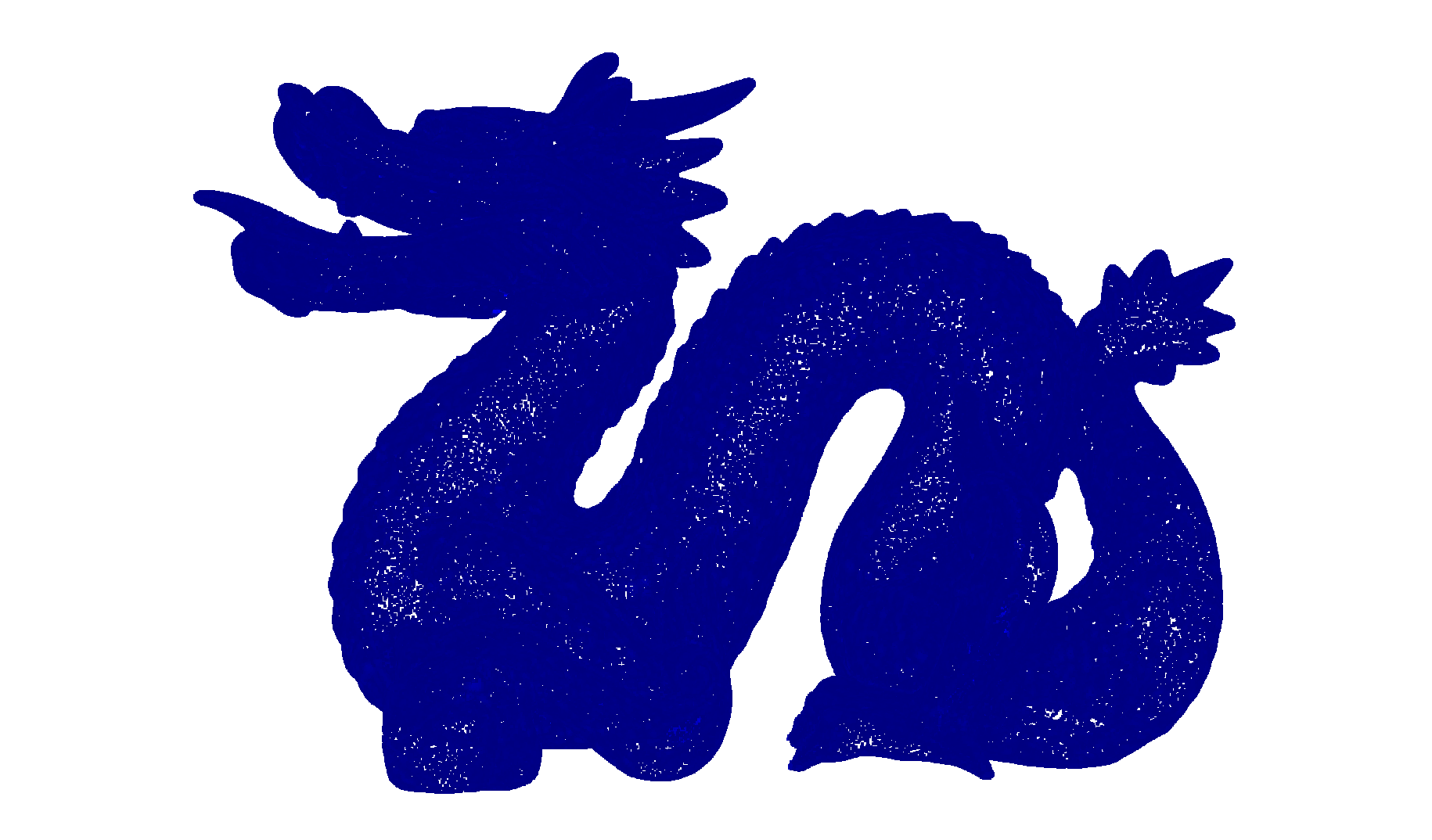}\vspace{0pt} 
    \includegraphics[width=\linewidth]{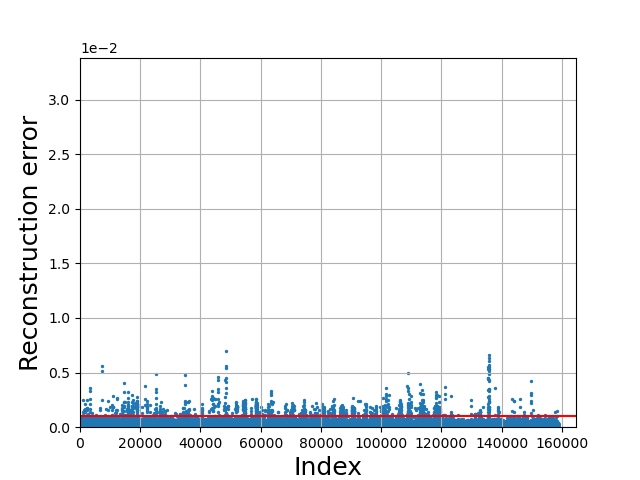}\vspace{0pt} 
    \includegraphics[width=\linewidth,trim={250 10 250 10},clip]{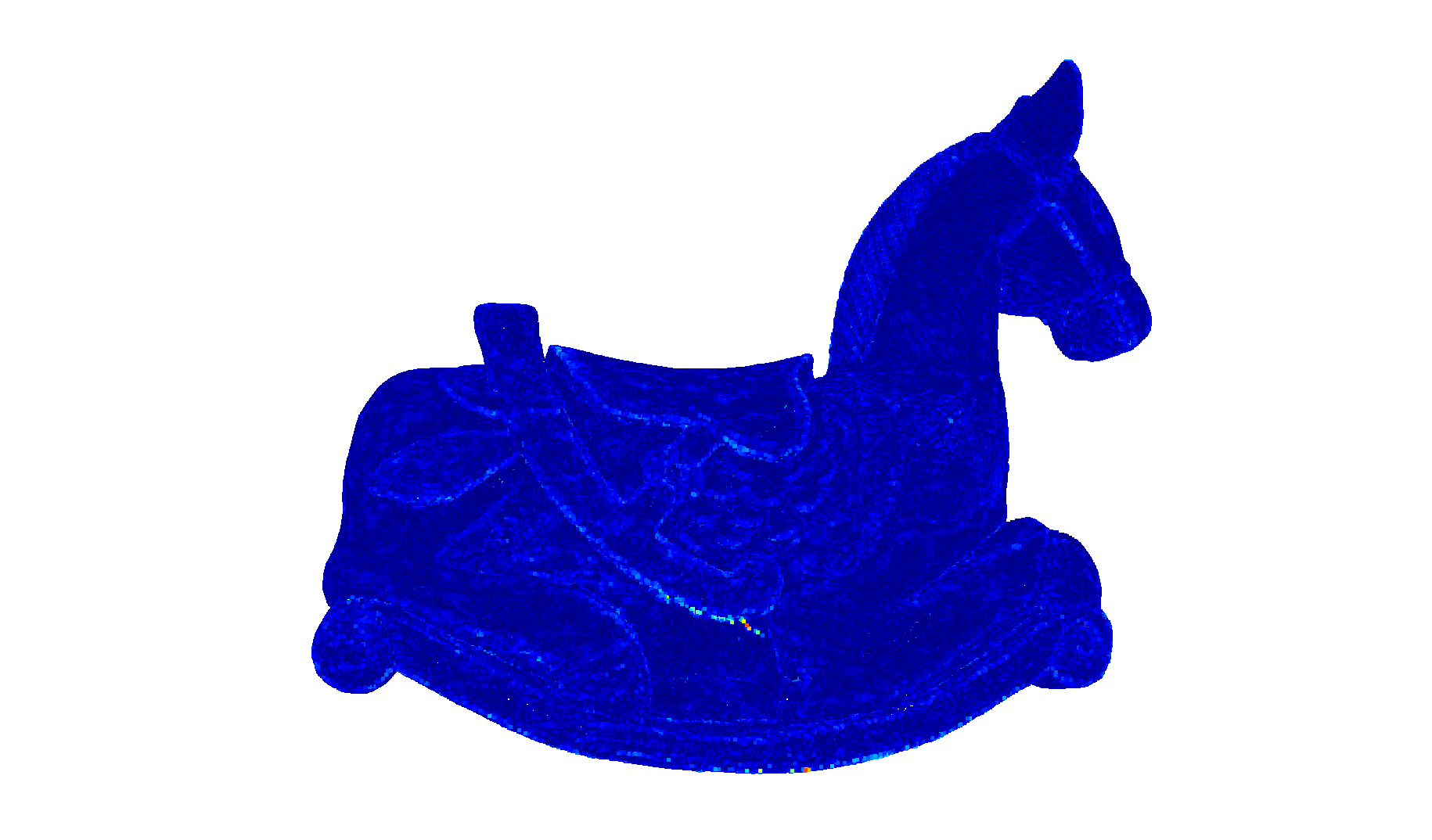}\vspace{0pt}
    \includegraphics[width=\linewidth]{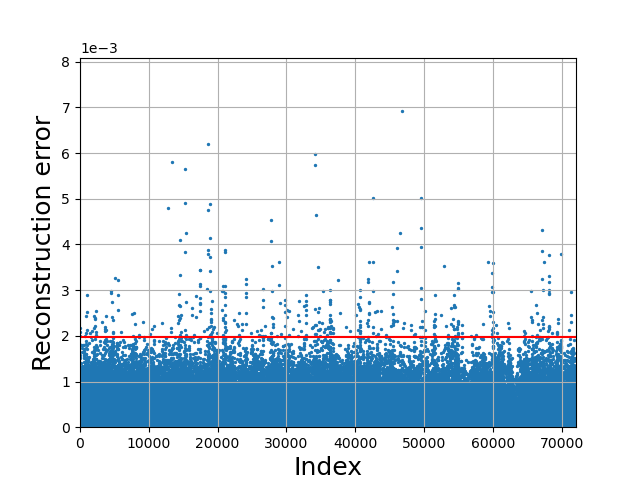}\vspace{0pt} 
    \end{minipage}
}
\\
\includegraphics[width=\linewidth,trim={30 0 30 0},clip]{fig/jet-colorbar.png}
\caption{Reconstruction errors for the Buddha, Dragon, and Isidore horse 3D point clouds. The jet color ramp is used to depict the reconstruction error for each facet. The red horizontal line is the upper reconstruction error bound in~\eqref{eq-error-bound}, which is also depicted in the plot of reconstruction error for other methods.}
\label{lfs-recon-err-cmp}
\end{figure}

The surface Delaunay ball radius $R$ can provide an upper bound for the local facet size $l$ during the Delaunay refinement process. 
For simplicity and ease of analysis, we choose to use $R$ instead of $l$ in our reconstruction error analysis. Note that the surface reach, representing the smallest local feature size, corresponds to areas where the local curvature radius is smaller than the local thickness/separation, or vice versa. By assuming that the surface reach corresponds to a curvature radius, the surface reach area will achieve the maximum reconstruction error. Thus, we can analyze the maximum reconstruction error as a function of the estimated reach and the radius of the corresponding surface Delaunay ball as
\begin{equation}
 \epsilon \le \frac{R_{\mathrm{min}}^2}{2I_R},
 \label{eq-error-bound}
\end{equation}
where $R_{\mathrm{min}}$ is the radius of the surface Delaunay balls for the smallest facet and $I_R$ is the estimated reach. This formula suggests an estimation of the upper error bound for the reconstruction. Fig.~\ref{lfs-recon-err-cmp} verifies the consistency with respect to the formula and compares the per-facet reconstruction error with other methods. The red line depicts the maximum analytical error, i.e., the upper bound in~\eqref{eq-error-bound}. Furthermore, the per-facet reconstruction error scatter plots show that our local maxima reconstruction errors are uniformly distributed compared to other methods.

\subsection{Time and Memory} 

\begin{figure}[t!]
\centering
\def\tabularxcolumn#1{m{#1}}
\scalebox{1.0}{
\begin{tabularx}{\linewidth}{ccXXc}
\setlength{\tabcolsep}{0pt}
\renewcommand{\arraystretch}{0} 
\begin{tabular}[p]{l ccc}
\rotatebox{90}{ \footnotesize{\quad  Ellipsoid}} &
\includegraphics[width=0.3\columnwidth,trim={300 150 300 150},clip]{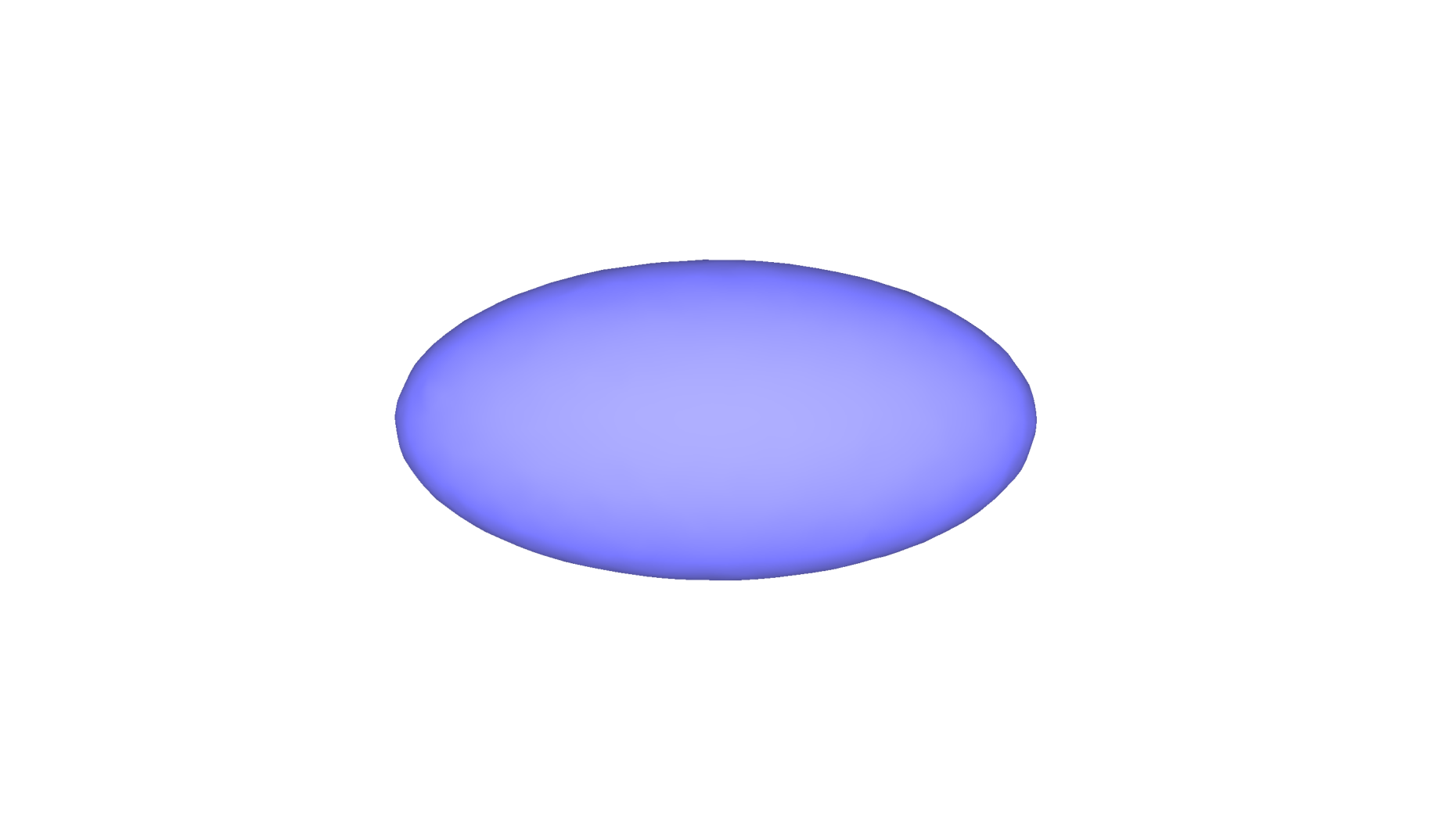}\vspace{0pt} &
\includegraphics[width=0.3\columnwidth]{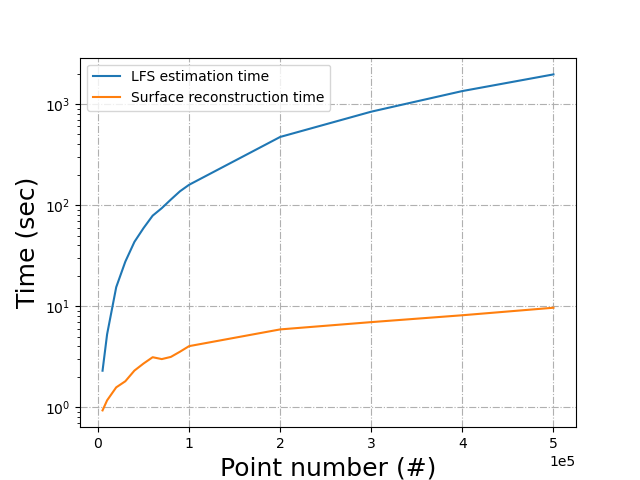}\vspace{0pt} &
\includegraphics[width=0.3\columnwidth]{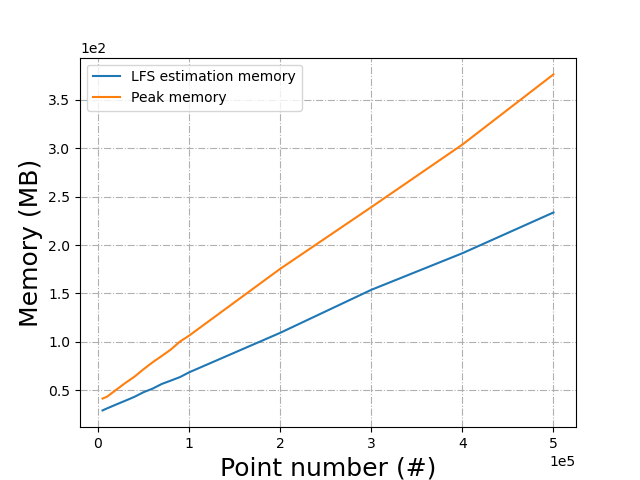}\vspace{0pt} \\
\rotatebox{90}{ \footnotesize{\quad \quad  Hippo}} &
\includegraphics[width=0.3\columnwidth,trim={300 100 300 100},clip]{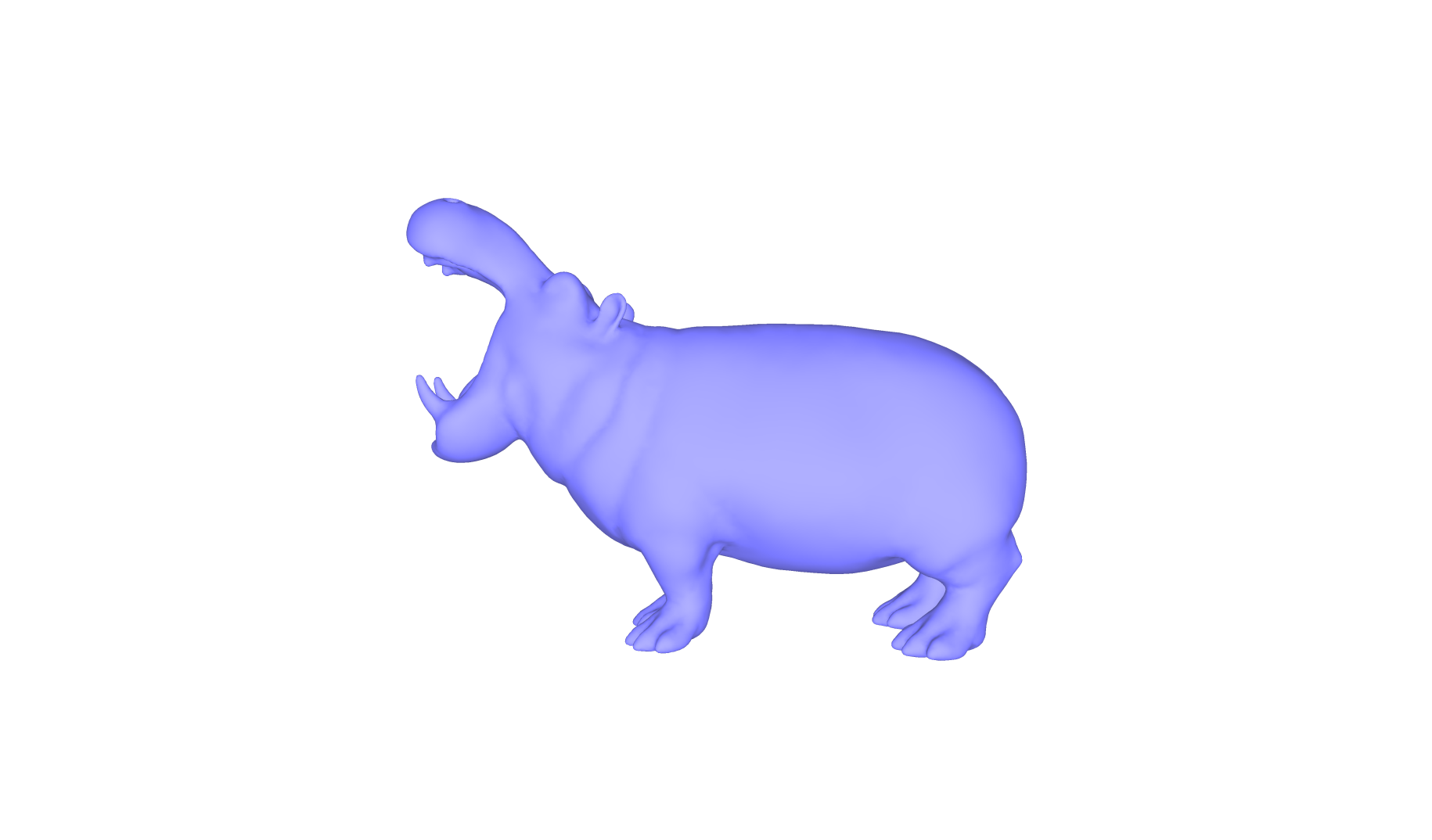}\vspace{0pt} &
\includegraphics[width=0.3\columnwidth,trim={0 0 0 30},clip]{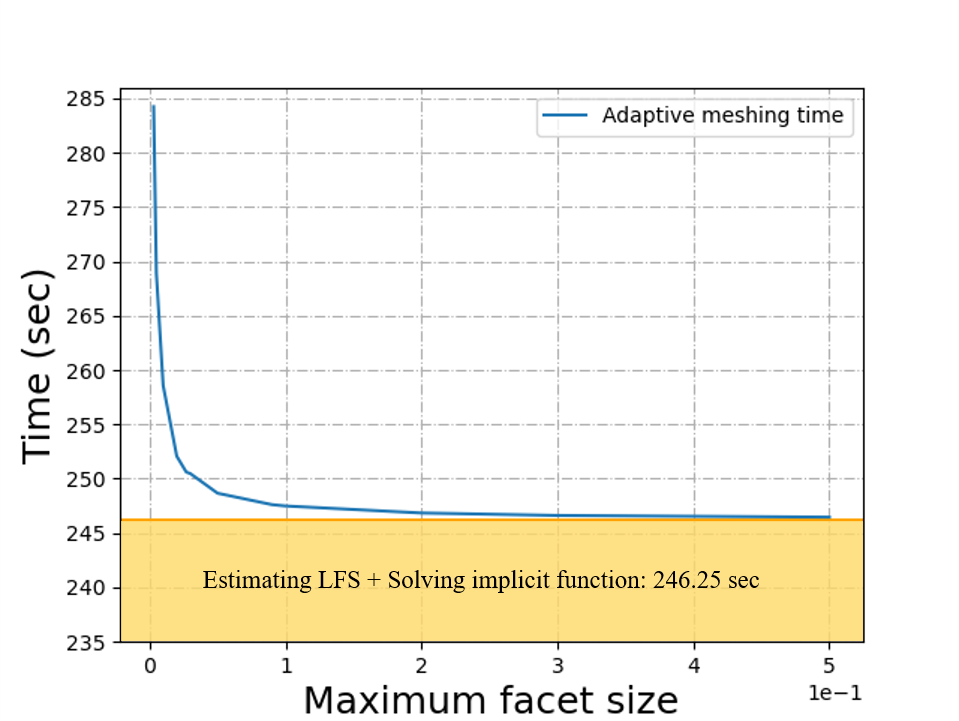}\vspace{0pt} &
\includegraphics[width=0.3\columnwidth,trim={0 0 0 30},clip]{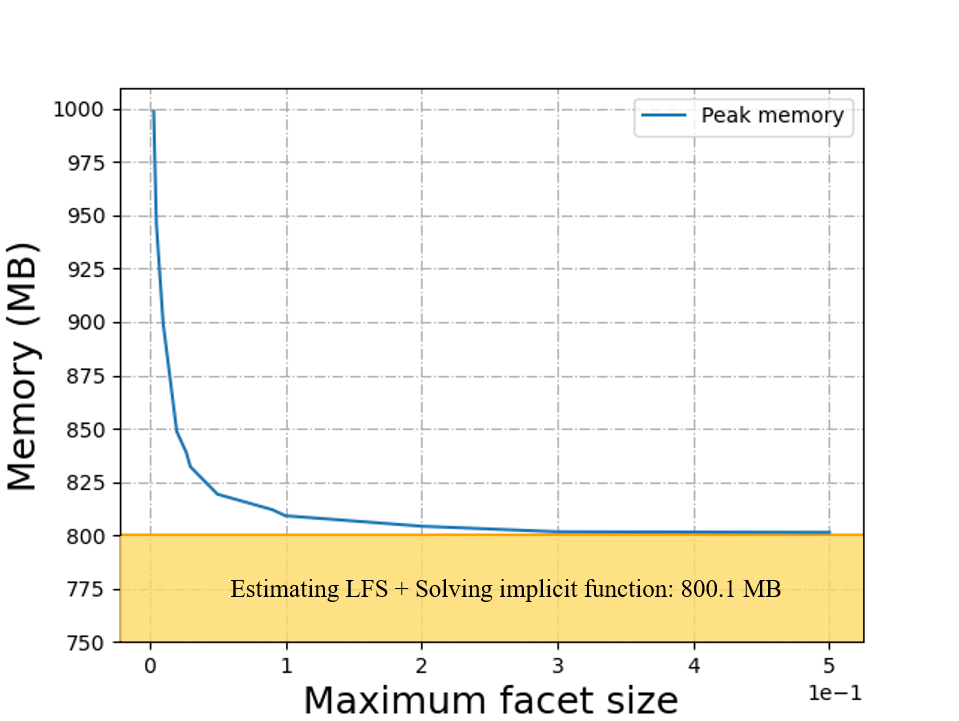}\vspace{0pt}
\end{tabular}
\end{tabularx}
}
    \caption{Computation times and memory consumption. 
    Top: ellipsoid with increasing density of the input 3D point cloud. 
    Bottom: experiments on the hippo 3D point cloud (58,188 points) with varying maximum facet size parameter ($\textrm{size}_\textrm{max}$).} 
    \label{time-memory}
\end{figure}

We now measure the computational time and the memory consumption through a series of experiments. We first generate a series of point clouds sampled on an ellipsoid using an increasing density. The memory usage is almost linear with respect to the number of input points. We also measure time and memory consumption on the hippo point cloud under different user-defined parameters (maximum facet size $\textrm{size}_\textrm{max}$). The computation time and memory usage for estimating LFS and solving the implicit function are constant because the input point cloud is unchanged. However, the total time and peak memory depend on $\textrm{size}_\textrm{max}$. Fig.~\ref{time-memory} plots the computation time (in seconds) and memory usage (in MBytes) against the number of input points or $\textrm{size}_\textrm{max}$.

\subsection{Limitations} 

\begin{figure}[ht!]
\centering
\subfigure[input]{
    \begin{minipage}[h]{0.35\columnwidth}
    \includegraphics[width=\linewidth,trim={0 0 0 0},clip]{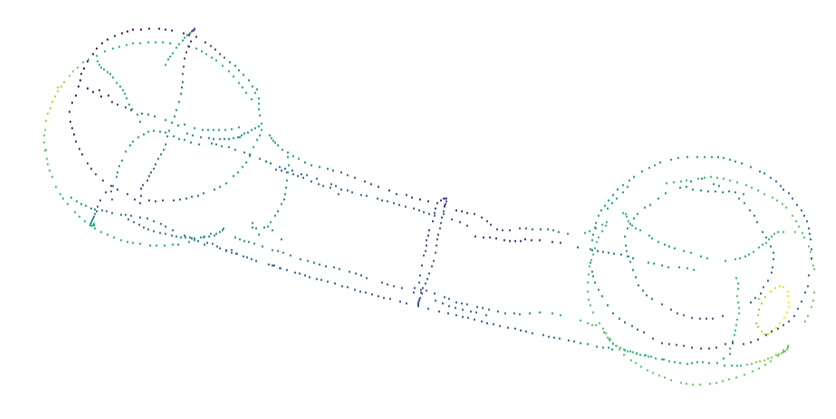}\vspace{0pt} 
    \end{minipage}
}
\subfigure[ours]{
    \begin{minipage}[h]{0.35\columnwidth}
    \includegraphics[width=\linewidth,trim={0 0 0 0},clip]{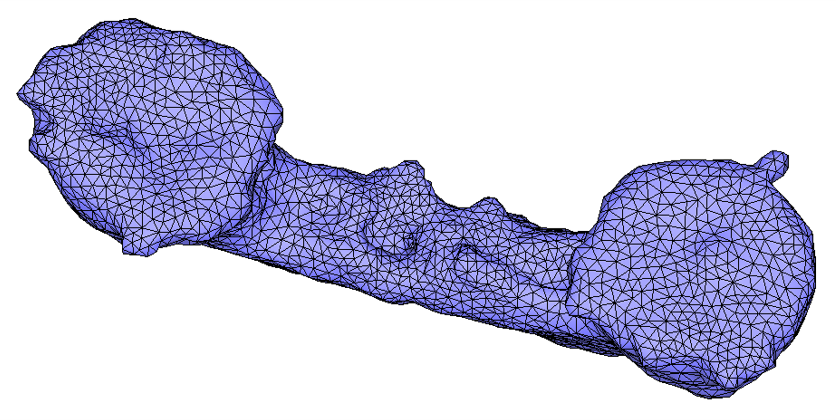}\vspace{0pt} 
    \end{minipage}
}
 \caption{Failure cases. The reconstructed mesh is not smooth.} 
\label{failure-cases}
\end{figure}

The primary limitation of our method lies in its performance with sketch 3D point clouds (extremely sparse, see Fig.~\ref{failure-cases}). Our mesh solver relies on sign guesses to propagate vertex signs across the multi-domain, which can be challenging to determine for sketch point clouds. Unlike Poisson-based solvers, we do not diffuse normals. In addition, our method cannot handle boundaries if the surface is open. Regarding our future work, we plan to extend our approach to address these two issues, enhancing its versatility and effectiveness in handling sketch 3D point clouds and reconstructing boundaries.

\bibliographystyle{IEEEtran}
\bibliography{reference}

\end{document}